\documentclass[a4paper,11pt]{book}
\author{Daniel Weber}
\title{Teaching Unknown Objects by Leveraging Human Gaze and Augmented Reality in Human-Robot Interaction}
\date{\today}
\usepackage[T1]{fontenc}
\usepackage[utf8]{inputenc}

\usepackage[square,numbers]{natbib} 
\usepackage{notoccite} 
\usepackage[nottoc]{tocbibind}

\usepackage[french,german,english]{babel}

\usepackage{lmodern} 

\usepackage{fourier} 
\DeclareMathAlphabet{\mathcal}{OMS}{zplm}{m}{n}
\SetMathAlphabet{\mathcal}{bold}{OMS}{zplm}{b}{n}
\usepackage[DIV=default,BCOR=1cm,headlines=2.1,headinclude]{typearea}	

\usepackage{setspace} 

\usepackage{graphicx}
\usepackage[table]{xcolor}
\graphicspath{{images/} {figures/}}

\usepackage{booktabs}
\usepackage{lipsum}
\usepackage{microtype}
\usepackage{url}

\usepackage{fancyhdr}

	\renewcommand{\headrulewidth}{0.4pt}
	\renewcommand{\footrulewidth}{0.0pt}
\fancypagestyle{plain}{
	\fancyhf{}
	\renewcommand{\headrulewidth}{0pt}
	\renewcommand{\footrulewidth}{0pt}
	\fancyfoot[EL,OR]{\thepage}
}
\fancypagestyle{addpagenumbersforpdfimports}{
	\fancyhead{}
	\renewcommand{\headrulewidth}{0pt}
	\fancyfoot{}
	\fancyfoot[EL,OR]{\thepage}
}

\usepackage{listings}
\usepackage[hyperfootnotes=false]{hyperref}	
\makeatletter
\hypersetup{pdfborder={0 0 0},
	colorlinks=true,
	linkcolor=black,
	citecolor=black,
	urlcolor=black,
    pdfstartview=,
    pdftitle={\@title},
    pdfauthor={\@author},
    pdfkeywords={Augmented Reality, Eye Tracking, Human-Robot Interaction, Machine Learning, Object Detection, Robotics, Teaching Unknown Objects}}
\makeatother
\ifpdf
\usepackage[final]{pdfpages}
\else
\usepackage{calc}
\usepackage{breakurl}
\usepackage[nlwarning=false]{hypdvips}
\usepackage{backref}
\renewcommand*{\backref}[1]{}
\fi
\usepackage{bookmark}

\makeatletter
\renewcommand\@pnumwidth{20pt}
\makeatother

\makeatletter
\def\cleardoublepage{\clearpage\if@twoside \ifodd\c@page\else
    \hbox{}
    \thispagestyle{empty}
    \newpage
    \if@twocolumn\hbox{}\newpage\fi\fi\fi}
\makeatother \clearpage{\pagestyle{plain}\cleardoublepage}

\usepackage{color}
\usepackage{tikz}
\usepackage[explicit]{titlesec}
\definecolor{gray75}{gray}{0.75}
\titleformat{\chapter}		
[hang]					
{\normalfont\Huge\bfseries}	
{\thechapter} 
{20pt}					
{\begin{tabular}[t]{@{\color{gray75}\vrule width 2pt\hspace{20pt}}p{0.84\textwidth}}\raggedright#1\end{tabular}} 
[]						
\titleformat{name=\chapter,numberless}		
[hang]					
{\normalfont\Huge\bfseries}	
{} 
{0pt}					
{\raggedright#1} 		
[]						

\newcounter{myparts}
\newcommand*\partlabel{}
\titleformat{\part}[display]  
	{\normalfont\bfseries\Huge} 
	{\gdef\partlabel{\thepart\ }}     
 	{0pt} 
 	  {\ifpdf\setlength{\unitlength}{20mm}\else\setlength{\unitlength}{0mm}\fi
	  \addtocounter{myparts}{1}
	  \begin{tikzpicture}[remember picture,overlay]
    \node[anchor=north west,xshift=-65mm,yshift=-6.9cm-\value{myparts}*20mm] at (current page.north east) 
      {\begin{tikzpicture}[remember picture, overlay]
        \draw[fill=black] (0,0) rectangle(62mm,20mm);   
        \node[anchor=north west,yshift=-6.1cm-\value{myparts}*\unitlength,xshift=-60.5mm,minimum height=30mm,inner sep=0mm] at (current page.north east)
        {\parbox[top][30mm][t]{55mm}{\raggedright \color{white}Part \partlabel \rule{0cm}{0.6cm}}};  
        \node[anchor=north east,yshift=-6.1cm-\value{myparts}*\unitlength,xshift=-63.5mm,text width=\textwidth,minimum height=30mm,inner sep=0mm] at (current page.north east)
              {\parbox[top][30mm][t]{\textwidth}{\raggedleft \rule{0cm}{0.6cm}\color{black}#1}};
       \end{tikzpicture}
      };
   \end{tikzpicture}
   \gdef\partlabel{}
  } 
\titlespacing*{\part}{11.06cm}{26.4pt-\parskip-\parskip}{0pt}

\usepackage{amsmath}
\usepackage{amsfonts}
\usepackage{amssymb}
\usepackage{mathtools}
\makeatletter
\def\resetMathstrut@{%
  \setbox\z@\hbox{%
    \mathchardef\@tempa\mathcode`\(\relax
      \def\@tempb##1"##2##3{\the\textfont"##3\char"}%
      \expandafter\@tempb\meaning\@tempa \relax
  }%
  \ht\Mathstrutbox@1.2\ht\z@ \dp\Mathstrutbox@1.2\dp\z@
}
\makeatother

\counterwithout*{footnote}{chapter}

\usepackage{graphicx}
\usepackage{multirow}
\usepackage{subcaption}
\usepackage{mathtools}
\usepackage{amssymb}  
\usepackage[flushleft]{threeparttable} 
\usepackage{booktabs}
\usepackage{siunitx}
\usepackage{array}
\usepackage{balance}
\usepackage{algpseudocode, algorithm}   
\usepackage[printonlyused]{acronym}
\usepackage{bibentry} \nobibliography*
\usepackage{doi}
\usepackage{enumitem}
\usepackage[bottom,splitrule]{footmisc}
\usepackage{grffile}    
\usepackage{misc/jigsaw_v0.4}
\usepackage{bm}     
\usepackage{xurl}   
\usepackage{footnotehyper} 
\usepackage{chngcntr}   

\usepackage[titles]{tocloft}    
\cftsetindents{figure}{1.5em}{3em}
\cftsetindents{table}{1.5em}{3em}

\newcommand*{\figref}[1]{Figure~\ref{#1}}

\newcommand*{\tblref}[1]{Table~\ref{#1}}
\newcommand*{\secref}[1]{Section~\ref{#1}}

\newcommand*{\pluseq}{\mathrel{+{=}}}
\newcommand*{\round}[1]{\ensuremath\left\lfloor#1\right\rceil}
\DeclarePairedDelimiter\abs{\lvert}{\rvert}%
\DeclarePairedDelimiter\norm{\lVert}{\rVert}%
\makeatletter
\let\oldabs\abs
\def\abs{\@ifstar{\oldabs}{\oldabs*}}
\let\oldnorm\norm
\def\norm{\@ifstar{\oldnorm}{\oldnorm*}}
\makeatother

\newcommand*{\pz}{\phantom{0}}

\newcommand*{\AP}[1]{\ensuremath{\text{AP}_{#1}}}
\newcommand*{\mAP}[1]{\ensuremath{\text{mAP}_{#1}}}
\newcommand*{\AR}[1]{\ensuremath{\text{AR}_{#1}}}
\newcommand*{\mAR}[1]{\ensuremath{\text{mAR}_{#1}}}

\newcommand*{\Image}{\mathcal{I}}
\newcommand*{\FeatureMap}{\mathcal{M}}
\newcommand*{\ActivationMap}{\mathcal{A}}
\newcommand*{\VisualDissimilarity}{d}
\newcommand*{\SpatialDissimilarity}{F}
\newcommand*{\Weight}{w}
\newcommand*{\TransMat}{\mathcal{T}}
\newcommand*{\WeightN}{w'}
\newcommand*{\TransMatN}{\mathcal{T}'}

\newcommand*{\chref}[1]{Chapter~\ref{#1}}
\newcommand*{\subsecref}[1]{Subsection~\ref{#1}}
\newcommand*{\algoref}[1]{Algorithm~\ref{#1}}

\newcommand\blfootnote[1]{%
	\begingroup
	\renewcommand\thefootnote{}%
	\footnote{#1}%
	\addtocounter{footnote}{-1}%
	\endgroup
}

\newcommand*{\Description}[1]{}
\begin{document}

\frontmatter
\input{src/head/titlepage.tex}
\cleardoublepage
\thispagestyle{empty}

\vspace*{\fill}
\vspace{-\headheight}	
\vspace{-\headsep}		


\begin{center}
    To my wife, my parents, my sister, and my grandmother, whose unwavering love and endless support are greater than my words could ever be.
\end{center}

\vspace*{\fill}

\setcounter{page}{0}
\chapter*{Acknowledgments}
\markboth{Acknowledgments}{Acknowledgments}
\addcontentsline{toc}{chapter}{Acknowledgments}

In the course of my doctoral research in recent years, I have had the privilege of engaging in many exciting collaborations and fruitful discussions with numerous people, all of whom I have benefited from and whom I would like to express my sincere gratitude to.
The successful completion of my research, culminating in this dissertation, would not have been possible without their support.

First of all, I would like to thank my two supervisors, Prof.~Dr.~Enkelejda Kasneci and Prof.~Dr.~Andreas Zell, for hosting me at the chairs Human-Computer Interaction and Cognitive Systems, respectively, and whose guidance has profoundly shaped both my research and my academic pursuits.
The former, in particular, has been a consistent source of support, both in research and in general matters.
Furthermore, special and sincere appreciation also go to Jun.-Prof.~Dr.~Michael Krone and Dr.~Shahram Eivazi for their immediate willingness to evaluate my work. 
I am also very grateful to Vita Serbakova and Margot Reimold for always assisting me with the universities' bureaucracy, and to Uli Ulmer for providing all the exceptional technical support and consistently being reliably eager to help.
In addition, I would also like to thank the Deutsche Forschungsgemeinschaft (DFG, German Research Foundation) for funding me and my project, as well as Michaela Bitzer, who has always been a reliable contact person at the university in these matters.

Throughout my journey, I have been fortunate to work with a number of former and present colleagues, whose experiences have greatly enriched my own.
Foremost, I would like to express my heartfelt gratitude to Dr.~Thiago Santini, Dr.~David Geisler, Dr.~Wolfgang Fuhl, and Dr.~Nora Castner.
I can't thank you enough for your warm and gracious welcome and for giving me an incredibly delightful start by taking me into your care.
I deeply appreciate your patience in answering my countless questions and I truly enjoyed all our amusing conversations and funny moments both within our office and during the numerous ``non-smoking smoking breaks''.
Special thanks also extend to the rest of my esteemed colleagues from both the Human-Computer Interaction team as well as the Cognitive Systems group, including Dr.~Thomas Kübler, Dr.~Efe Bozkir, Dr.~Nuri Benbarka, Dr.~Julian Jordan, Dr.~Benedikt Hosp, Dr.~Hong Gao, Björn Severitt, Mario Laux, and many others, for their support and assistance whenever I was in need.
This also applies in particular to Benjamin Kiefer, Martin Meßmer, Timon Höfer, and Valentin Bolz, who deserve special recognition.
Your moral support and exhilarating gallows humor have been instrumental in making my academic journey so enjoyable and memorable.
Especially the latter has been an invaluable presence in my life as a long-time and incredibly close friend.

Finally, and above all, I would like to thank my parents Brunhilde and Waldemar, my sister Marina, together with Sebastian and Lena, and my grandmother Elfriede, from the bottom of my heart.
Your support, encouragement, and unconditional love mean more to me than you could possibly imagine.
Likewise, I want to express my profound gratitude to my wife, Carmen, who has always believed in me, encouraged me, lifted me up, helped me through challenging times and accompanied me through all the ups and downs.
Your support has been an immeasurable blessing in my life.
Thank you for always being there for me -- I am deeply indebted to you.

\bigskip
\hfill Daniel~Weber

\cleardoublepage
\chapter*{Abstract}
\markboth{Abstract}{Abstract}
\addcontentsline{toc}{chapter}{Abstract} 

Robots are becoming increasingly popular in a wide range of environments due to their exceptional work capacity, precision, efficiency, and scalability.
This development has been further encouraged by advances in \ac{AI}, particularly \ac{ML}.
By employing sophisticated neural networks, robots are given the ability to detect and interact with objects in their vicinity.
However, a significant drawback arises from the underlying dependency on extensive datasets and the availability of substantial amounts of training data for these object detection models.
This issue becomes particularly problematic when the specific deployment location of the robot and the surroundings, including the objects within it, are not known in advance.
The vast and ever-expanding array of objects makes it virtually impossible to comprehensively cover the entire spectrum of existing objects using preexisting datasets alone.

The goal of this dissertation was to teach a robot unknown objects in the context of \ac{HRI} in order to liberate it from its data dependency, unleashing it from predefined scenarios.
In this context, the combination of eye tracking and \ac{AR} created a powerful synergy that empowered the human teacher to seamlessly communicate with the robot and effortlessly point out objects by means of human gaze.
This holistic approach led to the development of a multimodal \ac{HRI} system that enabled the robot to identify and visually segment the \acp{OOI} in three-dimensional space, even though they were initially unknown to it, and then examine them autonomously from different angles.
Through the class information provided by the human, the robot was able to learn the objects and redetect them at a later stage.
Due to the knowledge gained from this \ac{HRI} based teaching process, the robot's object detection capabilities exhibited comparable performance to state-of-the-art object detectors trained on extensive datasets, without being restricted to predefined classes, showcasing its versatility and adaptability.

The research conducted within the scope of this dissertation made significant contributions at the intersection of \ac{ML}, \ac{AR}, eye tracking, and robotics.
These findings not only enhance the understanding of these fields, but also pave the way for further interdisciplinary research.
The scientific articles included in this dissertation have been published at high-impact conferences in the fields of robotics, eye tracking, and \ac{HRI}.

\acresetall 
\begin{otherlanguage}{german}
\cleardoublepage
\chapter*{Zusammenfassung}
\markboth{Zusammenfassung}{Zusammenfassung}
\addcontentsline{toc}{chapter}{Zusammenfassung}

Roboter finden aufgrund ihrer außergewöhnlichen Arbeitsleistung, Präzision, Effizienz und Skalierbarkeit immer mehr Verwendung in den verschiedensten Anwendungsbereichen.
Diese Entwicklung wurde zusätzlich begünstigt durch Fortschritte in der Künstlichen Intelligenz (KI), insbesondere im Maschinellem Lernen (ML).
Mit Hilfe moderner neuronaler Netze sind Roboter in der Lage, Objekte in ihrer Umgebung zu erkennen und mit ihnen zu interagieren.
Ein erhebliches Manko besteht jedoch darin, dass das Training dieser Objekterkennungsmodelle, in aller Regel mit einer zugrundeliegenden Abhängig von umfangreichen Datensätzen und der Verfügbarkeit großer Datenmengen einhergeht.
Dies ist insbesondere dann problematisch, wenn der konkrete Einsatzort des Roboters und die Umgebung, einschließlich der darin befindlichen Objekte, nicht im Voraus bekannt sind.
Die breite und ständig wachsende Palette von Objekten macht es dabei praktisch unmöglich, das gesamte Spektrum an existierenden Objekten allein mit bereits zuvor erstellten Datensätzen vollständig abzudecken.

Das Ziel dieser Dissertation war es, einem Roboter unbekannte Objekte mit Hilfe von \ac{HRI} beizubringen, um ihn von seiner Abhängigkeit von Daten sowie den Einschränkungen durch vordefinierte Szenarien zu befreien.
Die Synergie von Eye Tracking und \ac{AR} ermöglichte es dem als Lehrer fungierenden Menschen, mit dem Roboter zu kommunizieren und ihn mittels des menschlichen Blickes auf Objekte hinzuweisen.
Dieser holistische Ansatz ermöglichte die Konzeption eines multimodalen \ac{HRI}-Systems, durch das der Roboter Objekte identifizieren und dreidimensional segmentieren konnte, obwohl sie ihm zu diesem Zeitpunkt noch unbekannt waren, um sie anschließend aus unterschiedlichen Blickwinkeln eigenständig zu inspizieren.
Anhand der Klasseninformationen, die ihm der Mensch mitteilte, war der Roboter daraufhin in der Lage, die entsprechenden Objekte zu erlernen und später wiederzuerkennen.
Mit dem Wissen, das dem Roboter durch diesen auf \ac{HRI} basierenden Lehrvorgang beigebracht worden war, war dessen Fähigkeit Objekte zu erkennen vergleichbar mit den Fähigkeiten modernster Objektdetektoren, die auf umfangreichen Datensätzen trainiert worden waren.
Dabei war der Roboter jedoch nicht auf vordefinierte Klassen beschränkt, was seine Vielseitigkeit und Anpassungsfähigkeit unter Beweis stellte.

Die im Rahmen dieser Dissertation durchgeführte Forschung leistete bedeutende Beiträge an der Schnittstelle von \ac{ML}, \ac{AR}, Eye Tracking und Robotik.
Diese Erkenntnisse tragen nicht nur zum besseren Verständnis der genannten Felder bei, sondern ebnen auch den Weg für weitere interdisziplinäre Forschung.
Die in dieser Dissertation enthalten wissenschaftlichen Artikel wurden auf hochrangigen Konferenzen in den Bereichen Robotik, Eye Tracking und \ac{HRI} veröffentlicht.

\end{otherlanguage}

\cleardoublepage
\pdfbookmark{\contentsname}{toc}
\tableofcontents

\cleardoublepage
\phantomsection
\listoffigures

\cleardoublepage
\phantomsection
\listoftables

\cleardoublepage
\phantomsection
\chapter*{List of Abbreviations}
\markboth{List of Abbreviations}{List of Abbreviations}
\addcontentsline{toc}{chapter}{List of Abbreviations}

\begin{acronym}[DGA-GBVS]
    \acro{AGV}{Automated Guided Vehicle}
    \acro{AI}{Artificial Intelligence}
    \acro{AMR}{Autonomous Mobile Robot}
    \acro{AR}{Augmented Reality}    
    \acro{DGA-GBVS}{Dual Gaze-Assisted GBVS}    
    \acro{FOV}{Field of View}
    \acro{GA-GBVS}{Gaze-Assisted GBVS}    
    \acro{GBVS}{Graph-Based Visual Saliency}    
    \acro{HRC}{Human-Robot Collaboration}    
    \acro{HRI}{Human-Robot Interaction}    
    \acro{IoU}{Intersection over Union}    
    \acro{KNN}{k-Nearest Neighbor}    
    \acro{LLM}{Large Language Model}
    \acro{mAP}{mean average precision}    
    \acro{mAR}{mean average recall}    
    \acro{ML}{Machine Learning}
    \acro{OMD}{Objects in Multiperspective Detail}    
    \acro{OOI}{Object of Interest}
    \acroplural{OOI}[OOIs]{Objects of Interest}
    \acro{PbD}{Programming by Demonstration}
    \acro{RANSAC}{random sample consensus}  
    \acro{RLHF}{Reinforcement Learning from Human Feedback}
    \acro{ROI}{Region of Interest}
    \acroplural{ROI}[ROIs]{Regions of Interest}
    \acro{ROS}{Robot Operating System}   
    \acro{SVM}{Support Vector Machine}    
    \acro{UAV}{Unmanned Aerial Vehicle}    
    \acro{UWP}{Universal Windows Platform}    
    \acro{VR}{Virtual Reality}
\end{acronym}



\mainmatter
\acresetall 
\cleardoublepage
\chapter{List of Publications}
\label{ch:publist}

\section{Publications Relevant to This Thesis} \label{sec:publist_relevant}
\begin{enumerate}
	\item[\cite{weber2020distilling}]
	\textbf{Daniel Weber}, Thiago Santini, Andreas Zell, and Enkelejda Kasneci. Distilling Location Proposals of Unknown Objects through Gaze Information for Human-Robot Interaction. In \textit{2020 IEEE/RSJ International Conference on Intelligent Robots and Systems (IROS)}, pages 11086–11093. IEEE, 2020. \href{https://doi.org/10.1109/IROS45743.2020.9340893}{doi:10.1109/IROS45743.2020.9340893}.

 	\item[\cite{geisler2020exploiting}]
	David Geisler, \textbf{Daniel Weber}, Nora Castner, and Enkelejda Kasneci. Exploiting the GBVS for Saliency aware Gaze Heatmaps. In \textit{ACM Symposium on Eye Tracking Research and Applications}, pages 1–5, 2020. \href{https://doi.org/10.1145/3379156.3391367}{doi:10.1145/3379156.3391367}.
	
	\item[\cite{weber2022exploiting}]
	\textbf{Daniel Weber}, Enkelejda Kasneci, and Andreas Zell. Exploiting Augmented Reality for Extrinsic Robot Calibration and Eye-based Human-Robot Collaboration. In \textit{Proceedings of the 2022 ACM/IEEE International Conference on Human-Robot Interaction (HRI)}, pages 284–293. IEEE, 2022. \href{https://doi.org/10.1109/HRI53351.2022.9889538}{doi:10.1109/HRI53351.2022.9889538}.
	
	\item[\cite{weber2022gaze}]
	\textbf{Daniel Weber}, Wolfgang Fuhl, Andreas Zell, and Enkelejda Kasneci. Gaze-based Object Detection in the Wild. In \textit{2022 Sixth IEEE International Conference on Robotic Computing (IRC)}, pages 62–66. IEEE, 2022. \href{https://doi.org/10.1109/IRC55401.2022.00017}{doi:10.1109/IRC55401.2022.00017}.
	
	\item[\cite{weber2023multiperspective}]
	\textbf{Daniel Weber}, Wolfgang Fuhl, Enkelejda Kasneci, and Andreas Zell. Multiperspective Teaching of Unknown Objects via Shared-gaze-based Multimodal Human-Robot Interaction. In \textit{Proceedings of the 2023 ACM/IEEE International Conference on Human-Robot Interaction (HRI)}, pages 544–553, March 2023. \href{https://doi.org/10.1145/3568162.3578627}{\nolinkurl{doi:10.1145/3568162.3578627}}.

	\item[\cite{weber2023leveraging}]
	\textbf{Daniel Weber}, Valentin Bolz, Andreas Zell, and Enkelejda Kasneci. Leveraging Saliency-Aware Gaze Heatmaps for Multiperspective Teaching of Unknown Objects. In \textit{2023 IEEE/RSJ International Conference on Intelligent Robots and Systems (IROS)}, 2023. (Accepted for publication).
\end{enumerate}

\section{Further Publications}
\begin{enumerate}
	\item[\cite{fuhl2023gaze}] 
	Wolfgang Fuhl, \textbf{Daniel Weber}, and Shahram Eivazi. The Gaze and Mouse Signal as Additional Source for User Fingerprints in Browser Applications. In \textit{Proceedings of the 18th International Joint Conference on Computer Vision, Imaging and Computer Graphics Theory and Applications - HUCAPP}, pages 117–124. INSTICC, SciTePress, 2023. \href{https://doi.org/10.5220/0011607300003417}{doi: 10.5220/0011607300003417}.
		
	\item[\cite{fuhl2023pistol}] 
	Wolfgang Fuhl, \textbf{Daniel Weber}, and Shahram Eivazi. Pistol: Pupil Invisible Supportive Tool to Extract Pupil, Iris, Eye Opening, Eye Movements, Pupil and Iris Gaze Vector, and 2D as well as 3D Gaze. In \textit{Proceedings of the 18th International Joint Conference on Computer Vision, Imaging and Computer Graphics Theory and Applications - HUCAPP}, pages 27–38. INSTICC, SciTePress, 2023. \href{https://doi.org/10.5220/0011607200003417}{doi: 10.5220/0011607200003417}.
	
	\item[\cite{fuhl2023groupgazer}] 
	Wolfgang Fuhl, \textbf{Daniel Weber}, and Shahram Eivazi. GroupGazer: A Tool to Compute	the Gaze per Participant in Groups with Integrated Calibration to Map the Gaze Online to a Screen or Beamer Projection. In \textit{Proceedings of the 18th International Joint Conference on Computer Vision, Imaging and Computer Graphics Theory and Applications - HUCAPP}, pages 109–116. INSTICC, SciTePress, 2023. \href{https://doi.org/10.5220/0011607000003417}{doi: 10.5220/0011607000003417}.

    \item[\cite{fuhl2023watch}] 
    Wolfgang Fuhl, Björn Severitt, Nora Castner, Babette Bühler, Johannes Meyer, \textbf{Daniel Weber}, Regine Lendway, Ruikun Hou, and Enkelejda Kasneci. Watch out for those bananas! Gaze Based Mario Kart Performance Classification. In \textit{Proceedings of the 2023 Symposium on Eye Tracking Research and Applications}, pages 1–2. ACM, 2023. \href{https://doi.org/10.1145/3588015.3590136}{doi: 10.1145/3588015.3590136}.

\end{enumerate}

\section{Scientific Contribution}
\begin{figure}
    \vspace{1ex}
    \centering
    \includegraphics[width=\linewidth]{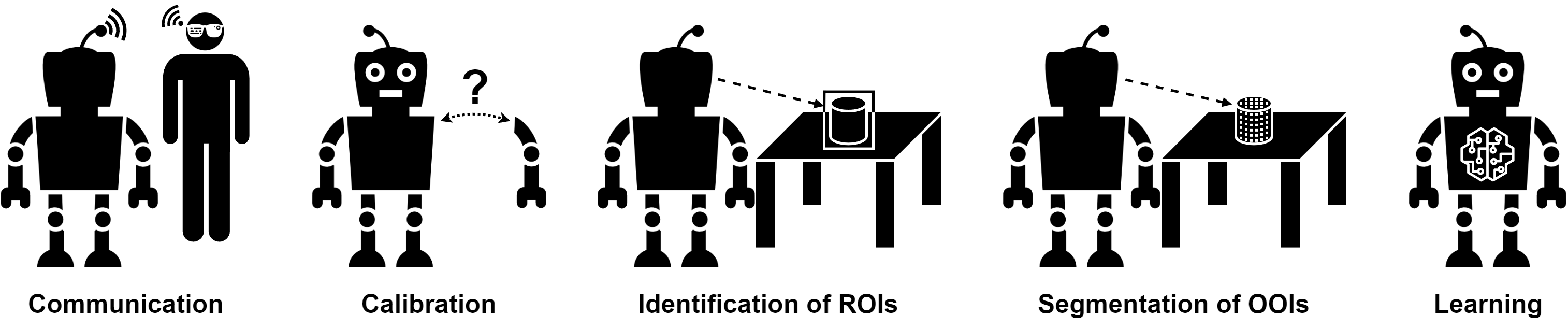}
    \caption{
    This thesis unites realms of research, that were previously running predominantly in parallel.
    The contributions cover a spectrum of domains including \ac{HRI}, robotics, computer vision, machine learning, and various others.
    The figure provides an overview, visualizing some of the most significant contributions.}
    \label{fig:contributions_sketch}
\end{figure}

This work explores a holistic perspective on \ac{HRI}, where interdisciplinary synergies from the research domains of \ac{AR}, \ac{ML}, eye tracking and robotics empower a human to teach a robot novel objects, facilitating its adaptability to cope with unfamiliar environments.
The most noteworthy contributions, as illustrated in \figref{fig:contributions_sketch}, encompass (1) the development of a natural and intuitive interaction channel between humans and robots utilizing \ac{AR} and eye tracking, (2) a practical extrinsic robot calibration method, (3) multiple approaches for the identification of \acp{ROI}, (4) the segmentation of unknown objects through gaze-based \ac{HRI}, and (5) robot learning by means of automatically recorded and labelled data.
These contributions have been published at renowned conferences in the field of robotics, eye tracking, and \ac{HRI}, and have paved the way for further research on robotic teaching.

The structure of this dissertation is as follows:
\chref{ch:introduction} commences by emphasizing the differences between humans and robots and outlining how robots benefit humans.
Following the various application areas of robots, their deficiencies and learning capabilities are explained.
Subsequently, diverse approaches are described how humans can teach robots.
These include, among others, the usage of \ac{AR} and eye tracking, which in particular offer great potential for teaching unknown objects and which are further elaborated on in the context of \ac{HRI}.
Then, the general setting is presented, and the objectives are defined.
At the end of the chapter, the utilized hardware is described and the fundamental terminology for subsequent evaluations is introduced.
\chref{ch:motivation_results} presents the major contributions of the six scientific publications listed in \secref{sec:publist_relevant}.
This encompasses the motivation, the methodology, as well as the main contributions and results of each paper.
Finally, \chref{ch:discussion} concludes with a discussion of the findings, achievements and limitations of this work and provides an outlook on future research.
The research conducted in the context of this dissertation was funded by the Deutsche Forschungsgemeinschaft (DFG, German Research Foundation) under Germany's Excellence Strategy ``Machine Learning: New Perspectives for Science'' -- EXC number 2064/1 -- Project number 390727645.

\cleardoublepage

\chapter{Introduction} \label{ch:introduction}
Humans and robots possess distinct capabilities and characteristics due to their inherent nature and design.
These include, but are not limited to, emotional intelligence, intuition and creativity, learning and adaptation, physical and cognitive abilities, as well as physical and mental limitations.

Presumably one of the more obvious differences is intuition and creativity.
Gut feelings, intuition, and a capacity for creativity allow humans to think outside the box.
This quality facilitates the development of innovative ideas and solutions to problems.
Robots, on the other hand, lack this genuine intuition and creativity as they are governed by logical patterns and unequivocal algorithms~\cite{gizzi2019creative}.
Although advances in robotics and \ac{AI} have enabled them to emulate certain aspects of intuition and creativity, they are still far from covering the entire spectrum of human intuition and creativity.

Closely related are the cognitive abilities.
Humans are adept at interpreting ambiguous information, comprehending context, and making decisions based on multiple factors, which stem from their abilities in complex reasoning and critical thinking.
While robots can accomplish certain cognitive tasks, such as identifying patterns or analyzing data, accurately and fast, their cognitive capabilities are limited and rarely extend beyond familiar circumstances~\cite{carff2009human}.

Emotional intelligence empowers humans to express feelings and to convey, recognize, and understand emotions.
In this way, we are able to empathize and connect with other human beings, develop social bonds, and build relationships.
In contrast, robots are devoid of any concept of real emotion.
Even though they can simulate or react to predefined emotions, their emotional abilities are exclusively artificial~\cite{loffler2018multimodal}.

In terms of physicality, humans and robots differ in both capabilities and underlying limitations.
Humans have versatile and adaptable physical abilities, such as dexterity, fine motor skills, and a broad portfolio of movements. 
However, they also experience physical and mental limitations owing to factors like fatigue, inconsistency regarding repetitions, and biological constraints.
Robots, on the other hand, are proficient in precise and repetitive actions, and are endowed with higher levels of strength and endurance in comparison to humans \cite{werner2017design}.
They are often engineered with specific physical characteristics required to accomplish a particular task.
Apart from a dependency on power supply and maintenance, this tailored design and the absence of an intrinsic awareness of their surroundings often leaves them inflexible and inadequate to operate under unfamiliar conditions~\cite{boteanu2015towards}.

Some of these mentioned attributes render robots superior to humans for certain tasks, and others inferior (see \figref{fig:robot_abilities}).
Either way, this repertoire of characteristics can benefit humans, which is why robots continue to enter a growing array of application fields, serving as a valuable complement to human capabilities.

\begin{savenotes}
\newcommand*{\fntexti}{\copyright{} %
    Haophuong21 / %
    \href{https://commons.wikimedia.org/wiki/File:Robot-cong-nghiep-the-he-moi.jpg}{Wikimedia Commons} / %
    \href{https://creativecommons.org/licenses/by-sa/4.0/deed.en}{CC BY-SA 4.0} / %
    Cropped from original. %
    See \cite{img:wikimedia:industrial_robot}. %
}
\newcommand*{\fntextii}{\copyright{} %
    \href{https://commons.wikimedia.org/wiki/User:Mbrickn}{Mbrickn} / %
    \href{https://commons.wikimedia.org/wiki/File:Stuck_Starship_Robot.jpg}{Wikimedia Commons} / %
    \href{https://creativecommons.org/licenses/by/4.0/deed.en}{CC BY 4.0}. %
    See \cite{img:wikimedia:stuck_starship_robot}. %
}
\begin{figure}
    \centering
    \includegraphics[width=.53\textwidth,trim={2.7cm 0cm 1.3cm 0cm},clip]{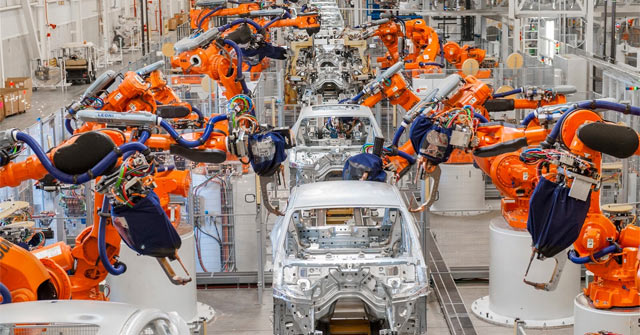}
    \hfill
    \includegraphics[width=.45\textwidth,trim={0cm 0cm 0cm 0cm},clip]{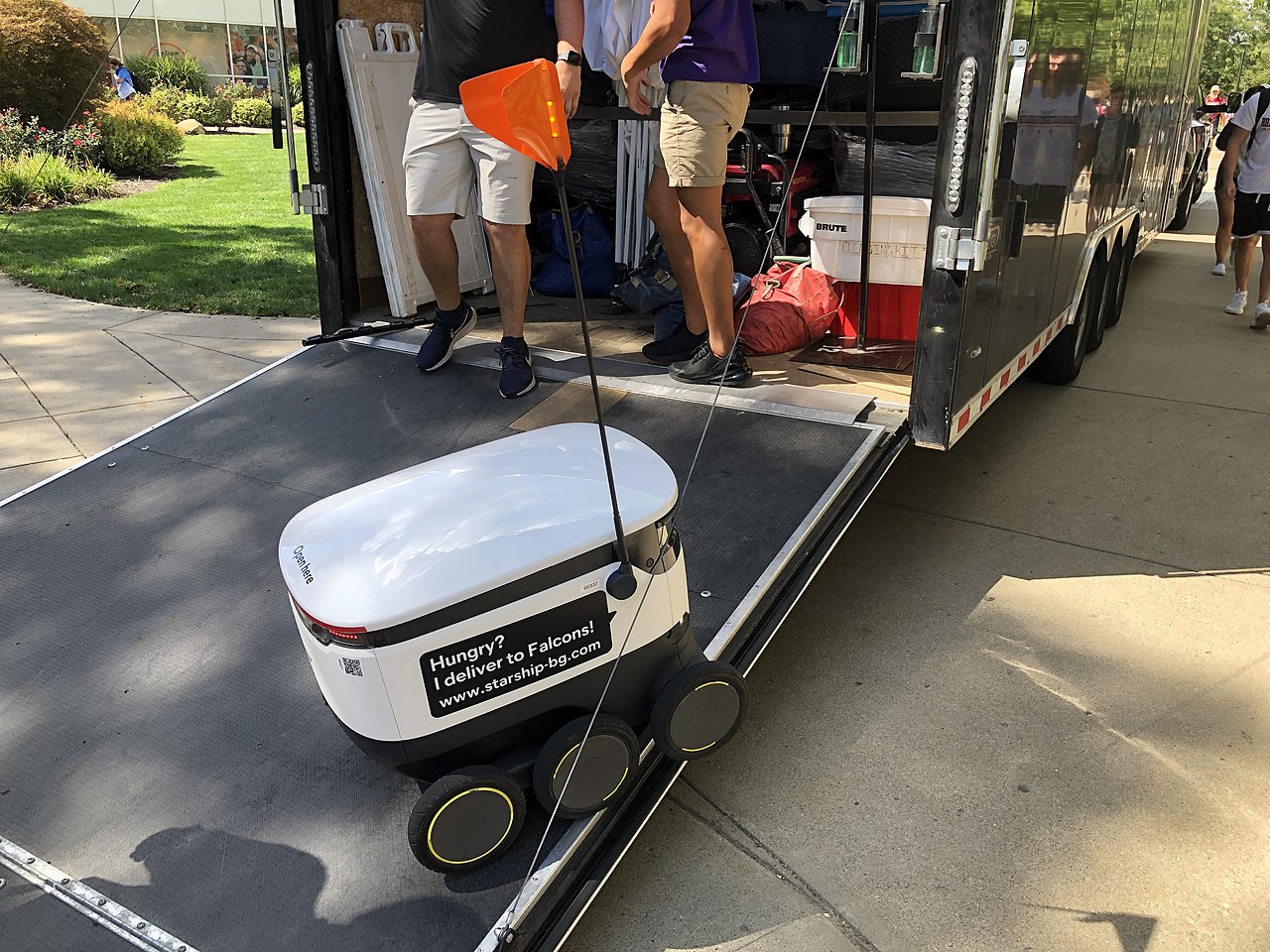}
    \caption[
    Robots excel over humans in terms of physical abilities, precision, and speed, rendering them exceptionally well-suited for tasks such as factory automation, as shown in the left image.
    However, robots face limitations in their natural ability to autonomously adapt to unpredictable situations or operate in unfamiliar environments due to their lack of inherent abstraction capabilities.
    These can lead to failures under real-life conditions, as evident in the image on the right, where an autonomous delivery vehicle got stuck due to an unexpected obstacle.
    ]{
    Robots excel over humans in terms of physical abilities, precision, and speed, rendering them exceptionally well-suited for tasks such as factory automation, as shown in the left image\footnote{\fntexti}.
    However, robots face limitations in their natural ability to autonomously adapt to unpredictable situations or operate in unfamiliar environments due to their lack of inherent abstraction capabilities.
    These can lead to failures under real-life conditions, as evident in the image\footnote{\fntextii} on the right, where an autonomous delivery vehicle got stuck due to an unexpected obstacle.
    }
    \label{fig:robot_abilities}
\end{figure}
\end{savenotes}

\section{Applications Areas of Robots} \label{sec:application_areas}
Robots' excellent precision, working capacity, efficiency, and ability to operate in arduous and hazardous environments creates an increasing demand for robots in a variety of sectors.
In the manufacturing industry, they are deployed extensively for tasks such as assembly \cite{marvel2018multi,kyrarini2019robot}, painting \cite{asadi2018pictobot, zhang2020accurate}, welding \cite{pires2003welding, pires2006welding}, packaging \cite{dubey2006packaging, do2012dual}, and quality control \cite{herakovic2010robot, burghardt2017monitoring}. 
To this end, they are designed to perform these tasks repetitively, fast, and with high precision.
Furthermore, in warehouses and distribution centers, robots are used to sort \cite{song2021robot,liu2022rapid}, and \acp{AMR} and \acp{AGV} are used to handle materials \cite{li2019task,polten2021scheduling,zhang2023application} to reduce human labor.
Within the medical and healthcare sector, robots play an essential role conducting diagnoses \cite{abolmaesumi2001user,najarian2011advances}, assisting in surgeries \cite{rosen2011surgical,peters2018review}, and supporting rehabilitation \cite{krebs2003rehabilitation,colombo2018rehabilitation}.
While robotic exoskeletons provide physiotherapy and mobility assistance for people with disabilities, surgical robots enable minimally invasive surgical operations.
Agricultural robots, also known as agribots or agrobots, are utilized for agricultural purposes such as weeding \cite{van2006mobile,utstumo2018robotic,gokul2019gesture}, planting seeds \cite{haibo2015study,ruangurai2015automated}, disease and insect detection \cite{pilli2015eagrobot, schor2016robotic, schor2017development, rey2019xf}, plant monitoring \cite{dos2015towards,bietresato2016evaluation,vidoni2017byelab}, spraying pesticides \cite{singh2005autonomous,sammons2005autonomous,mahmud2019multi} and harvesting crops \cite{tanigaki2008cherry,baeten2008autonomous,bulanon2010fruit,feng2018design}.
They reduce the resources and labor costs required, protect humans and the environment from unnecessary use of chemical substances, and enable precision farming techniques.
Furthermore, robots are deployed in space as part of space exploration missions to investigate distant planets, moons, and asteroids \cite{hirzinger1994rotex,yim2003modular,moosavian2007free,bogue2012robots}.
Among others, their tasks include gathering data, conducting experiments and facilitating research.
In defense and security, application fields of robots include surveillance \cite{rybski2000enlisting,ghouse2017military,ghute2018design}, reconnaissance \cite{chemel1999cyclops,voth2004new,usha2017military}, support and rescue \cite{choi2019development,ismail2020military,ismail2021military}, as well as the detection and disposal of underground mines \cite{nonami2003development,aoyama2007development,farooq2016wirelessly}.
In these contexts, \acp{UAV} also play a crucial role because they can operate in impassable terrain: For example, they are highly suitable for search missions in maritime environments \cite{varga2022seadronessee}.
In the domestic and personal sphere, robots are primarily used as vacuum cleaners \cite{chiu2009design,vaussard2014lessons,kang2014robust}, lawn mowers \cite{hicks2000survey,verne2020adapting,liao2021designing} or social companions \cite{bechade2019towards,ghafurian2021social}.

The ever-expanding applications and benefits of robots have been substantially fertilized by the advances in the research field of \ac{AI}, especially \ac{ML}.
These advancements have unlocked capabilities in robots for acquiring and learning certain behaviors to act in specific situations in a similar manner to humans or as desired by humans.
However, the way robots learn fundamentally differs from that of humans.
This foundational distinction expands upon the differences between humans and robots discussed earlier in this chapter and further emphasizes the unique nature of robot learning.

\section{Robot Deficiencies and Learning} \label{sec:robot_deficiencies_learning}
Humans possess a broad spectrum of learning styles, such as observational learning \cite{bandura2008observational}, conceptual learning \cite{feldman2003simplicity} or learning by trial and error \cite{young2009learning}, feedback \cite{balzer1989effects} or by means of a teacher \cite{hanushek2006teacher}.
This variety empowers them to link different concepts, understand and interpret complex contexts, as well as to perceive subtle clues and approach problems from unconventional angles.
By extracting underlying principles, building contextual understanding and reusing past experiences, humans can adapt their knowledge to new circumstances and varying environments, respond to unexpected changes and apply their skills to other domains.
The deficiency in the capacity for such generalizations and the lack of adaptability of behaviors and strategies are among the key shortcomings of robots.
Robots suffer from an inability to think beyond established patterns, as their learning behavior primarily relies on programmed algorithms or machine learning techniques; the latter usually demands data-driven training to acquire new skills.
Even in a familiar environment with a well-defined setting, a substantial amount of training data is required.
On top of that, any change in circumstances may necessitate retraining.
When robots are challenged with unpredictable or not predefined conditions that they have not been explicitly prepared for, their performance is negatively affected.
However, deployed in close proximity to humans, the robot can be assisted by the human in dealing with such unfamiliar scenarios~\cite{carff2009human}.
This implies that \ac{HRI} holds the potential to mitigate the previously outlined disparities to some extent.
Through interactive teaching by a human, knowledge can be imparted to the robot and obstacles can be collectively overcome.

In many of the robot applications described in \secref{sec:application_areas}, the identification of the relevant \acp{OOI} plays a crucial role, as illustrated in \figref{fig:robots_ooi}.
\begin{savenotes}
\newcommand*{\fntexti}{\copyright{} %
    UKM Elisabeth Deiters-Keul / %
    \href{https://commons.wikimedia.org/wiki/File:Kommissionierroboter_in_der_Apotheke_der_Universit\%C3\%A4tsklinik_M\%C3\%BCnster.jpg}{Wikimedia Commons} / %
    \href{https://creativecommons.org/licenses/by-sa/3.0/deed.en}{CC BY-SA 3.0} / %
    Cropped from original. %
    See \cite{img:wikimedia:consignment_robot}. %
}
\newcommand*{\fntextii}{\copyright{} %
    \href{https://www.tevel-tech.com}{Tevel Aerobotics Technologies} / %
    Cropped from original. %
    See \cite{img:tevel}. %
}
\begin{figure}
    \centering
    \includegraphics[width=.49\textwidth,trim={1cm 0.75cm 1cm 0.78cm},clip]{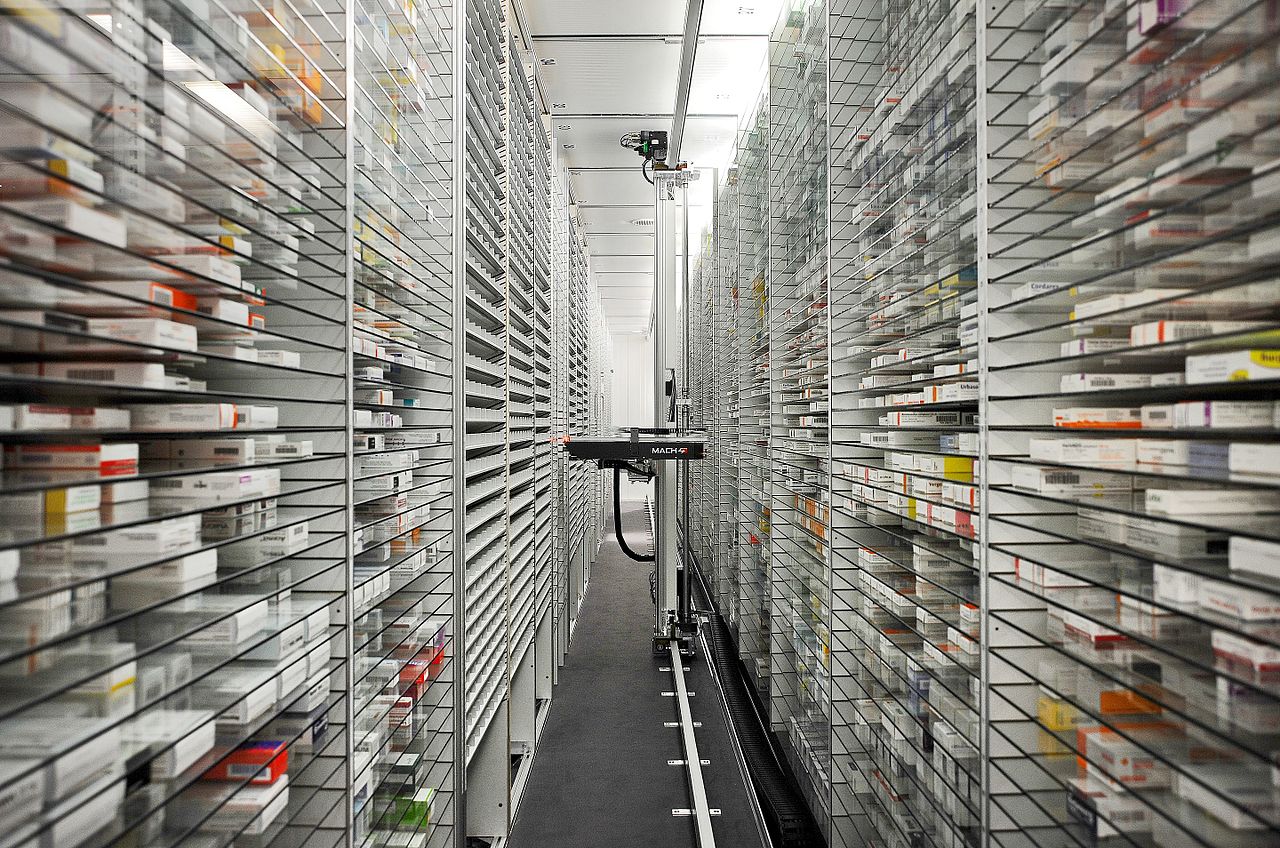}
    \hfill
    \includegraphics[width=.49\textwidth,trim={0cm 1.5cm 0cm 1.5cm},clip]{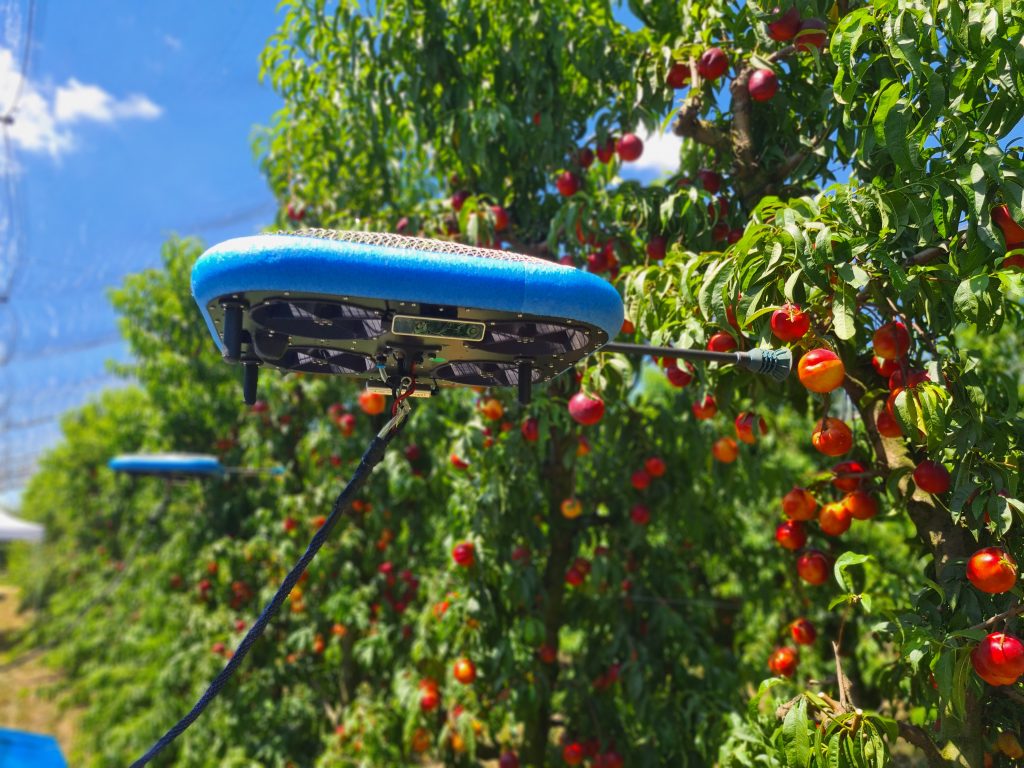}
    \caption[
    In the left image, a consignment robot in a pharmacy retrieves packages of medicines from a storage shelf, and in the right image, a drone harvests apples. In both applications, the robot must detect the respective \ac{OOI} prior to performing the action.
    ]{
    In the left image\footnote{\fntexti}, a consignment robot in a pharmacy retrieves packages of medicines from a storage shelf, and in the right image\footnote{\fntextii}, a drone harvests apples. In both applications, the robot must detect the respective \ac{OOI} prior to performing the action.
    }
    \label{fig:robots_ooi}
\end{figure}
\end{savenotes}
The capability to detect them must be either acquired beforehand -- often data-driven -- or taught during runtime, for example by means of \ac{HRI} or \ac{HRC}.
The former is the most popular and most widespread variant.
Based on an available data set, the robot is trained and is then capable of interacting with the respective objects, enabling potential subsequent actions like grasping.
However, such an approach is based on two underlying assumptions.
First and foremost, it is imperative that the future \acp{OOI} are known in advance.
Secondly, the existence and accessibility of appropriate training data encompassing all \acp{OOI} is indispensable.
These aspects are the crux of the issue at hand.
The plethora of objects and thus potential interaction entities is -- at least for all practical matters -- basically unlimited.
Consequently, the existence of appropriate data cannot be assumed without further ado.
The problem that arises from such data dependency has already been hinted at above.
If the setting is not precisely predefined and limited to fixed set of \acp{OOI}, this will impair the performance of the robot and thus prevent its successful deployment.
Herein lies the necessity of the contributions of this work, as it becomes increasingly important to find alternative concepts that disengage from dataset dependency in robot learning.
This dissertation revolves around one such an approach that entails the human teaching the robot the unknown objects, assisting it to cope in environments where novel objects are encountered.

\section{Towards Teaching Robots}
In general, robots can learn from humans in different ways.
One possibility, for example, is through demonstration, which is particularly popular for teaching assembly tasks \cite{skubic2000acquiring,dillmann2004teaching,wang2018facilitating,zhu2018robot}.
Here, the robot typically observes the actions performed by a human and then replicates them through imitation.
In a related framework of \ac{PbD}, a human teacher assumes the role of manually guiding the robot \cite{aleotti2004leveraging,calinon2007active,alexandrova2014robot}, either by physically manipulating its limbs or utilizing an interface.
The robot meticulously captures and stores the teacher's actions, enabling it to generalize and autonomously execute similar tasks in the future.
Another teaching technique is \ac{RLHF}, which integrates valuable human insights into reinforcement learning \cite{griffith2013policy,christiano2017deep}.
By incorporating evaluative feedback, such as rewards or punishments, provided by a human collaborator, the robot utilizes this information to refine its learning process.
Through iteration, the robot progressively enhances its performance based on the guidance received through the human's assessments.
Instructions in natural language offer another avenue for robots to learn from humans \cite{lauria2001training,muthugala2018review}.
The robot processes the verbal commands of a human teacher and can thus learn new tasks, such as grasping \cite{she2014teaching} or navigating \cite{fasola2013using}, or improve its existing skills.
Moreover, \ac{AR} and \ac{VR} technologies offer immersive environments that facilitate the training of robots by humans \cite{pettersen2003augmented,liu2018interactive,tahriri2015optimizing,roman2018low}.
Within these augmented or virtual spaces, the human can interact with virtual objects or simulated events and the robot learns by assimilating the transferred information.
Naturally, there are various other methods beyond the scope of this dissertation that cannot all be discussed in detail.
For more information and examples, the readers are referred to~\cite{trafton2006children, lockerd2004tutelage,xu2017robot}. 

With regard to the intended teaching of novel objects, \ac{AR} demonstrates significant potential.
This technology stands out as a promising avenue to realize natural interaction and collaboration between human and robot.

\section{Human-Robot Interaction Employing \texorpdfstring{\acs*{AR}}{AR} and Eye Tracking}
The combination of robotics with augmented reality \cite{green2008human,green2010evaluating,suzuki2022augmented} and robotics with human eye tracking \cite{admoni2017social,aronson2018eye,scalera2021human} led to the intensification of the interaction between humans and robots.

\ac{AR} refers to the technology that integrates computer-generated content into the real world.
By overlaying digital information, such as virtual objects, onto the physical world, the latter is augmented.
According to \citeauthor{azuma1997survey} \cite{azuma1997survey}, \ac{AR} can be defined as a system that exhibits the following three properties: It combines real and virtual world, allows real-time interaction, and accurately registers virtual and real objects in 3D.
The primary advantage of \ac{AR} lies in its inherent ability to blend digital components into the user's individual perception of the physical world.
Rather than merely displaying data, \ac{AR} embeds immersive sensory modalities that are perceived as natural parts of the environment and enhance the user's overall experience.
\ac{AR} functions across a wide range of domains, such as medicine \cite{roberts1986frameless,bajura1992merging,sielhorst2008advanced,wen2014hand}, military \cite{julier2000information,livingston2002augmented,foxlin2004flighttracker}, manufacturing \cite{reinhart2003integrating,tang2003comparative,lai2020smart}, entertainment and games \cite{cavallaro1997foxtrax,azuma2006performance,cavallaro2011augmenting}, education \cite{yuen2011augmented,billinghurst2012augmented,kesim2012augmented,chang2013integrating}, navigation and path planning \cite{rekimoto1997navicam,narzt2006augmented,chung2016augmented}, as well as tourism \cite{vlahakis2002archeoguide,ingram2003trust,martinez2004designing,fritz2005enhancing}.
Through integration of \ac{AR} into \ac{HRI} scenarios, the human gains awareness of the robot's state and intentions.
\ac{AR} offers a more immersive and contextually rich interaction by visualizing supplementary information of the robot that is not directly observable by the human \cite{collett2006developer}.
In summary, \ac{AR} bridges the digital world of the robot and the analog world of the human to facilitate communication between them.

Owing to the growing prevalence of \ac{AR} glasses, such as the Microsoft HoloLens~2\footnote{\url{https://www.microsoft.com/en-us/hololens} (accessed: June 16, 2023)}, the Magic Leap~2\footnote{\url{https://www.magicleap.com/magic-leap-2} (accessed: June 16, 2023)} or the recently announced Apple Vision Pro\footnote{\url{https://www.apple.com/apple-vision-pro} (accessed: June 16, 2023)}, eye tracking can be adroitly employed alongside \ac{AR}.
Eye tracking is the process of measuring a person's eye movements and focus.
This technology enables the analysis and comprehension of eye behavior, such as the direction of the gaze \cite{tafaj2012bayesian,santini2016bayesian}, the fixation duration on specific points \cite{eivazi2017optimal}, and gaze patterns \cite{kubler2014subsmatch}.
It is applied in numerous fields and research areas, spanning psychology \cite{mele2012gaze,orquin2018threats,rahal2019understanding}, marketing \cite{dos2015eye,zamani2016eye,wedel2017review}, driving \cite{horng2004driver,devi2008driver,singh2011eye,tafaj2013online,braunagel2015driver,braunagel2017ready,khan2019gaze,bozkir2019assessment,carr2020role}, sports \cite{roca2013perceptual,panchuk2015eye,huttermann2018eye,hosp2021expertise}, education \cite{busjahn2014eye,ashraf2018eye,sumer2018teachers,strohmaier2020eye,goldberg2021attentive,sumer2021multimodal,hahn2022eye,gao2022evaluating,byrne2023leveraging}, and medicine \cite{wilson2011gaze,almansa2011association,voisin2013investigating,castner2020deep,hosp2021differentiating,hosp2021states,arsiwala2023gaze}.
In the realm of robotics, especially with regard to \ac{HRI}, eye tracking can be used, for example, to allow robots to proactively anticipate and perform tasks based on the eye movements of their human partner \cite{huang2016anticipatory} or to empower people with disabilities to control assistive robots \cite{cio2019proof}.
In this \ac{HRI} context, eye tracking plays a pivotal role, as it enables the robot to perceive the focus of the human's attention.
By tracking the user's eye movements, the system can ascertain where the user is looking, and subsequently, the robot can react in accordance with the interpreted user intentions. 
This seamless and natural interaction enables the human to effortlessly issue commands or express interest by simply looking at objects or specific locations of interest.

These benefits offered by \ac{AR} and eye tracking were leveraged to close the existing gap of robots being confined to predefined scenarios, mentioned in the concluding remarks of \secref{sec:robot_deficiencies_learning}.
The interdisciplinary contributions of this dissertation furthered this holistic perspective, ultimately fulfilling the objectives detailed in the following.

\section{Setting and Objectives} \label{sec:objectives}
\begin{figure}[ht]
    \centering
    \includegraphics[width=.9\linewidth]{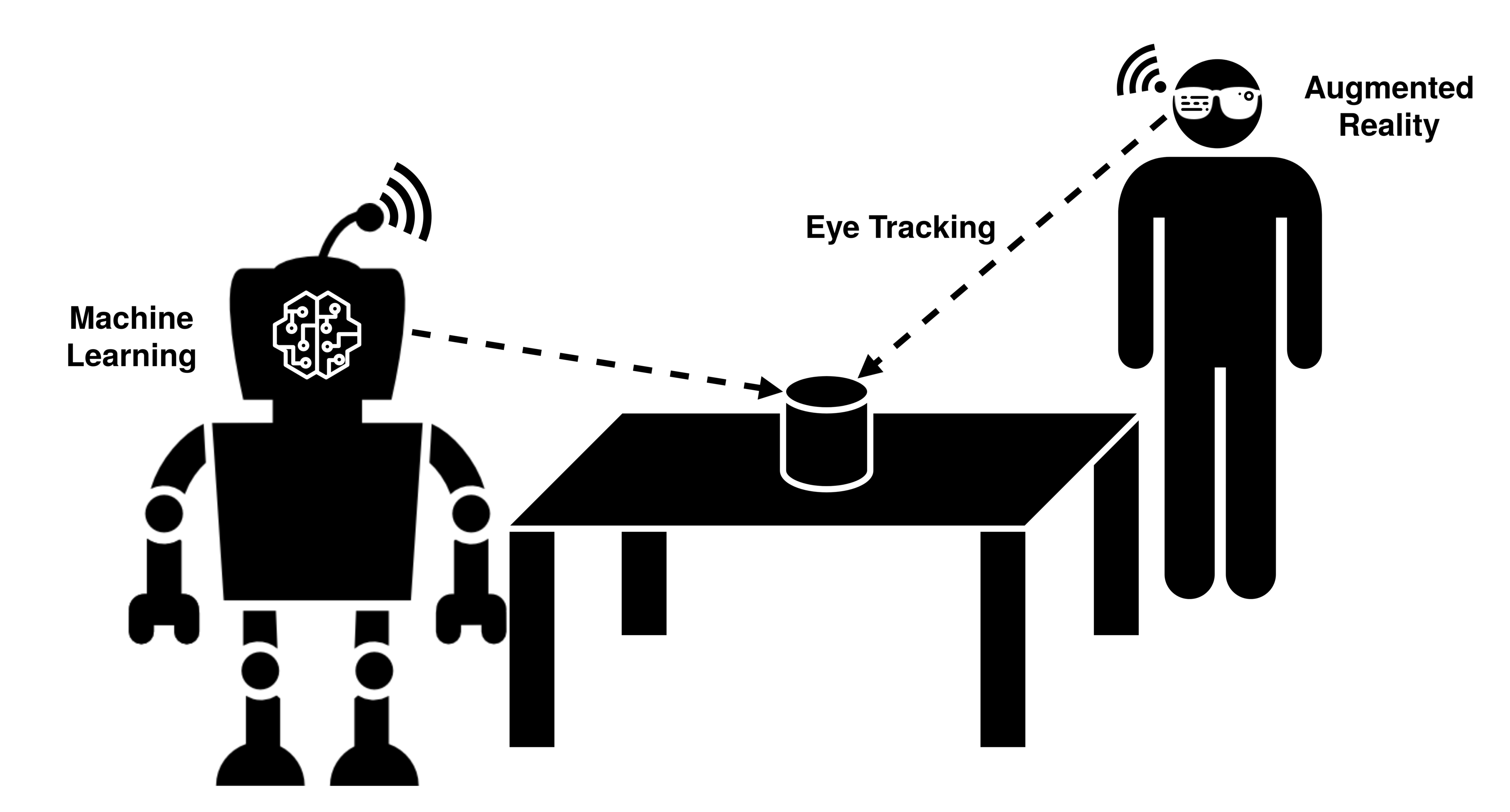}
    \caption{The human and the robot both stand in front of a table. The human selects the \ac{OOI} using gaze. The robot must identify the object correctly and then learn it by means of the class information which the human provides. The communication takes place through an \ac{AR} interface and Wi-Fi.}
    \label{fig:project_sketch}
\end{figure}

As briefly covered in \secref{sec:robot_deficiencies_learning}, the central goal of this dissertation was to develop a framework for teaching a robot unknown objects, with a strong focus on flexibility in non-predefined scenarios.
For this purpose, the human directs the robot's attention towards the \ac{OOI} by looking at it and pointing it out via gaze.
Subsequently, after the robot has segmented the object visually, it records it from slightly different angles and finally learns and later redetects it based on the class information provided by the human during the teaching process.
The general setting is visualized in \figref{fig:project_sketch}.

In order to accomplish this overall goal, a detailed series of intermediate steps had to be completed.
More specifically, this thesis addresses each of the following challenges and provides methods to approach them in realistic settings:

\begin{enumerate}[label=\text{C\arabic*.},ref=\text{C\arabic*}, parsep=3pt plus 2pt minus 1pt, itemsep=\parsep]

    \item\label{item:p1} The human gaze has to be mapped from the human's frame of reference to the robot's frame of reference to convey the information regarding where the human is looking and directing their gaze.
    
    \item\label{item:p2} The identification and localization of the \ac{OOI} on the robot side must be accomplished. To this end, the robot needs to visually segment the object, even though the object is not yet known to it at this point. This segmentation process allows the robot to distinguish and delimit the object from its surroundings.

    \item\label{item:p3} In order for the human and the robot to communicate with each other and exchange data at all, a bidirectional connection needs to be established between them using an \ac{AR} interface. The implementation of this dedicated interface is essential, as it serves the dual purpose of enabling communication and tracking the human gaze.

    \item\label{item:p4} Prior to the human teaching the robot, the human must ascertain whether both, human and robot, are sharing their attention on the same object. Meaning, the human needs to receive feedback regarding the segmented \ac{OOI} in order to verify and intervene if necessary.

    \item\label{item:p5} As part of the teaching process, the human must impart the class information to the robot.

    \item\label{item:p6} To facilitate the robot's learning process, it is crucial to develop a procedure by which the robot can examine the object from various perspectives, as previously described, and generate autonomously labeled training data. Subsequently, this data can be utilized to train the robot using \ac{ML} models.
\end{enumerate}

All of the above intermediate challenges were successfully completed through pronounced interdisciplinary efforts and all contributed to the higher holistic objective.
Their solutions will be explained in more detail in \chref{ch:motivation_results} and \chref{ch:discussion}.
Upon completion, the robot's capability to redetect taught objects could be assessed and evaluated.

\section{Hardware and Evaluation Fundamentals}

\subsection{Hardware}
The \ac{HRI} teaching task described above requires a variety of different hardware.
This hardware will be described in more detail below, as it reoccurs throughout the dissertation.

\begin{figure}[b!]
    \centering
    \includegraphics[width=.47\textwidth,trim={0cm 7cm 0cm 7cm},clip]{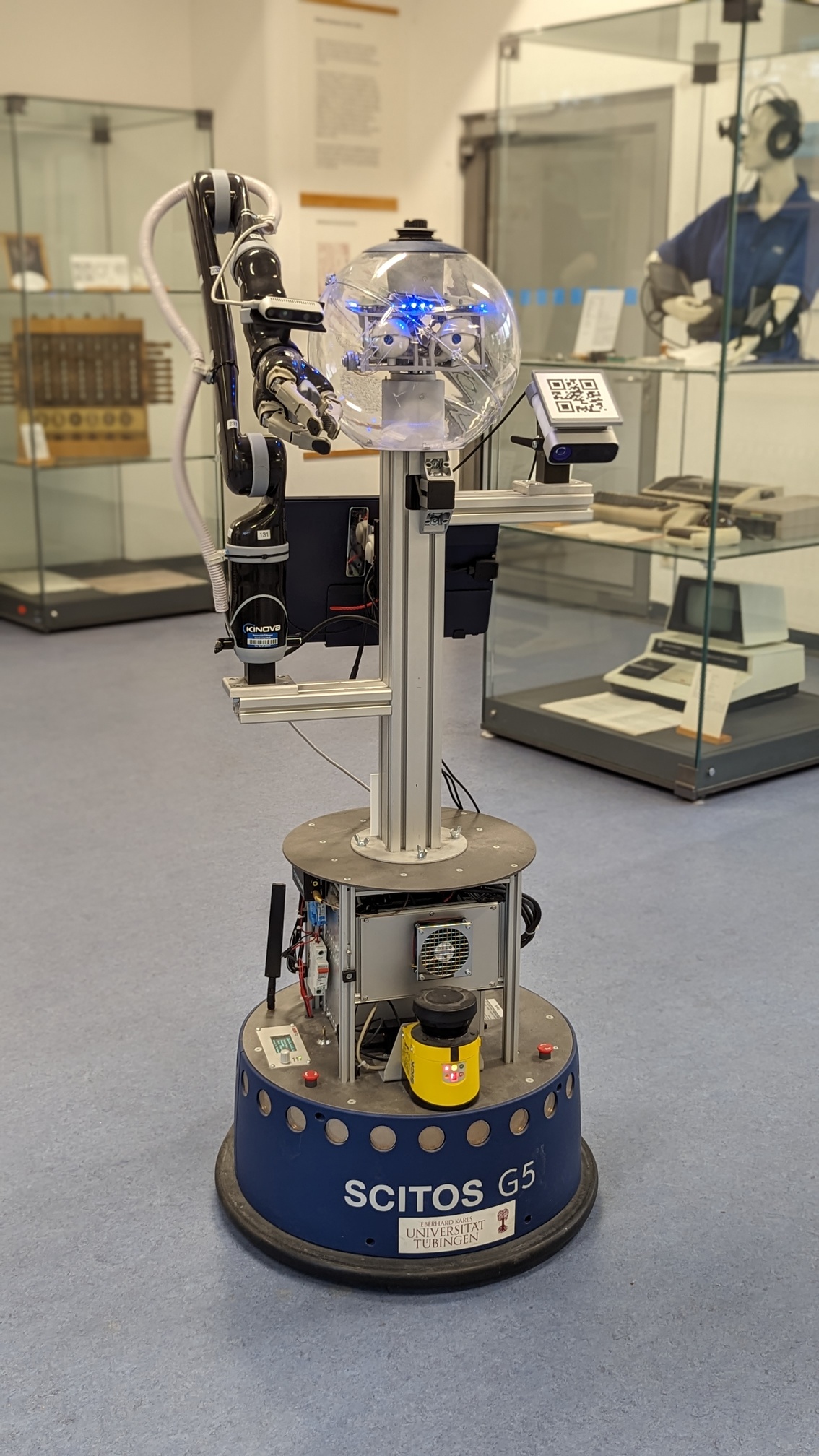}
    \hfill
    \includegraphics[width=.51\textwidth,trim={0cm 4.5cm 0cm 1cm},clip]{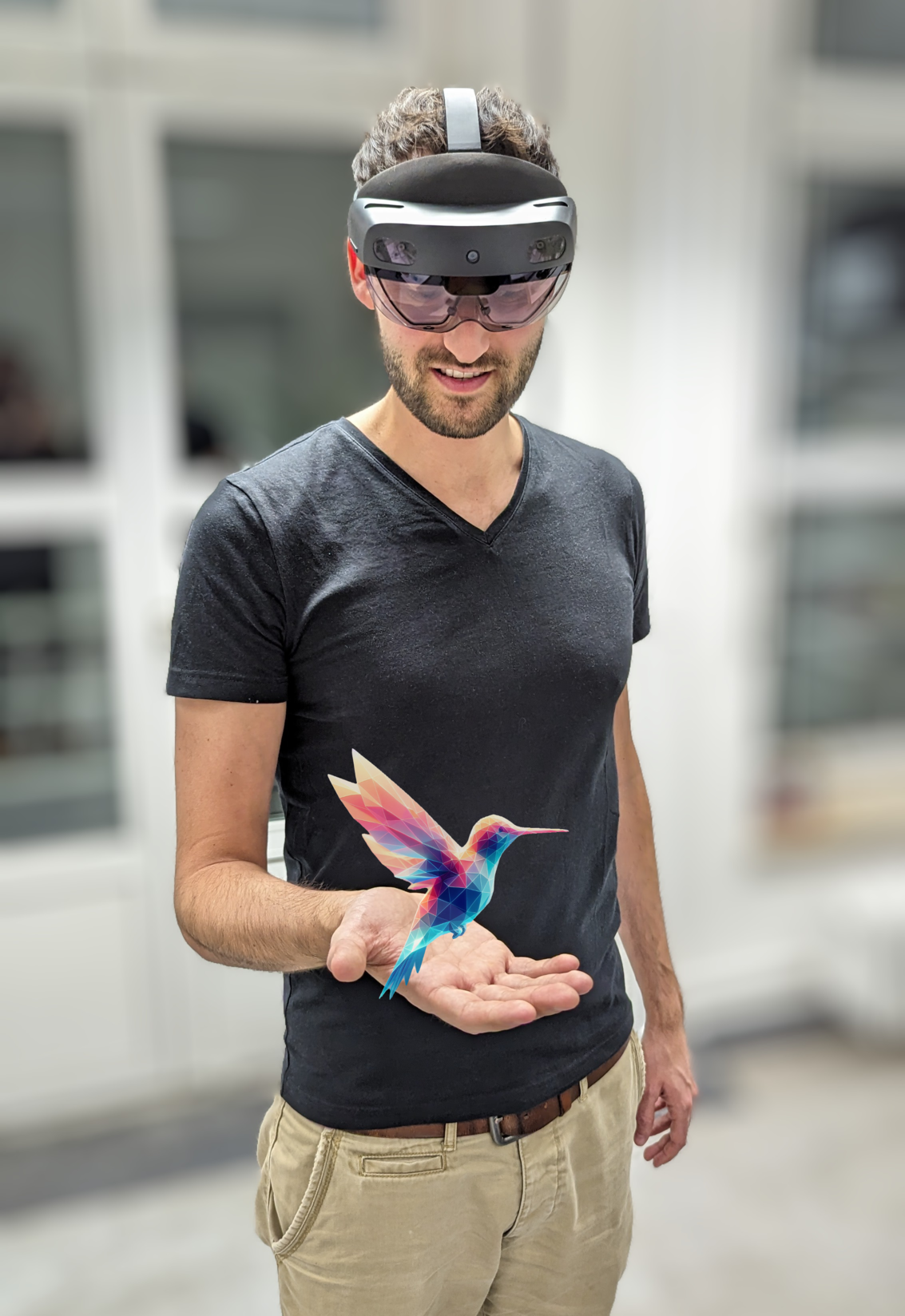}
    \caption[The left image depicts the Scitos~G5 robot, and the right image illustrates a person interacting with a virtual object while wearing the HoloLens~2.
    Both devices were used throughout this dissertation.]
    {The left image depicts the Scitos~G5 robot, and the right image\protect\footnotemark{} illustrates a person interacting with a virtual object while wearing the HoloLens~2.
    Both devices were used throughout this dissertation.}
    \label{fig:HL2andScitos}
\end{figure}%
\footnotetext{The hummingbird was generated with DALL-E~3.}%

In the first part of \chref{ch:motivation_results}, for the general investigation of the potential of eye tracking in determining \acp{ROI}, standard eye trackers, decoupled from any \ac{AR} functionality, are used.
In the \ac{AR}-related part of \chref{ch:motivation_results}, from \subsecref{subsec:HRI22} and on, the hardware remains the same and is pictured in \figref{fig:HL2andScitos}.
The mobile robot employed throughout is the Scitos~G5 developed by MetraLabs\footnote{\url{https://www.metralabs.com/mobiler-roboter-scitos-g5} (accessed: June 16, 2023)}, with the robotic middleware suite \ac{ROS} \cite{quigley2009ros} as framework to control it.
Additionally, the robot is further equipped with a Kinova Jaco2\footnote{\url{https://www.kinovarobotics.com/product/gen2-robots} (accessed: July 10, 2023)} robot arm.
The human partner wears a pair of \ac{AR} glasses, specifically the HoloLens~2.
This \ac{AR} device manufactured by Microsoft is equipped with a built-in eye tracker.
The eventual user interface developed for the HoloLens~2, as a crucial component of this work, is implemented within the game development environment Unity\footnote{\url{https://unity.com} (accessed: June 16, 2023)}.
The actual communication between the HoloLens~2 and \ac{ROS} is handled by the \ac{UWP} version of ROS\# \cite{bischoff2019rossharp}, which is a collection of open-source software libraries and tools designed to facilitate communication and data transfer between Unity applications and \ac{ROS}.

\subsection{Evaluation Fundamentals and Terminology}
The proficiency of the teaching pipeline introduced in this dissertation is determined by the extent to which the robot was able to redetect the objects it was taught by the human.
For better understanding of this pipeline and its evaluation, an overview of the common terminology frequently appearing in the subsequent chapters is provided here.

Among these terms is \ac{IoU}.
The \ac{IoU}, also known as the Jaccard index, is a measure for the similarity of two sets.
It is defined by the intersection of the two sets divided by their union.
In the research field of computer vision, the \ac{IoU} is widely applied to compare and quantify the similarity of two bounding boxes in order to assess the accuracy of object detection algorithms and models.
It can assume values ranging from $0$ to $1$, where $0$ signifies that the two bounding boxes do not overlap, and $1$ indicates that they are identical in terms of position and size.
Thus, a high \ac{IoU} score suggests a better alignment or larger overlap.
Commonly, values of $0.5$ and above are considered acceptable, albeit slightly generous, an \ac{IoU} of $0.9$ is rather strict, and $0.7$ a reasonable compromise in between.

Further relevant terms revolve around the MS~COCO metrics~\cite{lin2014microsoft}, which are common evaluation metrics in object detection.
The central focus lies on the average precision and the average recall across a variety of \acs{IoU} thresholds.
In this context, the precision is defined as the ratio of correct predicted bounding boxes to the total number of predictions, while recall represents the fraction of correct predictions among the relevant bounding boxes.
Meaning, precision and recall specify how many of the obtained predictions are correct and how many of the relevant items were detected, respectively.
An object detection model can then be evaluated for different thresholds of the model's confidence scores, resulting in pairs of precision and recall values.
Based on these pairs, a precision-recall curve can be constructed.
The average precision then results from the area under this curve.
In practice, however, MS~COCO determines the average precision by averaging $101$ interpolated precision values at equidistant recall values using a step size of $0.01$ between $0$ and $1$, denoted by $[0:0.01:1]$.
The abbreviations $ \AP{50} = \text{AP}^{\text{IoU}=0.5} $ and $ \AP{75} = \text{AP}^{\text{IoU}=0.75} $ refer to the average precision values at the \ac{IoU} thresholds of $0.5$ and $0.7$, respectively.
Furthermore, $ \AP{} = \text{AP}^{\text{IoU}=0.5:0.05:0.95} $ is the average precision averaged across all \acs{IoU} thresholds in $[0.5:0.05:0.95]$.
Analogously, this notation extends to the average recall.
In this case, the maximum recall is ascertained allowing $1$, $10$, and $100$ detections per image, respectively, averaged over \acp{IoU}, and is represented by the abbreviations $ \AR{1} = \text{AR}^\text{max=1} $, $ \AR{10} = \text{AR}^\text{max=10} $, and $ \AR{100} = \text{AR}^\text{max=100} $.
Each of the aforementioned metrics are calculated independently for each individual class.
Additionally, the mean across all given classes is denoted by the \ac{mAP} and the \ac{mAR}.

\cleardoublepage

\chapter{Major Contributions}
\label{ch:motivation_results}

In this chapter, the relevant research contributions towards the objectives and setting introduced in \chref{ch:introduction} are further detailed.
First, the motivation sets up the respective research question.
Then, the methodology shows the processes involved with addressing the research question and subquestions, followed by the work's main contributions and results.
An overview of the subsequent papers published at high-impact conferences in the fields of robotics, eye tracking and \ac{HRI} is listed in \chref{ch:publist}.

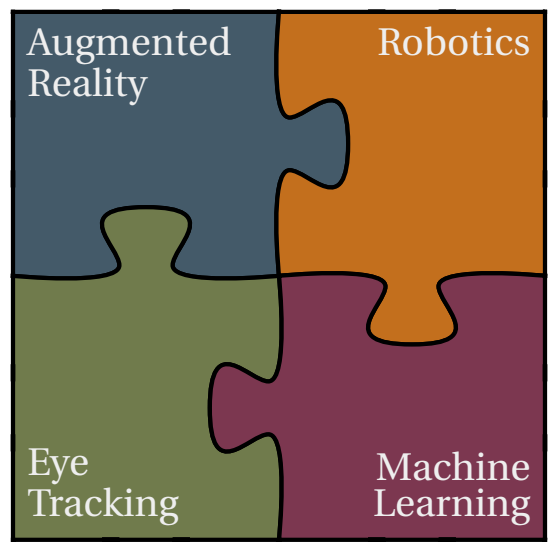
\begin{figure}[b!]
	\centering
	\begin{tikzpicture}
        \newcommand*{\jsscale}{3.5}
        \newcommand*{\textformat}{\fontsize{15.8}{0}\selectfont} 
        \definecolor{colorText}{HTML}{EEEEEE}
        \definecolor{color1}{HTML}{2F4858}
        \definecolor{color2}{HTML}{BC5F04}
        \definecolor{color3}{HTML}{606C38}
        \definecolor{color4}{HTML}{6D213C}
        \matrix [nodes=draw]{
        \pic [fill=color1!90, draw=black, ultra thick, scale=\jsscale,%
                pic text={\textformat Augmented\\ \textformat Reality},%
                pic text options={text=colorText,align=left,anchor=north west,shift={(-1.7,1.7)}}%
                ]{piece={1}{-1}{0}{0}};&
        \pic [fill=color2!90, draw=black, ultra thick, scale=\jsscale,%
                pic text={\textformat Robotics\\},%
                pic text options={text=colorText,align=right,anchor=north east,shift={(1.7,1.7)}}%
                ]{piece={-1}{0}{0}{1}}; \\
        \pic [fill=color3!90, draw=black, ultra thick, scale=\jsscale,%
                pic text={\textformat Eye\\ \textformat Tracking},%
                pic text options={text=colorText,align=left,anchor=south west,shift={(-1.7,-1.7)}}%
                ]{piece={0}{1}{-1}{0}}; &
        \pic [fill=color4!90, draw=black, ultra thick, scale=\jsscale,%
                pic text={\textformat Machine\\ \textformat Learning},%
                pic text options={text=colorText,align=right,anchor=south east,shift={(1.7,-1.7)}}%
                ]{piece={0}{0}{1}{-1}};\\
        };
	\end{tikzpicture}
	\caption{The distinct research fields of augmented reality, eye tracking, machine learning, and robotics are intertwined as essential parts of a larger \ac{HRI} system.
    }
	\label{fig:overview}%
\end{figure}

This research has an impact on multiple fields, which are illustrated in \figref{fig:overview}.
These fields interlock and combine to form a system that enables two-way, human-robot interaction, ultimately the human to teach the robot unknown objects in a natural and feasible way.

This chapter is divided into two parts.
The first part of this chapter, \secref{sec:results_A}, investigates the general potential of the human gaze in determining \acp{ROI}.
These defined \acp{ROI} based on human gaze offer great potential towards the goal of teaching a robot unknown objects, since the robot needs to identify which object a human is looking at in the first place, that is, what belongs to the object and what does not.
In this part, the major focus lies on the domains eye tracking and machine learning.
The accompanying publications are included in \chref{app:A}.

The aforementioned findings are applied in the second part of the chapter, \secref{sec:results_B}.
Here, the other two domains from \figref{fig:overview}, \acl{AR} and Robotics, become involved.
All parts together finally empower the robot's ability to perceive and learn unknown objects by collaborative interaction with the human.
The accompanying publications are included in \chref{app:B}.

\section{Investigating the Potential of Gaze in Determining Regions of Interest}
\label{sec:results_A}

The first subsection \subsecref{subsec:IRC22} examines how gaze can be potentially used for unknown object detection without considering the stimulus (image of the observed scene). The second subsection \subsecref{subsec:geisler2020exploiting} introduces the concept of saliency-aware gaze heatmaps.

\subsection{Gaze-based Object Detection in the Wild} \label{subsec:IRC22}
\begin{quote}
\textbf{Daniel Weber}, Wolfgang Fuhl, Andreas Zell, and Enkelejda Kasneci. Gaze-based Object Detection in the Wild. In \textit{2022 Sixth IEEE International Conference on Robotic Computing (IRC '22)}.
\end{quote}

\subsubsection{Motivation}
One of the first challenges before a human can teach a robot unknown objects in a \ac{HRI} setting, as described in \secref{sec:objectives}, is to let the robot perceive and locate the object the human is looking at.
This semantic awareness applies not only to learning, but to generally all desired interactions of a robot with an unknown object of interest.
However, detecting an object that is not known is a non-trivial problem.
At the same time, the human gaze can assist and provide additional information.
The main motivation of this work was to investigate how far the information capacity of the gaze spans, and whether it is possible to detect objects without any context from the scene image.
Thereby, object detection was conceived in the same way as face detection, where the primary task is to detect whether a face is present or not and, if so, to determine its position.
Usually, no classification takes place.
As this is not possible without a scene image, also in relation to objects, we have limited ourselves as well to the binary detection task and subsequent localization.

\subsubsection{Principal Methodology}
The foundation for the investigations was a self-reported data set.
This data set comprised of multiple participants, who were instructed to move freely inside and outside the venue while wearing a head-mounted eye tracker.
The gaze and the \ac{FOV} were recorded without any specifications of how long the participants should look at the encountered objects.
Afterward, the \acp{OOI} were labelled.

For the object detection task, the gaze data was divided into ranges of gaze points using temporal windows and then classified whether an object has been observed.
If the classification resulted in the detection of an object of interest, a regression of the bounding box parameters followed.
As input feature to the \ac{ML} models, the spatial distribution of the gaze points in the respective time window encoded as 2D and 3D heatmaps was used.

\subsubsection{Main Findings}
From the methodological point of view, the extension of the 2D~heatmap encoding into the three-dimensional space was essential.
As a result, in combination with a \ac{KNN} approach, a classification accuracy of \SI{92}{\%} was achieved.
Overall, multiple different \ac{ML} models were investigated as a proof of concept.
In addition to KNNs, this also included Bagged Trees, \acp{SVM}, and Gaussian Process.
In a detailed evaluation, the performance of these models were analyzed using different time window sizes and grid sizes for the heatmap features.
Apart from the high classification accuracy, the regression of the bounding box parameters yielded an average absolute error for the position of around \SI{6}{\%}.
However, the determination of the bounding box dimensions proved to be more difficult than the position.
The best average absolute error of the height and width of the bounding boxes ranged between \SI{10}{\%} and \SI{15}{\%}.

Besides the classification accuracy and the average absolute error of the bounding box parameters, the speed and resource consumption of the object detection using the gaze heatmap features were investigated.
For speed, the execution time for 1000 different inputs with a batch size of one was measured.
Both heatmap features required only a fraction of the time needed by conventional image-based object detectors.
In numbers, this translates to $8$ to $58$~seconds and $64$ to $611$~seconds for the 2D and 3D~heatmap features, respectively.
In contrast, only the smallest of the conventional image-based object detectors was able to stay with $164$~seconds below $3$~minutes, whereas the majority took significantly more than $10$~minutes.
For instance, the very popular Faster \mbox{R-CNN}~\cite{ren2015faster} with a ResNet-50-FPN backbone~\cite{he2016deep} needed the most time with over $8705$~seconds for the $1000$ different inputs.
This made it several orders of magnitude slower than the models that used the proposed heatmap features.

Regarding resource efficiency, the difference was even more pronounced.
While the use of the heatmap features only occupied a few hundred kilobytes to a few megabytes of memory for a single input, the smallest comparison model YOLOv5n \cite{jocher2022yolov5-short} allocated around \SI{270}{MB}.
The most resources were claimed by Faster \mbox{R-CNN}, which allocated over \SI{1.7}{GB} of memory.
Therefore, the conventional image-based object detectors were also several orders of magnitude more inefficient in terms of memory.

The dataset created in the context of this work was unique at the time of publication, hence it was made publicly available to the research community\footnote{\url{https://cloud.cs.uni-tuebingen.de/index.php/s/QPzJC48xDGsjnZK}}.

Overall, the work has shown that the gaze carries important information that can be useful and harnessed for object detection even if the class of the \ac{OOI} is unknown.
Consequently, combined with a scene image, it can unfold even more impact, which will be elaborated on in the following section.

\subsection{Exploiting the \texorpdfstring{\acs*{GBVS}}{GBVS} for Saliency aware Gaze Heatmaps} \label{subsec:geisler2020exploiting}
\begin{quote}
David Geisler, \textbf{Daniel Weber}, Nora Castner, and Enkelejda Kasneci. Exploiting the GBVS for Saliency aware Gaze Heatmaps. In \textit{ACM Symposium on Eye Tracking Research and Applications (ETRA '20)}.
\end{quote}

\subsubsection{Motivation}
Although, gaze can assist locating unknown \acp{OOI}, the use of pure gaze data in the form of heatmaps -- or similar representations -- has its limitations. 
Firstly, humans can sometimes unconsciously fixate on objects, which is why objects do not necessarily have to be observed entirely, as the human already perceives them beforehand.
Secondly, the eye tracking signal is always subject to some degree of error, so the resulting estimated gaze is never perfectly accurate.
For this reason, this work combined visual information and gaze signal.
Algorithms such as \ac{GBVS} \cite{harel2006graph} extract visually attentive areas of the stimuli, which are likely to attract the attention of a human observer.
The resulting saliency maps indicate regions of particular interest.
In combination with the recorded human gaze, deficiencies on both ends can be compensated, leading to better overall estimation of the observed \acp{ROI}.

\subsubsection{Principal Methodology}
Fusing gaze and visual information was accomplished by incorporating the gaze into the salience maps determined by the \ac{GBVS}.
The original GBVS algorithm is composed of three main steps.
Namely, 1.) the extraction of feature maps, comprising low-level features such as luminance, orientation, or color, based on the input image, 2.) the ensuing computation of activation maps, and finally 3.) their normalization and the aggregation of all the activation maps obtained from all the different feature maps.

The first and last steps have been retained in this work as in the original.
In the second step, the idea is to weight a pixel more saliently the more it differs from its surroundings.
A graph is constructed that connects each pixel to all other pixels in the feature map generated in the first step.
In other words, the fully-connected graph represents the image, specifically the feature map, whereby the nodes can be interpreted as the pixels.
The weight assigned to an edge of the graph is the product of the spatial and visual differences of the two pixels connected by the respective edge.
Finally, the desired activation map can be treated as a state vector within a Markov chain operating on this graph.
More precisely, the activation map is derived from the state of equilibrium (or stationary distribution), with the edge weights defining the Markov transition matrix.
The stationary distribution is given as a solution to an eigenvector problem, that is, as the eigenvector of the transition matrix corresponding to the eigenvalue of $\bm{1}$ (identity).
In practice, this eigenvector is typically determined by iteratively multiplying the Markov transition matrix by a probability vector with an initially uniform distribution.
However, rather than choosing the initial activation map to be uniformly distributed, it is initialized by means of the observed gaze points, namely the measured visual activation ascertained from the gaze signal.
This tailored initialization allows for the integration of eye tracking data in the saliency calculation process.

\subsubsection{Main Findings}
Due to the challenging nature of conducting a quantitative evaluation within this particular context, the evaluation primarily relied on experimental demonstration.
Three different types of stimulus were analyzed: A painting, a short video, and a text-rich website.
The salience maps obtained using the aforementioned method were visualized by overlaying them in the form of a heatmap onto the respective stimulus.
In this way, saliency enhanced gaze heatmaps were examined alongside the standard gaze heatmaps and the fixation maps, which visually depict the areas where the observer's eyes were fixated or focused.

In the case of the website stimulus, it was immediately apparent that the approach described takes direct account of the displayed text as opposed to the pure gaze heatmap.
The gaze signal was guided to the individual lines and letters, making the entire heatmap appear sharper and more meaningful.
As a result, it was easier to judge whether certain regions attract the intended level of visual attention and to assess the ease of visual accessibility for observers in perceiving the presented information.
This finding is particularly relevant in sectors such as web design and advertising.

A more challenging stimulus was the painting, where the foreground contrasted less distinctly from the background.
Also in this context, the strength of the proposed method emerged.
The saliency-aware gaze heatmap exhibited a better balance between areas that were more difficult to perceive due to their complexity and thus were observed for longer periods of time, and areas that were less complex and hence observed only briefly.
In contrast, the conventional gaze heatmap did not consider the region's accessibility to the observer and misleadingly suggested higher levels of interest in certain areas than they truly deserved, simultaneously undermining the prominence of other genuinely relevant regions.

In summary, the findings revealed that the saliency-aware gaze heatmaps effectively guide the eye-tracking signal towards salient regions, producing a more accurate attention pattern.

\section{Perceiving and Multiperspective Teaching of Unknown Objects} \label{sec:results_B}
This section elaborates on the intermediate steps that were necessary to achieve the main objective stated in \secref{sec:objectives}.
All the following papers worked towards this goal, and their respective motivations must therefore always be considered in this overall context.
The first two subsections, \subsecref{subsec:IROS20} and \subsecref{subsec:HRI22}, focus on the perception and localization of unknown \acp{OOI}.
The latter two subsections, \subsecref{subsec:HRI23} and \subsecref{subsec:IROS23}, deal with the learning process in which the human teaches the robot unknown objects within a \ac{HRI} scenario.

\subsection{Distilling Location Proposals of Unknown Objects through Gaze Information for Human-Robot Interaction} \label{subsec:IROS20}
\begin{quote}
\textbf{Daniel Weber}, Thiago Santini, Andreas Zell, and Enkelejda Kasneci. Distilling Location Proposals of Unknown Objects through Gaze Information for Human-Robot Interaction. In \textit{2020 IEEE/RSJ International Conference on Intelligent Robots and Systems (IROS '20)}.
\end{quote}

\subsubsection{Motivation}
This work marked the beginning of the \ac{HRI} teaching pipeline, in which now a real robot was deployed.
However, instead of using \ac{AR} glasses, as in the later course of the research, the gaze was still estimated and recorded with a regular head-mounted eye tracker.

Prior to learning unknown objects later in the project, the robot must first understand which (unknown) object the current \ac{OOI} is.
That means that the localization problem must be solved beforehand.
This time, unlike in \secref{sec:results_A}, not directly on the basis of the gaze data and the corresponding stimulus from the \ac{FOV} of the human, but rather from the robot's perspective.
In order to ensure further processing later on, it was primarily a matter of narrowing down the \ac{ROI} by means of a bounding box around the \ac{OOI}.
The straightforward utilization of modern neural network-based object detectors was thereby precluded due to their reliance on pre-existing training data, which cannot be assumed in general and excludes the detection of objects not contained therein.
Moreover, it would contradict the fact that the \ac{OOI} are truly unknown.

\subsubsection{Principal Methodology}
Consequently, the approach presented in this work was algorithm-based.
It can be divided into three building blocks.

The first block aims at mapping the human's gaze into the robot's frame of reference.
This can be achieved either by directly locating the robot in the camera frame of the human's eye tracker or vice versa, or alternatively by indirect co-location.
For the latter, at least four common points must be known in the respective \acp{FOV} of the robot and the human.
If this is the case, the respective position of the human and the robot can be determined by means of trilateration and accordingly also the mutual position.
To ensure the presence of common anchor points, fiducial markers were used, as their detection is robust and efficient.
In practice, the transformation from the human reference frame to the one of the robot is achieved by finding a homography that performs a perspective transformation between the image plane of the eye tracker and the image plane of the robot's body camera.
The homography can then be used to translate the human gaze points to the corresponding coordinates within the robot's camera image.

The second block requires the robot to predict candidate bounding boxes that are likely to contain an object.
This task was facilitated by utilizing location proposal methods, which were applied to the image captured by the robot's scene camera.
Such methods are commonly employed to effectively reduce the search space, thereby accelerating the detection process and diminishing the computational costs involved.
In the present case, selective search \cite{uijlings2013selective} was resorted to due to the fact that it is class-independent.
This property makes it suitable for unknown \ac{OOI}, just like in our case, where the class is not known in advance.
Typically, the output of such location proposal methods consists of thousands of bounding box candidates, which is why the cardinality of the output set was reduced in the third block.

In the third and final building block, the robot's set of proposed candidate locations was distilled using gaze information from the human partner, previously mapped from the human's frame of reference to that of the robot.
The intention was to significantly reduce the number of candidates while simultaneously increasing their relevance.
As a filtering mechanism, the requirement that the human gaze must fall within the bounding box associated with the respective \ac{OOI} was leveraged.
This ensured that the resulting subset contained only bounding boxes that had an intersection with the object tagged by the gaze point.
The distillation was tailored in such a way that the hierarchical order of the proposed bounding boxes was preserved.
The order of the proposals hints at the likelihood of them containing an object.

\subsubsection{Main Findings}
A qualitative analysis showed that eye tracking, marker recognition, and gaze mapping operated in real time.
Therefore, the proposed method proved to be suitable for real-time \ac{HRI}.
One prerequisite, however, was that a sufficient number of fiducial markers were visible to reliably estimate the mutual position of human and robot and to map the gaze as accurately as possible.
As long as this premise was fulfilled, the human was not constrained in his movements and was able to move freely.

In the scope of a quantitative analysis, the position indices of the bounding boxes were evaluated object-wise before and after distillation.
The position index denoted the position of a bounding box in terms of the hierarchical order in which it appeared in the set of location proposals.
The smaller the position index of a bounding box pertaining to the respective \ac{OOI}, the faster it can be found by iteration and the fewer communication with the robot is necessary to select it.
However, a low position index is not the sole significant factor.
The quality of the bounding box, that is, its accuracy, is equally crucial.
Ideally, a bounding box perfectly enclosing the \ac{OOI} appears in the first position set of proposals.
Regarding the accuracy of the boxes, the \ac{IoU}, also sometimes called the Jaccard index, was assessed as a performance metric.
This metric quantified the extent to which the bounding boxes align with the ground truth.
The output of the state-of-the-art object detector FCOS \cite{tian2019fcos}, which was pre-trained on the MS~COCO dataset \cite{lin2014microsoft}, served as a ground truth.
In general, an object is considered as correctly detected if the \ac{IoU} of the corresponding bounding box exceeds a threshold value of $0.5$ \cite{everingham2010pascal, zitnick2014edge}. 
Once the \ac{IoU} threshold reaches $0.7$, the detection can be deemed reasonably good.

The experiments have revealed that the overwhelming majority of boxes within the full set of location proposals possess an \ac{IoU} value of less than $0.1$, rendering them irrelevant.
Instead of having to search through over $2000$ proposals, the distillation method significantly reduced this number to an average of about $126$.
At the same time, the position index of the best existing box was improved from $1315$ to $61.5$.
This trend was also evident with regard to the precision, that is, the fraction of the relevant boxes (boxes with an \ac{IoU} of at least $0.7$) among the boxes distilled by the method.
The distillation process increased the precision from less than \SI{2}{\%} to almost \SI{40}{\%}.
Among the first three bounding boxes after distillation, at least one box consistently exhibited an \ac{IoU} value of $0.7$ or higher, with an average \ac{IoU} of over $0.81$, which equates an accuracy of almost \SI{90}{\%} compared to the best box in the full set of proposals.
Considering the first 15 boxes, the accuracy was even further enhanced to almost \SI{98}{\%}.

Overall, the conducted proof of concept demonstrated functionality and validity in the sense that the distillation process increased the precision by a factor of approximately $21$ and was able to locate objects comparably well as the neural network-based object detector FCOS, although pre-training was completely dispensed with.
Moreover, the proposed method had the capability to detect objects that FCOS was not specifically trained on and were therefore undetectable within the FCOS framework by nature.

\subsection{Exploiting Augmented Reality for Extrinsic Robot Calibration and Eye-based Human-Robot Collaboration}\label{subsec:HRI22}
\begin{quote}
\textbf{Daniel Weber}, Enkelejda Kasneci, and Andreas Zell. Exploiting Augmented Reality for Extrinsic Robot Calibration and Eye-based Human-Robot Collaboration. In \textit{Proceedings of the 2022 ACM/IEEE International Conference on Human-Robot Interaction (HRI '22)}.
\end{quote}

\subsubsection{Motivation}
In order to make the interaction between human and robot as pleasant as possible, it is crucial to establish a communication way between the two parties that is feasible and natural.
For this reason, \ac{AR} in the form of the HoloLens~2, a pair of \ac{AR} glasses manufactured by Microsoft, was integrated into this work.
In the previous work, outlined in \subsecref{subsec:IROS20}, successful transmission of gaze data from the human to the robot was accomplished.
However, it necessitated an overlapping \ac{FOV} with fiducial markers within it, and even more pertinent, the communication was limited to a unidirectional exchange.
Both of these problems ought to be solved in this work by means of \ac{AR}.
The latter is especially important because another aspect needed to be addressed.
Whereas up to this point the localization of objects was restricted to the 2D images of the robot camera, the robot and the human actually operate in three-dimensional space.
Therefore, in this work, the 3D positions of the unknown \acp{OOI} should be determined.
The \acs{AR}-based two-way communication was then intended to enable the robot to provide the human with feedback regarding the detected \acp{OOI}.

\vspace*{\baselineskip}
\subsubsection{Principal Methodology}
The HoloLens~2 constitutes the junction between the digital world of the robot and the analog world of the human.
All interactions take place wirelessly via an implemented \ac{AR} interface.
The interface offers gesture and speech navigation capabilities, enabling users to issue control commands to the robot or access its camera stream, among other functionalities.
Conversely, the robot can superimpose detected objects directly in the human's \ac{FOV}.

In order for the two interaction participants to be aware of each other's position, a calibration must first be carried out.
Simultaneously, this serves to determine the transformation between the robot and its body camera.
During this procedure, virtual counterparts of the robot and its camera are positioned according to the physical instances by means of QR~codes.
The QR~codes, however, are only needed during the calibration and do not impose any burden during runtime.
After calibration, the HoloLens~2 acts as a bridge between human and robot, ensuring that the robot maintains awareness of the human's position at all times.
In this way, the \ac{AR} interface facilitates real-time provision of human gaze information to the robot, unaffected by the movements of either the human or the robot, and without imposing any restrictions on their \ac{FOV}.

The localization strategy of the \acp{OOI} in three-dimensional space is based on a sequence of well-known computer vision techniques enhanced by gaze data.
The starting point forms the point cloud that originates from the robot's body camera.
Due to the preceding calibration, the position of the camera and thus the point cloud is known to both the human and the robot.
The point cloud first undergoes a pass through filter, followed by a voxel grid filter, aimed at diminishing complexity by reducing the number of points within the cloud.
These two processing stages increase the computing time significantly.
Due to the extrinsic camera calibration, the orientation of the table on which the objects are placed is known, enabling the identification and extraction of the corresponding plane using \ac{RANSAC} \cite{fischler1981random}.
Finally, by conducting Euclidean clustering on the remaining points of the point cloud and incorporating the gaze information to identify the points belonging to the \ac{OOI}, a segmented representation of the \ac{OOI} is obtained.

\subsubsection{Main Findings}
The calibration method described above demonstrated high practicality in the conducted trails, especially due to its efficiency and minimal time expenditure.
Depending on the experience of the user, a single calibration cycle typically took only between 15 and 40~seconds.
Hence, the method highlighted its suitability for spontaneous recalibrations, enabling system modifications, such as adjustments to the camera's tilt, during runtime.
Quantitatively, it proved to be reliable and accurate, deviating from the reference method by only a few millimeters on average. 
However, due to the absence of a real ground truth, it could not be conclusively decided which method was more accurate, as the deviations from each other were within the margin of error.
In terms of speed, the presented \ac{AR}-based calibration was clearly superior.

\begin{figure}[htb]
	\centering
	\includegraphics[width=\linewidth]{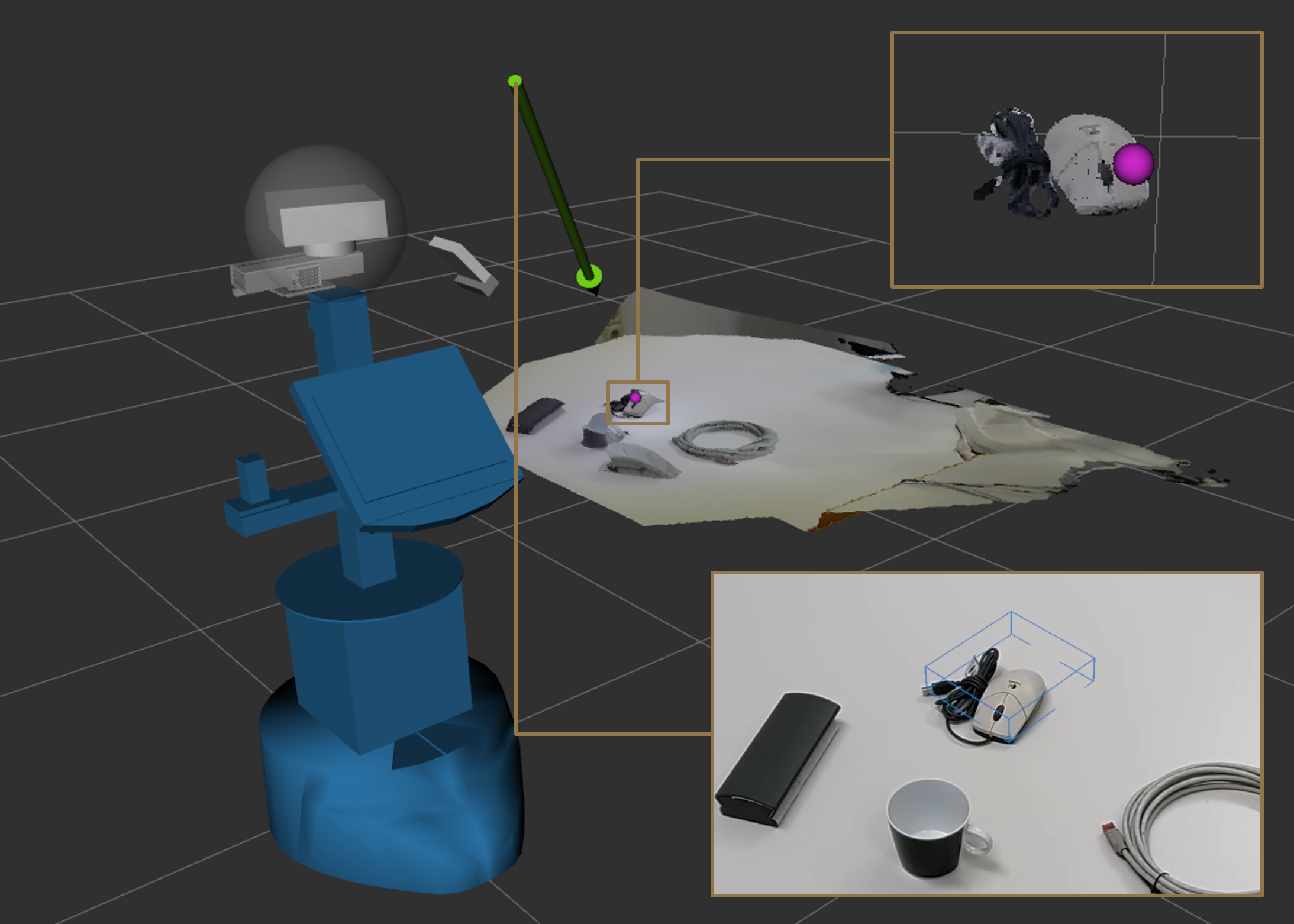}
	\caption{Visualization of the robot and the human observing a scene. The human’s gaze vector is represented by the green arrow. The intersection of the gaze ray with the environment is depicted as a purple sphere. The top right shows the object segmented by the robot and the bottom right shows the feedback (blue bounding box) provided to the human, displayed in the human’s \ac{FOV}.
    }
	\label{fig:HRI22}%
\end{figure}

The permanent bidirectional communication channel introduced by the AR interface enabled continuous real-time segmentation of the respective \ac{OOI} observed by the human.
A 3D bounding box derived from the segmented point cloud can finally be overlaid in the human's \ac{FOV} to indicate the specific object that the robot believes the human is focusing on.
A comprehensive visualization is shown in \figref{fig:HRI22}.

The quality of the segmented objects was assessed in two different ways: In 2D and in 3D.
The neural network-based 3D object detectors VoteNet \cite{qi2019deep} and Frustum ConvNet \cite{wang2019frustum}, which were originally intended as baselines, only achieved a \ac{mAP} of \SI{27.8}{\%} and less than \SI{1}{\%} respectively for the test objects and were therefore discarded.
Note that in 3D, an object is typically already considered to be detected at an \ac{IoU} threshold of $0.25$ \cite{qi2018frustum, song2015sun}.
Remarkably, even with a threshold twice as large, the method proposed in this work achieved a flawless recall rate of \SI{100}{\%}.
Moreover, the mean 3D \ac{IoU} of all test objects reached a value of almost $0.7$, further emphasizing the quality of the results.

In addition to the evaluation in 3D, a 2D assessment was carried out to mitigate potential susceptibility to bias due to self-labeled 3D ground truth.
The 3D bounding boxes were projected onto the 2D image plane of the robot's camera and then compared to the output of FCOS, which served as the ground truth.
The achieved mean \ac{IoU} of $0.81$ considerably surpassed the 2D detection threshold of $0.5$.

In summary, all test objects were successfully detected, and the system exhibited intuitive access to natural communication and \ac{HRI}.

\subsection{Multiperspective Teaching of Unknown Objects via Shared-gaze-based Multimodal Human-Robot Interaction} \label{subsec:HRI23}
\begin{quote}
\textbf{Daniel Weber}, Wolfgang Fuhl, Enkelejda Kasneci, and Andreas Zell. Multiperspective Teaching of Unknown Objects via Shared-gaze-based Multimodal Human-Robot Interaction. In \textit{Proceedings of the 2023 ACM/IEEE International Conference on Human-Robot Interaction (HRI '23)}.
\end{quote}

\subsubsection{Motivation}
After the previous work in \subsecref{subsec:HRI22} had enabled the robot to segment unknown \ac{OOI} by means of human gaze, this work aimed to further enhance the robot's capabilities by addressing the key aspect of learning.
In order to enable the robot to detect the objects independently and without help, the human should teach the robot within the context of a \ac{HRC} scenario.
The human should act in the pivot role of the teacher and, after visually indicating the object to the robot by gaze pointing and then verbally providing the corresponding class name, impart the knowledge to facilitate the robot's understanding of the \ac{OOI}.
In pursuit of this, the robot was supposed to capture various images of the \ac{OOI} from multiple different angles. 
These images were then to be automatically labeled and utilized for training purposes.
Eventually, the robot's proficiency in detecting the recently learned objects was tested.

\subsubsection{Principal Methodology}
The segmentation principles were borrowed from the work presented in the previous \subsecref{subsec:HRI22}.
The AR interface introduced therein was further extended to support multimodal \ac{HRI} and the transmission of class information.
To this end, once the human and robot have directed their attention towards the same object, the bounding box of the \ac{OOI} is displayed through the HoloLens~2.
The human user can then conveniently approve the bounding box using speech commands or gestures.
Subsequently, the class name can be specified and submitted to the robot via a virtual keyboard or again via speech recognition.

For robust training purposes, the robot necessitates an abundant amount of data.
Consequently, it autonomously generates a comprehensive dataset comprising images recorded from multiple perspectives.
For this purpose, the robot utilized a second camera attached to its arm to perform a circular movement around the \ac{OOI} and examines it from all sides.
Thereby, he proceeds according to \algoref{alg:HRI23}.
The camera mentioned therein always refers to the one attached to the robot arm and not to the body camera used for segmentation.
\begin{algorithm}[hbt]
   \caption{Multiperspective Recording} \label{alg:HRI23}
   \newcommand*{\PCdD}{\ensuremath{pc_{\text{2D}}}}
   \newcommand*{\PCtD}{\ensuremath{pc_{\text{3D}}}}
   \begin{algorithmic}[1]
        \Require Class $c$, point cloud $\PCtD$, and bounding box $box$ of segmented \ac{OOI}
        \State $p \gets \operatorname{calcCircularPath}(box)$
        \State $w \gets \operatorname{getReachableWaypoints}(p)$
        \State $t \gets \operatorname{calcTrajectory}(w)$
        \While{moving along trajectory $t$}
            \State $\PCtD ' \gets \operatorname{transform}(\PCtD)$     \Comment{Transform 3D point cloud to camera frame}
            \State $\PCdD \gets \operatorname{project3DToPixel}(\PCtD ')$ \Comment{Project onto image plane of camera}
            \State $roi \gets \operatorname{min/max}(\PCdD)$         \Comment{Calculate 2D bounding box}
            \State Store in folder $c$: RGB image, depth image, camera parameters, $roi$
        \EndWhile
    \end{algorithmic}
\end{algorithm}

The core idea resolves around the autonomous labeling of each image with the \ac{ROI} as the robot moves the camera around the object.
The \ac{ROI} is determined by transforming the segmented 3D point cloud into the respective current camera frame, which changes dynamically as the robot arm moves.
The 3D points are then projected onto the 2D image plane of the camera using the intrinsic camera parameters, and the 2D bounding box is derived from the boundary points.
Finally, the robot stores the \acp{ROI}, RGB and depth images as well as the intrinsic and extrinsic camera parameters.
The extrinsic parameters specify the camera's position in relation to the object. 
The entire progress of this multi-perspective acquisition of training data is visually reported to the human by means of the \ac{AR} interface.

In terms of the object detection architecture, the robot was equipped with state-of-the-art models such as Faster \mbox{R-CNN}~\cite{ren2015faster}, which were fed with the obtained RGB images.
Although, in principle, numerous datasets are available in the field of computer vision, it cannot be presupposed that they contain a particular \ac{OOI}.
Nevertheless, this offers an opportunity to build on.
Therefore, the training of the robot's object detector was rooted in the extension of existing knowledge ad hoc in the situation through teaching assuming a general awareness of objectness in the form of pretraining on irrelevant objects.
By applying a transfer learning approach, only the classification and regression heads were reinitialized and trained, while the feature layers remained frozen.
This training strategy allowed successful training even on non-high-end hardware, such as the robot, while mitigating the risk of overfitting.

\subsubsection{Main Findings}
As part of the evaluation, the robot was taught ten different classes, each with two objects, using the teaching pipeline described above. 
Per teaching run, that is, per object, the robot acquired a large amount of labelled multiperspective training data in a short period of time (about $1$ minute).
After training, the robot's gained knowledge was examined on a separate test set, which comprised two other objects from each of the ten classes, distinct from the objects used during training.

In total, several different object detectors were tested, including Faster \mbox{R-CNN} and FCOS \cite{tian2019fcos}, among others.
It showed that the robot was able to generalize from the training objects to unseen objects of the same respective classes.
A thorough evaluation using Faster R-CNN revealed that the robot successfully detected the majority of objects, achieving an impressive \mAP{50} of almost \SI{70}{\%}.
In contrast, the same object detector, when trained on the entire MS~COCO \cite{lin2014microsoft} dataset, achieved an \mAP{50} of over \SI{80}{\%} on the six classes that were also part of MS~COCO.
However, its performance on all ten test classes was only slightly around \SI{50}{\%}, which is considerably less compared to the results obtained through the multiperspective teaching pipeline.

The complete training data generated by the robot consists of more than $3100$ viewpoints and has a considerable density of information as it encompasses various essential components mentioned above.
Therefore, especially due to the inclusion of the camera poses, it becomes particularly appealing for other prominent research areas, such as Neural Radiance Fields~\cite{mildenhall2020nerf, yu2021pixelnerf, lin2021barf, deng2022depth}.
For this reason and in order to ensure reproducibility, this data, along with the validation and test set, were collected into the \ac{OMD} dataset, which was made publicly available to the research community\footnote{\url{https://cloud.cs.uni-tuebingen.de/index.php/s/2oRPs2o3FZkdBHW}}.
In addition, the complete code base of the \ac{HRI} system, featuring the \ac{AR}~interface, ROS nodes, and learning policy, was made publicly available as well\footnote{\url{https://github.com/dnlwbr/Multiperspective-Teaching}}.

All in all, the introduced novel teaching pipeline employing multimodal \ac{HRI} demonstrated its practical efficacy as an intuitive and natural method for teaching the robot new, yet unknown objects using few instances.
Furthermore, it enabled the robot to detect classes that lack a dedicated training dataset.

\subsection{Leveraging Saliency-Aware Gaze Heatmaps for Multiperspective Teaching of Unknown Objects} \label{subsec:IROS23}
\begin{quote}
\textbf{Daniel Weber}, Valentin Bolz, Andreas Zell, and Enkelejda Kasneci. Leveraging Saliency-Aware Gaze Heatmaps for Multiperspective Teaching of Unknown Objects. In \textit{2023 IEEE/RSJ International Conference on Intelligent Robots and Systems (IROS '23)}. (Accepted for publication).
\end{quote}

\subsubsection{Motivation}
Although the robot was successfully taught through \ac{HRI} in the previous work from \subsecref{subsec:HRI23}, this approach exhibited a drawback.
The outcome heavily relied on the quality of the segmentation discussed in \subsecref{subsec:HRI22}, as it was instrumental in determining the \acp{ROI}.
Furthermore, even minor inaccuracies in the transformation calibration between the robot, arm, and wrist camera can propagate throughout the system and accumulate to discrepancies in the result.
A more direct and robust approach with fewer system components was therefore intended to reduce the complexity and thus the susceptibility to errors of the entire system.
Rather than relying solely on the segmentation by \ac{HRC}, a more straightforward approach involved the human observer devoting a slightly longer duration of time focusing the \ac{OOI} and then to consider a series of the resulting gaze points.

\subsubsection{Principal Methodology}
Albeit, building upon the preceding achievements, the underlying paradigm that the \acp{OOI} need to be fully identified prior to the robot's data acquisition was abandoned.

The human looks at the object and initiates the teaching procedure by means of speech recognition.
For a duration of $10$~seconds, the gaze data is logged while the \ac{AR} interface provides an audible countdown.
Based on the gathered gaze points, the robot approximates the position and size of the \ac{OOI} and records it from different angles as before.
Instead of the segmented point cloud being transformed into the camera frame and then mapped onto the 2D image plane, it is the gaze points that undergo this procedure.
Theoretically, the bounding box labels could be determined directly from the edges of the transformed and projected gaze points.
However, as mentioned above, even slight imprecision in the hardware calibration or in the tracking of the human's position or gaze can cause an offset in the ensuing \ac{ROI}.
To compensate for these inaccuracies, the gaze points were refined and guided towards the \ac{OOI} using saliency.
This step reflected the findings of \subsecref{subsec:geisler2020exploiting} in the form of the \ac{GA-GBVS}, which is additionally extended to the \ac{DGA-GBVS}.
For each perspective captured by the robot, feature maps are extracted from the corresponding RGB image.
These feature maps are then used along with the gaze data to generate activation maps, which are eventually normalized and merged into a saliency-aware gaze heatmap.

By means of Otsu's method\cite{otsu1979threshold}, a threshold value is set to delimit relevant points and to binarize the heatmap.
This also sharpens the edge of the heatmap points, which then define the \ac{ROI}.
Note that relying solely on saliency maps, without considering gaze points, is insufficient for reliably determining the \ac{ROI}.
This is because there may be other salient areas within the stimulus that are not part of the intended \ac{OOI}.

Finally, the robot can proceed analogously to \subsecref{subsec:HRI23}, leveraging the acquired and labeled perspectives to learn the taught objects.

\subsubsection{Main Findings}
In order to compare the approach with the results published in \cite{weber2023multiperspective} and discussed in \subsecref{subsec:HRI23}, the evaluation was carried out accordingly.
Exactly the same ten objects were taught to the robot via the new \ac{HRI} pipeline, whose detection capability was then evaluated on the \ac{OMD} dataset.
Faster \mbox{R-CNN} served as the backbone.

In general, the described method consistently outperformed the previous approach in almost all classes.
This basically applied to all common object detector metrics.
Regarding the \acl{mAP}, the previous values for \mAP{50}, \mAP{75}, and \mAP{} were improved from \SI{66.9}{\%}, \SI{31.4}{\%}, and \SI{33.6}{\%} to \SI{73.6}{\%}, \SI{38.1}{\%}, and \SI{39.5}{\%}, respectively.
A model trained on the full MS~COCO dataset, as in \subsecref{subsec:HRI23}, was again able to perform quite well on the known classes, but overall, including the unknown classes, only achieved a \mAP{50}, \mAP{75}, and \mAP{} of \SI{50.7}{\%}, \SI{33.3}{\%}, and \SI{30.5}{\%}, respectively.
Furthermore, this model was outperformed, even for some of the known classes, by the method presented in this work. 

A similar pattern emerged with regard to the \acl{mAR}.
The \mAR{1}, \mAR{10}, and \mAR{100} were increased from \SI{43.7}{\%}, \SI{50.1}{\%}, and \SI{50.4}{\%} to \SI{50.1}{\%}, \SI{54.4}{\%}, and \SI{54.4}{\%}, respectively, in comparison to the scheme presented in \subsecref{subsec:HRI23} and thereby also surpassed the model trained on the entire MS~COCO set.
The latter only achieved a recall of \SI{35.2}{\%} in each of the three metrics.
Likewise, this model was exceeded even for some of the known classes by the new proposed teaching pipeline.

Additionally, the curves of precision and recall as functions of \ac{IoU}, generally revealed superior values compared to the previous approach from \subsecref{subsec:HRI23}, especially at high \ac{IoU} thresholds.
This improvement indicated more accurate bounding boxes, enhancing the learning and detection capabilities of the entire \ac{HRI} system.

\cleardoublepage

\chapter{Discussion \& Outlook}
\label{ch:discussion}

This chapter discusses the publications presented in the previous \chref{ch:motivation_results} and aligns them with the objectives of the thesis outlined in \secref{sec:objectives}.
The central aspect is the development of a \ac{HRI} teaching pipeline, as elaborated therein.
The findings regarding the potential of gaze in determining \acp{ROI}, based on the publications \cite{geisler2020exploiting} and \cite{weber2022gaze} from the list in \chref{ch:publist}, are discussed in \secref{sec:discussion_A}.
\secref{sec:discussion_B} delves into the multiperspective teaching approach for unknown objects by means of \ac{HRI}, utilizing \ac{AR} and eye-tracking techniques.
Here the insights of \cite{weber2020distilling}, \cite{weber2022exploiting}, \cite{weber2023multiperspective}, and \cite{weber2023leveraging} are examined.
Finally, a conclusion is drawn, providing a concise summary of the findings, which is followed by an outlook of possible future steps and directions to consider.

\section{The Potential of Gaze in Determining Regions of Interest} \label{sec:discussion_A}
In the general setting, without the robot and \ac{AR}, the focus lies not on the teaching per se, but on the preceding identification of the \acp{OOI}.
The results from publication~\cite{weber2022gaze} have demonstrated that the gaze by itself can already yield information about the position of objects and their approximate size up to a certain degree.
The image of the scene does not necessarily have to be included, but due to the omission of image information, time windows with a fixed size were used.
Especially the classification of the time windows, that is, whether an object was within focus or not, was convincing with \SI{92}{\%} for the best combination of parameters.
Conversely, it was also found that without using the scenery, the estimation of the size of the bounding box belonging to the \acp{OOI} was quite challenging and produced mixed results.
However, this is not particularly surprising, as humans do not observe objects completely in every detail in order to perceive them appropriately.
Furthermore, since the subjects were allowed to move freely, different gaze points from subsequent timestamps could belong to the same point in the environment and vice versa.
This further complicated the bounding box regression.

A clearer picture emerges with regard to the estimation time and the required resources.
The introduced heatmap feature, resulted in a tremendous speed boost in comparison to standard object detectors such as FCOS \cite{tian2019fcos} or RetinaNet \cite{lin2017focal}, while requiring only a fraction of the resources.
This discrepancy can be attributed to the notable difference in size between heatmaps and images when employed as input features for machine learning models.
This disparity can be attributed to the significant difference in sizes between heatmaps and images when used as input features for the \ac{ML} models.
Depending on the number of grid cells into which the \ac{FOV} was divided to build the heatmap input features, even the three-dimensional variant had a relatively small size, containing only $50^3 = 125\,000$ values.
They are, therefore, easier to process compared to images, which, at the resolution of $1088 \times 1080 \times 3$, had to handle a much larger number of $3\,525\,120$ values.
The low hardware requirements are particularly attractive for the operation on mobile robots, as their hardware is limited and not arbitrarily expandable.
Even though this is a really powerful advantage, in practice one would rather seek to make use of any information available and look for a balanced combination to also enhance the regression of the bounding box parameters.
To exploit the full potential of the robot, visual information or even depth data could therefore be incorporated in addition to the heatmap features.
Such a combination remains the subject of future research.

The experiments in \cite{geisler2020exploiting} revealed that gaze signal can be effectively combined with saliency maps.
These maps identify and highlight salient areas and thus process visual information.
As described in \subsecref{subsec:geisler2020exploiting}, the joint fixation-saliency maps were superior to the standard fixation heatmaps.
The eye-tracking signals could be combined with the \ac{GBVS} algorithm and inaccuracies in the gaze data could thus be corrected.
This observation was true both for simple stimuli like a website, where the text prominently contrasts with the background, and for more intricate stimuli such as paintings.
The downside of this approach, which is also its main limitation, is the runtime and memory consumption.
As the size of the input image increases, the size of the Markov transition matrix grows quadratically.
The initialization of this matrix has a complexity of $\mathcal{O}\left(n^2\right)$.
The combined effect of these two factors restricts the \ac{GBVS} to extremely low input resolutions, resulting in a reduced level of acuity.
Consequently, the original implementation of the \ac{GBVS} algorithm sets the internal resolution to a maximum edge length of $32$ pixels \cite{harel2006saliency}.
Nevertheless, by sparsifying the matrix through the omission of very small values, the complexity can be reduced to $\mathcal{O}(n)$.
Moreover, in the eventual application within the \ac{HRI} teaching scenario, the blurring effect is less relevant compared to the original work.
This is because the goal is not to generate a comprehensive saliency map, but rather to refine the gaze map in light of the stimulus.

With the outcome of these general experiments, a step was made towards the objectives outlined in \secref{sec:objectives}.
The observations, which highlight the benefit of gaze data in determining \acp{ROI} and that saliency can mitigate inaccuracies in gaze signals, hold particular relevance to the challenges \ref{item:p2} and \ref{item:p6}.

\section{Multiperspective Teaching of Unknown Objects via Human-Robot Interaction} \label{sec:discussion_B}
While the previous studies were detached from \ac{AR} and \ac{HRI}, in \cite{weber2020distilling} the robot came into play.
The main takeaway was that the presented approach successfully enabled the robot to detect unknown objects, that the human was looking at, within the 2D image of its body camera.
Remarkably, this method performed comparably well to the state-of-the-art object detector FCOS \cite{tian2019fcos} without requiring any pretraining, meaning that the deployment was not tied to predefined objects.
By leveraging gaze information, the precision of the selective search algorithm witnessed a substantial increase by a factor of over 20.
It should be noted, that although the devised method yielded a distilled and this more relevant output compared to the original, it was not univocal, as it still consisted of a set of multiple proposals.
However, this characteristic is not actually a disadvantage, but rather presents opportunities as it allows the human to select the best and most appropriate bounding box in cases where the first one was not suitable.
This is especially helpful when the \ac{OOI} is difficult to distinguish from the background or when it comprises varying colors that do not clearly suggest to the robot whether it represents a single object or multiple distinct objects.
Furthermore, the possibility of making decisions through \ac{HRI} closely resembles the interaction and learning process between humans, rendering it a natural approach.
Eventually, the challenges \ref{item:p1} (gaze mapping) and \ref{item:p2} (unknown object localization) could be solved within the realm of an ordinary tracker without any \ac{AR} functionality.
Nevertheless, the former relied on the detection of fiducial markers in order to map the gaze from the human's frame of reference to that of the robot.
Due to the fact that the mapping process required a sufficient number of adequately sized markers to be present in both the robot's and the human's field of vision, the human's mobility was restricted, and the system became more cumbersome and error-prone.
The latter was in turn not suitable for unknown object localization in three-dimensional space.

Both mentioned problems were further improved and fully resolved by the publication \cite{weber2022exploiting} from the list in \chref{ch:publist}.
By using the HoloLens~2, the fiducial markers became obsolete.
The position and orientation of the device and thus of the human could be tracked in real time without restricting the movements of the human or the robot in any way.
Since the robot and the human were thus constantly aware of each other's position, the gaze vector and gaze point could be directly transformed from one coordinate system to the other.
This solved \ref{item:p1} entirely.
The necessary information was exchanged through the dedicated communication channel provided by the specifically implemented \ac{AR} interface.
Furthermore, this interface was responsible for the registration of the human gaze data.
Consequently, \ref{item:p3} could be marked as completed.
Additionally, the involvement of speech and gesture control, which was absent in the preceding work \cite{weber2020distilling}, allowed for a more natural and intuitive interaction.
This enhancement significantly improved the user-friendliness.

Another milestone was that the unknown object localization was lifted from 2D to 3D.
This was indispensable for the later teaching process.
The possibility of the human using gaze to guide the robot's attention towards the \ac{OOI}, which the robot then segments in real time, also laid the foundation for the ensuing research.
Gaze pointing offers an intuitive and less ambiguous alternative to pointing with a finger and, unlike speech, it can be utilized prior to the robot knowing the object.
With this, \ref{item:p2} could ultimately be deemed resolved.
Moreover, due to the synergy of the \ac{AR} and segmentation components, a virtual three-dimensional bounding box around the object segmented by the robot can be visualized within the field of view of the HoloLens~2.
Through this feedback mechanism, the robot can provide direct indications to the human regarding its estimation of the human's attentional focus. 
Consequently, the person can immediately identify whether the robot's assessment aligns with his or her own perception.
Hence, \ref{item:p4} was also solved within the scope of publication \cite{weber2022exploiting}.

Even though the presented algorithm-based approach clearly outperformed the state-of-the-art neural network-based 3D object detectors, it has a limitation to be considered.
The segmentation is not able to distinguish objects that are close to each other.
As the semantic understanding of objects for the localization task is not yet evolved, situations may arise where objects positioned in close proximity to each other are interpreted as a single large object.
However, this could be addressed by means of additional \ac{HRI} in which the human advises the robot of the approximate size of the \ac{OOI}.

For the follow-up work, \ref{item:p5} and \ref{item:p6} remained to be eliminated.
These were addressed in \cite{weber2023multiperspective} from the list in \chref{ch:publist}, focusing on the teaching aspect of the pipeline.
With regard to \ref{item:p5}, the previously implemented \ac{AR} interface was extended to enable the human to convey the class information to the robot, thereby enhancing the collaboration between the human and the robot.
The virtual bounding box derived from the robot's segmentation feedback serves two purposes.
Firstly, it verifies the robot's attention, and secondly, once the shared attention of the human and the robot is ensured, the human can select the \ac{OOI} encompassed by the virtual bounding box.
This interactive selection mechanism appears intuitive and natural, as it can be conducted either by gestures or by speech.
The same modalities can then be used to determine the class of the \ac{OOI}.
For this purpose, the \ac{AR} interface implements a virtual keyboard, solving \ref{item:p5}.

Last, \ref{item:p6} is addressed by means of the robotic arm technique and the transfer learning approach described in \subsecref{subsec:HRI23}.
The autonomous and efficient way for the robot to examine the \ac{OOI} in more detail and acquire labelled training data completed the teaching pipeline.
As a result, the robot was able to successfully redetect the objects it had been taught through \ac{HRI}, fulfilling the overarching objective outlined in \ref{sec:objectives}.
In particular, it must be emphasized that the objects used in the test phase were distinct from the objects, that the robot was taught with.
The robot was thus able to generalize to unseen object entities of the previously learned classes.
This reveals the capabilities of the robot's object detection system, representing a big step towards the deployment of robots in unfamiliar and non-predefined environments.

Naturally, the results achieved through \ac{HRI} teaching could hardly match those attained by training on extensive datasets containing thousands or even millions of images.
At least in the assessment of the individual classes.
Assuming the initial issue of not having adequate datasets for all classes, the strengths of teaching through \ac{HRI} became apparent.
Considering all classes, rather than only those contained within the corresponding datasets, the evaluations revealed that the teaching approach outperformed the baseline with an \ac{mAP} of \SI{33.6}{\%} compared to \SI{30.5}{\%}.
This highlights the teaching approach's enhanced flexibility and adaptability.
Consequently, in practice, one would neither want to forego prior knowledge, if it is available, nor the flexibility of \ac{HRI} teaching.
Hence, a combination of pretraining and \ac{HRI} teaching emerges as the best strategy as prior knowledge can be used, but still be expanded when unknown objects occur.

Either way, the performance of supervised machine learning methods relies on the quality of the input data during training.
In the current form of the \ac{HRI} teaching pipeline, the output of the object segmentation from publication \cite{weber2022exploiting} described in \subsecref{subsec:HRI22} is thus a carrying factor.
When exposed to challenging sensor data, such as with extremely dark or highly reflective object surfaces, the depth determination might be inaccurate, leading to insufficient segmentation of the \ac{OOI}.
This in turn can affect the labels of the training data gathered by the robot and hinder the robot's learning progress.
While this issue does not fully impede the robot's ability to learn unknown objects, it can be alleviated by increasing the number of objects from the respective class during the training phase.

Alternatively, in \cite{weber2023leveraging}, an alternative approach was attempted to determine the labels on the image data collected by the robot.
Hence, challenge \ref{item:p2} was solved in a different way, not with upstream segmentation, but based on a series of gaze points collected over a period of time.
In combination with the saliency of the individual training images, the \acp{ROI} could be calculated and used the corresponding labels.
By encoding the three-dimensional gaze points as saliency-aware 2D gaze heatmap, the human gaze could be aligned with the \ac{OOI} and imprecision in the signal could be rectified.
Compared to the segmentation in publication \cite{weber2022exploiting}, an even greater amount of gaze information from the human has been incorporated.
As a result, the estimated \acp{ROI} are more accurate and the labels of higher quality.
This is also reflected in the robot's detection performance.
In fact, the results have improved compared to the preceding \ac{HRI} variant in \cite{weber2023multiperspective}, for example, the \ac{mAP} has further increased from \SI{33.6}{\%} to \SI{39.5}{\%}.
Even in terms of the class-specific analysis, it was possible to keep up with the model that was trained on the entire MS~COCO dataset.
Furthermore, one of the limitations of the segmentation in \cite{weber2022exploiting}, wherein objects in close proximity were erroneously interpreted as a single object, is remedied.
Although the advantages outweigh the disadvantages, there is also a small drawback.
Specifically, after the robot has examined the \ac{OOI}, an intermediate processing step is necessary, because due to the runtime of the \ac{GA-GBVS} and \ac{DGA-GBVS} techniques already mentioned earlier, the labels cannot be produced during the recording phase itself.
In practice, however, this is to a certain extent negligible, as the calculations can take place in parallel with the \ac{HRI} teaching of other new objects.

\section{Conclusion and Outlook}
In conclusion, the presented interdisciplinary research broke new ground by fusing fields that were previously running predominantly in parallel.
More specifically, this dissertation marks a significant step towards teaching robots their unknown environment, particularly when the required training data is limited or unavailable.
Instead of relying entirely on data based pre-training, \ac{HRI} was consulted to foster the robot's object detection abilities.
By engaging in such collaborative settings, the robot could successfully be taught unknown objects within its environment by its human peer.
The robot became capable of independently detecting these objects without further external assistance, enhancing its adaptability to non-predefined scenarios.
Along this line, a variety of innovative steps were solved at the intersection of robotics, human eye tracking, \ac{AR} and \ac{ML}.
This includes, in addition to successfully teaching the robot, natural and human-like interaction by means of eyes and speech.
The novel \ac{AR}-based extrinsic calibration required for this purpose is characterized by its speed, ease of use, as well as competitive accuracy in comparison to classical approaches.
To summarize, it can be concluded that all challenges, initially outlined at the beginning of this dissertation, were successfully addressed, resulting in the attainment of the overarching goal of teaching unknown objects through \ac{HRI}.
This accomplishment not only validated the intended outcome, but also aligned with the broader desire of closing the yawning gap of data dependency in robot learning.

Despite this, there are still several opportunities for potential enhancements and future research.
For instance, as of now, no form of optimization has taken place with regard to the object detection backbone.
Naturally, the hyperparameters of the neural network models could be fine-tuned. 
However, this is also accompanied by an increased risk of overfitting to a specific setting or environment.
This, in turn, would negate the advantage of modularity inherent in the presented system, which in principle works with any object detector, and would limit the flexibility of the entire system.
It would therefore be more beneficial to develop a model that takes all robot sensors into account, rather than solely relying on the RGB images.
Especially, the depth data could contain valuable additional information that can be leveraged.
By incorporating the full range of sensor data, such a model would unlock enhanced capabilities and maximize the system's potential.

Another direction to consider is the further intensification of the interaction between the human and the robot.
The potential that \ac{HRI} offers has not yet been fully exhausted, and humans continue to hold significant value in providing further support.
Especially within situations where an object class has been detected incorrectly, the human can play an even more pivotal role in instructing and correcting the robot.
In this regard, new opportunities also arise due to the ascent of increasingly powerful \acp{LLM}.
Prominent exemplars such as \mbox{GPT-4} \cite{openai2023gpt4} and PaLM~2 \cite{anil2023palm}, used in ChatGPT and Google Bard, respectively, along with Meta's LLaMA \cite{touvron2023llama} hold the potential to advance communication, streamlining the conveyance of intricate scenarios and subject matters to the robot.
Large visual models or large multimodal models, like SAM \cite{kirillov2023segment}, may also, in the future, facilitate tasks such as identifying the \ac{ROI} of the unknown \acp{OOI} or even serve as a backbone for the teaching process itself.
For the latter, it is important to note that the comprehension of objects in these models is rooted in extensive amounts of textual and image data.
As a result, training these large-scale models, comprising hundreds of billions of parameters, necessitates supercomputers equipped with exceptionally high-performance hardware.
Therefore, training these models directly on the limited hardware of a robot is currently entirely precluded.
For the former, even with a robust generalization due to the substantial volume of training data, it must be ensured that totally unfamiliar objects, not previously included in the training data, are detectable.

Furthermore, the presented system still has to prove its operational capability in alternative contexts outside the office environment.
Thus far, the used objects have exhibited a considerable degree of diversity and the tests encompassed multifaceted conditions, while the \ac{HRI} teaching process has been conducted within the confines of controlled office scenarios, rather than real-world environments where the robot would be confronted with a plethora of disturbing and irrelevant objects within its \ac{FOV}.

Even though there is still further research required to refine the system's general applicability, it has already indicated some promising aptitude for certain applications.
For instance, it was possible to design a prototypical extension that enabled the robot to acoustically name the classes of objects that the human was looking at.
This might already be advantageous in domestic settings for handicapped or elderly people with speech deficits or limited mobility, in order to point out an object of their desire to other people in the vicinity.

In all, the emergence of the presented research findings in robotics, \ac{HRI}, \ac{AR}, \ac{ML}, and eye tracking will prospectively fuel the steady expansion of potential applications and highlight the growing demand for \ac{HRI} based teaching.

\addtocontents{toc}{\vspace{\normalbaselineskip}}
\cleardoublepage
\bookmarksetup{startatroot}

\appendix
\addtocontents{toc}{\vspace{\normalbaselineskip}}

\setcounter{footnote}{0}

\counterwithin{figure}{section}
\counterwithin{table}{section}
\counterwithin{equation}{section}

\newif\ifpaper
\papertrue

\chapter{Investigating the Potential of Gaze in Determining Regions of Interest}
\label{app:A}

The following publications are enclosed in this chapter:

\begin{enumerate}
	
	\item[\cite{weber2022gaze}]\label{app:itm:weber2022gaze} 
	\textbf{Daniel Weber}$^*$, Wolfgang Fuhl$^*$, Andreas Zell, and Enkelejda Kasneci. Gaze-based Object Detection in the Wild. In \textit{2022 Sixth IEEE International Conference on Robotic Computing (IRC)}, pages 62–66. IEEE, 2022. \href{https://doi.org/10.1109/IRC55401.2022.00017}{doi:10.1109/IRC55401.2022.00017}.
	
	\item[\cite{geisler2020exploiting}]\label{app:itm:geisler2020exploiting} 
	David Geisler, \textbf{Daniel Weber}, Nora Castner, and Enkelejda Kasneci. Exploiting the GBVS for Saliency aware Gaze Heatmaps. In \textit{ACM Symposium on Eye Tracking Research and Applications}, pages 1–5, 2020. \href{https://doi.org/10.1145/3379156.3391367}{doi:10.1145/3379156.3391367}.
	
\end{enumerate}

\blfootnote{
	\hspace{-16.5pt}{\scriptsize $^*$ equal contribution}\\
	{\scriptsize The publications templates have been slightly adapted to match the formatting of this dissertation.
	The ultimate versions are accessible via the digital object identifier at the respective publisher.
	Publication~\hyperref[app:itm:geisler2020exploiting]{\protect\NoHyper\cite{geisler2020exploiting}\protect\endNoHyper} is \textcopyright~2020~ACM.
	Publication~\hyperref[app:itm:weber2022gaze]{\protect\NoHyper\cite{weber2022gaze}\protect\endNoHyper} is \textcopyright~2022~IEEE and reprinted, with permission, from~\hyperref[app:itm:weber2022gaze]{\protect\NoHyper\cite{weber2022gaze}\protect\endNoHyper}.
	In reference to IEEE copyrighted material which is used with permission in this thesis, the IEEE does not endorse any of University of Tübingen’s products or services. Internal or personal use of this material is permitted. If interested in reprinting/republishing IEEE copyrighted material for advertising or promotional purposes or for creating new collective works for resale or redistribution, please go to \url{http://www.ieee.org/publications_standards/publications/rights/rights_link.html} to learn how to obtain a License from RightsLink. If applicable, University Microfilms and/or ProQuest Library, or the Archives of Canada may supply single copies of the dissertation.}
}

\clearpage
\setcounter{footnote}{0}
\section{Gaze-based Object Detection in the Wild}
\label{app:sec:paper03}
\blfootnote{\hspace{-14pt}Funded by the Deutsche Forschungsgemeinschaft (DFG, German Research Foundation) under Germany’s Excellence Strategy -- EXC number 2064/1 -- Project number 390727645.}

\ifpaper
\subsection{Abstract}
In human-robot collaboration, one challenging task is to teach a robot new yet unknown objects enabling it to interact with them.
Thereby, gaze can contain valuable information.
We investigate if it is possible to detect objects (object or no object) merely from gaze data and determine their bounding box parameters.
For this purpose, we explore different sizes of temporal windows, which serve as a basis for the computation of heatmaps, i.e., the spatial distribution of the gaze data.
Additionally, we analyze different grid sizes of these heatmaps, and demonstrate the functionality in a proof of concept using different machine learning techniques.
Our method is characterized by its speed and resource efficiency compared to conventional object detectors.
In order to generate the required data, we conducted a study with five subjects who could move freely and thus, turn towards arbitrary objects.
This way, we chose a scenario for our data collection that is as realistic as possible.
Since the subjects move while facing objects, the heatmaps also contain gaze data trajectories, complicating the detection and parameter regression.
We make our data set publicly available to the research community for download.
\subsection{Introduction}
Recent research has shown that eye tracking has becoming increasingly relevant for a variety of applications. These include even dynamic real-world scenarios, such as driving \cite{braunagel2016necessity}, 
medicine \cite{harezlak2018application}, 
and sports \cite{hosp2021soccer}. 
Especially the combination with computer vision problems \cite{shanmuga2015eye}, 
has in turn great potential for the employment of eye tracking in other fields, such as robotics \cite{weber2020distilling}. 
In the field of robotics, the focus is often on the interaction with the environment, for example, detecting and grasping objects \cite{mahler2019learning}. 
In such settings, however, the interaction entities are often unknown due to the enormous amount of potentially existing objects.
For this purpose, a semantic understanding of scenes must be present.
In conveying this understanding, humans can play an important role and provide assistance to the robot.
One modality that has proven to be particularly suitable and helpful for such human-robot collaboration (\acs{HRC}) settings is the human gaze \cite{weber2022exploiting}.
Gaze allows objects to be intuitively selected by the human and communicated (e.g., gaze pointing) to the interaction partner (e.g., robot). 
An additional advantage of the gaze modality is that it is far more unambiguous than gestures and, unlike speech, can also be used effortlessly in the case of unknown objects whose class name may not be known at all.

In this work, we address the problem of unknown object detection in real-world scenarios based on gaze.
This is an essential challenge for \acs{HRC}, as an example.
After all, if the robot could detect an unknown object by the fact that the human is looking at it, this paves the way for further interaction possibilities.
We refer to object detection in a similar manner to face detection.
In face detection, the task is to estimate whether there is a face or not.
In our task, the challenge is to find out whether the current gaze pattern belongs to a perceived object or not.
While there is work investigating unknown object detection on static imagery, there is little research addressing unknown object detection on videos and settings in the wild.
Along this line, 
\cite{xiao2018salient} used fixations to infer the saliency of objects.
A gaze map was used by \cite{shi2017gaze}, who combined it with candidate regions to segment objects.
In the work by \cite{luo2019interested}, gaze points were grouped into clusters to determine whether a cluster belonged to an object of interest and whether it was looked at intentionally or unintentionally.
However, all these related works used multiple gaze points on one image, which is only possible if the stimulus (image of the observed scene) is static or if, for instance, eye tracking data from multiple people is used, as in \cite{shanmuga2015eye}.
Contrary to all aforementioned related works, we present a method capable of using gaze data from a single person in dynamic scenes, i.e., with non-static stimuli, to detect unknown objects.

Our way to meet this challenge is by considering and analyzing gaze data across multiple frames and constructing a heatmap from it. 
In contrast, \cite{weber2020distilling} significantly reduced the amount of candidate bounding boxes of unknown objects on a static image using only one gaze point.
In another recent work in a \acs{HRC} scenario, \cite{weber2022exploiting} achieved segmentation of unknown objects and calculated corresponding bounding boxes in 3D space in real time.
Although only one gaze point was required here, the scene image including depth information was needed.
Some other approaches dispense with the gaze altogether, but focus rather on single-class images \cite{pang2020multi}, 
or use additional information, e.g., from a depth sensor \cite{bao2015saliency}.
While robots typically have many sensors, they often have limited computing power.
Additionally, there is often only one object of interest at a time, obviating the need to detect all objects at once. 
By completely omitting image data and employing gaze data instead, we can accomplish the task of detecting unknown objects of interest and still saving large amounts of required computer resources.

In this work, we build on existing work and pave the way for successful human-robot interaction through the following  main contributions:
\begin{itemize}
    \item We present a method for detecting unknown objects in a scene without stimulus, based solely on gaze information.
    \item We only use heatmaps instead of scene images, enabling thus for a significantly faster approach than image-based object detection, while at the same time requiring considerably less computational resources.
    \item We make our unique data set, which contains both gaze data and bounding boxes of the observed objects, publicly available to the research community for download at \url{https://cloud.cs.uni-tuebingen.de/index.php/s/QPzJC48xDGsjnZK}.
\end{itemize}
\subsection{Method}
In this work, we follow two goals.
First, we classify which gaze points or ranges of gaze points belong to an object, and we assign temporal windows to the gaze points, which belong to an annotated bounding box. This creates a classification problem in which the gaze points windows with an associated bounding box are assigned to class one and gaze points windows without a bounding box are assigned to class zero. 

The second goal is to regress the bounding box parameters on the gaze points. These parameters are the width and height, as well as the x and y position. For this task, we also assigned the gaze points to temporal windows. For the regression, we used only temporal windows with associated bounding box, since all others have no parameters for the regression.

We decided to use a spatial distribution as a feature since this worked best in our initial evaluations. This spatial distribution is a heatmap as previously proposed by \cite{fuhl2021gaze} to classify gaze position data. To create such a heatmap, the gaze position data of a temporal window are used, and the individual gaze positions are assigned to cells in the heatmap (grid).
Each time window results in one heatmap.
After the assignment, the heatmap is divided by the sum over all values to obtain a distribution. As an extension to the approach in \cite{fuhl2021gaze}, we extended the 2D heatmap to 3D. This was possible because the software used for gaze determination generates 3D gaze points~\cite{PISTOL} based on a $k$-nearest neighbor regression. In the case of the 3D heatmap, a cell is assigned to each gaze point based on its spatial position with the difference to the 2D heatmap that the depth or distance of the gaze points is additionally considered along the z-axis.
The assignment procedure is illustrated in \figref{paper03:fig:heatcreation}.
\begin{figure}[tb]
	\centering
	\includegraphics[width=.94\linewidth]{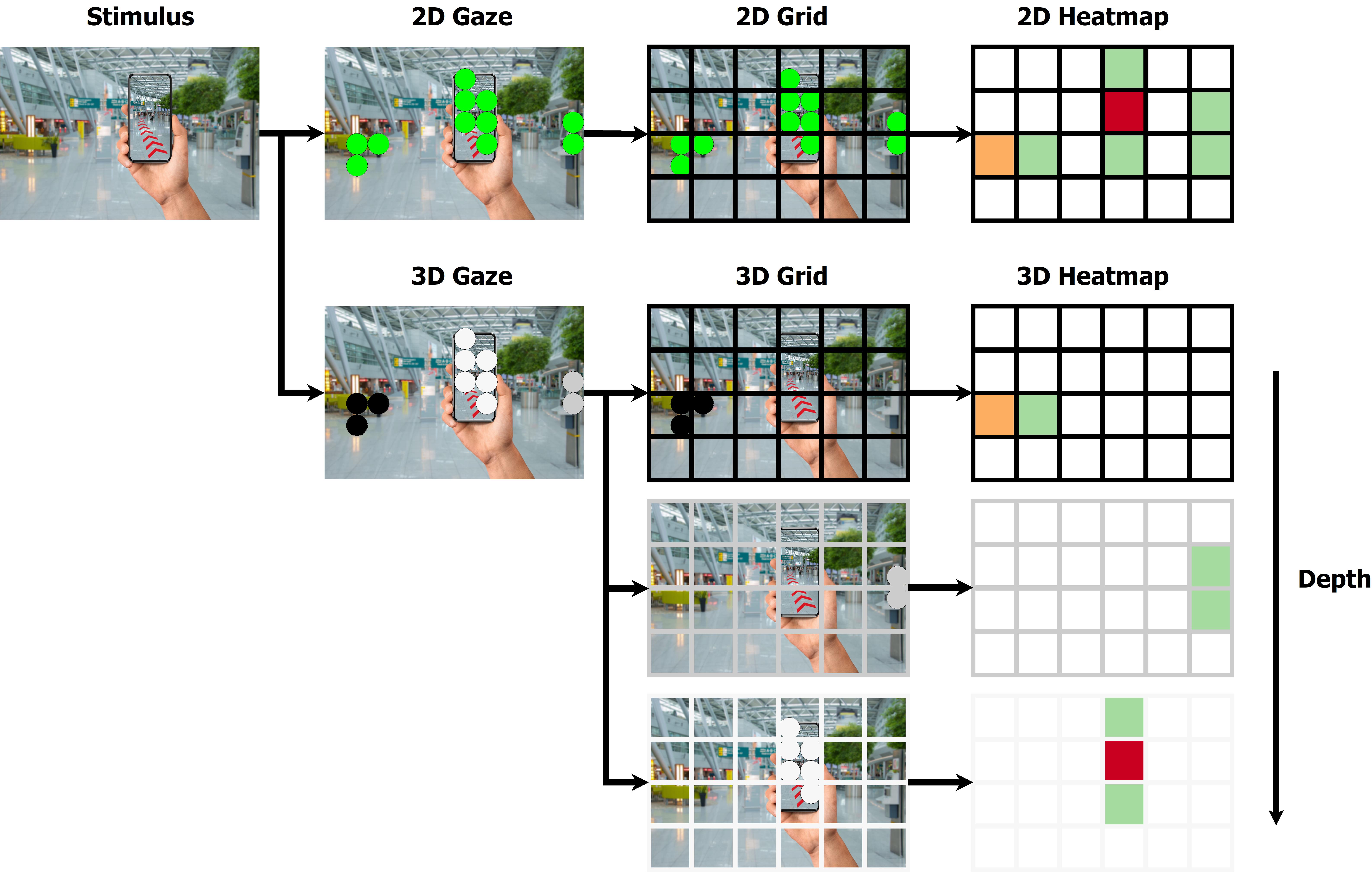}
	\caption{Creation of a 2D or 3D heatmap based on the gaze information and the stimulus resolution.}
	\label{paper03:fig:heatcreation}
\end{figure}

A formal description of the generation of the heatmap in 3D is given in Equation~\ref{paper03:eq:heat}.
\begin{equation}
    \operatorname{heat}\left(\, \round{\frac{p_x}{R_x} \cdot G_x},\, \round{\frac{p_y}{R_y} \cdot G_y},\, \round{\frac{p_z}{R_z} \cdot G_z}\, \right) \pluseq 1.
          \label{paper03:eq:heat}
\end{equation}
The gaze positions in x, y, and z coordinates in an Euclidean coordinate system are denoted by $p_x$,~$p_y$,~and~$p_z$, respectively. The constants $R_x$,~$R_y$,~and~$R_z$ represent the maximum resolution of the stimulus in x and y direction and the maximum depth supported by the software Pistol~\cite{PISTOL}. By dividing the gaze points by the maximum resolution, these ranges are normalized between~ $ 0 $ and~$ 1 $. Subsequently, these values are multiplied by the number of grid cells ($G_x$,~$G_y$,~and~$G_z$) and rounded to the nearest integers, denoted by~``$\round{\cdot}$''. These new values correspond to the index in the heatmap and the selected cell is incremented by one, denoted by~``$\pluseq$''. In the case of a 2D heatmap, the cell for depth (z~coordinate) is fixed at one.

Equation~\ref{paper03:eq:norm} describes the normalization of the heatmap in 3D and 2D since for the 2D case there would be only one depth.
\begin{equation}
    \operatorname{heat}(x,y,z) = \frac{\operatorname{heat}(x,y,z) }{\sum_{i=1}^{G_x} \sum_{j=1}^{G_y} \sum_{k=1}^{G_z} \operatorname{heat}(i,j,k)}.
          \label{paper03:eq:norm}
\end{equation}
The variables~$x$,~$y$,~and~$z$ are the indexes to the heatmap corresponding to the x-axis, y-axis, and z-axis.
Finally, the one-dimensional vector resulting from the flattening of the heat map can be used as an input feature for various machine learning techniques.

\subsection{Study Design \& Data Acquisition}
In this section, we describe the dataset we used. In order to evaluate our approach, a  dataset was required which contains not only eye tracking information but also, in addition to the gaze points, the bounding boxes of the objects that the participants were looking at.
Since, to the best of our knowledge, no such dataset exists or is publicly available, we collected a novel data set.
At the beginning, a calibration was performed with each participant, following the procedure described in \cite{PISTOL}.
Subsequently, the subjects were allowed to move freely around the site.
In this course, they should look at arbitrary objects they encountered.
There was no specification as to how long they were supposed to look at the objects.
To evaluate gaze accuracy, the participants were asked to look at the calibration marker again at the end of each recording.
All recordings were conducted with the Pupil Invisible eye tracker, a head-mounted eye tracker developed by Pupil Labs, 
whose scene camera provides RGB images with a resolution of 1088 $\times$ 1080.
Each participant captured three recordings (each recording was about five minutes long, including calibration and evaluation), resulting in 14 valid videos in total.
This led to a total length of about one hour of recording, consisting of $ 102\,620 $ frames of which $ 27\,946 $ contained objects.

Finally, we labeled the obtained data with DarkLabel \cite{DarkLabel}.
\figref{paper03:fig:labeledBoxes} shows individual example moments from the recordings.
\begin{figure}[b]
	\centering
	\includegraphics[width=.325\linewidth]{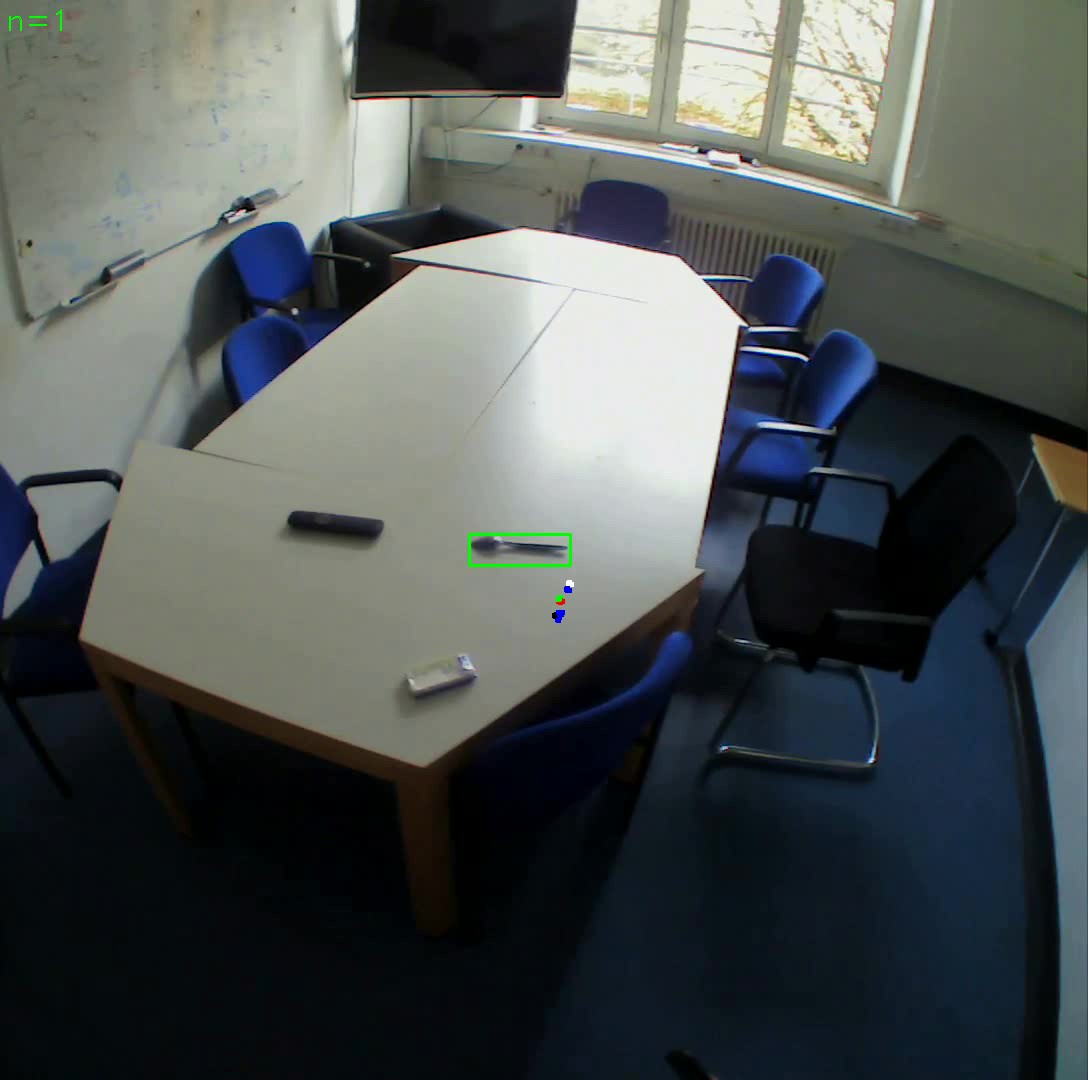}\hfil
	\includegraphics[width=.325\linewidth]{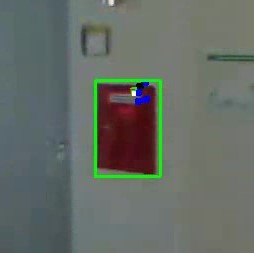}\hfil
	\includegraphics[width=.325\linewidth]{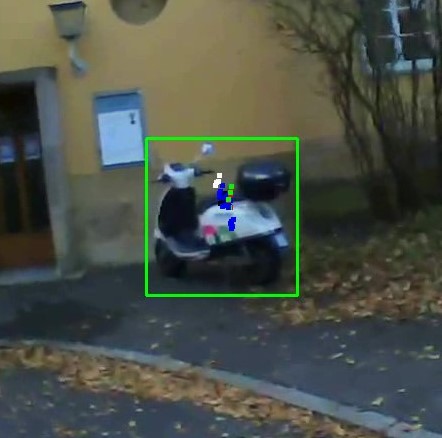}
	\caption{The images, some of them zoomed in, show exemplary moments of our data, where the objects that were consciously observed are labeled with a bounding box.
	}
	\label{paper03:fig:labeledBoxes}
\end{figure}
Due to the errors related to the gaze estimation, the gaze points are, especially for small objects, not always on the labeled object, even though the participant was actually looking at it.
In fact, even for a human, it is not always easy to determine the target object, and sometimes only possible considering the context and the observation of an image sequence.
This demonstrates quite clearly the difficulties and challenges associated with this task.
Our final, publicly available dataset only contains the gaze information and bounding boxes, yet no stimuli-related information.

\subsection{Evaluation}
In this section, we evaluate the classification of the gaze points with respect to the affiliation to an object, and we try to extract the position and the size of the object from those. 
To this end, we applied a variety of different, well-established machine learning methods and list here a selection comprising the best of them.
In the classification experiments, we always specify the mean accuracy of a 5-fold cross validation. For the regression experiments, the mean error as a percentage of the image resolution from a 5-fold cross validation is given. We evaluated different heatmap grid sizes as well as different time window sizes.
We conducted our evaluations on a computer system with Windows 10, an AMD Ryzen 9 3950X 16-core processor with 3.50 GHz, and 64 GB DDR4 Ram. All machine learning methods were implemented on the Matlab version 2021b and for reproducibility we restrict ourselves to Matlab's default parameters.

\begin{table}[b]
	\caption{Best and worst classification results of the 2D and 3D heatmap features. The mean is denoted by $ \mu $ and the standard deviation by $ \sigma $.}
	\label{paper03:tbl:evalClassification}
	\centering
	\begin{tabular}{cc|ccc}
		\multirow{2}{*}{Feature} & \multirow{2}{*}{ML} & \multicolumn{2}{c}{Accuracy} & \multirow{2}{*}{$\mu\pm\sigma$} \\
		& & Worst & Best & \\
		\midrule
		\multirow{3}{*}{2D heatmap} & \acs{KNN} & 68 & 88 & $83.2\pm3.8$ \\
		& Bagged Trees & 79 & 89 & $86.3\pm1.9$ \\
		& Gaussian \acs{SVM} & 73 & 84 & $79.4\pm2.6$ \\
		\midrule
		\multirow{3}{*}{3D heatmap} & \acs{KNN} & 73 & 92 & $87.8\pm3.3$ \\
		& Bagged Trees & 80 & 89 & $86.8\pm1.3$ \\
		& Gaussian \acs{SVM} & 72 & 83 & $76.5\pm3.6$ \\
	\end{tabular}
\end{table}
\begin{figure}
    \centering
    \includegraphics[width=.9\linewidth]{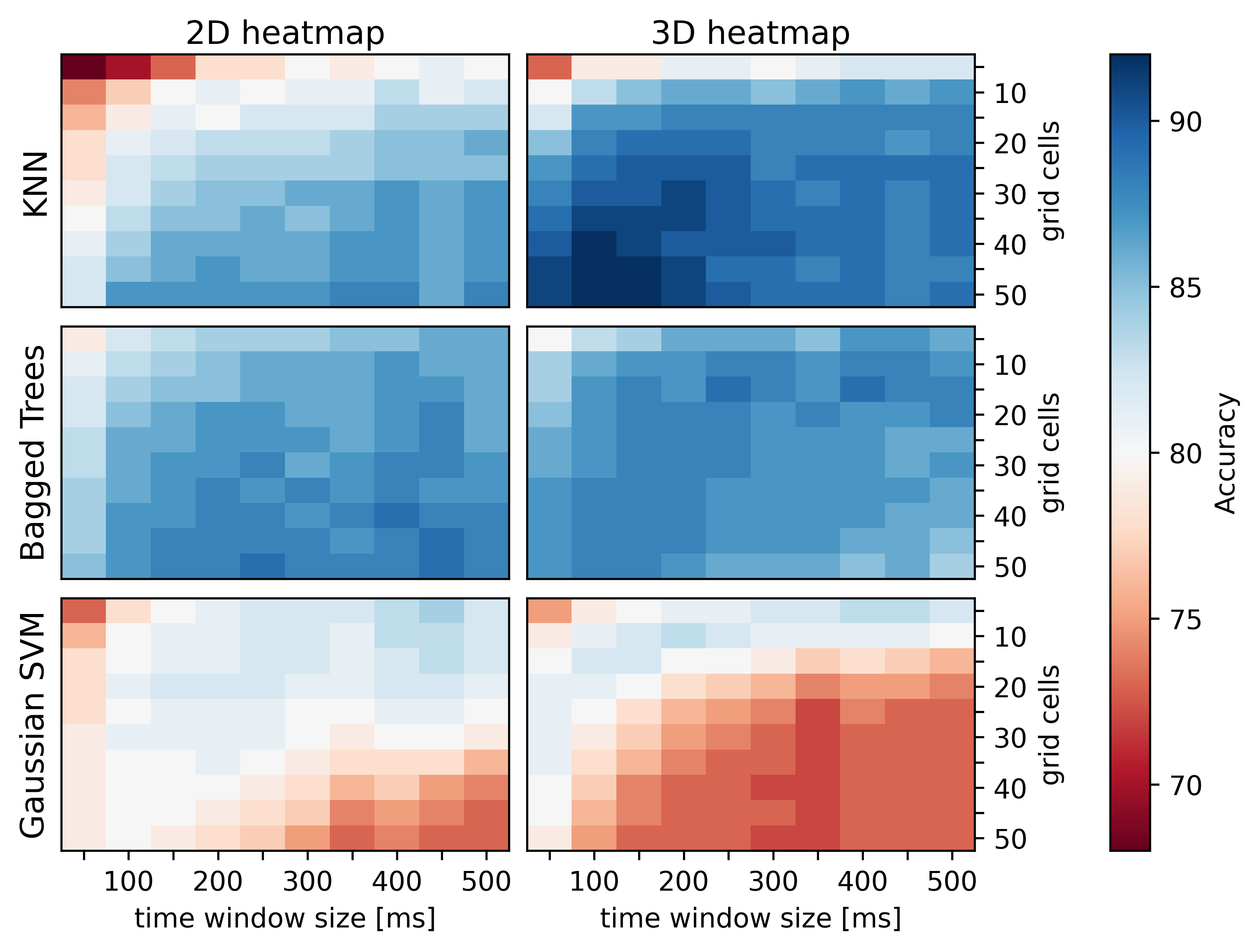}
    \caption{Classification results of the 2D and 3D heatmap features for different time window sizes (in ms), number of grid cells, and machine learning methods illustrated in a heatmap. The results are the average accuracy of a 5-fold cross validation.}
    \label{paper03:fig:evalClassification}
\end{figure}

The assignment of classes (object or no object) to time windows was done based on the presence of an annotated object in the time window.
This means that if there was an annotated object in the time window, the class was set to one, and zero otherwise.
In the regression, only time windows with an existing annotated object were used. Here, the parameters of the annotated object closest to the central timestamp of the time window were chosen. This was assigned because, in most cases, our subjects moved while looking at an object. Thus, there are usually different positions and sizes of bounding boxes in a time window.

\figref{paper03:fig:evalClassification} and \tblref{paper03:tbl:evalClassification} show a summary of the results of our classification experiment.
Comparing the results of the three methods (\acs{KNN}, bagged trees, and Gaussian \acs{SVM}) for the 2D heatmap feature, the approach based on bagged trees achieves the best results. Looking at the progression over the grid and time window size, we can see that the \acs{KNN} and the bagged trees perform best with a high number of grid cells and large time windows. In contrast, the Gaussian \acs{SVM} performs best at a small number of grid cells but still large time windows.
Moving on to the 3D heatmaps, the accuracy of the \acs{KNN} method improves by 4 percent to 92 percent, which is also significantly better than the bagged trees.

The best results of our regression experiment are shown in \tblref{paper03:tbl:evalRegression}.
\begin{table}
    \caption{Best regression error results as the average absolute error of a 5-fold cross validation in percentage. The columns X and Y denote the position of the bounding box, W is the width, and H is the height of the bounding box.}
    \label{paper03:tbl:evalRegression}
    \centering
    \begin{tabular}{cc|cccc}
        \multirow{2}{*}{Feature} & \multirow{2}{*}{ML} & \multicolumn{4}{c}{Error}  \\
        & & X & Y & W & H \\
        \midrule
        \multirow{3}{*}{2D heatmap} & Gaussian Process & 6.1 & 6.8 & 12.2 & 15.1 \\
        & Bagged Trees & 6.4 & 6.9 & 12.0 & 14.3 \\
        & Gaussian \acs{SVM} & 6.4 & 6.9 & 13.4 & 15.5 \\
        \midrule
        \multirow{3}{*}{3D heatmap} & Gaussian Process & 5.8 & 6.0 & \pz9.9 & 11.6 \\
        & Bagged Trees & 6.4 & 6.7 & 10.5 & 12.3 \\
        & Gaussian \acs{SVM} & 6.2 & 6.2 & 11.0 & 12.9 \\
    \end{tabular}
\end{table}
Looking at the individual methods (Gaussian process regression, bagged trees, and Gaussian \acs{SVM}), we see that all methods perform similarly well.
As expected, based on the spatial heatmap feature, the position estimation is the most accurate. In contrast, the regression of the bounding box size, using only gaze data an no stimuli, is even more difficult than the position estimation and therefore less accurate.
Comparing the results for the 2D and the 3D heatmap feature, the position results remain about the same, with some overall improvement.
In terms of bounding box size, the best results improve significantly for all of the three methods.
All in all, the Gaussian process method combined with the 3D heatmap feature performs best.

\figref{paper03:fig:QualEval} shows a qualitative extract of the Gaussian process regression in comparison to the ground truth.
Naturally, the position is more accurate than the bounding box size, since humans tend not to observe the entire object when looking at it. Overall, however, both can be determined quite well.
\begin{figure}
	\centering
	\includegraphics[width=0.49\linewidth]{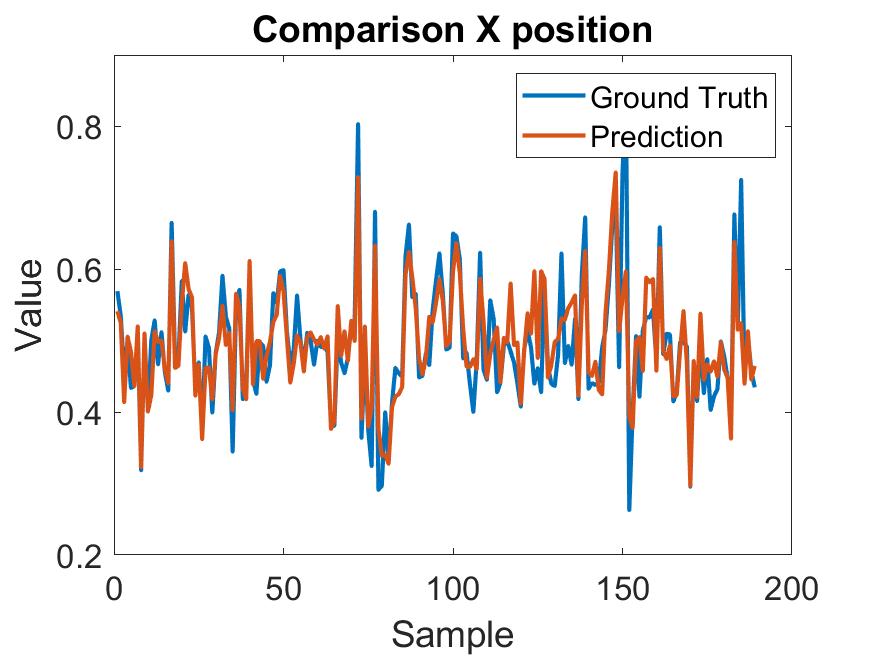}
	\includegraphics[width=0.49\linewidth]{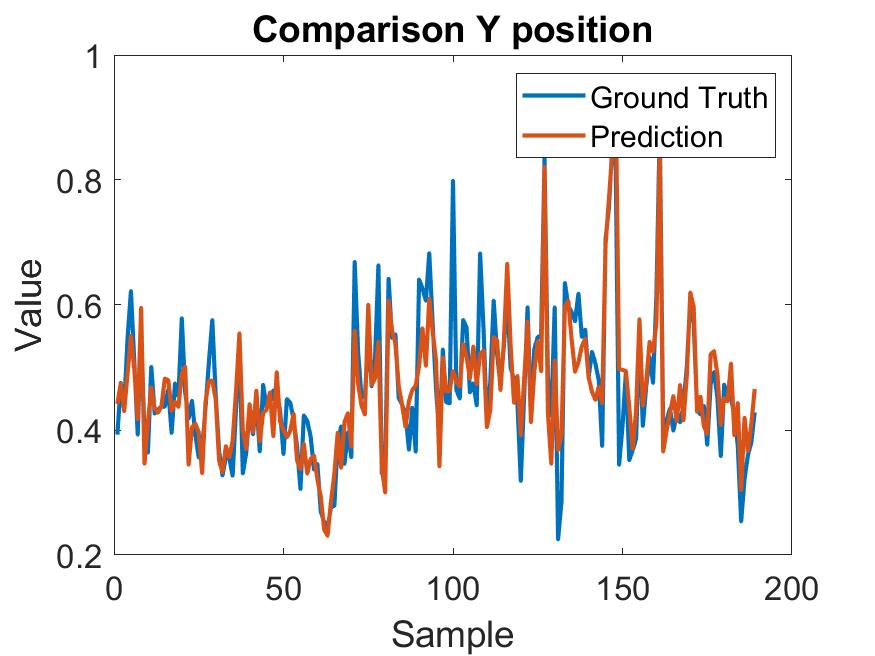}
	\par\bigskip
	\includegraphics[width=0.49\linewidth]{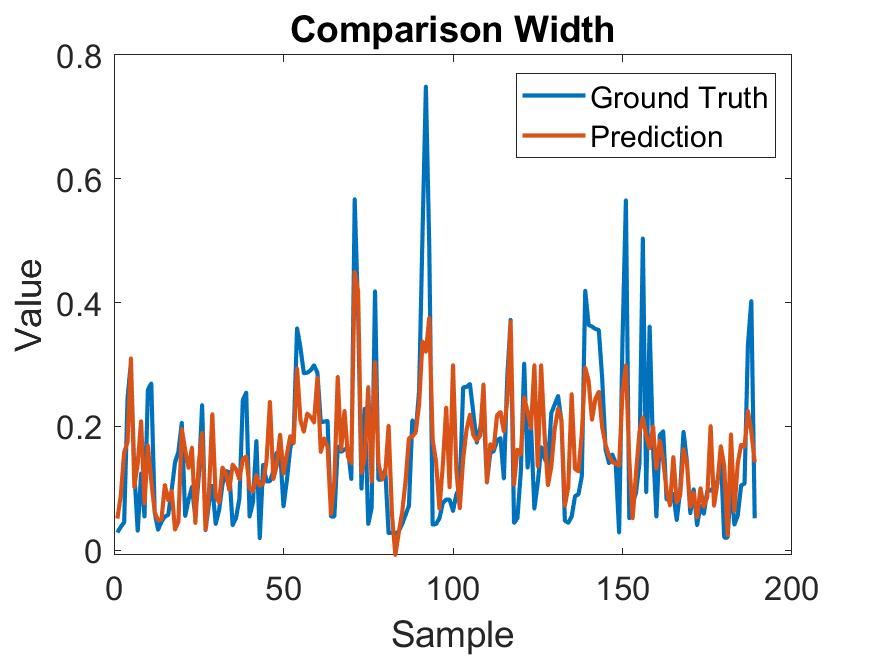}
	\includegraphics[width=0.49\linewidth]{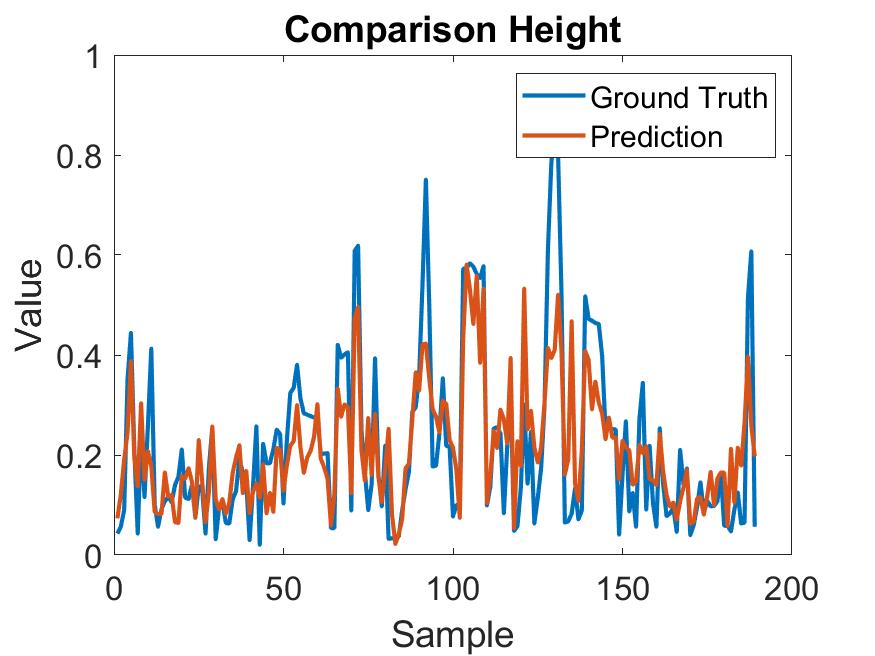}
	\caption{Qualitative evaluation of the bounding box parameter regression.
	The results are from the Gaussian Process Regression with a time window size of 100, a grid cell number of 15 and the 3D heatmap feature.}
	\label{paper03:fig:QualEval}
\end{figure}

Hereafter, we will investigate the runtime and memory requirements.
It should be borne in mind that classical object detectors pursue a slightly different goal than we do.
Whereas in their case all objects are
to be detected, we are primarily interested in the existence of an object of interest, that is, the one that the human is looking at.
Since classical object detectors only use scene images and do not obtain information about human gaze behavior, they cannot know whether a human is looking at an object, nor which object.
Thus, it would be
a matter of chance whether the statement is correct.

With the regression task, the detection of all objects would be possible. Here, however, we encounter a different real-world problem, outside of laboratory conditions, which also makes our method so appealing.
Since we are in a wild world, the objects of interest are extremely diverse and their number tremendous.
The vast majority of objects in our dataset, such as doorknobs, light switches, and fire extinguishers, are simply not part of any publicly available data sets, such as Microsoft COCO~\cite{lin2014microsoft} or ImageNet~\cite{deng2009imagenet}, that are typically used for training.
Since the methods differ too much in this respect, we need a benchmark that covers more the commonalities.
Therefore, in the remainder of this section, we will establish a baseline comparison in terms of speed and computing resources.
As a baseline, we use state-of-the-art object detectors.
These include Faster \mbox{R-CNN}~\cite{ren2015faster}, FCOS~\cite{tian2019fcos}, and RetinaNet~\cite{lin2017focal}, each with a ResNet-50-FPN backbone~\cite{he2016deep},
SSDlite320~\cite{liu2016ssd} and Faster \mbox{R-CNN} both with a MobileNetV3 Large backbone~\cite{howard2019searching},
as well as SSD300~\cite{liu2016ssd} with a VGG16 backbone~\cite{simonyan2014very}.
These are supplemented by various YOLOv5~\cite{jocher2022yolov5-short} variants.
In order to test the speed, we measured the runtime of all methods on the CPU for 1000 individual predictions, i.e. 1000 different inputs with a batch size of one.
The resource consumption was determined by measuring the amount of memory required for a single input.
For our method with the heatmap input features, we used a time window size of 250\,ms.
For the classic object detectors, the $ 1088 \times 1080 \times 3 $ RGB images were used as input.
The summary of the results are shown in \tblref{paper03:tbl:evalResources}.
\begin{table}
    \caption{Comparison of the required resources for the different input features. The time column indicates the execution time for 1000 different inputs at a batch size of one in seconds. The memory column specifies the required memory of a single input in kilobytes. For the 2D and 3D heatmap features, the results shown are from a time window size of 250\,ms and a grid cell number of 30.}
    \label{paper03:tbl:evalResources}
    \centering
    \begin{tabular}{cccc}
        Feature & ML & Time [s] & Memory [KB] \\
        \midrule
        \multirow{3}{*}{2D heatmap} & \acs{KNN} & 10.8 & \pz424 \\
        & Bagged Trees & 57.8 & 1134 \\
        & Gaussian \acs{SVM} & \pz8.6 & \pz406 \\
        \midrule
        \multirow{3}{*}{3D heatmap} & \acs{KNN} & 276.9 & 3045 \\
        & Bagged Trees & \pz64.7 & 1467 \\
        & Gaussian \acs{SVM} & 610.6 & 3650 \\
        \midrule
        \multirow{11}{*}{RGB Image} & F. R-CNN~\cite{ren2015faster} (RN50) & 8\,705.3 & 1\,745\,456 \\
        & F. R-CNN~\cite{ren2015faster} (MN) & 1\,205.6 & \pz\,545\,400 \\
        & FCOS~\cite{tian2019fcos} & 4\,723.2 & \pz\,995\,416 \\
        & RetinaNet~\cite{lin2017focal} & 5\,184.5 & 1\,390\,580 \\
        & SSD300~\cite{liu2016ssd} & \pz900.8 & \pz\,529\,744 \\
        & SSDlite320~\cite{liu2016ssd} & \pz163.7 & \pz\,293\,788 \\
        & YOLOv5n~\cite{jocher2022yolov5-short} & \pz200.6 & \pz\,270\,168 \\
        & YOLOv5s~\cite{jocher2022yolov5-short} & \pz486.3 & \pz\,312\,104 \\
        & YOLOv5m~\cite{jocher2022yolov5-short} & 1\,127.6 & \pz\,421\,904 \\
        & YOLOv5l~\cite{jocher2022yolov5-short} & 2\,174.5 & \pz\,622\,536 \\
        & YOLOv5x~\cite{jocher2022yolov5-short} & 3\,677.9 & \pz\,940\,508 \\
    \end{tabular}
\end{table}

The fastest are the Gaussian \acs{SVM} and the \acs{KNN} with the 2D heatmap feature.
The Bagged Trees are slower, but the runtime increases proportionally less as the number of grid cells increases.
Consequently, the runtime for the 3D heatmap feature is in the range of one minute for the 1000 predictions while the runtime for \acs{KNN} and Gaussian \acs{SVM} increases considerably from a few seconds to several minutes.
Nonetheless, it is immediately apparent that the runtime is in general significantly lower compared to the object detectors using the RGB images as input features.
While only the smaller models like YOLOv5n and SSDlite remain under three minutes, the other models are much slower. 
In particular, the computation time required by the popular Faster \mbox{R-CNN (RN50)} exceeds that of the Bagged Trees by a factor of over 100.

A similar picture emerges with respect to the RAM allocated for one single prediction.
The memory requirements of the bagged trees are larger for small inputs, but do not increase as much in proportion to the number of grid cells as for the \acs{KNN} and the Gaussian \acs{SVM}.
Overall, the heatmap features require only a few 100\,KB to a few MB.
This is substantially less than the most frugal neural network YOLOv5n, which needs around 270\,MB.
Faster \mbox{R-CNN} with the \mbox{ResNet-50} backbone requires the most memory with over 1.7\,GB.
Again, the factor is more than 100 times larger than for the Gaussian \acs{SVM} with the maximum number of 50 grid cells.
Compared to the Bagged Trees, it even exceeds 860 times.

In summary, our method is several orders of magnitude faster than conventional object detectors while requiring only a fraction of their resources.

\subsection{Conclusion}
In this work, we addressed object detection in the wild by means of gaze data.
Our results show that it is possible to detect objects and determine their bounding box based solely on gaze information.
Additionally, we have used a variety of machine learning methods to show that they work for solving such challenges.
Besides, the functionality of several machine learning methods proves that our heatmap feature, which we have extended to 3D, can be used efficiently for this problem.
In comparison to classical object detectors that use image input features, we have shown that object detection by means of our heatmap features is significantly faster while only requiring a fraction of the computational resources.
This is of major relevance due to the fact that robots usually have only limited computing capacity at their disposal and cannot be equipped with powerful graphics units as they consume a lot of power.

However, a significant amount of work remains for the future as we plan to extend our proof of concept to a real robot by making the gaze of the human collaborator accessible to it.
Our approach can serve as a foundation for future applications in the field of human-machine interaction and \acs{HRC}, where robots can learn new objects from humans through instant knowledge sharing.
Hence, we hope our methods and dataset can help to advance researchers in this challenging context.

\fi

\clearpage
\setcounter{footnote}{0}
\section{Exploiting the GBVS for Saliency aware Gaze Heatmaps}
\label{app:sec:geisler2020exploiting}

\ifpaper
\begin{figure}[h]
    \begin{minipage}[t]{0.3333333\textwidth}
        \begin{center}
            \includegraphics[trim={9cm 0 9cm 0},clip,width=\linewidth]{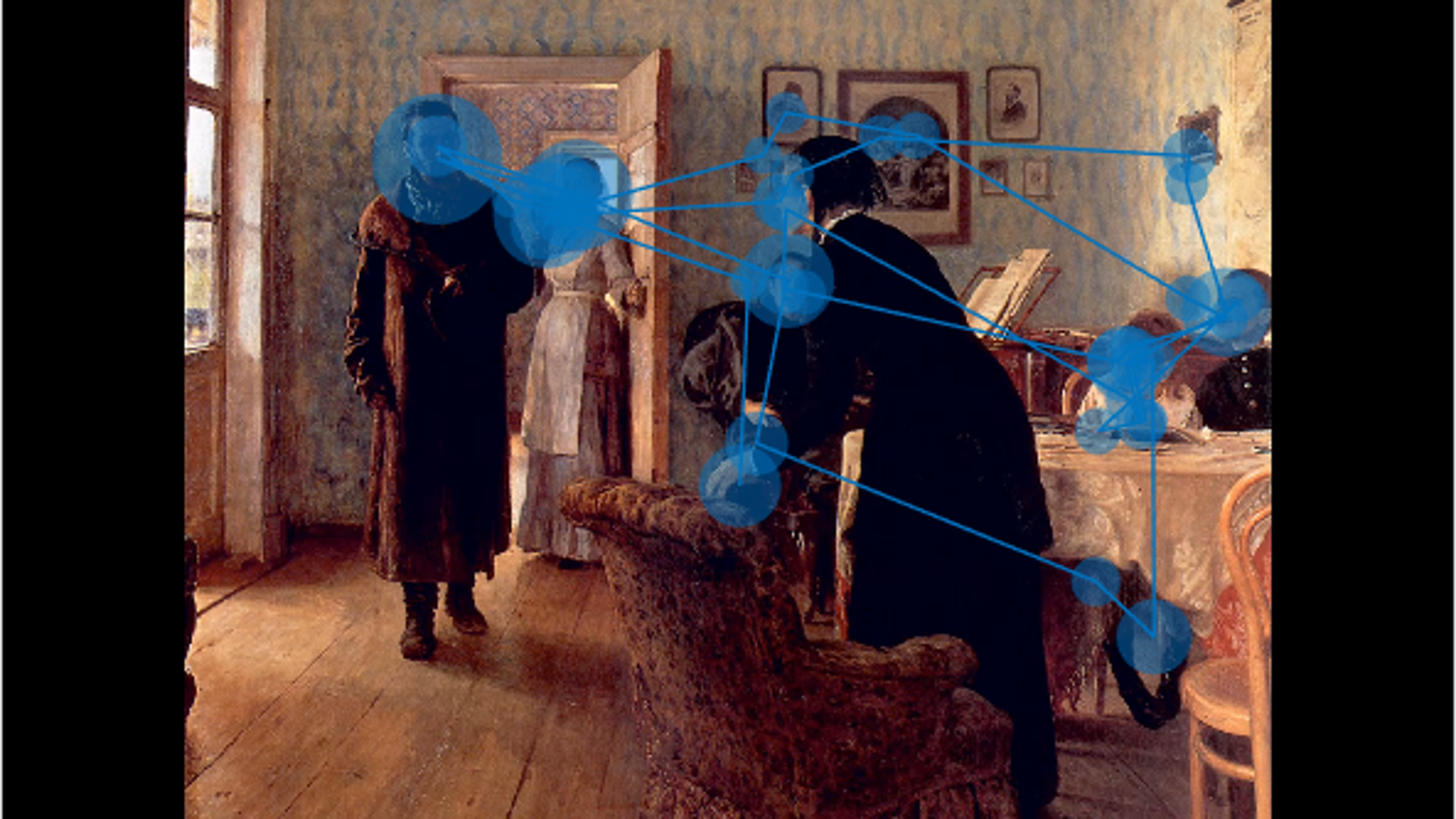}
            (a) Fixation sequence on the painting \emph{An Unexpected Visitor} from Ilya Repin.
        \end{center}
    \end{minipage}\hfill%
    \begin{minipage}[t]{0.3333333\textwidth}
        \begin{center}
            \includegraphics[trim={9cm 0 9cm 0},clip,width=\linewidth]{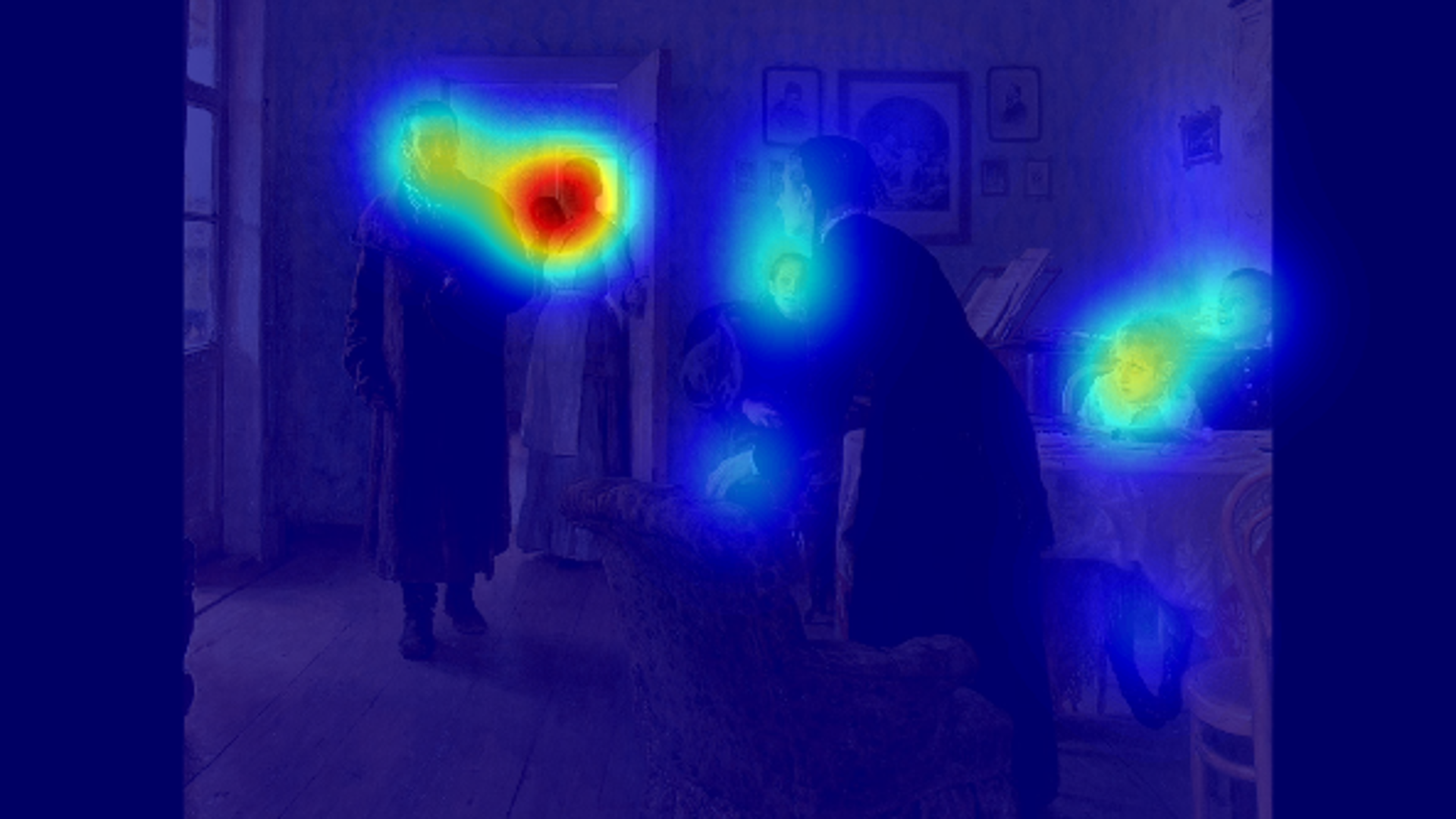}
            (b) Regular gaussian like fixation heatmap.
        \end{center}
    \end{minipage}\hfill%
    \begin{minipage}[t]{0.3333333\textwidth}
        \begin{center}
            \includegraphics[trim={9cm 0 9cm 0},clip,width=\linewidth]{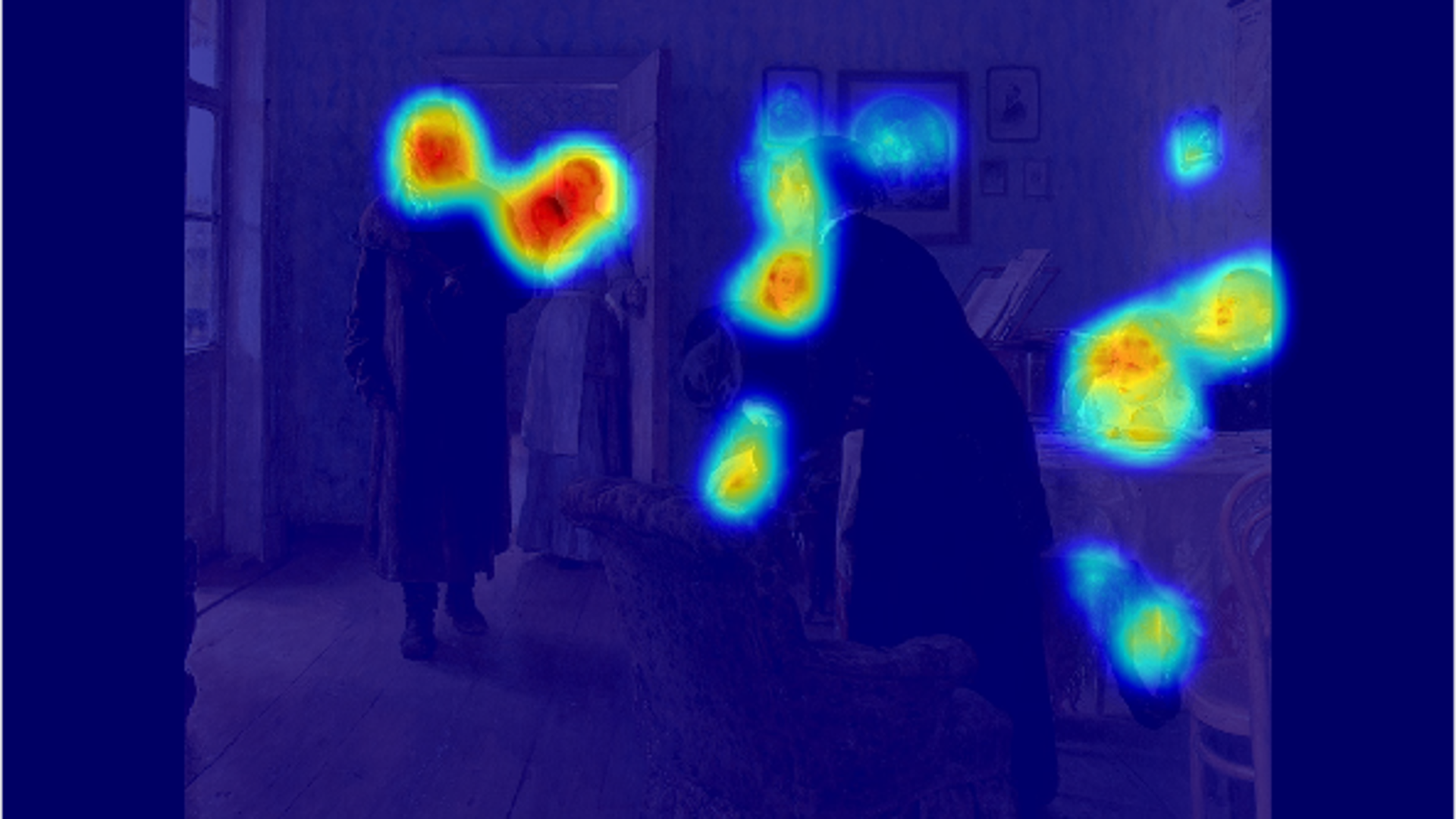}
            (c) \acs{GBVS} attention map with incorporated gaze signal.
        \end{center}
    \end{minipage}%
    \caption{(a)~shows the sequential fixation signal, where the size of the circles encodes the fixation time. (b)~shows the corresponding gaussian like fixation heatmap. (c)~shows the output of the proposed approach, where the tracked fixations are incorporated into the \acs{GBVS} attention map calculation.}
    \label{geisler2020exploiting:fig:teaser}
\end{figure}
\subsection{Abstract}
    Analyzing visual perception in scene images is dominated by two different approaches: 1.)~Eye Tracking, which allows us to measure the visual focus directly by mapping a detected fixation to a scene image, and 2.)~Saliency maps, which predict the perceivability of a scene region by assessing the emitted visual stimulus with respect to the retinal feature extraction. One of the best-known algorithms for calculating saliency maps is \acs{GBVS}. In this work, we propose a novel visualization method by generating a joint fixation-saliency heatmap. By incorporating a tracked gaze signal into the \acs{GBVS}, the proposed method equilibrates the fixation frequency and duration to the scene stimulus, and thus visualizes the rate of the extracted visual stimulus by the spectator.
\subsection{Introduction}\label{geisler2020exploiting:sec:introduction}
Our eyes move around to perceive and understand the scene in order to compensate for our limited- but clearest- foveal vision. When viewing a scene, we frequently focus our attention, known as a fixation, before shifting to another area with a rapid eye movement known as a saccade. The selectivity of the focused scene locations is a highly optimized and developed process, mainly driven by two factors: 1.) visual scene features, extracted by the retina (bottom-up), and 2.) the interpretation of the extracted features regarding their semantic value by higher cognitive processes, and subsequent identification of the next fixation target (top-down)~\cite{van2019real,itti1998model,itti2000saliency}. Modeling and understanding this reciprocal process is a long-term core topic in cognitive psychology and the computer vision community~\cite{kotseruba201840}.

While retinal feature extraction can be modeled using saliency maps such as \acs{GBVS} \cite{harel2006graph}, the selectivity of visual attention can be measured using eye tracking. Saliency maps are predominantly bio-physiologically inspired algorithms to predict potential fixation targets~\cite{zhao2013learning}. Therefore, regions emitting a strong, recognizable visual stimulus are identified and emphasized by replicating the retinal visual stimulus processing. Eye tracking, on the other hand, often tracks the pupil center and extrapolates the line of sight to a scene image. A subsequent typical visualization is to illustrate the extracted fixations as a heatmap overlay on the scene image. This is used to investigate the visual attention on the scene, and to identify areas of particular interest.

However, fixation heatmaps are subject to some limitations. For instance, slight shifts in the eye-tracking signal make it difficult to identify the scene parts that attracted the visual attention and which information of the scene was actually perceptible. In addition, depending on the implementation, long or frequent fixations on the same scene region may lead to a high density in the fixation heatmap. Hence, it is assumed that these regions are particularly relevant to the spectator since a comparatively high amount of visual information was extracted. However, frequent or long fixations may also be caused by difficult scene conditions such as low contrasts. Thus, the visual information may be harder to extract, and therefore requires longer or more frequent fixations to be perceived.

In this work, we propose to incorporate detected fixations from an eye tracking signal into the calculation of the \acs{GBVS} attention map. The resulting heatmap equilibrates the measured visual attention to the retinal-perceivable stimulus, and thus visualizes the density of perceived information in the scene more accurately as pure fixation or saliency heatmaps.

\paragraph{Structure of the Paper:}
Section~\ref{geisler2020exploiting:sec:related_work} gives a short introduction into state-of-the-art eye tracking visualization and saliency methods. Section~\ref{geisler2020exploiting:sec:method} contains a comprehensive description of how the proposed approach takes place in the \acs{GBVS} algorithm. Section~\ref{geisler2020exploiting:sec:evaluation} shows the exemplary application of the proposed visualization to different types of stimuli. The final sections~\ref{geisler2020exploiting:sec:conclusion} state the limitations of the presented approach and the final remarks.
\subsection{Related Work}\label{geisler2020exploiting:sec:related_work}
The eye tracking community is a research powerhouse. Continuous improvements in tracking accuracy, precision, and availability over the last decades made eye tracking to one of the most eminent sensors in numerous research fields: Psychology, HCI, medicine, neuroscience, marketing, and many more. In particular, the success of recent years in the field of vision-based eye tracking has boosted the technology in terms of affordability, convenience, and usability for a broad community~\cite{santini2017calibme,santini2017eyerectoo,santini2019get,BRM2019HospB}. However, the ability to conduct comprehensive eye-tracking studies led to increasing demand for sophisticated methods for visualization and qualitative evaluation of the acquired data~\cite{blascheck2015challenges}.

An initial exploratory step in eye tracking studies is often to examine the spatial location, duration, and frequency of fixations as a heatmap over the stimulus~\cite{bylinskii2015eye}. This can be efficiently calculated over a large amount of data and gives a first impression of the distribution of visual attention on the stimulus~\cite{duchowski2012aggregate,dao2014heatmap}. But, in order to gain deeper insights into the data, an extensive repertoire of different visualization techniques is available, such as various saccade metrics~\cite{kubler2016novel,32683c26f5f049368a8dae9c18ccfab6,fb24eebc85f04f019ea615bdc87da319}, AOI hierarchies~\cite{548feb0e41cf4592b1f37e7d9eeb6265,kurzhals2016visual}, or extensive interactive visualizations including stimulus and time domains~\cite{kurzhals2015gaze,kurzhals2016fixation,03456eff1f104efc8b26336458435a54,TCKWGJRWE2015}, etc. A comprehensive overview can be found in the survey of~\cite{fd80e03966dc41ceb55cc78e6aa9836a} and~\cite{blascheck2017visualization}.

Saliency maps assess the stimulus by modeling the retinal signal processing to determine whether a scene area is particularly prominent in its immediate neighborhood, and therefore more likely to be perceived. The stimulus is evaluated by its intensity and its opponent color spaces: Driven by the neuronal circuit of the photoreceptors. Additionally, further feature spaces can be formed, such as edge orientation and difference formation of sequential images~\citep{geisler2017saliency,harel2006graph,hou2007saliency,itti1998model,bian2008biological,zhang2013saliency}. While these bottom-up approaches mainly reproduce the feature extraction of the retina perception, newer deep-learning-based approaches show great success in modeling the whole processes, from the retinal feature extraction up to the semantic interpretation of the visual cortex and higher cognitive levels. They include the recognition and evaluation of abstract forms regarding their object-relatedness and semantic relevance~\citep{zhao2015saliency,john2014traffic,li2015visual,lee2016deep,liu2016dhsnet,kummerer2014deep,chen2017look,dubey2015makes,zou2015harf}.

The proposed approach combines the worlds of fixation heatmaps and salience maps, as a novel visualization technique. The resulting heatmap provides insights into the extracted information rate of the scene and extends the existing visualization techniques towards a more stimulus-driven paradigm.
\subsection{Method}\label{geisler2020exploiting:sec:method}
Similar to most saliency map approaches, the \acs{GBVS} algorithm is divided into 3 consecutive steps~\cite{harel2006graph}:
\begin{enumerate}
    \item Extraction of a feature map $M_t$ on a given image $I_t$.
    \item Calculation of an activation map $A_t$ based on $M_t$.
    \item Normalization and combination of the activation map $A_t$.
\end{enumerate}
Our approach amends step (2) by injecting the gaze signal $g_t$ into the calculation of the activation map $A_t$. Steps 1 and 3 remain unchanged to the \acs{GBVS} publication~\cite{harel2006graph} and not further discussed here. The subscript $t$ indicates the time domain since the gaze signal is given as a time series of consecutive fixation points. However, it also simplifies the handling with dynamic stimuli, such as videos. In the following, we assume that for each $t$ exists a corresponding gaze signal $g_t$, as well as a stimulus $I_t$, respectively a feature map $M_t$.

The \acs{GBVS} interprets the activation map as a state vector of a Markov model. The transition between two states is defined by a dissimilarity score over the feature map $M_t$. Thus, a random walk over the Markov model empowers those states that are dissimilar in the respective feature map. Analogous to the original \acs{GBVS}, the dissimilarity between the two states $i$ and $j$ in the feature map $M_t$ is defined as follow:
\begin{equation}\label{geisler2020exploiting:equ:d}
    d_t\left(i,j\right) = \left|\log{\frac{M_t\left(i\right)}{M_t\left(j\right)}}\right|,
\end{equation}
where $M_t\left(i\right)$ is the $i$-th value of the corresponding feature map $M_t$. The transition weight $w_t\left(i,j\right)$ between the two states $i$ and $j$ is defined as the product of their dissimilarity score $d_t\left(i,j\right)$ and a distance weight $F_w\left(i,j\right)$:
\begin{equation}\label{geisler2020exploiting:equ:w}
    w_t\left(i,j\right) = d_t\left(i,j\right) \cdot F\left(i,j\right).
\end{equation}
The distance weight adds a local sensitivity to the dissimilarity score. Thus, states that are dissimilar to their immediate neighborhood are emphasized while the impact of the dissimilarity score is attenuated with increasing distance. $F\left(i,j\right)$ is defined as an exponentially weighted square distance between the states $i$ and $j$ in their spatial dimension in the input image $I_t$:
\begin{equation}\label{geisler2020exploiting:equ:F}
    F\left(i,j\right) = \exp{\left(-\frac{
            \left(x\left(i\right)-x\left(j\right)\right)^2+
            \left(y\left(i\right)-y\left(j\right)\right)^2
        }{
            2\cdot\sigma
        }\right)},
\end{equation}
where $x\left(i\right)$ and $y\left(i\right)$ is the $x$- and $y$-coordinate of the $i$-th state in the respective input image $I_t$. The free parameter $\sigma$ controls the shape of the exponential distance weight. The larger $\sigma$ is chosen, the more weight is given to the dissimilarities of more remote states.

The final Markov transition matrix $T_t$ is then assembled as follows:
\begin{equation}\label{geisler2020exploiting:equ:T}
    T_t = \left(\begin{array}{cccc}
                           1 & w_t\left(0,1\right) & \ldots & w_t\left(0,n\right) \\
         w_t\left(1,0\right) &                   1 & \ddots & w_t\left(1,n\right) \\
                      \vdots &              \ddots & \ddots &              \vdots \\
         w_t\left(n,0\right) & w_t\left(n,1\right) & \ldots &                   1
    \end{array}\right),
\end{equation}
where $n$ is the number of elements in the feature map $M_t$ respective the input image $I_t$.

The activation map $A_t$ is then calculated by $k$ repeated multiplication with the transition matrix $T_t$:
\begin{equation}\label{geisler2020exploiting:equ:Ak}
    A^{\left(k\right)}_t = T_t \cdot A^{\left(k-1\right)}_t.
\end{equation}

\paragraph{Incorporate Gaze:}
Up to this step, the procedure follows the original \acs{GBVS} algorithm. However, instead of initializing $A^{\left(0\right)}_t$ equally distributed, the gaze position is encoded as initial activation map:
\begin{equation}\label{geisler2020exploiting:equ:v0}
    A^{\left(0\right)}_t = 
                       q \cdot A^{\left(k\right)}_{t-1} + 
        \left(1-q\right) \cdot \left(F\left(0,g_t\right),\ldots,F\left(n,g_t\right)\right),
\end{equation}
where $F\left(i,g_t\right)$ is the exponential weighted square distance between the recorded gaze position $g_t$ and the spatial location of the $i$-th element in the activation map. In other words, the activation map is initialized by the measured visual activation from the eye tracking signal. Additionally, parameter $q\in\left[0,1\right]$ controls the influence of the previously calculated activation map $A_{t-1}$ into the initialization of $A^{\left(0\right)}_t$. Thus, for $q>0$, $A^{\left(0\right)}_t$ encodes the recently measured visual attention, but also the history of previous predicted attention areas. This smooths the resulting activation map $A^{\left(k\right)}_t$ in the temporal domain, and makes noise in the gaze signal less significant. However, it also poses the risk to generate a distorted activation map. For instance, on a dynamic stimulus: the previous predicted attentive area in frame $I_{t-1}$ is located somewhere in frame $I_t$. Yet, $A^{\left(0\right)}_t$ provides values at this area and the Markov model will adapt it to the next salient region -- which may not have ben actually focused on. Nevertheless, this effect only occurs if the content of the scene changes significantly, for instance on scene cuts in movies, or opening a new web page while browsing.

When generating static heat maps (such as Figure~\ref{geisler2020exploiting:fig:teaser}), it is common to ignore the temporal domain completely. In this case, $q$ is set to zero. The overall heatmap $\mathbf{A}^{\left(k\right)}$ is then the weighted sum over $A^{\left(k\right)}_t$:
\begin{equation}
    \mathbf{A}^{\left(k\right)} = \sum\limits_{t} A^{\left(k\right)}_t \cdot b_t,
\end{equation}
where the weighting $b_t$, for instance, can be chosen in relation to the fixation time.

\paragraph{Parameters:} On regular gaussian like gaze heatmaps, $\sigma$ models the area of visual attention (foveal perception) and/or the expected noise of the eye tracking signal, and thus controls the acuity of the resulting heatmap. In the proposed approach, $\sigma$ controls the distribution of visual attention deduced from the fixation signal. But also how far the Markov model may adopt this distribution to the underlying stimulus in each iteration. The number of iterations is controlled by the parameter $k$. Whereby for $k=0$, $A^{\left(0\right)}_t=A^{\left(k\right)}_t$ corresponds to a regular gaussian like fixation heatmap of a single fixation point (respectively $\mathbf{A}^{\left(0\right)}$ overall fixation points). Figure~\ref{geisler2020exploiting:fig:sigma_k} shows how the initial gaze heatmap $A^{\left(0\right)}_t$ is gradually distorted to the stimulus for each additional iteration over equation~\ref{geisler2020exploiting:equ:Ak}.

\begin{figure*}
        \begin{minipage}[t]{1\linewidth}
            \begin{minipage}[t]{0.02\linewidth}
                \rotatebox{90}{\hspace{-0.05cm}{\small $k=0$}}
            \end{minipage}\hfill%
            \begin{minipage}[t]{0.108\linewidth}
                \includegraphics[width=\linewidth]{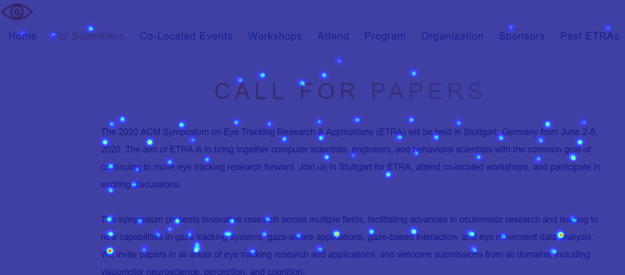}
            \end{minipage}\hfill%
            \begin{minipage}[t]{0.108\linewidth}
                \includegraphics[width=\linewidth]{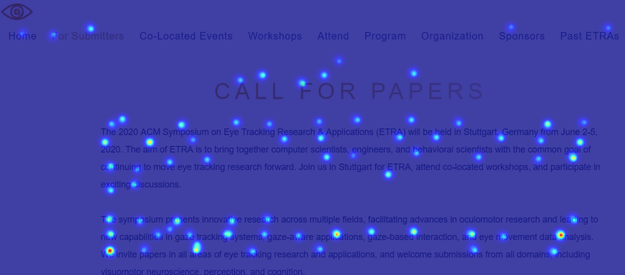}
            \end{minipage}\hfill%
            \begin{minipage}[t]{0.108\linewidth}
                \includegraphics[width=\linewidth]{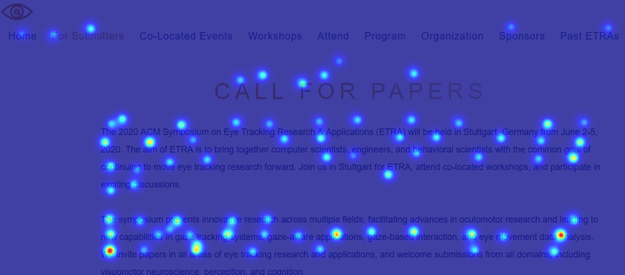}
            \end{minipage}\hfill%
            \begin{minipage}[t]{0.108\linewidth}
                \includegraphics[width=\linewidth]{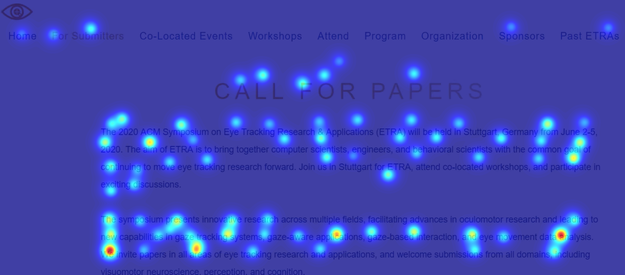}
            \end{minipage}\hfill%
            \begin{minipage}[t]{0.108\linewidth}
                \includegraphics[width=\linewidth]{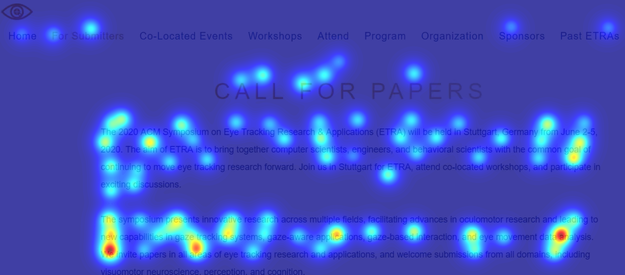}
            \end{minipage}\hfill%
            \begin{minipage}[t]{0.108\linewidth}
                \includegraphics[width=\linewidth]{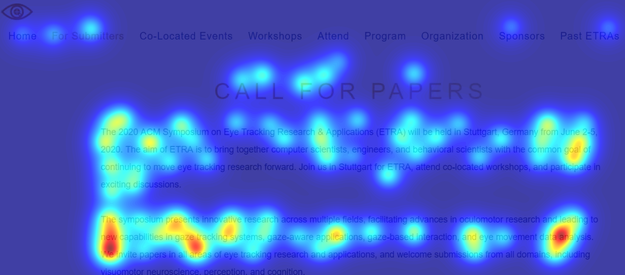}
            \end{minipage}\hfill%
            \begin{minipage}[t]{0.108\linewidth}
                \includegraphics[width=\linewidth]{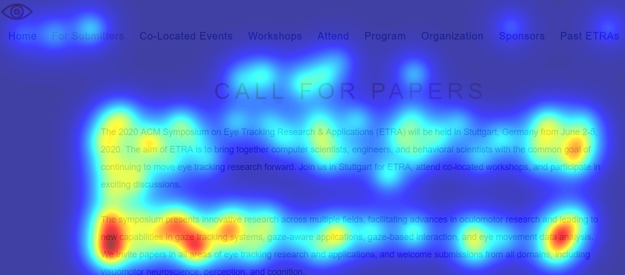}
            \end{minipage}\hfill%
            \begin{minipage}[t]{0.108\linewidth}
                \includegraphics[width=\linewidth]{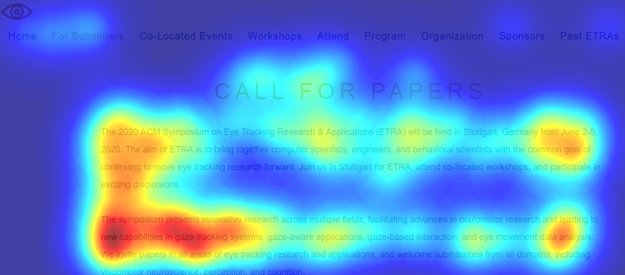}
            \end{minipage}\hfill%
            \begin{minipage}[t]{0.108\linewidth}
                \includegraphics[width=\linewidth]{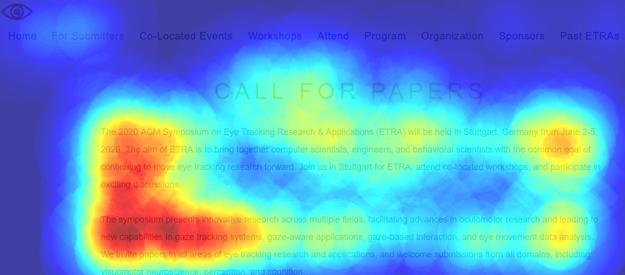}
            \end{minipage}
        \end{minipage}\\[0.1cm]
        \begin{minipage}[t]{1\linewidth}
            \begin{minipage}[t]{0.02\linewidth}
                \rotatebox{90}{\hspace{-0.05cm}{\small $k=1$}}
            \end{minipage}\hfill%
            \begin{minipage}[t]{0.108\linewidth}
                \includegraphics[width=\linewidth]{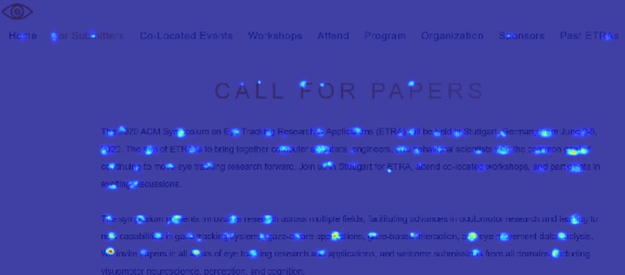}
            \end{minipage}\hfill%
            \begin{minipage}[t]{0.108\linewidth}
                \includegraphics[width=\linewidth]{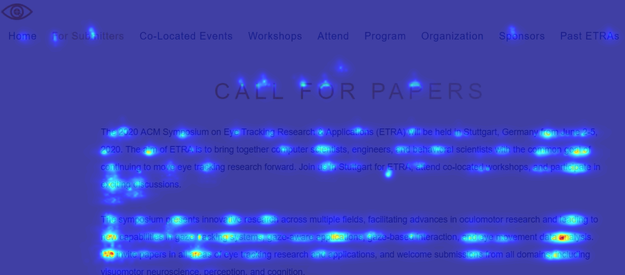}
            \end{minipage}\hfill%
            \begin{minipage}[t]{0.108\linewidth}
                \includegraphics[width=\linewidth]{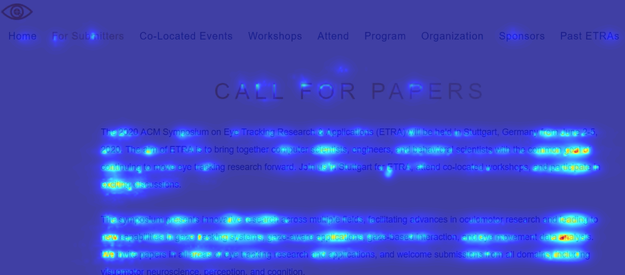}
            \end{minipage}\hfill%
            \begin{minipage}[t]{0.108\linewidth}
                \includegraphics[width=\linewidth]{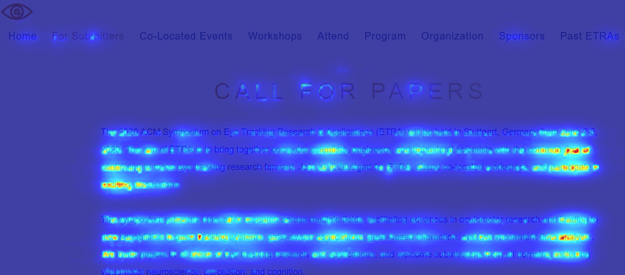}
            \end{minipage}\hfill%
            \begin{minipage}[t]{0.108\linewidth}
                \includegraphics[width=\linewidth]{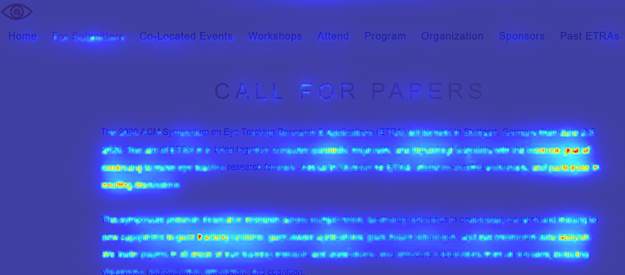}
            \end{minipage}\hfill%
            \begin{minipage}[t]{0.108\linewidth}
                \includegraphics[width=\linewidth]{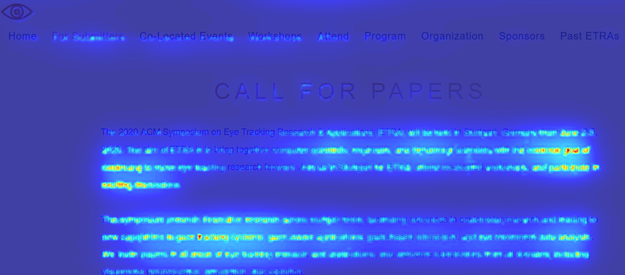}
            \end{minipage}\hfill%
            \begin{minipage}[t]{0.108\linewidth}
                \includegraphics[width=\linewidth]{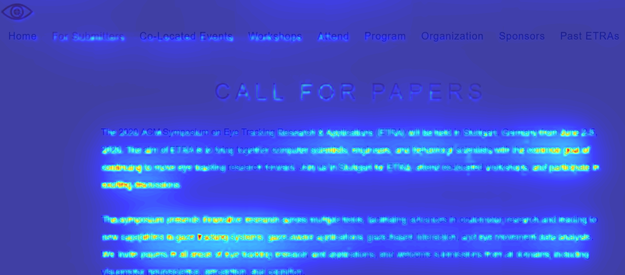}
            \end{minipage}\hfill%
            \begin{minipage}[t]{0.108\linewidth}
                \includegraphics[width=\linewidth]{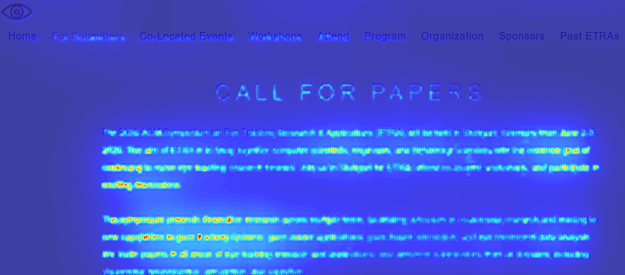}
            \end{minipage}\hfill%
            \begin{minipage}[t]{0.108\linewidth}
                \includegraphics[width=\linewidth]{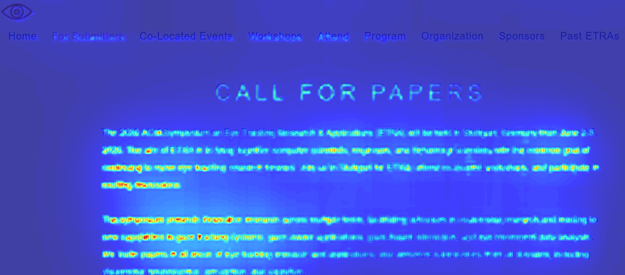}
            \end{minipage}
        \end{minipage}\\[0.1cm]
        \begin{minipage}[t]{1\linewidth}
            \begin{minipage}[t]{0.02\linewidth}
                \rotatebox{90}{\hspace{-0.05cm}{\small $k=2$}}
            \end{minipage}\hfill%
            \begin{minipage}[t]{0.108\linewidth}
                \includegraphics[width=\linewidth]{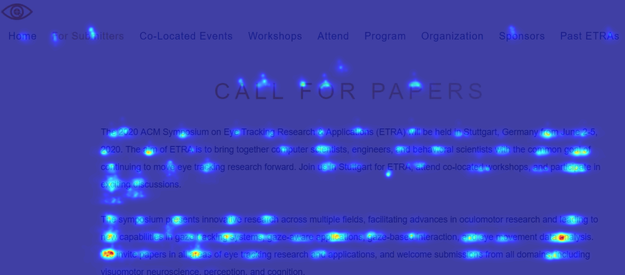}
            \end{minipage}\hfill%
            \begin{minipage}[t]{0.108\linewidth}
                \includegraphics[width=\linewidth]{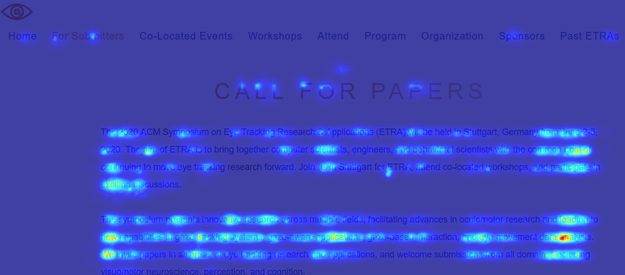}
            \end{minipage}\hfill%
            \begin{minipage}[t]{0.108\linewidth}
                \includegraphics[width=\linewidth]{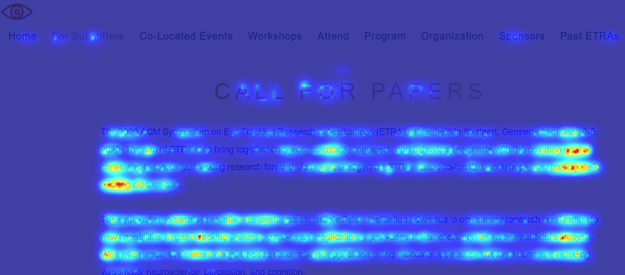}
            \end{minipage}\hfill%
            \begin{minipage}[t]{0.108\linewidth}
                \includegraphics[width=\linewidth]{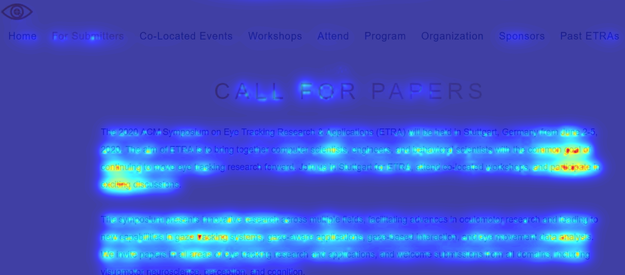}
            \end{minipage}\hfill%
            \begin{minipage}[t]{0.108\linewidth}
                \includegraphics[width=\linewidth]{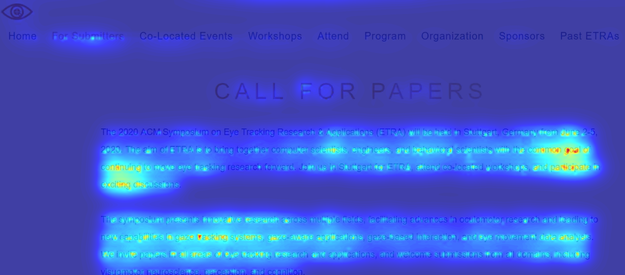}
            \end{minipage}\hfill%
            \begin{minipage}[t]{0.108\linewidth}
                \includegraphics[width=\linewidth]{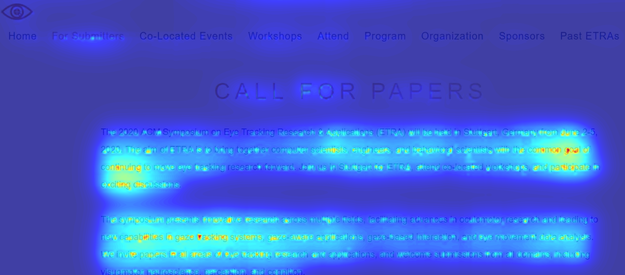}
            \end{minipage}\hfill%
            \begin{minipage}[t]{0.108\linewidth}
                \includegraphics[width=\linewidth]{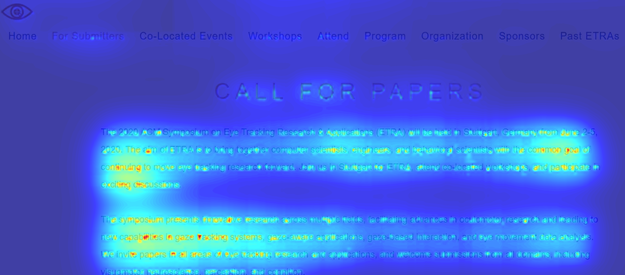}
            \end{minipage}\hfill%
            \begin{minipage}[t]{0.108\linewidth}
                \includegraphics[width=\linewidth]{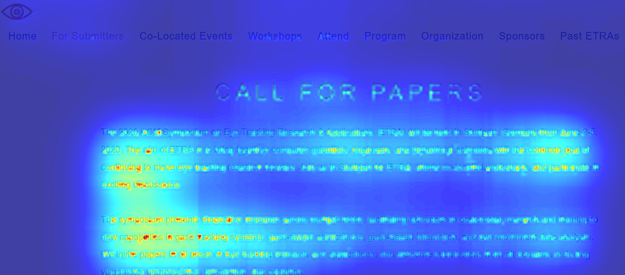}
            \end{minipage}\hfill%
            \begin{minipage}[t]{0.108\linewidth}
                \includegraphics[width=\linewidth]{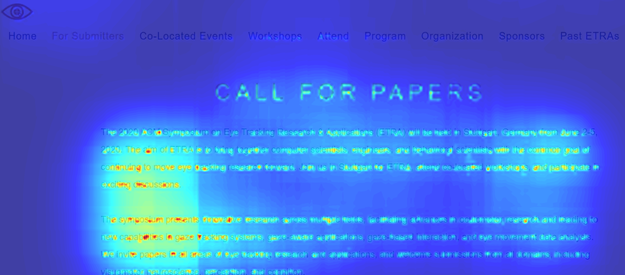}
            \end{minipage}
        \end{minipage}\\[0.1cm]
        \begin{minipage}[t]{1\linewidth}
            \begin{minipage}[t]{0.02\linewidth}
                \rotatebox{90}{\hspace{-0.05cm}{\small $k=4$}}
            \end{minipage}\hfill%
            \begin{minipage}[t]{0.108\linewidth}
                \includegraphics[width=\linewidth]{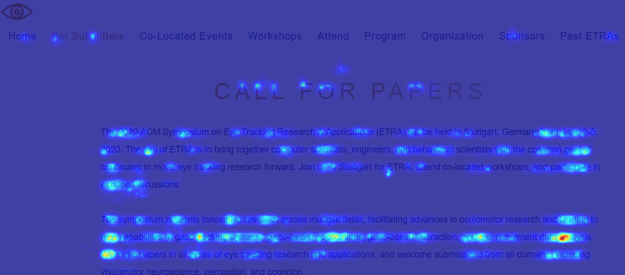}
            \end{minipage}\hfill%
            \begin{minipage}[t]{0.108\linewidth}
                \includegraphics[width=\linewidth]{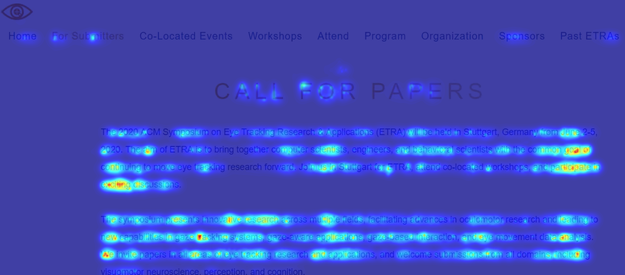}
            \end{minipage}\hfill%
            \begin{minipage}[t]{0.108\linewidth}
                \includegraphics[width=\linewidth]{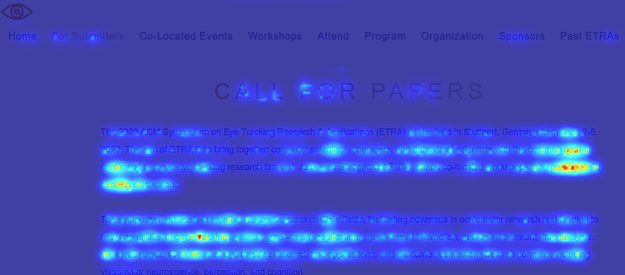}
            \end{minipage}\hfill%
            \begin{minipage}[t]{0.108\linewidth}
                \includegraphics[width=\linewidth]{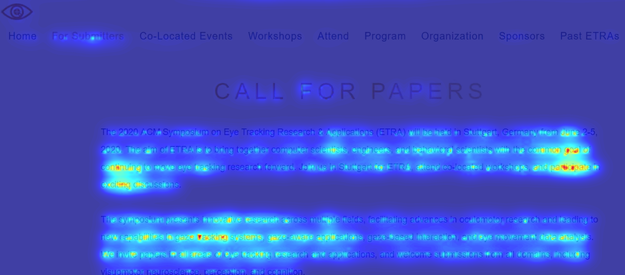}
            \end{minipage}\hfill%
            \begin{minipage}[t]{0.108\linewidth}
                \includegraphics[width=\linewidth]{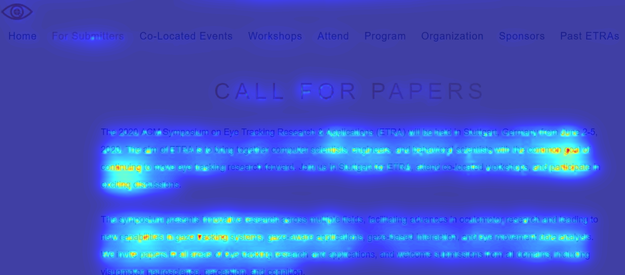}
            \end{minipage}\hfill%
            \begin{minipage}[t]{0.108\linewidth}
                \includegraphics[width=\linewidth]{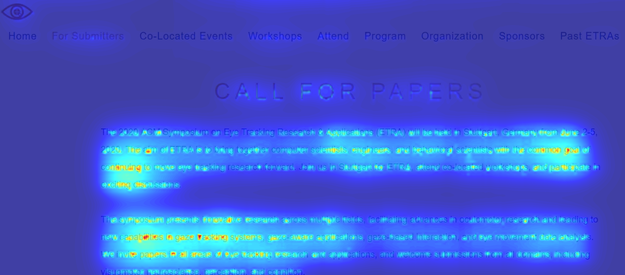}
            \end{minipage}\hfill%
            \begin{minipage}[t]{0.108\linewidth}
                \includegraphics[width=\linewidth]{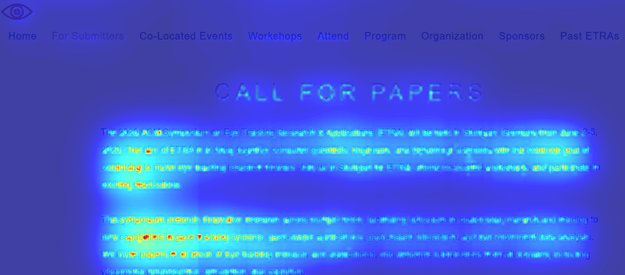}
            \end{minipage}\hfill%
            \begin{minipage}[t]{0.108\linewidth}
                \includegraphics[width=\linewidth]{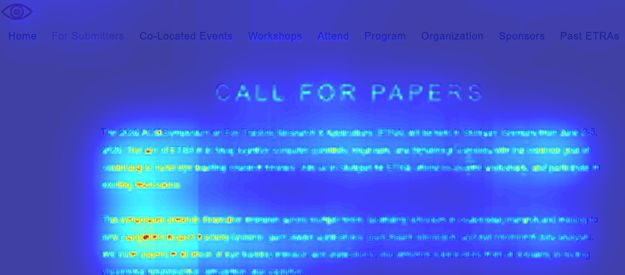}
            \end{minipage}\hfill%
            \begin{minipage}[t]{0.108\linewidth}
                \includegraphics[width=\linewidth]{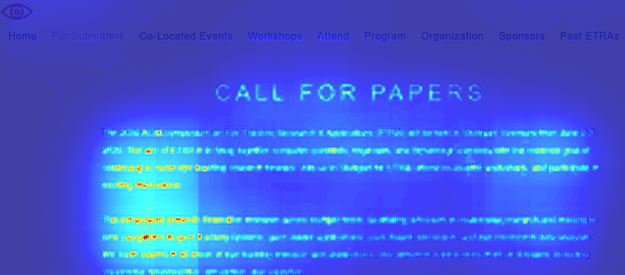}
            \end{minipage}
        \end{minipage}\\
        \begin{minipage}[t]{1\linewidth}
            \begin{minipage}[t]{0.02\linewidth}
                \hfill%
            \end{minipage}\hfill%
            \begin{minipage}[t]{0.108\linewidth}
                \centering \small $\sigma=1.0$
            \end{minipage}\hfill%
            \begin{minipage}[t]{0.108\linewidth}
                \centering \small $\sigma=2.0$
            \end{minipage}\hfill%
            \begin{minipage}[t]{0.108\linewidth}
                \centering \small $\sigma=4.0$
            \end{minipage}\hfill%
            \begin{minipage}[t]{0.108\linewidth}
                \centering \small $\sigma=8.0$
            \end{minipage}\hfill%
            \begin{minipage}[t]{0.108\linewidth}
                \centering \small $\sigma=16.0$
            \end{minipage}\hfill%
            \begin{minipage}[t]{0.108\linewidth}
                \centering \small $\sigma=32.0$
            \end{minipage}\hfill%
            \begin{minipage}[t]{0.108\linewidth}
                \centering \small $\sigma=64.0$
            \end{minipage}\hfill%
            \begin{minipage}[t]{0.108\linewidth}
                \centering \small $\sigma=128.0$
            \end{minipage}\hfill%
            \begin{minipage}[t]{0.108\linewidth}
                \centering \small $\sigma=256.0$
            \end{minipage}
        \end{minipage}
        %
        \caption{Influence of the parameters $\sigma$ (horizontal) and $k$ (vertical) on the adaptation of the fixations heatmap to the text stimulus. The input is the same as used in figure~\ref{geisler2020exploiting:fig:text}. For large $k$, the injected visual attention is increasingly distributed across the entire stimulus. The parameter $\sigma$ should be chosen depending on the desired level of detail of the visualization. Using a text stimulus, it is usually reasonable to choose a high degree of detail (here $\sigma \leq 8$), in order to visualize the perception rate of single words or lines.}
        \label{geisler2020exploiting:fig:sigma_k}
\end{figure*}

\paragraph{Implementation Details:}
The main limitation of \acs{GBVS} is runtime and memory consumption. The transition matrix $T_t$ grows in quadratic size with the input size $n$, and thus quickly exceeds the available memory (e.g. $>9.4\cdot10^{10}$ elements on a VGA resolution). Additionally, the initialization of $T_t$ requires a runtime complexity of $\mathcal{O}\left(n^2\right)$. Both together, limit the \acs{GBVS} to very low input resolutions, which leads to a loss of acuity. Thus, the standard parametrization of the \acs{GBVS} toolbox limits the internal resolution to an edge length of $32\text{px}$~\cite{harel2006saliency}.

On closer examination, however, it is apparent that most values in $T_t$ are extremely small and have no significant impact to the resulting activation map $A^{\left(k\right)}_t$. Thus, after applying a threshold $l$, the transition matrix $T_t$ becomes predominantly sparse. Furthermore, assuming that $M_t\in\left[0,1\right]$, the elements of $T_t$, which potentially exceed the threshold $l$ can be determined in relation to $\sigma$:
\begin{equation}\label{geisler2020exploiting:equ:l0}
    l < F\left(i,j\right),
\end{equation}
and resolves to:
\begin{equation}\label{geisler2020exploiting:equ:l1}
    \sqrt{-2\cdot\sigma\cdot\log{\left(l\right)}} \geq \sqrt{
        \left(x\left(i\right)-x\left(j\right)\right)^2+
        \left(y\left(i\right)-y\left(j\right)\right)^2
        },
\end{equation}
where the right term is the euclidean distance between the the $i$-th and $j$-th element in the feature map $M_t$. Thus, initializing $T_t$ only requires the calculation of $2\cdot\sqrt{-2\cdot\sigma\cdot\log{\left(l\right)}}$ elements per row, since all other elements are not exceeding the threshold $l$. This reduces the actual runtime from $\mathcal{O}\left(n^2\right)$ to $\mathcal{O}\left(n\right)$. Similar considerations can be made for the initialization of $A^{\left(0\right)}_t$ (although this is not a bottleneck). However, due to the sparseness of $T_t$ and $A^{\left(0\right)}_t$, solving equation~(\ref{geisler2020exploiting:equ:Ak}) is much faster~\cite{yuster2005fast}.
\subsection{Experimental Demonstration}\label{geisler2020exploiting:sec:evaluation}
Figures~\ref{geisler2020exploiting:fig:teaser}, \ref{geisler2020exploiting:fig:text}, and~\ref{geisler2020exploiting:fig:movie} demonstrate the application of the proposed visualization on different stimulus types: the \emph{An Unexpected Visitor} painting from Ilya Repin, the \emph{Call for Papers} website from ETRA 2020 as text, and a short video snippet of \emph{Big Buck Bunny} from the Peach open movie project~\cite{bbb}. The gaze signal was recorded by a Tobii Pro Spectrum at $1200\text{Hz}$. The fixation locations and duration were extracted using the fixation filter I-VT provided by Tobii Pro Lab and default parametrization~\cite{olsen2012tobii}. All stimuli were presented as full screen on the Monitor at $1920 \times 1080$ pixels.

\begin{figure}[b]
    \begin{minipage}[t]{0.33\linewidth}
        \begin{center}
            \includegraphics[width=\linewidth]{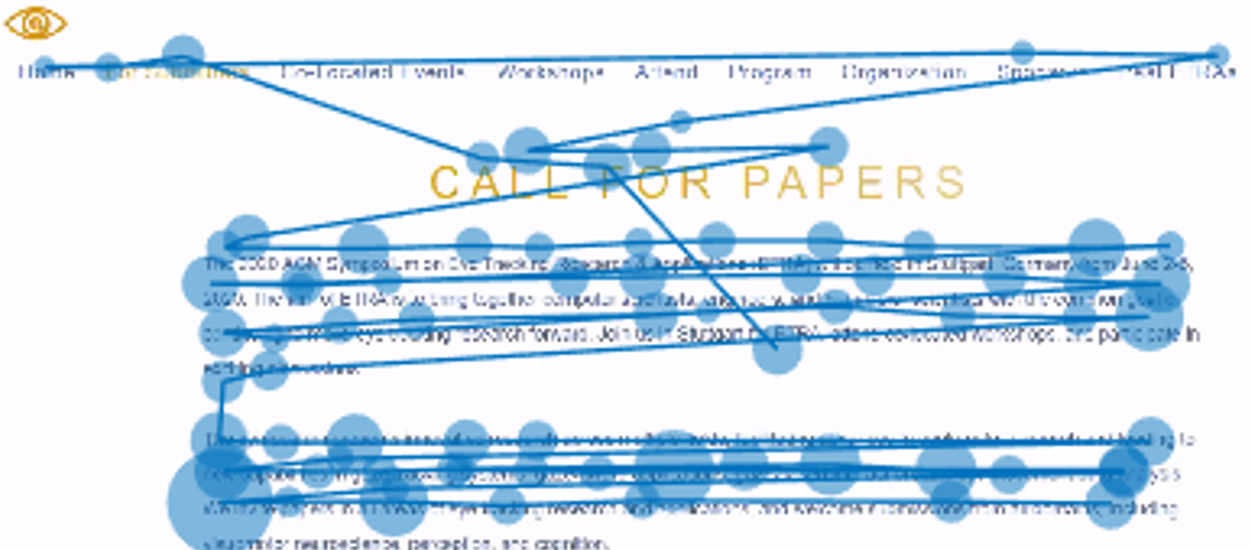}
            Input fixation map\\ \& stimuli
        \end{center}
    \end{minipage}\hfill
    \begin{minipage}[t]{0.33\linewidth}
        \begin{center}
            \includegraphics[width=\linewidth]{geisler2020exploiting/parameter/output_param_test_web_sigma_32_k_0_gbvs_heatmap.png}
            Gaze Heatmap\\ ($\sigma=32.0$)
        \end{center}
    \end{minipage}\hfill%
    \begin{minipage}[t]{0.33\linewidth}
        \begin{center}
            \includegraphics[width=\linewidth]{geisler2020exploiting/parameter/output_param_test_web_sigma_4_k_2_gbvs_heatmap.png}
            \acs{GBVS} Heatmap\\ ($\sigma=4.0$, $k=2$)
        \end{center}
    \end{minipage}
    \caption{The left image shows the recorded fixation sequence on the ETRA 2020s \emph{Call for Papers} website. The middle image shows a regular gaussian fixation heatmap ($\mathbf{A}^{\left(0\right)}$). The right image shows the output of the proposed method ($\mathbf{A}^{\left(k=2\right)}$).}
    \label{geisler2020exploiting:fig:text}
\end{figure}

On the text stimulus, it is recognizable how the Markov model depicts the measured visual attention to paragraphs, lines, down to single words and characters. Therefore, the acuity of the heatmap is increased, and consequently, interpretations about the perception rate to text passages are simplified. For instance, in the field of web design and advertising, the proposed model can help to analyze whether a certain area attracts the desired level of visual attention and whether the presented information was easily visual accessible to the spectator.

However, in this context, text reading is a relatively unambiguous challenge, since the text is very salient to its background. At the same time, the text is often the only element that attracts the visual attention of the reader. The strength of the proposed method of visual attention visualization is particularly evident in more complex stimuli as shown in figures~\ref{geisler2020exploiting:fig:teaser} and~\ref{geisler2020exploiting:fig:movie}. Considering the \emph{An Unexpected Visitor} painting, the fixations are mainly on the faces in the scene, but also on some miscellaneous areas, such as hands, the paintings in the background, or feet. However, the regular fixation heatmap has a particularly pronounced fixation cluster on the face of the woman in the background. This can be attributed to the fact that this face is particularly difficult to perceive due to its low contrast. Yet, the long and frequent fixations in this area lead to a suppression of all other fixations, which can lead to the interpretation that this area was of higher interest for the spectator. The \acs{GBVS} generated fixation heatmap incorporates not only the fixation duration and frequency but also how accessible the stimulus in the region is to the observer. The result is a much more balanced fixation heatmap, where all the fixated heads are clearly pronounced.

\begin{figure}
    \begin{minipage}[t]{0.33\linewidth}
        \begin{center}
            \includegraphics[width=\linewidth]{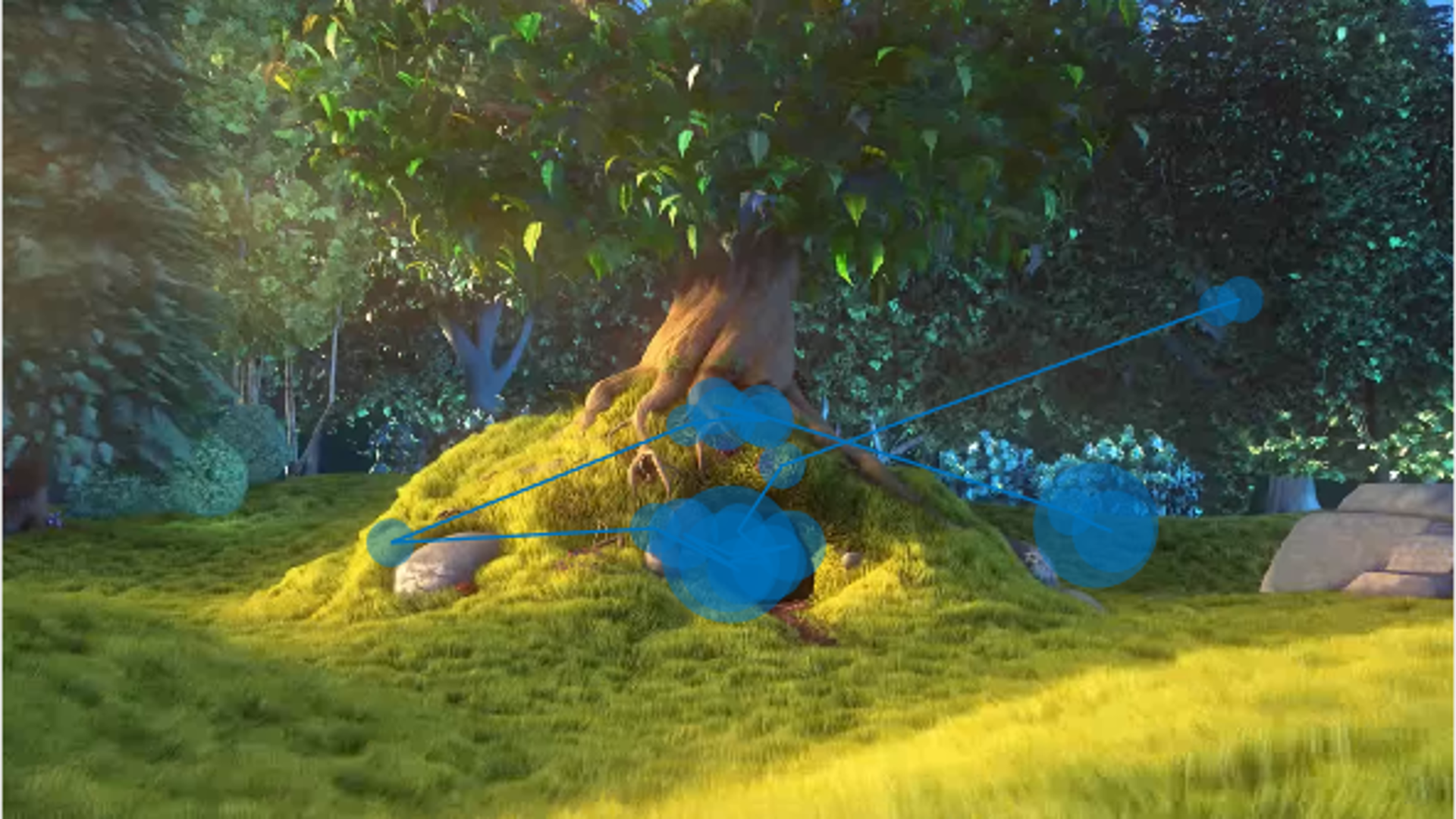}
            Input fixation map\\ \& stimuli
        \end{center}
    \end{minipage}\hfill
    \begin{minipage}[t]{0.33\linewidth}
        \begin{center}
            \includegraphics[width=\linewidth]{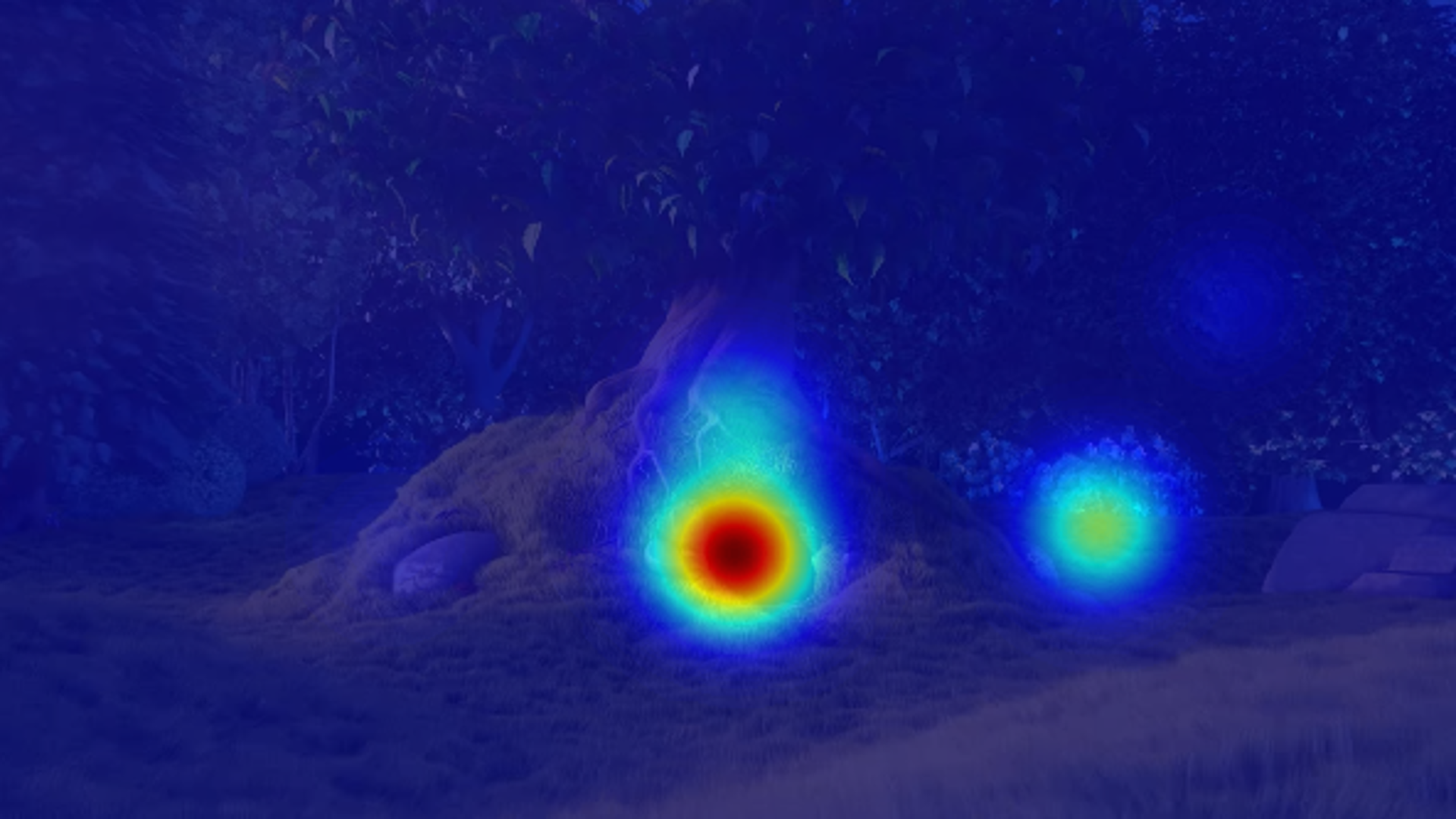}
            Gaze Heatmap\\ ($\sigma=16.0$)
        \end{center}
    \end{minipage}\hfill%
    \begin{minipage}[t]{0.33\linewidth}
        \begin{center}
            \includegraphics[width=\linewidth]{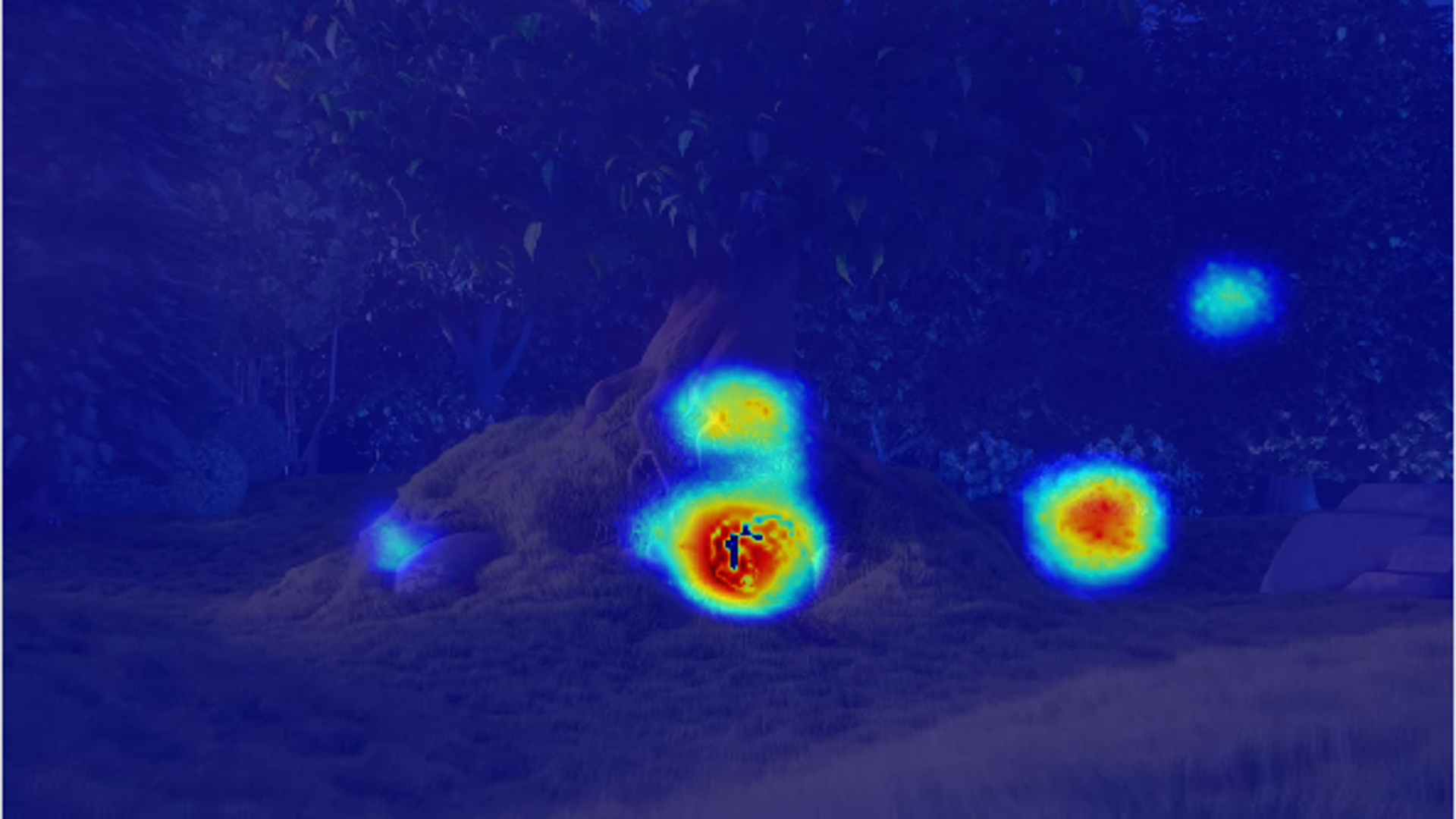}
            \acs{GBVS} Heatmap\\ ($\sigma=8.0$, $k=2$)
        \end{center}
    \end{minipage}
    \caption{The left image shows the recorded fixation sequence on a short snippet of the video clip \emph{Big Buck Bunny} from the Peach open movie project~\cite{bbb}. The middle image shows a regular gaussian fixation heatmap ($\mathbf{A}^{\left(0\right)}$). The right image shows the output of the proposed method ($\mathbf{A}^{\left(k=2\right)}$).}
    \label{geisler2020exploiting:fig:movie}
\end{figure}

\subsection{Final Remarks}\label{geisler2020exploiting:sec:conclusion}
The proposed method extends the well-known \acs{GBVS} saliency algorithm by incorporating the measured visual attention. The resulting heatmap visualizes a predicted perception rate of scene areas for an individual or multiple spectators. However, as the most bottom-up saliency algorithm, \acs{GBVS} uses exclusively intrinsic scene features to predict whether certain scene content is attractive for fixation. It turns out, this is very accurate for a free viewing scenario. Yet, various tasks may require the spectator to direct their visual attention to less saliency scene areas. The proposed algorithm might distort these fixation points to a close salient region and thus weigh the perception rate based on the wrong stimuli. This limitation can be compensated by using a small $\sigma$ and high-resolution scene images but requires a high accuracy of the fixation point.

In practice, however, it has been shown that the proposed visualization generates more intuitive heatmaps than pure fixation heatmaps. Thus, the presented visualization provides an ingenious overview of the scene areas with a distinctive high rate of visual awareness.
\fi

\chapter{Perceiving and Multiperspective Teaching of Unknown Objects}
\label{app:B}

The following publications are enclosed in this chapter:
\begin{enumerate}
	\item[\cite{weber2020distilling}]\label{app:itm:weber2020distilling} 
	\textbf{Daniel Weber}, Thiago Santini, Andreas Zell, and Enkelejda Kasneci. Distilling Location Proposals of Unknown Objects through Gaze Information for Human-Robot Interaction. In \textit{2020 IEEE/RSJ International Conference on Intelligent Robots and Systems (IROS)}, pages 11086–11093. IEEE, 2020. \href{https://doi.org/10.1109/IROS45743.2020.9340893}{doi:10.1109/IROS45743.2020.9340893}.
	
	\item[\cite{weber2022exploiting}]\label{app:itm:weber2022exploiting} 
	\textbf{Daniel Weber}, Enkelejda Kasneci, and Andreas Zell. Exploiting Augmented Reality for Extrinsic Robot Calibration and Eye-based Human-Robot Collaboration. In \textit{Proceedings of the 2022 ACM/IEEE International Conference on Human-Robot Interaction (HRI)}, pages 284–293. IEEE, 2022. \href{https://doi.org/10.1109/HRI53351.2022.9889538}{doi:10.1109/HRI53351.2022.9889538}.
	
	\item[\cite{weber2023multiperspective}]\label{app:itm:weber2023multiperspective} 
	\textbf{Daniel Weber}, Wolfgang Fuhl, Enkelejda Kasneci, and Andreas Zell. Multiperspective Teaching of Unknown Objects via Shared-gaze-based Multimodal Human-Robot Interaction. In \textit{Proceedings of the 2023 ACM/IEEE International Conference on Human-Robot Interaction (HRI)}, pages 544–553, March 2023. \href{https://doi.org/10.1145/3568162.3578627}{\nolinkurl{doi:10.1145/3568162.3578627}}.
	
	\item[\cite{weber2023leveraging}]\label{app:itm:weber2023leveraging} 
	\textbf{Daniel Weber}, Valentin Bolz, Andreas Zell, and Enkelejda Kasneci. Leveraging Saliency-Aware Gaze Heatmaps for Multiperspective Teaching of Unknown Objects. In \textit{2023 IEEE/RSJ International Conference on Intelligent Robots and Systems (IROS)}, 2023. (Accepted for publication).
\end{enumerate}

\blfootnote{
	\hspace{-16.5pt}
	{\scriptsize The publications templates have been slightly adapted to match the formatting of this dissertation.
	The ultimate versions are accessible via the digital object identifier at the respective publisher.
	Publication~\hyperref[app:itm:weber2023multiperspective]{\protect\NoHyper\cite{weber2023multiperspective}\protect\endNoHyper} is \textcopyright~2023~ACM.
	Publications~\hyperref[app:itm:weber2020distilling]{\protect\NoHyper\cite{weber2020distilling}\protect\endNoHyper} and \hyperref[app:itm:weber2022exploiting]{\protect\NoHyper\cite{weber2022exploiting}\protect\endNoHyper} are \textcopyright~2020~IEEE and \textcopyright~2022~IEEE, respectively, and reprinted, with permission, from~\hyperref[app:itm:weber2020distilling]{\protect\NoHyper\cite{weber2020distilling}\protect\endNoHyper} and \hyperref[app:itm:weber2022exploiting]{\protect\NoHyper\cite{weber2022exploiting}\protect\endNoHyper}.
	In reference to IEEE copyrighted material which is used with permission in this thesis, the IEEE does not endorse any of University of Tübingen’s products or services. Internal or personal use of this material is permitted. If interested in reprinting/republishing IEEE copyrighted material for advertising or promotional purposes or for creating new collective works for resale or redistribution, please go to \url{http://www.ieee.org/publications_standards/publications/rights/rights_link.html} to learn how to obtain a License from RightsLink. If applicable, University Microfilms and/or ProQuest Library, or the Archives of Canada may supply single copies of the dissertation.}
}

\clearpage
\setcounter{footnote}{0}
\section{Distilling Location Proposals of Unknown Objects through Gaze Information for Human-Robot Interaction}
\label{app:sec:paper01}

\ifpaper
\subsection{Abstract}
Successful and meaningful human-robot interaction requires robots to have knowledge
about the interaction context -- e.g., which objects should be interacted
with.
Unfortunately, the corpora of interactive objects is -- for all practical
purposes -- infinite. This fact hinders the deployment of robots with
pre-trained object-detection neural networks other than in pre-defined
scenarios.
A more flexible alternative to pre-training is to let a human teach the
robot about new objects after deployment.
However, doing so manually presents significant usability issues as the user
must manipulate the object and communicate the object's boundaries to the
robot.
In this work, we propose streamlining this process by using automatic
object location proposal methods in combination with human gaze
to distill pertinent object location proposals.
Experiments show that the proposed method 1) increased the precision by a
factor of approximately 21 compared to location proposal alone, 2) is able
to locate objects sufficiently similar to a state-of-the-art pre-trained
deep-learning method (FCOS) without any training, and 3) detected objects
that were completely missed by FCOS.
Furthermore, the method is able to locate objects for which FCOS was not
trained on, which are undetectable for FCOS by definition.

\subsection{Introduction}
In today's modern world, interaction between human and machines is omnipresent, e.g.\ in the figure of Alexa and Siri.
Moreover, the significant progress in augmented reality is also pushing the boundaries of the cooperation between human and robots.  For instance, this emerging kind of human-robot interaction (\acs{HRI}) has already helped to optimize manufacturing steps in production as well as been applied in factories \cite{nee2012augmented} and for assembly guidance \cite{rentzos2013augmented}.
The great majority of such technological developments has been strongly fueled by machine learning methods, such as neural networks.
The collection of huge databases allows us to train and continuously improve (deep) neural networks in order to fulfill challenging tasks.
However, these use cases typically operate under the assumption that there are sufficient data sets for training available.
But what if the training data is biased (e.g., geographically~\cite{devries2019}) or not labeled, for example in many production processes -- such as, the assembly of a recently developed electric engine of a car?
Furthermore, in some application scenarios such as search and rescue work with unmanned aerial vehicles (\acsp{UAV}) or the classification of medical images, there might be very few or no available training instances.
For example for \acsp{UAV}, aerial footage is simply difficult to obtain \cite{bejiga2017convolutional}
, whereas for medical images, storage is often prohibited due to patient privacy \cite{cho2015much}.

In addition, labeling data is a costly process due to the amount of human effort involved.
Drawing a high quality bounding box in an image, including quality and coverage verification, can take a human from 7 up to 42 seconds per object \cite{papadopoulos2017extreme, su2012crowdsourcing}.
With multiple objects in a scene this can quickly add up to prohibitive amounts.

In this paper we address this challenge by connecting findings from two research areas: eye tracking and robotics.
On the human side, we use the gaze modality to enable the exchange of information for a specific problem on the robot side, namely the detection of unknown objects.
Our goal is to enable the deployment of a robot in a non-predefined scenario and to explain an interaction context to the robot, e.g., the class of an object after detection.
That is, rather than using a neural network for object detection, we resort to the human gaze and want the robot to detect which object the human is looking at, even though its class is not yet known (see Figure~\ref{paper01:fig:teaser}). Moreover, interaction requires online operation -- in contrast to post processing.
\begin{figure}[t]
	\centering
	\includegraphics[width=\linewidth]{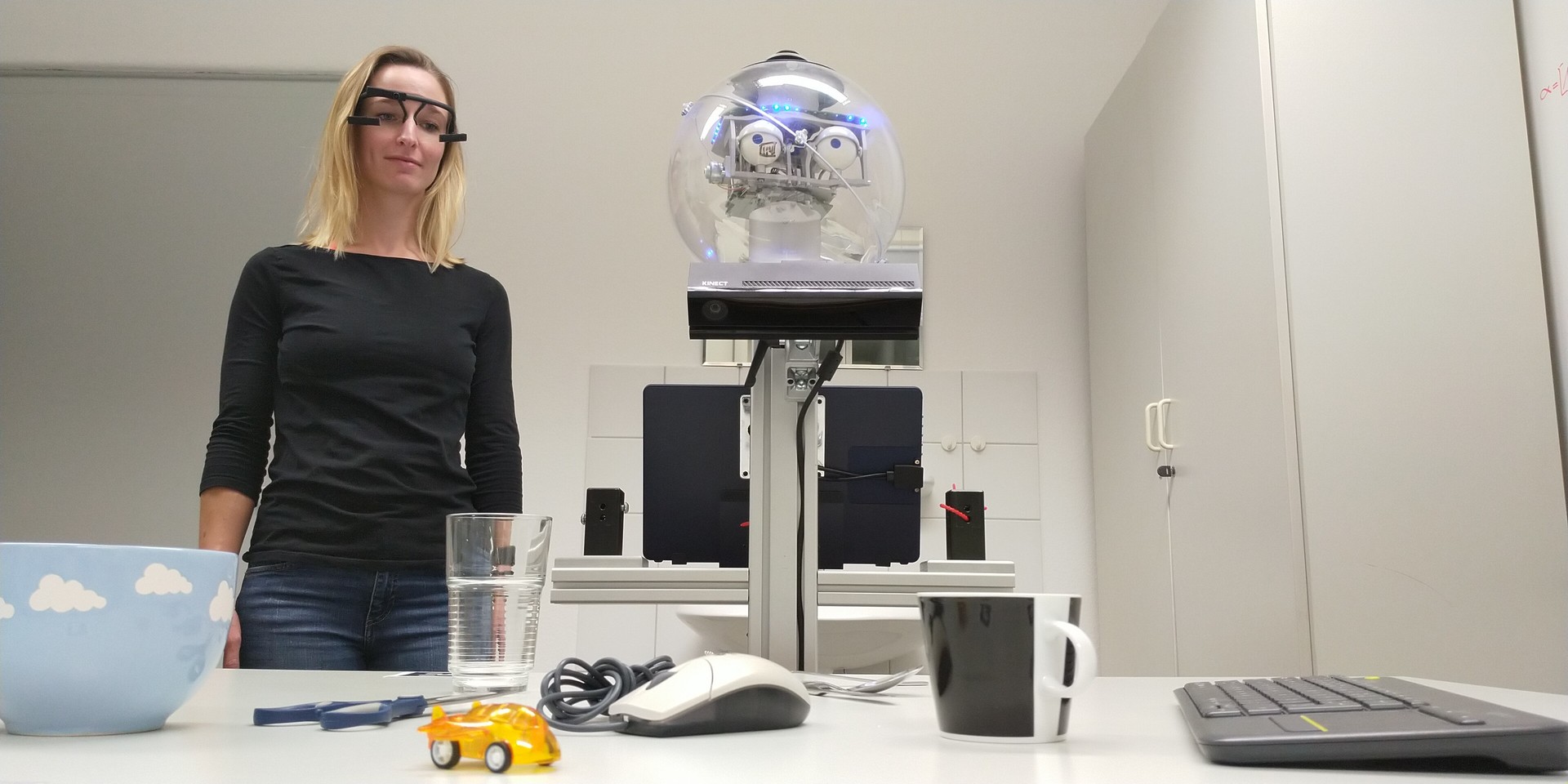}
	\caption{Without specialized pre-training, the robot does not know the
		objects in front of it. Nonetheless, through our proposed approach, the robot
		is capable of detecting these unknown objects based on gaze-based
		human-robot interaction without any training instances.}
	\label{paper01:fig:teaser}
\end{figure}
To the best of our knowledge, this is the first work to combine the well known technique of selective search \cite{uijlings2013selective}, which outputs thousands of class-independent object location proposals, with human gaze information to separate useful and useless areas of interest in a scene image.
Thus, the proposed approach enables us to detect and process objects in an image without training but still in an efficient way.
In summary, our most important contributions are:
\begin{enumerate}
	\item A novel method towards the deployment of robots in non-predefined scenarios.
	\item We are the first to connect eye tracking and robotics to detect unknown objects without the usage of neural networks, alleviating training-data dependency.
	\item As a proof of concept we conduct an experiment and demonstrate the validity and feasibility of our method.
\end{enumerate}

\subsection{Related Work}

Mapping gaze data from a head-mounted eye tracker with moving point of view, i.e. coordinate system, to a known reference frame is a well-known and open problem in current research.
Most works, such as \cite{huang2016anticipatory, peysakhovich2018aruco}, that were confronted with this issue solved it by using fiducial markers.
Even though \cite{kalash2016gaze} additionally tested feature matching and achieved reasonable results, markers provided better stability and reliability at significantly less computational cost in all of their test cases.
Apart from that, their purpose was to match a picture of an image displayed on a screen to a planar reference image, which was very similar to the one displayed on the screen.
As described in \cite{macinnes2018wearable}, feature matching reaches its limit when applied to a three-dimensional target object.
Accordingly, in our case it is more difficult to find and match features than with a simple painting or a poster, especially when the viewing perspectives differ significantly.
In \cite{aronson2019semantic} the authors succeeded in mapping the gaze by utilizing velocity features. However, this was limited to the user looking at one of several pre-defined key points.

In recent years, object recognition has been one of the most intensively researched areas in computer vision.
The availability of better hardware led to the emergence of deep neural networks as a go-to solution for object detection.
YOLO \cite{redmon2016you},
Mask R-CNN \cite{He2017Mask}, SSD \cite{liu2016ssd} and FCOS \cite{tian2019fcos}  are great examples of the extensive use of neural networks that constantly have been pushing the boundaries of object detection.
These networks typically rely on fully supervised learning methods and the existence of large annotated data sets, such as PASCAL VOC \cite{everingham2010pascal}, Microsoft COCO \cite{lin2014microsoft} and Imagenet \cite{deng2009imagenet}.
This means that they do not generalize well and lack reliability on unknown domains \cite{wilber2017bam}.

Moreover, with increasing climate-related public awareness, there has been some research focusing on energy efficiency of neural networks
\cite{chen2016eyeriss}
and its environmental impact \cite{strubell2019energy}.
\cite{canziani2016analysis}
analyzed the power consumption of popular image classification models.
Consequently, we follow the recommendation of \cite{strubell2019energy} and
prioritize a simple non-deep-learning approach instead.

A few works have already investigated the combination of eye tracking and computer vision tasks.
The authors of \cite{toyama2012gaze} performed gaze guided object recognition by matching features around human fixations to features from known objects in a database.
After a database was created, it was possible to classify an image, but not to determine the position of the object within the image.
\cite{papadopoulos2014training} concentrated on annotating images with bounding boxes.
They utilized fixation points to extend existing training data with gaze information.
Subsequently, a model was trained that predicted bounding boxes from the fixations while viewing an image.
A strategy for superpixel segmentation with eye tracking data was proposed by \cite{xiao2018salient}.
Just like the previous method, training data was already required right from the start.
In addition, both methods require multiple gaze points. In contrast, our method is able to operate with as few as one gaze point, thus being applicable in an online fashion.

In this paper, we build on existing work and benefit from collaborative working with a robot.
In this context, eye tracking can play an important role and connect humans and
robots in a natural and intuitive manner, offering an additional communication
channel available even when traditional channels, such as speech and
gestures~\cite{stiefelhagen2004natural}, might not be available for \acs{HRI} -- e.g.,
during microsurgery~\cite{fuhl2016nonintrusive}.
We use the human gaze to enable a robot to interact with its unknown environment by letting it recognize objects we are looking at.
Thereby, we bridge the gap between existing approaches for object detection and data independence with eye tracking.

\subsection{Method}
\label{paper01:sec:method}

In this work, we propose finding pertinent and accurate location proposals of
unknown objects through gaze information.
This process can be thought of as three building blocks: 1) estimating the human
partner's gaze in the robot's frame of reference, 2) generating location
proposals for unknown objects, and 3) distilling the location proposals using the
gaze information.
Throughout this section, we assume the robot to be equipped with at least one
camera.

\subsubsection{Gaze Estimation}
\label{paper01:subsec:gaze}

The most straightforward and inexpensive way of estimating the partner's gaze in
the robot's frame of reference is by estimating the gaze directly through the
robot's sensors -- e.g., through appearance or model-based~
remote gaze estimation methods~\cite{%
zhang2018revisiting,%
palinko2016robot}.
However, this poses a key limitation as the partner must be facing the robot,
severely limiting the perspectives from which gaze-based \acs{HRI} can happen.

This limitation can be alleviated through multiple remote eye trackers
distributed around the environment or the usage of a head-mounted eye tracker.
However, in both cases, it is necessary to map the estimated gaze from the eye
tracker frame of reference to the robot's.
This transformation can be achieved in multiple ways, for example by 1) directly finding the
eye tracker's pose in the robot's camera or vice versa, or 2) indirect
co-location, by finding at least four corresponding points in images of the eye
tracker's and robot's cameras\footnote{By finding the plane defined by the these
four points, one can estimate the pose of each camera relative to the plane and,
thus, the pose of one camera relative to the other.}.

In this work, we favor the usage of a head-mounted eye tracker due to the
reduced costs (i.e., only a single eye tracker is required) and user
constraints. Moreover, we employ fiducial markers~\cite{garrido2014automatic
} for co-location as these
provide a robust and inexpensive solution to the gaze mapping issue that can
be employed in traditional \acs{HRI} scenarios such as in factories, care facilities,
or individual homes.

\subsubsection{Unknown Object Location Proposal}
\label{paper01:subsec:proposal}

Location (or region) proposal methods consist of determining candidate object
locations (e.g., bounding boxes, or segmentation masks) that \emph{might}
contain an object.
This task can be realized, for example, through
segmentation~\cite{carreira2010constrained}, randomly-sampled boxes
classification~\cite{alexe2012measuring}, jumping
windows~\cite{vedaldi2009multiple}, and selective
search~\cite{uijlings2013selective}.
Such methods are typically used as an alternative to exhaustive search for
object detection to reduce the search space, speeding up the detection and
reducing the associated computing costs.

The cardinality of the proposed locations set is, naturally, image-dependent but
tends to be in the order of thousands. Normally, each location proposal is run
through a pre-trained classifier to detect whether an object is present in it.
However, many of these methods, such as the ones proposed by
\cite{
	uijlings2013selective,alexe2012measuring},
have a particularly interesting property: The proposed locations are
\emph{class-independent}.
In other words, within the proposed regions there are objects that a computer
vision system might not have been trained to identify~--~i.e., unknown objects.
This begs the question: Can we identify pertinent locations from the set of
proposals for interaction or to teach a robot about new objects in a natural way?

\subsubsection{Distillation Through Gaze Information}
\label{paper01:subsec:distillation}

In this work, we approach the task of identifying location proposals that are
pertinent for a human-robot interaction from the full set of class-independent
proposals by using gaze information from the human partner.
This distillation process can be activated through multimodal interactions
-- e.g., through touch or voice. Nevertheless, we also envision an automatic
approach in which the robot notices the human's gaze continously attending to a region
where no known object has been identified yet.

In order to obtain an initial set of candidate bounding boxes, we resort to selective search \cite{uijlings2013selective}.
Selective search uses the segmentation method from Felzenszwalb and Huttenlocher \cite{felzenszwalb2004efficient} to analyze the intensity of the pixels of the image and perform segmentation. The segmented parts and groups of adjacent segments are then used to calculate and propose regions of interest.
In other words, this algorithm-based approach combines the high recall of exhaustive search with the image guided sampling process of segmentation and outputs bounding boxes in a hierarchical order.
The benefits here are two-fold: the method can capture all possible object
locations and the region proposals are guided by the structure of the image, such as color, texture, size and shape, leading to a reduced number of proposed locations.
In this paper, we will refer to the position with respect to the order in which the boxes appear in the output set of region proposals as position \emph{index}.

Although the number of bounding boxes is reduced in comparison to an exhaustive search approach, this does not effect the high recall we need to ensure that we can find a suitable box for each object.
Moreover, it is possible to further distill the regions into a smaller and more-pertinent set of proposals: Since we know that the gaze coordinate has to lie within the searched bounding box, we can employ this information as a filtering mechanism.
Let $ \smash{(x_1^{(i)},y_1^{(i)}) \in \mathbb{N}^2} $ be the lower left and $ \smash{(x_2^{(i)},y_2^{(i)}) \in \mathbb{N}^2} $ be the upper right corners of
the bounding box $ B_i \in B$, where $ B $ is the full set of class-independent proposals.
By tracking our gaze point $ g = (x, y) \in \mathbb{N}^2 $, we can distill a smaller subset $ B_g \subset B $ of pertinent proposals from $ B $:
\[ B_g \coloneqq \left\{\, B_i \in B \,\Big\vert\, x_1^{(i)} \leq x \leq x_2^{(i)},\ y_1^{(i)} \leq y \leq y_2^{(i)} \,\right\}. \]
This subset $ B_g $ contains only bounding boxes that have an intersection with the object marked by the gaze point.
As we will see later, to achieve satisfactory results, we are dependent on a high gaze-tracking
accuracy and a robust gaze mapping.

Note that getting multiple (but hierarchically-sorted) bounding boxes proposals
is not a disadvantage but an advantage in our use case.
As previously mentioned in \cite{uijlings2013selective}, an object can consist of different colors, multiple objects can have the same color, or the object could be indistinguishable from its background.
In Figure~\ref{paper01:fig:ColourOfObjects} one can see that this could lead to problems if the detection
fails in terms that the only proposed bounding box is not correct or the object is not detected at all.
\begin{figure}
	\centering
	\begin{subfigure}{0.25\linewidth}
		\centering
		\includegraphics[width=.98\linewidth]{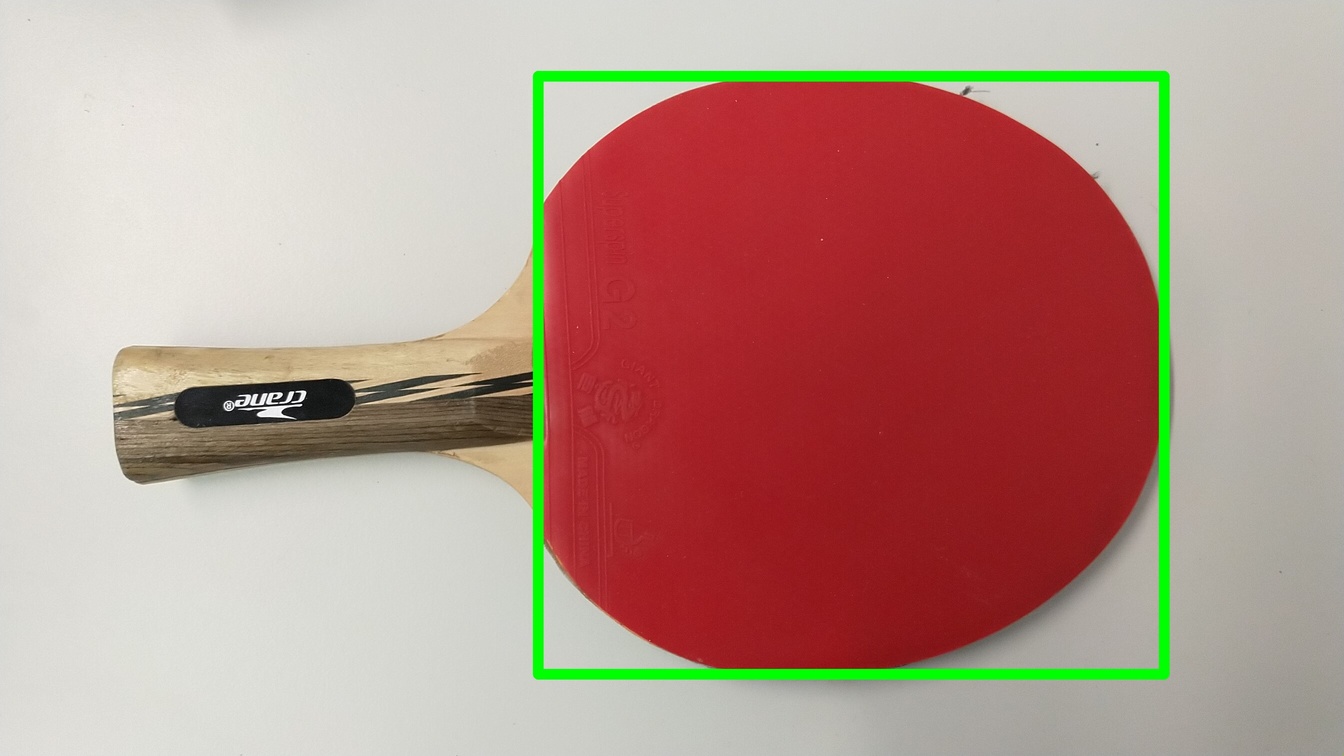}
		\caption{table tennis racket}
		\label{paper01:subfig:ttracket}
	\end{subfigure}%
	\begin{subfigure}{0.25\linewidth}
		\centering
		\includegraphics[width=.98\linewidth]{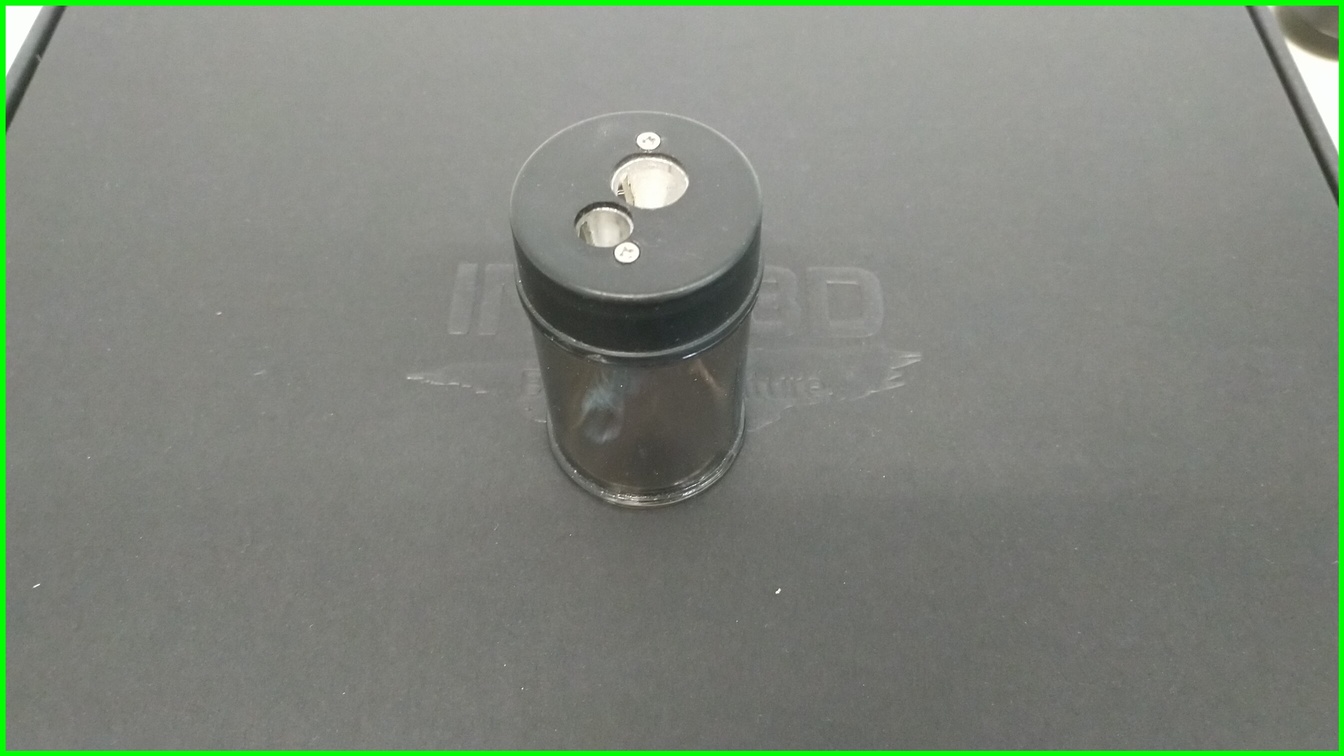}
		\caption{sharpener}
	\end{subfigure}%
	\begin{subfigure}{0.25\linewidth}
		\centering
		\includegraphics[width=.98\linewidth]{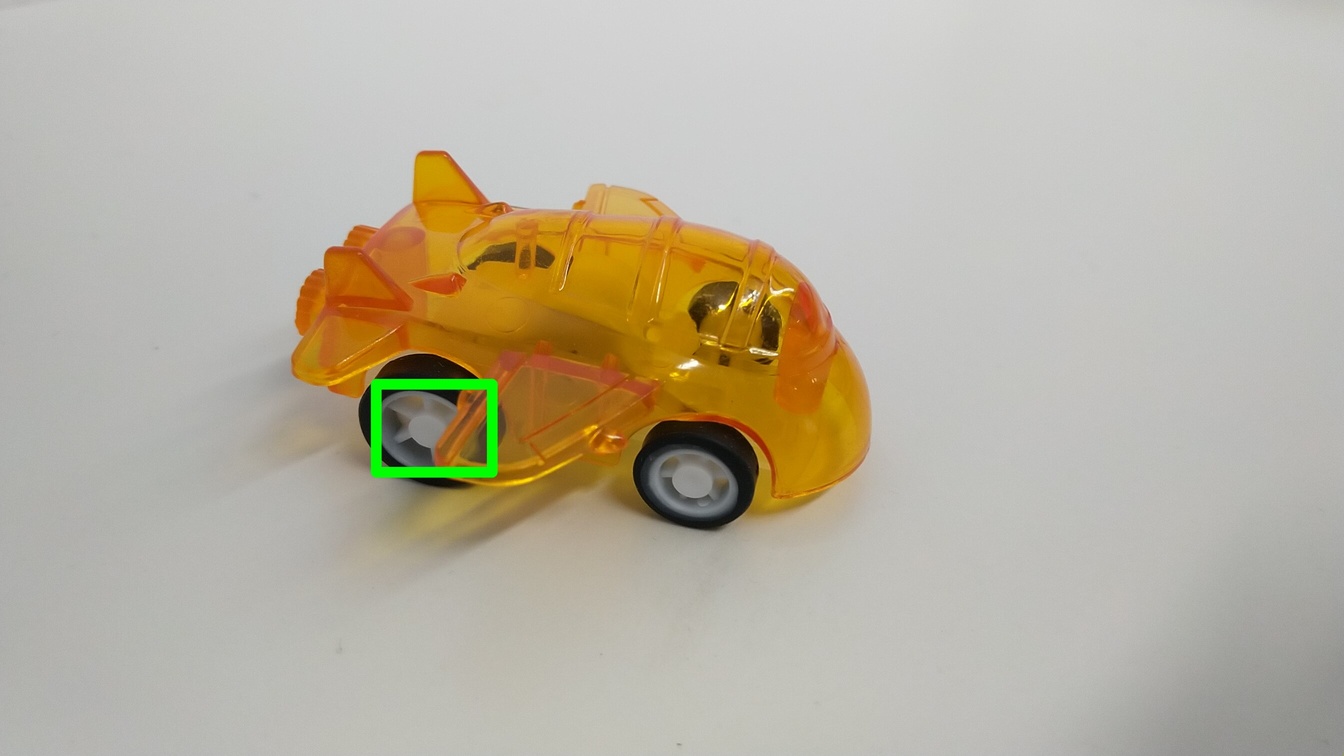}
		\caption{toy}
	\end{subfigure}%
	\begin{subfigure}{0.25\linewidth}
		\centering
		\includegraphics[trim=550 375 125 5, clip, width=.98\linewidth]{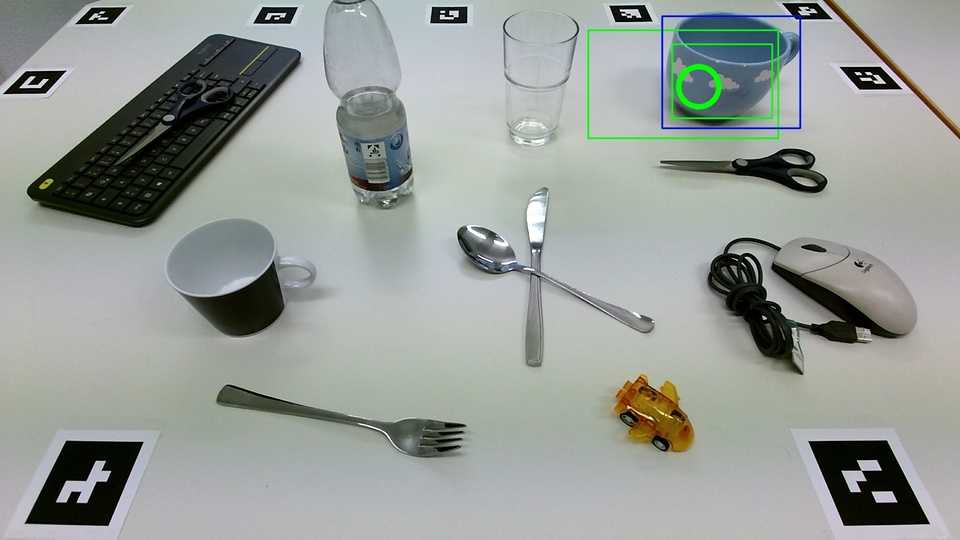}
		\caption{cup}
		\label{paper01:subfig:interaction}
	\end{subfigure}
	\caption{Objects can vary in shape and size, have different backgrounds and
	can consist of multiple colors. This may cause errors regarding the
	detection. The green boxes in the figure indicate proposed regions. In~(\subref{paper01:subfig:ttracket}) the red part is proposed earlier, meaning the corresponding bounding box has a lower position index than the whole racket.
	(\subref{paper01:subfig:interaction}) shows the first three proposals we receive for the blue cup. The first two (green) are not as accurate as the third (blue). Through interaction it is possible to communicate the preferred bounding box.}
	\label{paper01:fig:ColourOfObjects}
\end{figure}

Moreover, we strive for a more human-like learning process, in the sense of an interaction between robot and human, similar to that of a human with another human.
Multiple proposals also mean that we can decide to choose the second or third proposed and more accurate box instead of the first one (see Figure~\ref{paper01:subfig:interaction}).
Interaction between robot and human makes these decisions possible and brings us closer to a natural learning process.

\subsection{Experimental Setup}
\label{paper01:sec:experiment}

In order to showcase a working proof of concept of the proposed application, we collected a session for a participant (one of the system's designers) with the whole system working in real-time\footnote{Eye tracking and gaze mapping working at about $ 30 $ frames per second.}.
This session serves as basis for our initial evaluation of the system.

On a table, we placed different objects, including partially overlapping objects
to some extent. To have a wide appearance range, we selected objects with
distinct sizes, colors, and shapes.
In Figure~\ref{paper01:fig:scitos}, one can see the robot and his view in front of the table with all objects he is supposed to detect.
For the sake of simplicity and for later evaluation, we have used ordinary office and household items that are all part of the Microsoft COCO data set \cite{lin2014microsoft}.

\begin{figure}
	\centering
	\includegraphics[width=\linewidth]{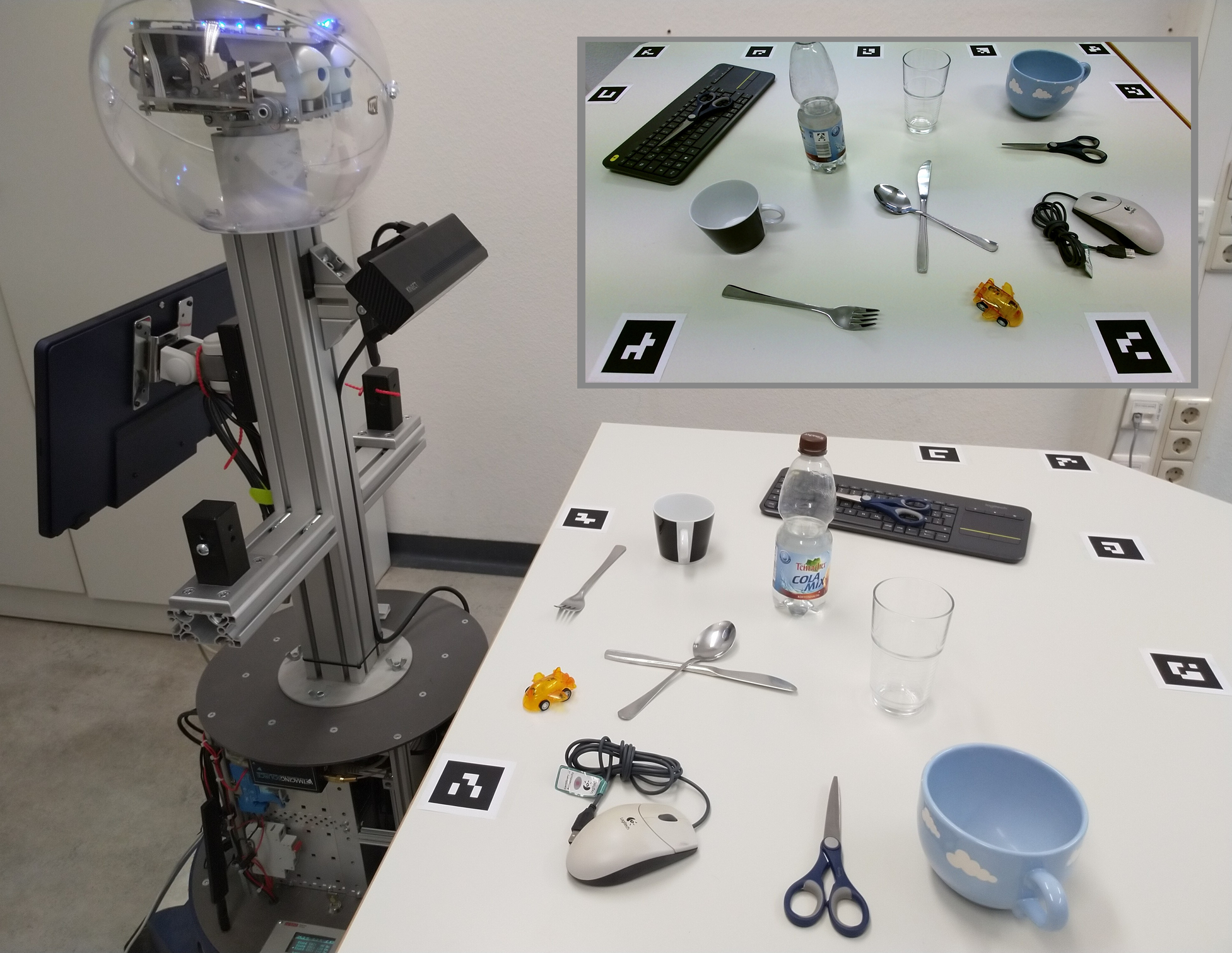}
	\caption{With a Microsoft Kinect~v2 the robot sees different objects on a table: Keyboard, scissors, cups, bottle, fork, knife, spoon, mouse and a small toy car.}
	\label{paper01:fig:scitos}
\end{figure}

As hardware, we used the first generation of Pupil Core \cite{PupilLabs}, a head mounted eye tracker developed by Pupil Labs. Although Pupil Labs provides a software solution called \emph{Pupil Capture} and \emph{Pupil Player}, we decided to utilize \emph{EyeRecToo}~\cite{santini2017eyerectoo}, an open-source software for real-time pervasive head-mounted eye-tracking. The main reasons were the calibration method \emph{CalibMe}~\cite{santini2017calibme}, the robust detection of ArUco markers, and slippage robustness~\cite{santini2019get}.
\emph{EyeRecToo}'s pupil tracking pipeline was set to use \emph{PuRe}~\cite{santini2018pure} / \emph{PuReST}~\cite{santini2018purest}.
Our robot counterpart is a Scitos~G5 from MetraLabs~\cite{MetraLabs} equipped with a Microsoft Kinect for XBox~One.
We accessed the RGB channels of the Kinect~v2 using \emph{\acs{ROS}}~\cite{quigley2009ros},
\emph{libfreenect2}~\cite{libfreenect2}, and \emph{iai\_kinect2}~\cite{iai_kinect2}.
For the implementation, we make extensive use of the \emph{OpenCV} \cite{OpenCV:Bradski:2000} library.

\subsection{Evaluation}
\label{paper01:sec:evaluation}

To establish reference ground-truth values for the object locations, we have
employed the Fully Convolutional One-Stage Object Detector
(FCOS)~\cite{tian2019fcos} trained on Microsoft COCO \cite{lin2014microsoft}, using the ResNeXt-64x4d-101 backbone with
deformable convolutions.
This serves as a baseline representing a state-of-the-art object detection
for supervised learning.
Given an image viewed from the robot's perpective, the output of FCOS is
shown in Figure~\ref{paper01:subfig:fcos}.
It is worth noting that the bottle was detected twice; in this case, we
opted to ignore the smaller inaccurate bounding box.
Moreover, neither the knife that overlaps with the spoon nor the scissor
placed on the keyboard are recognized by FCOS, despite all of these classes
being present in the training data. Thus, we discuss these
separately.

\subsubsection{Qualitative Analysis}
Both eye tracking and marker detection work in real time, as well as the subsequent gaze mapping.
Therefore, our method is suitable for real-time human-robot interaction.
As long as the accuracy in all three steps is high enough, the robot knows at any time where we are looking at.
The human is even unrestricted in his movements.
Figure~\ref{paper01:fig:mapping} shows two attempts of pointing out an object to the robot. One was successful and the other one failed.
\begin{figure}[b!]
	\centering
	\begin{subfigure}{\columnwidth}
		\begin{subfigure}{0.5\columnwidth}
			\centering
			\includegraphics[width=\linewidth]{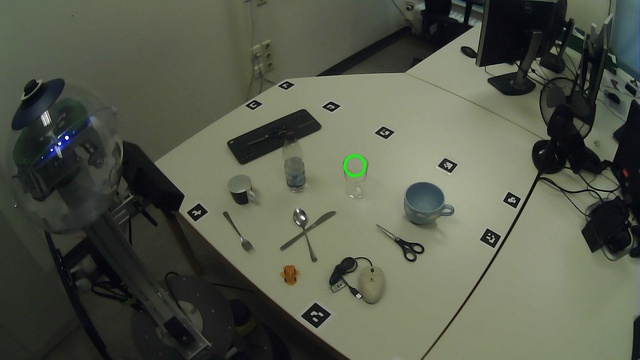}
		\end{subfigure}%
		\begin{subfigure}{0.5\columnwidth}
			\centering
			\includegraphics[width=\linewidth]{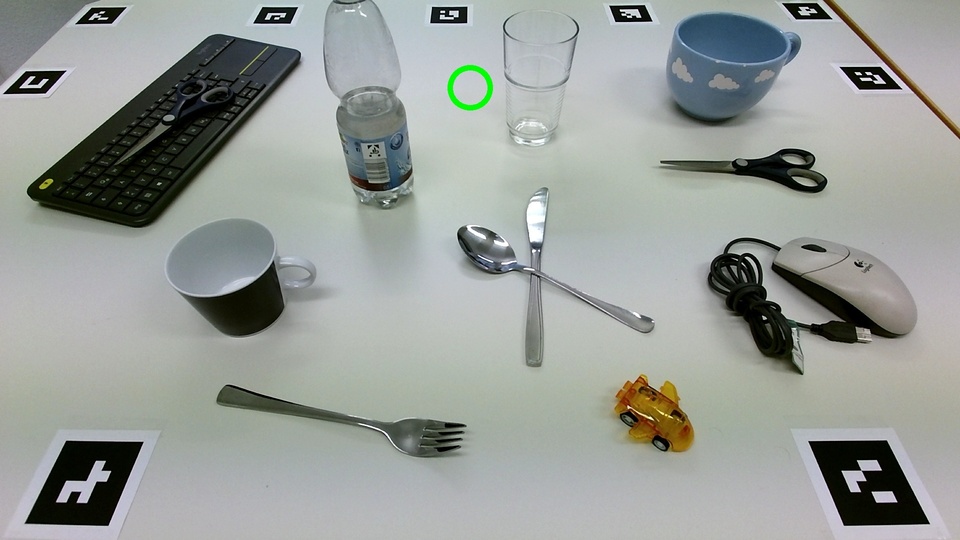}
		\end{subfigure}
		\caption{Failed}
		\label{paper01:fig:mapping-failed}
	\end{subfigure}
	\par\bigskip
	\begin{subfigure}{\columnwidth}
		\begin{subfigure}{0.5\columnwidth}
			\centering
			\includegraphics[width=\linewidth]{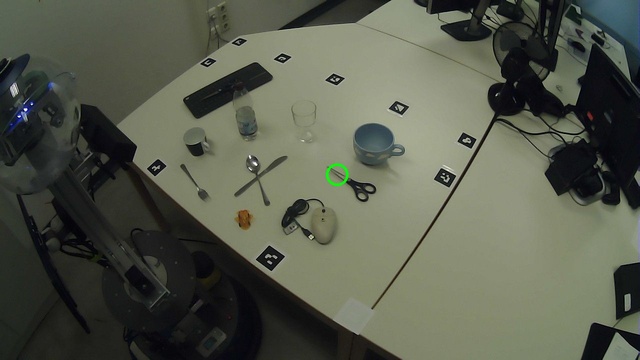}
		\end{subfigure}%
		\begin{subfigure}{0.5\columnwidth}
			\centering
			\includegraphics[width=\linewidth]{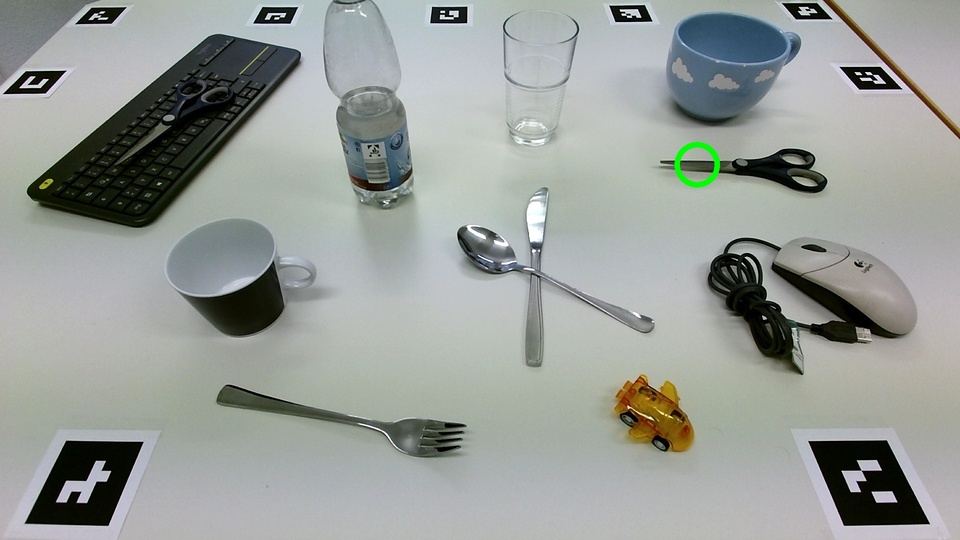}
		\end{subfigure}
		\caption{Successful}
		\label{paper01:fig:mapping-successful}
	\end{subfigure}
	\caption{A failed and a successful attempt of mapping the human gaze (left) on the robot's view (right).}
	\label{paper01:fig:mapping}
\end{figure}
Although the human from whom the gaze point in Figure~\ref{paper01:fig:mapping-failed} originated actually looked at the glass and his gaze was tracked correctly,
the gaze point in the robot's view is not on the glass, i.e. the mapping procedure was problematic in this case.
This exemplifies that enough markers have to be detected to guarantee reliable mapping and usability.
This could be ensured, for example, by using more accurate markers such as infrared tokens.
In addition, the tracking of the human gaze must work reliably to achieve
satisfactory usability.
Therefore, we have carefully calibrated the eye tracker to
achieve the desired accuracy. During interaction, however, the device is likely
to slip~\cite{niehorster2020impact} such that slippage robustness is paramount.

In contrast to the gaze mapping, the region proposal achieves real-time
operation only at lower frame rates.
The calculation of all the 2198 region proposals on our picture of the robot's view with a resolution of 1900x1080 took about 2.7 seconds with the ``quality'' method.
Nonetheless, this is not a problem, as
region proposal is not required for each frame but only sporadically.
Once a correct bounding box for the intended object has been found, it can be
tracked with well-known tracking algorithms like KCF \cite{henriques2014high} or
CSRT \cite{lukezic2017discriminative}.

\subsubsection{Quantitative Analysis}

To evaluate the efficiency of our method we compare the position indices of each bounding box within the complete hierarchical set of region proposals from the selective search algorithm with the indices we have distilled.
Of course these boxes should not only be easy and fast to find but have to be
accurate as well. For this reason, we need to investigate the similarity of the proposed
boxes w.r.t. the ground truth.
As measurement for accuracy, we calculate the Jaccard index
$ J(B_1, B_2) $, also known as \acf{IoU}.
This means that the closer the Jaccard index is to 1, the greater the similarity between the boxes.
For object detection, if the Jaccard index is more than $ 0.5 $, a detection is typically considered correct \cite{everingham2010pascal}.
Nevertheless, in general, a higher value is desirable.
\cite{zitnick2014edge} provides a comparison of different values of the Jaccard index and describes $ 0.5 $ as very loose, $ 0.9 $ as very strict and $ 0.7 $ as reasonable compromise in between.
Therefore, we set $ 0.7 $ as threshold and characterize bounding boxes with at least this value as ``sufficient''.
This allows us to analyze whether the selective search algorithm is a good choice and provides region proposals that are accurate enough, i.e. sufficient, for our use case.

In Table~\ref{paper01:tab:jaccard-index} the Jaccard index between the boxes predicted by FCOS and the best box in our set of proposals is listed for each item.
\begin{table}
	\centering
	\rowcolors{3}{white}{gray!25}
	\caption{Comparison between the full and our distilled set of bounding boxes.}
	\label{paper01:tab:jaccard-index}
	\resizebox{\textwidth}{!}{%
	\begin{threeparttable}
		\begin{tabular}{lcccccccccccccccc}
			\toprule
			\multirow{2}{*}{Item} & FCOS & \multicolumn{2}{c}{Best total} &\#Boxes& \multicolumn{3}{c}{First sufficient} & \multicolumn{3}{c}{Best among first 15} & \multicolumn{2}{c}{Recall} & \multicolumn{2}{c}{Precision} & \multicolumn{2}{c}{$ F_1 $ score} \\
						& Confidence & Index & \acs{IoU} & Dist. & Index & \acs{IoU} & Acc.\tnote{1} & Index & \acs{IoU} & Acc. & Full & Dist. & Full & Dist. & Full & Dist. \\
			\midrule
			Bottle		& 0.69	&\textbf{292}\tnote{2}& 0.943 &98& 1& 0.851&90.24\,\%&\textbf{12}& 0.943	& 100\,\% 	& 1 & 1 	& 0.012 & 0.265 & 0.023 & 0.419 \\
			Cup (black)	& 0.88	& 1748	& 0.863	&221& \textbf{3}& 0.828	& 95.94\,\%	& \textbf{3}& 0.828	& 95.94\,\%	& 1 & 1 	& 0.029 & 0.29	& 0.057 & 0.449 \\
			Cup (blue)	& 0.74	& 1199	& 0.918	&110& 3			& 0.870	& 94.77\,\%	& 15		& 0.899	& 97.93\,\%	& 1 & 1 	& 0.014 & 0.273 & 0.027 & 0.429	\\
			Fork		& 0.83	& 1873	& 0.945	&34& \textbf{1}& 0.939	& 99.37\,\%	& \textbf{1}& 0.939	& 99.37\,\%	& 1 & 1		& 0.013 & 0.824 & 0.025 & 0.903 \\
			Glass		& 0.82	& 1429	& 0.935	&110& 3			& 0.896	& 95.83\,\%	& 4			& 0.922	& 98.61\,\%	& 1 & 1 	& 0.020	& 0.391 & 0.038 & 0.562 \\
			Keyboard	& 0.55	& 1839	& 0.988	&189& 1			& 0.751	& 76.01\,\%	& 9			& 0.981	& 99.29\,\%	& 1 & 1 	& 0.046 & 0.529 & 0.087 & 0.692 \\
			Mouse		& 0.88	& 883	& 0.968	&231& 1			& 0.714	& 73.76\,\%	& 11		& 0.955	& 98.66\,\%	& 1 & 0.96	& 0.011 & 0.104 & 0.023 & 0.188 \\
			Scissor		& 0.68	& 1137	& 0.954	&89& 2			& 0.880	& 92.24\,\%	& 9			& 0.898	& 94.13\,\%	& 1 & 1 	& 0.018 & 0.449 & 0.036 & 0.620 \\
			Spoon		& 0.61	& 670	& 0.727	&118& \textbf{3}& 0.720	& 99.04\,\%	& \textbf{3}& 0.720	& 99.04\,\%	& 1 & 1 	& 0.002 & 0.034 & 0.004 & 0.066 \\
			Toy car		& 0.52	& 2079	& 0.946	&68& 3			& 0.712	& 75.26\,\%	& 5			& 0.909	& 96.09\,\%	& 1 & 1 	& 0.021 & 0.662 & 0.040 & 0.797 \\
			\midrule
			\O			& 0.72	&1314.9	& 0.919	&126.8& 2.1		& 0.816	&89.25\,\%	& 7.2		& 0.899	& 97.91\,\%	& 1 & 0.996 & 0.018 & 0.382 & 0.036 & 0.512 \\
			\bottomrule
		\end{tabular}
		\begin{tablenotes}
			\item[1] Accuracy compared to the best box in the full set (ratio of the two Jaccard indices).
			\item[2] Bold indices within the same line indicate identical boxes.
		\end{tablenotes}
	\end{threeparttable}}
\end{table}
Note that the knife and scissor placed on the keyboard are omitted from Table~\ref{paper01:tab:jaccard-index} because they are not recognized by FCOS, which means we do not have any reference values for these items.
We will discuss these items separately at the end of this section.
Besides, depending on whether we want to consider the mouse cable or not, the values in Table~\ref{paper01:tab:jaccard-index} naturally change.
Even though we can distill boxes for both cases, here we stick to the output of FCOS and only consider the mouse without cable.
One must keep in mind, however, that the composition of the mouse from two sub objects leads to different predictions being made.
For example, if the cable had not been bundled up, the relevant position index would indeed be smaller or the mean Jaccard index would be larger.
In our test, the use of a second gaze point allowed the delimitation to boxes containing the cable and the mouse body alone.
Compared to regular object detection, we are not bound to fixed ideas of objects but can vary the object's bounding box depending on the situation.
It is worth noting that this is a good example where our method surpasses all
pre-trained object detectors in terms of flexibility.
In this particular case, there is not only one correct box, but two.
\emph{Gaze allows us to resolve such ambiguities}.

We can observe that the Jaccard indices with few exceptions are all above 0.85 and the vast majority is even above 0.9 (see column 4).
This is highlighted visually in Figure~\ref{paper01:subfig:gt_best}, which shows the bounding box detected by FCOS and by the proposed approach.
\begin{figure}
	\centering
	\begin{subfigure}{0.5\columnwidth}
		\centering
		\includegraphics[width=.95\linewidth]{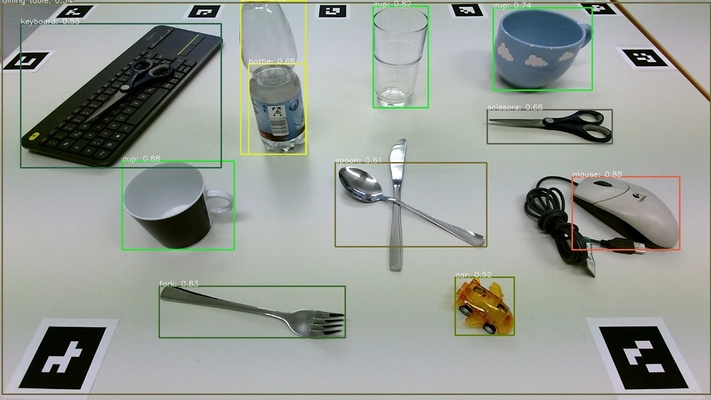}
		\caption{}
		\label{paper01:subfig:fcos}
	\end{subfigure}%
	\begin{subfigure}{0.5\columnwidth}
		\centering
		\includegraphics[width=.95\linewidth]{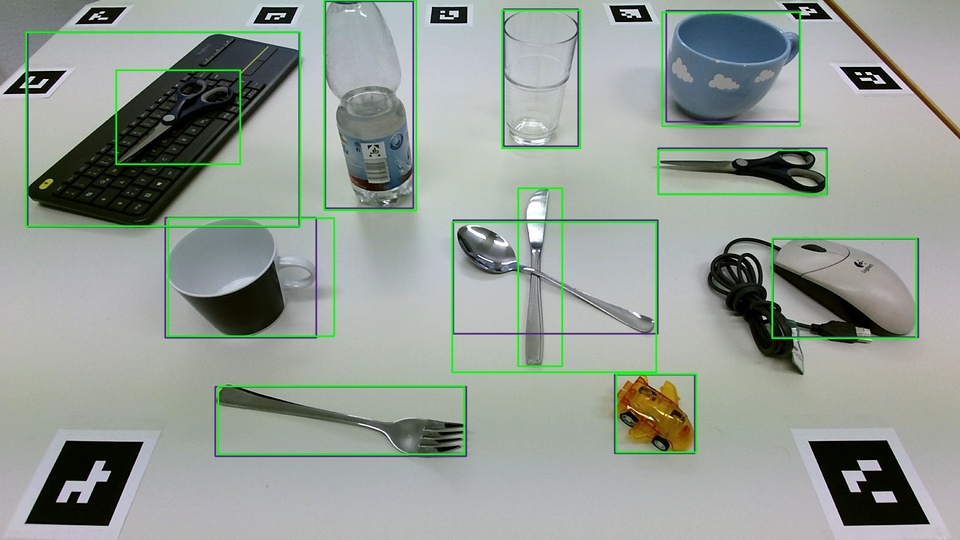}
		\caption{}
		\label{paper01:subfig:gt_best}
	\end{subfigure}
	\par\bigskip
	\begin{subfigure}{0.5\columnwidth}
		\centering
		\includegraphics[width=.95\linewidth]{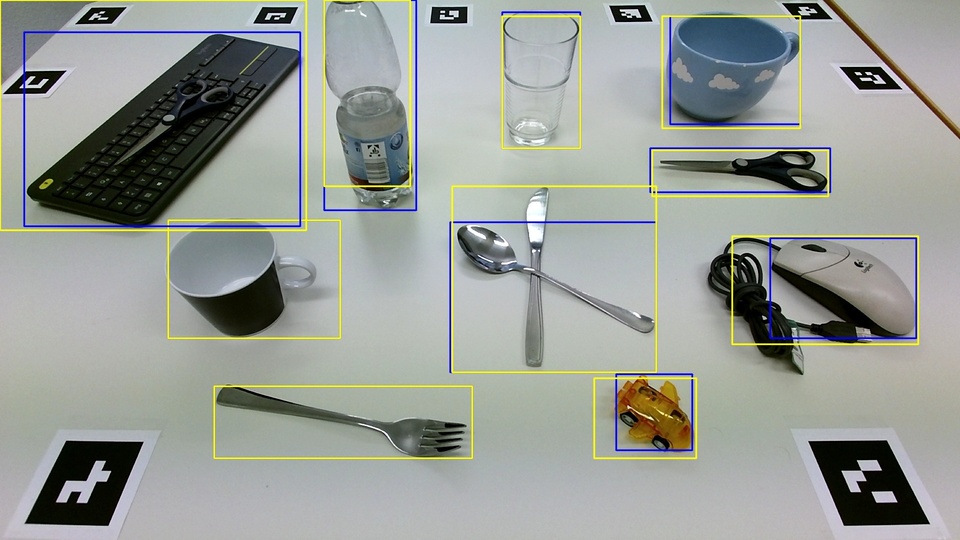}
		\caption{}
		\label{paper01:subfig:best15_sufficient}
	\end{subfigure}%
	\begin{subfigure}{0.5\columnwidth}
		\centering
		\includegraphics[trim=75 150 275 46, clip, width=.95\linewidth]{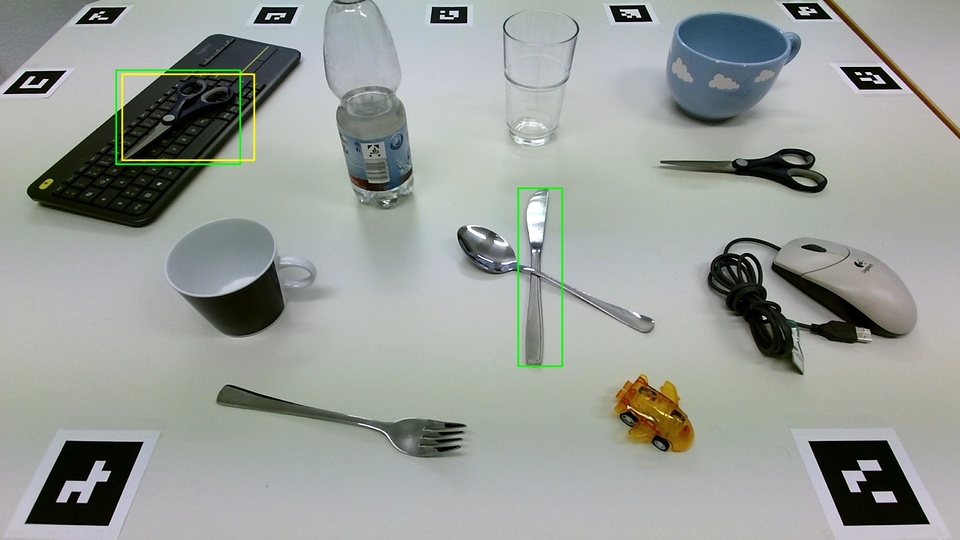}
		\caption{}
		\label{paper01:subfig:hard}
	\end{subfigure}
	\caption{
	(\subref{paper01:subfig:fcos}) shows the objects detected with FCOS. The confidence of the prediction can also be seen in Table \ref{paper01:tab:jaccard-index}.
	(\subref{paper01:subfig:gt_best}) shows a comparison of the total best bounding box (green) with the ground truth given by FCOS (purple).
	(\subref{paper01:subfig:best15_sufficient}) shows a comparison of the best bounding box among the first 15 proposals (blue) with earlier sufficient boxes (yellow). Note that these boxes are identical for the cup and the fork. The knife and scissors on the keyboard have been omitted as they are handled separately.
	(\subref{paper01:subfig:hard}) shows the best bounding boxes distilled for the knife and the scissor on the keyboard.
	}
\end{figure}
With a mean value of the Jaccard indices of $ 0.919 $, the proposed boxes are highly relevant.
This value can also be used as upper bound for the accuracy of our fast distillation.
Also worth noting are the massively high indices of the respective best box in each category. Whereas the position index of the best box of the bottle in the full set is the lowest at 292, the mean value of the position index is about $ 1315 $.
Through distillation, we managed to improve that value to an average of $ 61.5 $.
Figure~\ref{paper01:fig:violin} illustrates that the vast majority of the boxes in the full set of proposals has an Jaccard index below 0.1 and is therefore irrelevant.
\begin{figure}
	\centering
	\includegraphics[width=.85\linewidth]{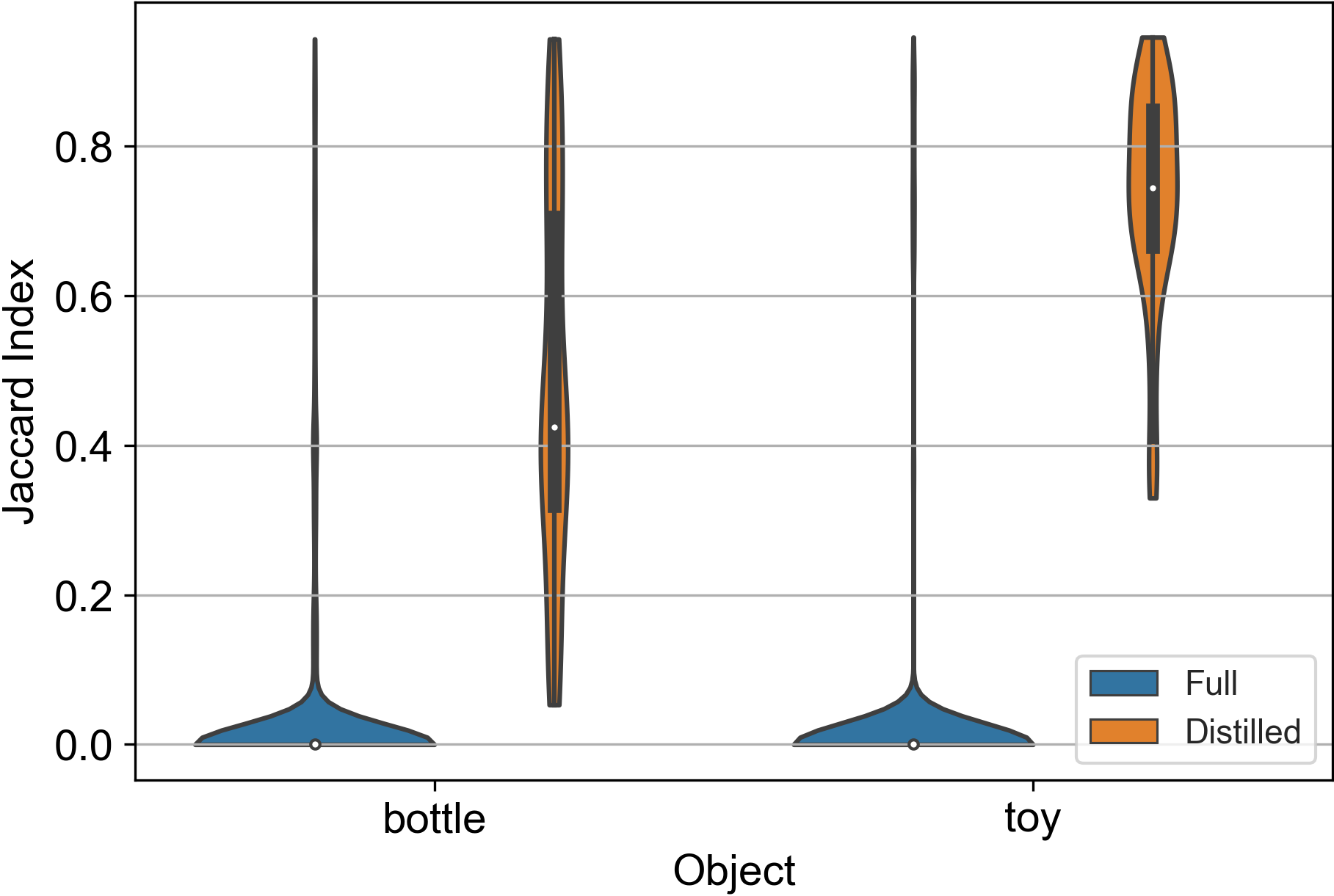}
	\caption{The violin plot shows the distribution of the Jaccard indices for the full and the distilled set of bounding boxes using the example of the bottle and the toy.}
	\label{paper01:fig:violin}
\end{figure}

As Table~\ref{paper01:tab:jaccard-index} and Figure~\ref{paper01:subfig:best15_sufficient} show, we do not have to find exactly the best boxes.
With our fast distillation method, we were able to provide a proposal among the first 15 boxes of each distilled subset with at least 94\,\% accuracy to the best. That is, with an average accuracy of even 97.91\,\%, we were almost as accurate as the best possible box, with a much smaller position index.
Therefore, we need much less communication with the robot to reach the desired box.

In addition, we have considered earlier sufficient boxes in the sense of boxes with a Jaccard index of at least $ 0.7 $.
Figure~\ref{paper01:subfig:best15_sufficient} shows these bounding boxes along with the best boxes among the first 15 described above.
Their position index is of course lower and, in our particular case, never higher than three.
As highlighted in Table~\ref{paper01:tab:jaccard-index}, it is often sensible to fall back to earlier boxes.
For instance, in the case of the blue cup, it is possible to reduce the position index from 15 to 3 while reducing the Jaccard index only by $ 0.03 $.
In contrast, the toy car's position index is acceptable either way, and a significant drop in accuracy results if the position index is lowered from 5 to 3.
In general, the average accuracy is about ten percent lower compared to the best possible box, but, as previously mentioned, it is found early since it is one of the first three proposals.

Considering that the sufficient boxes (Jaccard index $> 0.7$) are the relevant
ones, we define a) \emph{recall} as the ratio between relevant boxes retrieved
by our method and all relevant boxes, as well as b) \emph{precision} as the
ratio of boxes retrieved by our method that are relevant.
The per-object recall and precision are reported in Table~\ref{paper01:tab:jaccard-index} together with the $F_1$ score
$\left( 2 \cdot \frac{\text{precision} \cdot \text{recall}}{\text{precision} +
\text{recall}} \right)$, which is the harmonic mean of precision and recall.
Whereas the recall remained virtually the same, the precision increased significantly due to the distillation using the proposed method.
To be more specific, while on average not even $ 2\,\% $ of the boxes in the full set could be considered as sufficient, almost $ 40\,\% $ of the distilled boxes have a Jaccard index of at least $ 0.7 $.
Moreover, the resulting mean $ F_1 $ value is about $ 14 $ times higher for the distilled sets compared to the full set.

Finally, we would like to discuss the objects that were not recognized by FCOS.
The knife and the spoon lying on top of each other was a difficult task.
Both FCOS and our proposed method struggled on this part.
Although FCOS was not able to detect the knife at all, we at least managed to get a sufficient box with a high position index of 75.
Figure~\ref{paper01:subfig:hard} shows our best possible result.
However, we needed several attempts to match the human gaze point with the knife since the knife's width is relatively small.

Detecting the dark blue scissor on the black keyboard was on the other hand quite easy in terms of mapping the gaze point on the object.
Even though it was generally an even harder task with respect to the color of the background, unlike FCOS, the proposed method was able to find one sufficient and one best bounding box with the position index of $ 45 $ and $ 99 $, respectively. These bounding boxes were also relatively late to reach but far earlier than $ 635 $ and $ 1251 $, their indexes in the full set.

\subsubsection{Limitations}
Our method is based on a high accuracy in each partial step.
This is an issue if we have either bad gaze tracking or mapping, which could result by too small or too few markers, low-quality hardware, or external disturbances.
Even if the mapping part is accurate, a sloppy gaze estimation can lead to a gaze point that does not overlap with the object.
With an inaccurate gaze point in the robot's view, accurate bounding box proposals are difficult and sometime impossible to distill.
In this case, we have to repeat pointing out to the object.

Furthermore, with one exception (scissors on the keyboard), the proof of concept was carried out on a plain white table.
Although we would expect more candidate boxes in less homogeneous settings, our experiments suggest that there would still be highly relevant boxes due to the high recall that is in the nature of the method.

\subsection{Conclusion}
\label{paper01:sec:conclusion}

In this work, we have proposed and evaluated a novel method that enables the deployment of robots in non-predefined scenarios.
The proposed method combines automatic object location proposals with human gaze to distill pertinent location proposals.
Just by looking at an object and some human robot communication, we can find a bounding box with a Jaccard index of almost $ 0.9 $ compared to the ground truth.
These boxes can then be used to quickly extend the robot's object detection neural network.

Out of thousands of possible region proposals, we successfully distilled useful
object-independent bounding boxes, increasing the precision of the location
proposals by over 21 times with virtually no recall loss.
Despite challenging scenarios, our method was consistently applicable and does not need any training at all.
Relative to a state-of-the-art object detector (FCOS) trained on the Microsoft COCO data set, we achieved an average Jaccard index of almost $ 0.9 $ for at least one box out of the first $ 15 $ proposals.
Looking only at the first sufficient box of each object,
we observed an average accuracy of $ 89.25\,\% $ 
compared to the respective best possible box in the full set of proposals.

Since our gaze method significantly improved the position index of important bounding boxes compared to the large initial number of region proposals, it enables concentrating exclusively on relevant boxes.
This allows the robot to find the intended object more quickly and to generally reduce the necessary communication, improving the human-robot interaction.
In addition, we could find bounding boxes to objects that could not even been detected by FCOS.

In summary, our proposed method is therefore a broadly applicable and natural
way to achieve unknown-object detection by a robot in \acs{HRI} scenarios.
However, a significant amount of work remains for future work as we plan to extend our proof of concept by evaluating our method with more participants and additionally
investigate the impact of imperfect labels on the training of neural networks.

\subsection*{Acknowledgment}
This research was funded by the Deutsche Forschungsgemeinschaft (DFG, German Research Foundation) under Germany's Excellence Strategy -- EXC number 2064/1 -- Project number 390727645.
\fi

\clearpage
\setcounter{footnote}{0}
\section{Exploiting Augmented Reality for Extrinsic Robot Calibration and Eye-based Human-Robot Collaboration}
\label{app:sec:paper02}
\blfootnote{\hspace{-14pt}Funded by the Deutsche Forschungsgemeinschaft (DFG, German Research Foundation) under Germany’s Excellence Strategy -- EXC number 2064/1 -- Project number 390727645.}

\ifpaper
\subsection{Abstract}
For sensible human-robot interaction, it is crucial for the robot to have an awareness of its physical surroundings.
In practical applications, however, the environment is manifold and possible objects for interaction are innumerable.
Due to this fact, the use of robots in variable situations surrounded by unknown interaction entities is challenging and the inclusion of pre-trained object-detection neural networks not always feasible.
In this work, we propose deploying augmented reality and eye tracking to flexibilize robots in non-predefined scenarios.
To this end, we present and evaluate a method for extrinsic calibration of robot sensors, specifically a camera in our case, that is both fast and user-friendly, achieving competitive accuracy compared to classical approaches.
By incorporating human gaze into the robot's segmentation process, we enable the 3D detection and localization of unknown objects without any training.
Such an approach can facilitate interaction with objects for which training data is not available.
At the same time, a visualization of the resulting 3D bounding boxes in the human's augmented reality leads to exceedingly direct feedback, providing insight into the robot's state of knowledge.
Our approach thus opens the door to additional interaction possibilities, such as the subsequent initialization of actions like grasping.

\subsection{Introduction}
More and more robots are being used in environments within a close proximity to humans.
The possible applications of robots are diverse and possible interactions with humans are multifaceted.
Whether as a tour guide in museums \cite{thrun1999minerva} or as an assistant in supermarkets \cite{gross2008shopbot}, each interaction scenario involving robots has its own challenges.
Furthermore, successful technical advances in augmented reality (\acs{AR}) have promoted the interaction and collaboration between humans and robots.
Consequently, \acs{AR} has found application in factories \cite{nee2012augmented} and in imitating assembly processes that a human demonstrates \cite{blankemeyer2018intuitive}.

The long list of possible use cases results in at least as many tasks that need to be solved.
Among these tasks, the conveyance of the interaction context, such as the specification of an object to interact with, is particularly challenging.
Many tasks, especially object detection, can be accomplished through the benefit of machine learning methods, such as neural networks.
While advances in machine learning have had a major impact on the development of human-robot interaction, there are also some drawbacks.
Typically, many of these approaches require a sufficient amount of available training data, which cannot always be guaranteed.
This data dependency ties the deployment of robots to predefined scenarios and limits interaction with the environment, e.g. with unknown objects that cannot be detected.
For example, if a supermarket changes its assortment of products, the robot can usually only interact with the new items if it has learned them beforehand.
Our goal is to enable data-independent object detection for cases where no training data is available.

Another even more fundamental problem is the calibration of the robot.
In order for a robot to perceive a scene, its sensors, such as a fixed, but adjustable camera, must be properly targeted and its position relative to the robot base must be known.
Therefore, the scene or the purpose of the operation needs to be identified in advance, at least to a certain degree.
In addition, calibration of extrinsic robot parameters is often laborious \cite{elatta2004overview} since, in most cases, either the existence of a second sensor in the form of a laser scanner or another camera is assumed, or expensive external tools are used.
Both make subsequent adjustments in response to changing circumstances difficult.
On top of that, the authors of \cite{brvsvcic2015escaping} noted that robots in public attract the curiosity of people, especially children. In particular, children tend to touch the robot or exhibit abusive behavior when unobserved. This, in turn, can often lead to misalignments of the robot's sensors and require frequent recalibrations. A less time-consuming calibration method is beneficial in this case.

In this work, we attempt to fill this gap at the intersection of research fields of human-robot interaction, eye tracking, and augmented reality.
More specifically, we aim at a flexible deployment of robots, detached from predefined scenarios by leveraging collaboration with humans instead of training data.
Our contribution with this work is twofold:

On the one hand, we present a convenient method for determining the transformations between the robot and a sensor, in our case a camera, as well as between the human and the robot.
With our method, time does not have to be spent repeatedly for each calibration run, but only once during the initial setup.
Subsequent calibrations can then be performed in a matter of seconds, making the method particularly suitable for situations where frequent recalibrations are required.
The calibration can be executed at any time during runtime and allows both the human and the robot to move freely.

On the other hand, after utilizing said calibration, we fuse existing point cloud clustering methods with eye-tracking information to showcase the 3D detection of unknown objects.
More precisely, the robot and the human collaborate so that the robot detects which object the human is looking at without knowing the interaction context in advance.
Based on our calibration, we can establish a connection for continuous exchange of interaction information.
The human continually provides the robot with gaze data and the robot responds with the bounding box of the target object.
The human's perception is augmented by integrating the robot's feedback directly into the human's reality.
All of this without training and in an online fashion, not after the fact.

In summary, our most important contributions are as follows:
\begin{enumerate}
    \item We show and evaluate a calibration method via an augmented reality interface that is suitable for the deployment of robots in ever-changing scenarios and allows the robot's capabilities to be further extended by providing it with a new, additional real-time information channel --- the human gaze.
	\item We are the first to use augmented reality in a human-robot collaboration scenario to segment unknown objects in three-dimensional space without the use of neural networks. We also provide direct feedback to the human, enabling subsequent interactions.
\end{enumerate}

The remaining part of this paper is structured as follows. After a discussion of the related work,  in Section~\ref{paper02:sec:method} we describe and formalize our approach in detail.
Our results and the limitations of our approach are discussed in Section~\ref{paper02:sec:evaluation}.
Section~\ref{paper02:sec:conclusion} concludes this work and gives an outlook on our future activities.

\subsection{Related Work}
Employing gaze information to achieve human-robot interaction with unknown objects requires significant multidisciplinary efforts, which we will discuss in this section. From how
1) robots collaborate with humans, to
2) augmented reality in robotics,
3) robot calibration and
4) 3D object detection, to
5) mapping human gaze to a known frame of reference and
6) previous applications of eye tracking in the context of computer vision.

\subsubsection{Collaborative Settings}
In recent years, scenarios in which humans and robots work together side by side have gained attention.
Interaction with robots invites interesting possibilities for beneficial collaboration in human everyday life \cite{chandrasekaran2015human}.
In \cite{reardon2018come} a system was presented, that enables a robot to perform cooperative search with a human teammate, where the robot assists the human teammate in navigation to the
search target.
Collaboration between human and robot is also widespread in industrial environments, such as in assembly tasks \cite{gleeson2013gestures}, surface finishing applications \cite{wilbert2012robot} or welding work \cite{muller2017skill}.
In addition to the application in industry, robots have more and more of a social purpose.
Due to the lack of medical personnel and rising costs in the health sector, social robots are increasingly being used in the health care system \cite{olaronke2017state}.
They are typically deployed for surgical assistance \cite{kapoor2006constrained}, rehabilitation \cite{krebs2003rehabilitation}, elderly care \cite{broekens2009assistive}, and as companion robots \cite{robinson2013suitability}.

\subsubsection{Augmented Reality in Robotics}
With the increasing availability of various augmented reality glasses, the impact of \acs{AR} in research and industry has also grown.
In \cite{krupke2018comparison}, head orientation and pointing gestures were used to control an industrial robot arm for pick-and-place tasks. However, the arm was fixed in the room to facilitate coordinated transformation by means of a marker attached to the wall and the set of interaction objects was fixed.
An \acs{AR} device was also used by \cite{rosen2020mixed} in a multimodal communication setup to help a robot decide which object a human pointed to using gestures, gaze, and speech. In this setup, again, the objects were predefined and their positions were additionally measured accordingly in advance.
The authors of \cite{kastner2019augmented} visualized sensor data from a robot using \acs{AR} glasses. All sensors, though, were already calibrated, which additionally allowed for the utilization of a localization algorithm.
Following on from this, the same authors recently used a deep learning-based approach in \cite{kastner20203d} to determine the mutual position of the robot and \acs{AR} device. Nevertheless, this approach was not suitable for real time scenarios due to the limited computational capacity of the \acs{AR} glasses.
Within a manipulation frame, \cite{chu2018helping} used pre-trained 2D object detectors to determine 3D bounding boxes. This required a fiducial marker to be in the field of view at all times and was limited to a single object per pass. Such problems of ambiguity we will solve with gaze.

\subsubsection{Extrinsic Robot Calibration}
Modern robots are usually equipped with a large number of sensors, most frequently RGB-D cameras.
Ensuring their operability requires the most accurate calibration of extrinsic parameters, i.e. their position on the robot base.
A classical approach to this is the use of calibration patterns.
By observing the pattern, \cite{zhang2004extrinsic} determined the mutual position between a camera and a 2D laser range finder.
With only one image, but several markers, \cite{geiger2012automatic} succeeded in calibrating a camera with respect to a second camera or a laser scanner.
In both cases, however, the existence of a second sensor was a prerequisite and a common field of view of these two was mandatory.

In \cite{chen2018accurate}, a framework for parameter estimation using a motion capture system was built. 
While such systems, including Vicon \cite{Vicon} or OptiTrack \cite{OptiTrack} can be very accurate, they require careful calibration beforehand.
In addition, they are time-consuming to set up and expensive due to the amount of hardware involved, such as multiple cameras.
We try to close this gap with a fast and universally applicable method.

\subsubsection{3D Object Detection}
Due to the higher level of difficulty, many 3D object detectors are inspired by detection in 2D.
This includes the projection of the point cloud into bird's eye view \cite{zhou2018voxelnet} or cropping on frustums based on 2D bounding boxes \cite{qi2018frustum}, \cite{wang2019frustum}.
Few also operate on the point clouds directly \cite{qi2019deep}.
What they all have in common, however, is that they rely heavily on the availability of training data and focus predominantly on road scenes or furniture pieces.
An approach to instance segmentation of unseen objects was proposed by \cite{xie2020unseen}.
While they did not need real world images, they had to generate a large amount of synthetic data for which 3D CAD models were required.
As an alternative to neural networks, \cite{bao2015saliency} used a saliency-driven approach to detect unknown objects.
Nonetheless, the results were influenced to some extent by a parameter that depended on the size of the objects, and, due to the long calculation time, the system was not suitable for real-time applications.

\subsubsection{Gaze Mapping}
Mapping gaze data from a moving eye tracker to another coordinated frame is still an unsolved challenge and thus ongoing research \cite{peysakhovich2018aruco}.
One possible solution to this challenge is feature matching. For example, \cite{kalash2016gaze} achieved promising results with such a method, however it reaches its limit with diverging camera perspectives.
The authors also found that better robustness at less computational cost was achieved with fiducial markers \cite{kalash2016gaze}.
Such markers were used in recent works by \cite{peysakhovich2018aruco} and \cite{weber2020distilling}, among others.
One disadvantage of this approach is that fiducial markers have to be in the field of view of both cameras, restricting thus movements.
With our \acs{AR}-based approach, we overcome this problem and ensure stable gaze mapping despite free movement and thus independent of the field of view.

\subsubsection{Eye Tracking and Computer Vision}
Although not yet very popular, there are some works that have tried to solve computer vision problems with eye tracking.
In \cite{toyama2012gaze} for example, features in the neighborhood of human fixations were matched to features of known objects to determine the class of the respective object.
Statements about its position could not be made in this way.
The authors of \cite{xiao2018salient} reduced the number of superpixels for salient object detection with gaze data. 
In contrast to our approach, however, this required both multiple gaze points and training data.
With only one gaze point, \cite{weber2020distilling} managed to drastically reduce the number of candidate bounding boxes of a region proposal method, but this method is only applicable in 2D.

In this work, we build on existing research to improve human-robot interaction.
While speech and gesture are popular channels for communication, gaze is challenging~\cite{holmqvist2021eye} and often neglected.
In the following, we link eye tracking and augmented reality to address classical calibration problems as well as data-independent 3D object detection in a collaborative manner.

\subsection{Methods}
\label{paper02:sec:method}

In this work, we propose finding 3D positions of unknown objects by incorporating human gaze into the robot's segmentation process.
For this purpose, we first introduce the interface used to communicate with the robot.
Subsequently, we present an extrinsic robot calibration method, which is particularly characterized by its flexibility and ease of execution. In our case, we calibrate a camera's position relative to the robot's base, but in principle the method can be applied to any sensors.
Finally, we explain the segmentation process that applies said methods.

\subsubsection{Augmented Reality Interface}
\label{paper02:subsec:ARInterface}
All interaction with the robot is guided via an augmented reality interface and serves as a two-way communication channel between human and robot.
In this way, we can, for example, control the movement of the robot, access the robot's camera feed, or perform the extrinsic calibration between robot and its camera.
In addition, we can provide the robot with the human gaze data and display the results of the object detection.
We use the HoloLens~2 from Microsoft, a head-mounted pair of mixed reality glasses with a built-in eye tracker.
For the development of \acs{AR} applications, Microsoft provides an open-source cross-platform toolkit called Mixed Reality Toolkit~(MRTK).
The creation and development of our interface takes place in the game development environment Unity.
We use the versions MRTK~2.7.2 and Unity~2019.4.29.
For the actual communication between the HoloLens' \acf{UWP} and the robot operating system~(\acs{ROS}), we resort to the \acs{UWP} version of ROS\#~\cite{bischoff2019rossharp}, a set of open source software libraries and tools for communicating with \acs{ROS} from Unity applications.
On system startup, the robot launches ROS\#'s file\_server package as well as rosbridge\_server from the rosbridge\_suite.
As soon as the \acs{AR} interface is started on the HoloLens, it immediately establishes a connection with the robot via Wi-Fi.
Thereupon, ROS\# uses the rosbridge protocol to send JSON based commands via WebSockets, enabling the deployment of custom publishers and subscribers.
During runtime, the menu of our interface can be opened by looking at the user's palm.
Created virtual objects can then be selected by voice or gestures.
For example, menu buttons can be simply pressed with a finger or other virtual objects can be selected by looking at them and pinching the thumb and index finger together or saying ``select''.

\subsubsection{Calibration \& Gaze Estimation}
\label{paper02:subsec:calibration}
The incorporation of the human gaze into the robot's world requires the estimated gaze to be mapped from the reference frame of the human, provided by the HoloLens, into the robot's frame of reference.
For this purpose, the transformation can be computed either directly, if the pose of one device in the frame of the other is known, or through indirect co-location by finding corresponding points in the image of the two associated cameras \cite{weber2020distilling}.
The former is often difficult to realize in practice, while the latter has some disadvantages, namely limiting the view of both participants to an overlapping field of view.
Furthermore, for the robot to interact with objects in its field of view, the position and orientation of the robot's camera relative to its base must also be known.
The solution comes in the form of augmented reality, which we can employ as a bridge.
If we create virtual counterparts corresponding to the real poses of the respective frames, we become acquainted with the transformation between frames through the transformation between virtual elements.
More precisely, we determine the mutual position of the robot and the robot's camera sensor by aligning them with the corresponding virtual objects and calculating the transformation occurring in between in the virtual space of the HoloLens.
The authors of \cite{blankemeyer2018intuitive}, \cite{reardon2018come} and \cite{kastner2019augmented} did something similar to align the coordinate systems of a robot and that of a HoloLens. However, in their case, all the necessary robot sensors had been calibrated beforehand.
The advantage of our calibration method is that it splits the usual time-consuming extrinsic sensor calibration into multiple parts.
In case of frequent calibrations, only the fast part needs to be repeated.

For us, the approach described means we can intertwine both of our problems:
On the one hand, we can calibrate the position of the robot and the camera in relation to each other, and, on the other hand, we can establish a direct transformation between HoloLens and the robot, which means that the robot is aware of the gaze point at all times regardless of the field of view.
An overview of the underlying pipeline is shown in \figref{paper02:fig:pipeline}.
\begin{figure}
	\centering
	\includegraphics[width=.9\linewidth]{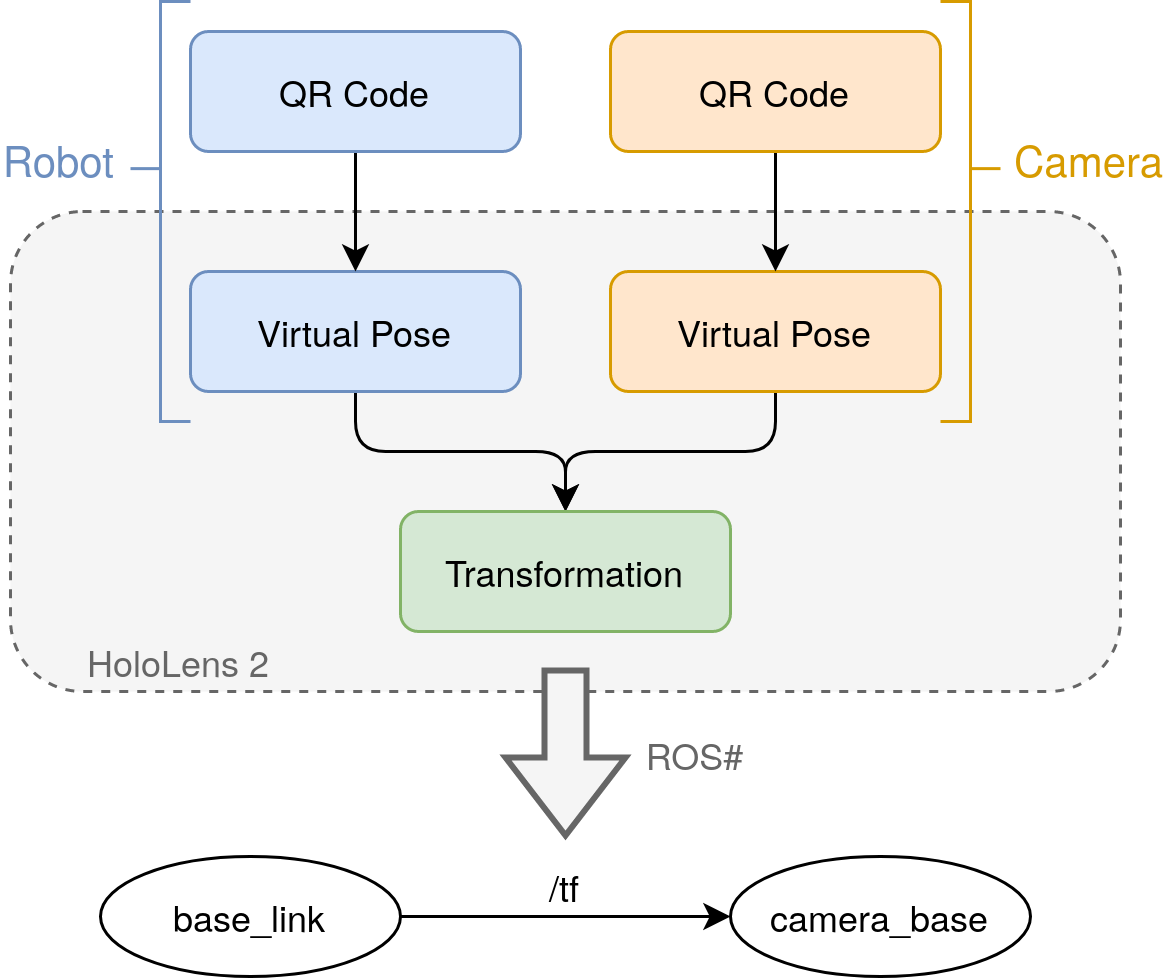}
	\caption{The QR codes specify the position of the virtual versions of the robot and the camera. The intervening transformation can be determined in the virtual world of the HoloLens~2 and is then published via ROS\#.}
	\label{paper02:fig:pipeline}
\end{figure}

We start by determining the poses of the two frames of interest.
This is, in our case, the so called base\_link on the robot side and the camera\_base frame on the camera side.
In principle, however, any frame can be used whose origin is known relative to a point on the housing.
To align real and virtual versions of the robot and its camera, we attach fiducial markers in the form of QR codes (see \figref{paper02:fig:ARCalibration}) as they allow for robust and inexpensive detection.
\begin{figure}
	\centering
    \includegraphics[width=\linewidth]{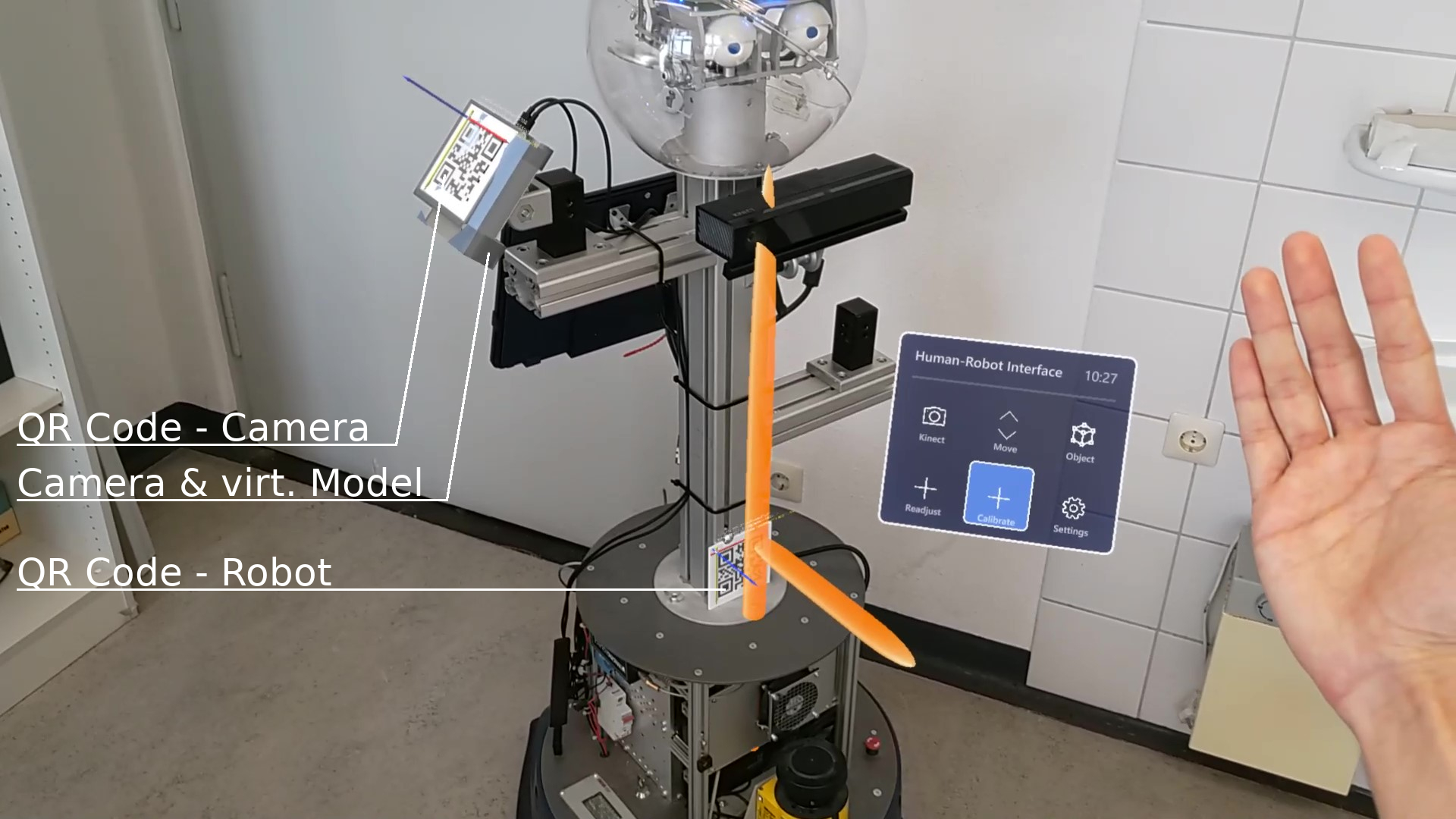}
	\caption{The \acs{AR} interface appears when looking at the open palm. The QR code on the camera positions the virtual camera model and the QR code on the robot's torso defines the robot's forward direction and center of rotation (orange).}
	\label{paper02:fig:ARCalibration}
\end{figure}
The HoloLens~2 is moreover capable of detecting QR codes at the system level in the driver.
However, we have to consider that there will be an offset between the pose of the markers and the actual frame.
So let $\{b\}$, $\{m_b\}$, $\{c\}$ and $\{m_c\}$ be the coordinate frames of the robot's base (base\_link), the QR code on the base, the camera (camera\_base), and the marker on the camera, respectively.
For two frames ${f_1}, {f_2} \in \left\{ \{b\}, \{m_b\}, \{c\}, \{m_c\} \right\}$, let the transformation from ${f_1}$ to ${f_2}$ be denoted by $^{f_1}T_{f2} \in \operatorname{SE}(3)$.
The connection between the frames can be illustrated by the following transformation graph:
\begin{center}
    \begin{tikzpicture}
        \node[] (B) at (-1,0) {$\{b\}$};
        \node[] (C) at (1,0) {$\{c\}$};
        \node[] (Mb) at (-3,0) {$\{m_b\}$};
        \node[] (Mc) at (3,0) {$\{m_c\}$.};
    
        \path [->, dashed] (B) edge node[above] {$\prescript{b}{}{T}_{c}$} (C);
        \path [->](Mb) edge node[above] {$\prescript{m_b}{}{T}_{b}$} (B);
        \path [->](Mc) edge node[above] {$\prescript{m_c}{}{T}_{c}$} (C);
    \end{tikzpicture}
\end{center}
The QR codes on the robot and camera can usually be attached to their housing so that they are either parallel or perpendicular to it.
Thus, their orientations and, hence, the rotations to the corresponding frames are known.
The same applies to the translation between $\{m_b\}$ and $\{b\}$, since the marker can be placed on the robot according to existing knowledge about other robot frames.
If, contrary to expectations, this is not possible, we also managed to approximately estimate the center of rotation of the robot, i.e. where the base\_link frame $\{b\}$ is located, as the geometric center of the virtual circle drawn by the marker on the camera as the robot rotates around its own axis.
The translation from $\{m_c\}$ to $\{c\}$ can be determined with the help of manufacturer information about the dimensions of the camera.
This means $\prescript{m_b}{}{T}_{b}$ and $\prescript{m_c}{}{T}_{c}$ are known.

We want to determine the transformation $\prescript{b}{}{T}_{c}$.
The idea is to add a frame $\{h\}$ corresponding to the coordinate system of the HoloLens to close the transformation graph:
\begin{center}
    \begin{tikzpicture}
        \node[] (B) at (-1,0) {$\{b\}$};
        \node[] (C) at (1,0) {$\{c\}$};
        \node[] (Mb) at (-3,0) {$\{m_b\}$};
        \node[] (Mc) at (3,0) {$\{m_c\}$};
        \node[] (H) at (0,1.2) {$\{h\}$};
    
        \path [->, dashed] (B) edge node[above] {$\prescript{b}{}{T}_{c}$} (C);
        \path [->](Mb) edge node[above] {$\prescript{m_b}{}{T}_{b}$} (B);
        \path [->](Mc) edge node[above] {$\prescript{m_c}{}{T}_{c}$} (C);
        \draw [->](H) -| node[above, pos=0.25] {$\prescript{h}{}{T}_{m_b}$} (Mb);
        \draw [->](H) -| node[above, pos=0.25] {$\prescript{h}{}{T}_{m_c}$} (Mc);
    \end{tikzpicture}
\end{center}
After the two QR codes have been detected by the HoloLens, they can be selected via our \acs{AR} interface and $\prescript{h}{}{T}_{m_b}$ and $\prescript{h}{}{T}_{m_c}$ can be estimated.
Finally, the transformation $\prescript{b}{}{T}_{c}$ from the base of the robot to the camera is given by the following equation:
\begin{equation*}
\label{paper02:eq:transformation}
    \prescript{b}{}{T}_{c} = \prescript{m_b}{}{T}_{b}^{-1} \prescript{h}{}{T}_{m_b}^{-1} \prescript{h}{}{T}_{m_c} \prescript{m_c}{}{T}_{c}.
\end{equation*}
The result can be published from the HoloLens using ROS\# to the transformation topic /tf, making it available to the robot.

Furthermore, we can use $\{h\}$ as a parent frame in which the robot's odometry frame is embedded.
This gives us a reference point for the gaze information that we can access via MRTK.
Associated with $\{h\}$, we can publish this data on a separate topic.
This includes the gaze vector and the hit point of the eye gaze ray with the target.

It should be noted that the fiducial markers are only needed while performing the calibration.
Once they have been detected and selected, the user is free from restrictions on the field of view.
In addition, contrary to the usual procedure, we do not determine the calibration parameters externally and then store them in configuration files.
This means that we can make changes to the camera, such as the tilt, even during runtime.
This is a great advantage for use under changing scenarios.

\subsubsection{Segmentation}
\label{paper02:subsec:segmentation}
We now address the problem of detecting unknown objects in the three-dimensional environment.
We tackle this task by enhancing existing segmentation methods on the robot side with gaze information from the human collaborator.
The segmentation process can be triggered either on demand by multimodal interaction, such as gestures or speech, or -- empowered by the calibration method -- continuously in real time.
The assistance that the robot receives from the human should be limited solely to the provision of the gaze information. Apart from that, the segmentation should only take place on the robot's side.
This makes sense due to the robot's higher resources and computing power compared to head-mounted devices like the HoloLens.

The segmentation process starts with a pass through filter where we assume that all relevant objects are between zero and three meters away from the camera, followed by a voxel grid filter with a leaf size of 0.03 along each axis that downsamples the point cloud we acquire from the robot's camera.
This is not mandatory, but it reduces the computation time drastically and allows a segmentation in real time.
In most cases, we can assume that the objects to be detected lie on a surface that is reasonably flat.
This could be, for example, a table, a shelf, or the floor itself.
We can take advantage of the parallelism between all these surfaces.
Due to our calibration, we know the orientation of the camera with respect to the robot standing on the floor.
This means that we can transform the upward vector from the HoloLens world frame into the camera frame and thus obtain the normal vector of the surface, that is parallel to the floor and on which the objects are located, in the frame of the camera.
We can then use \acs{RANSAC} to search for the largest plane in the robot's field of view, namely the said surface, that is perpendicular to the given normal vector.
Thereby, we set the maximum allowed deviation from the normal vector to $ 30 $ degrees.
All points belonging to this plane are finally removed from the point cloud.
In the next step, we let the gaze information flow in.
Since the human is looking at the object of interest, we know at least one point on its surface.
Starting from this point, we can cluster the point cloud using simple euclidean clustering.
That is, we first use a k-d tree to find the point in the point cloud that is closest to the gaze point.
Then we cluster the point cloud with respect to the Euclidean distance, a tolerance of \SI{5}{mm}, and a minimum cluster size of $ 500 $.
All points that belong to the same cluster as the nearest neighbor of the initial point result in the searched object.
Note that without the gaze information we would not be able to distinguish between clusters belonging to objects, clusters of parts of the environment, or noise.
\emph{This subtle gaze interaction resolves ambiguities and brings us closer to a natural learning process.}

Finally, we do not only obtain an instance segmentation of an object, but we can also calculate a 3D bounding box from it.
The box can be aligned properly in space again due to our calibration and the robot can share the result directly with the human via our \acs{AR} interface.
Thus, the bounding box can be displayed in the human's field of view, providing direct feedback and enabling a natural two-way communication component, as well as an initialization of further interactions of the robot with the object.
\subsection{Evaluation}
\label{paper02:sec:evaluation}
In our experiments, a Scitos~G5 from MetraLabs~\cite{MetraLabs} was employed as a robotic counterpart.
It has been equipped with an Azure Kinect DK, whose relative position to the robot we want to calibrate.
The camera also provides image data such as the point cloud on which we perform the object detection.
All components communicate with each other using \acs{ROS}~\cite{quigley2009ros}.

\subsubsection{Qualitative Analysis}
One of the advantages of our method is already evident when performing a single calibration run.
Whereas calibration methods based on data collection are time-consuming and difficult to automate \cite{elatta2004overview}, the entire procedure with our variant takes less than a minute.
Depending on the user's experience, a single run usually takes only between $ 15 $ and $ 40 $ seconds.
This is especially apparent when the camera needs to be adjusted more frequently, either because it has been unintentionally moved or because the setting has changed.

After calibration, the whole system, including gaze mapping and the object segmentation, runs in real time.
\figref{paper02:fig:RVIZ} shows a visualization in RVIZ.
\begin{figure}[b!]
	\centering
    \includegraphics[width=\linewidth]{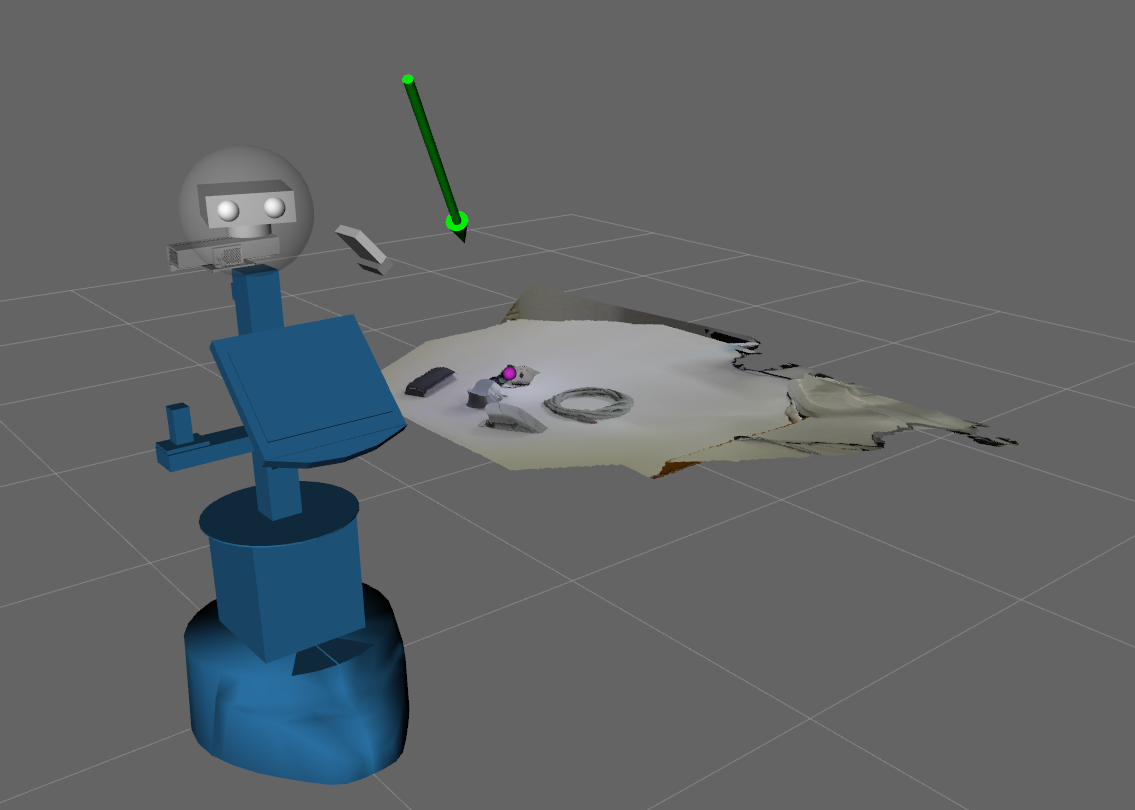}
    \caption{The robot model with the camera positioned relative to it. The human gaze vector is shown as a green arrow and the gaze hit point as a purple sphere.}
    \label{paper02:fig:RVIZ}
\end{figure}
The gaze ray vector as well as the coordinate of the hit point on the target are published with \SI{59}{Hz}.
Using the default configuration, the Azure Kinect provides the point cloud at $ 4 $ frames per second.
Subsequently, segmentation reduces the rate of the outgoing segmented cloud and thus also that of the bounding box to $ 2 $ frames per second.
Since the minimal fixation duration is, in most cases, at least \SI{200}{ms} \cite{holmqvist2011eye} and the recommended feedback delay time for manual pointing actions is approximately between \SI{350}{ms} and \SI{600}{ms} \cite{muller2007dwell}, an update every $ 0.5 $ seconds is sufficient.
Consequently, our method is suitable for human-robot interaction in real time.
 
\begin{figure}[t!]
	\centering
	\begin{subfigure}{0.5\columnwidth}
	    \centering
	    \includegraphics[width=\linewidth]{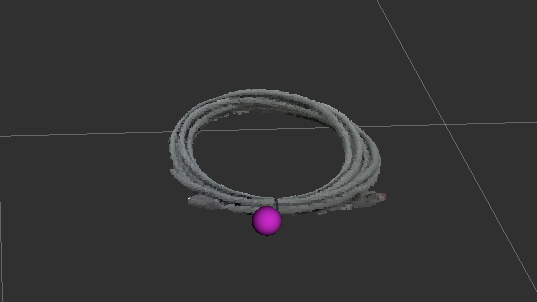}
	\end{subfigure}%
	\begin{subfigure}{0.5\columnwidth}
	    \centering
	    \includegraphics[width=\linewidth]{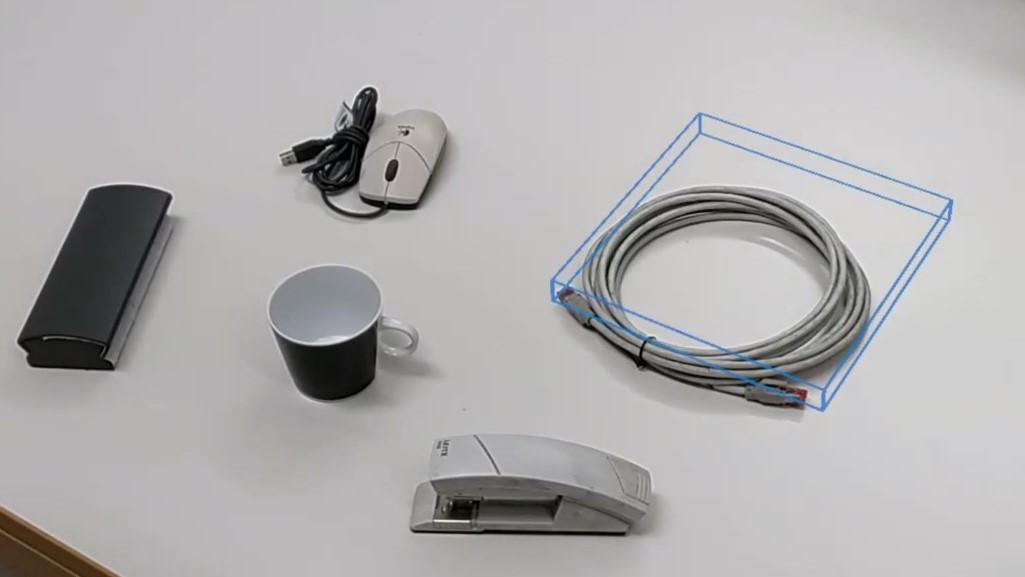}
	\end{subfigure}%
	\par\smallskip
	\begin{subfigure}{0.5\columnwidth}
	    \centering
	    \includegraphics[width=\linewidth]{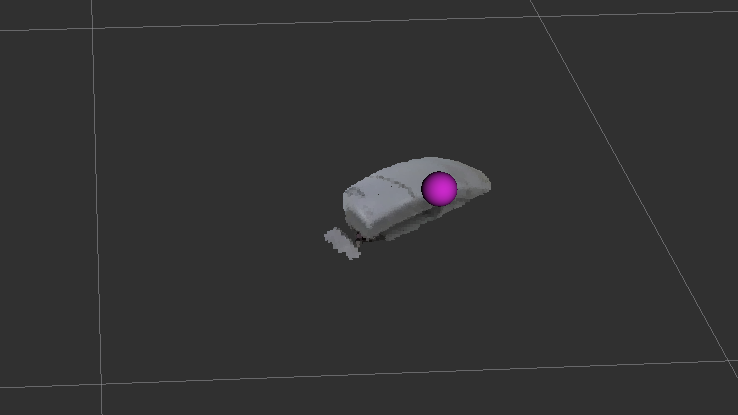}
	\end{subfigure}%
	\begin{subfigure}{0.5\columnwidth}
	    \centering
	    \includegraphics[width=\linewidth]{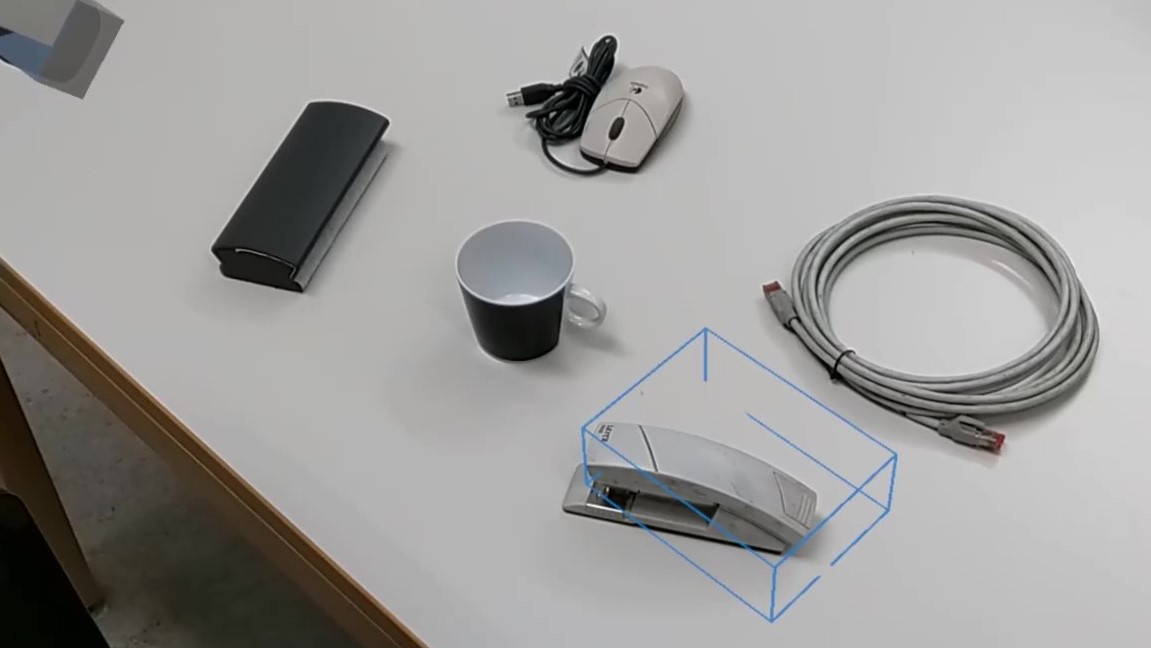}
	\end{subfigure}%
	\par\smallskip
	\begin{subfigure}{0.5\columnwidth}
	    \centering
	    \includegraphics[width=\linewidth]{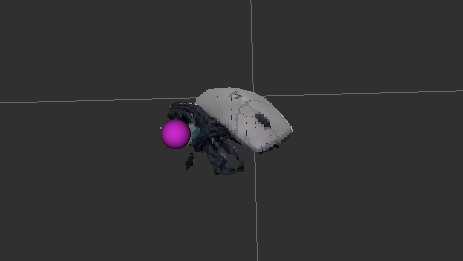}
	\end{subfigure}%
	\begin{subfigure}{0.5\columnwidth}
	    \centering
	    \includegraphics[width=\linewidth]{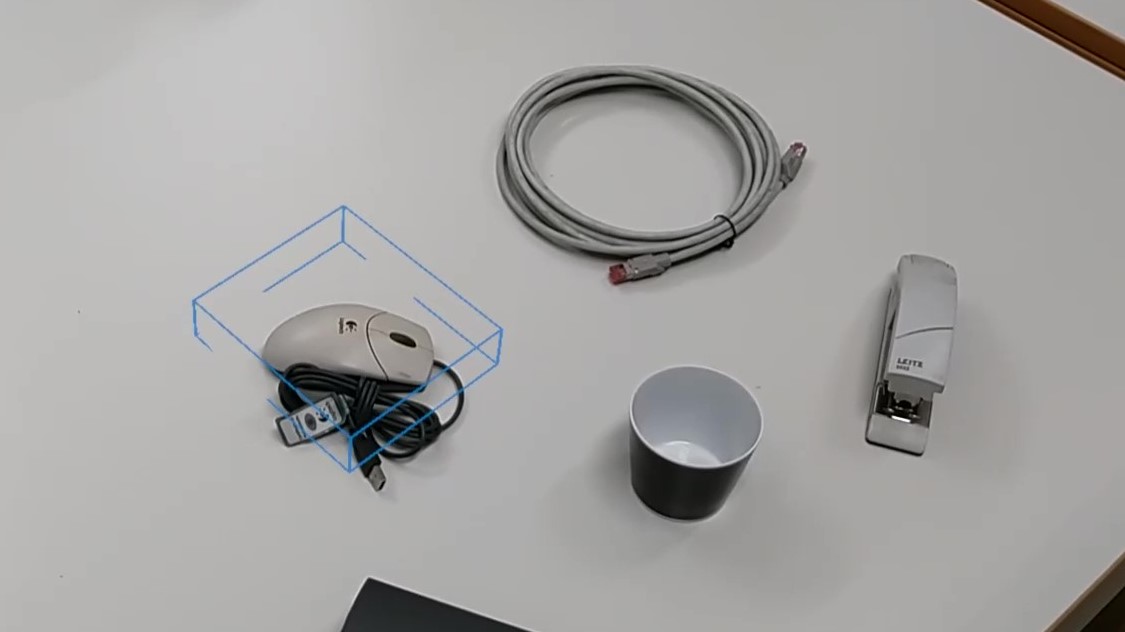}
	\end{subfigure}%
	\par
	\caption{The segmentation with the gaze point (left) and the resulting bounding box as seen from the human (right). The box is given in world coordinates, therefore tracking of already detected objects during movements of the robot is superfluous.}
	\label{paper02:fig:SegBox}
\end{figure}

Some final example results of segmented objects and the respective bounding box can be seen in \figref{paper02:fig:SegBox}.
For simplicity, we have chosen common household objects and office utensils, which we have placed on a table in front of the robot.
In principle, both humans and robots can move freely around the table, since the position of both is known in the HoloLens based parent frame.
However, to ensure that the robot's movements are tracked as precisely as possible, an additional localization procedure
is required, which is beyond the scope of this work, as solving such a problem has already been extensively researched, and possible solutions can be found in the literature \cite{montemerlo2002fastslam}, \cite{mur2015orb}.
Naturally, the current position can be manually repositioned at any time via our interface.

\subsubsection{Quantitative Analysis}
First we start with the evaluation of the calibration part.
To establish a reference ground truth, we utilize the OptiTrack motion capture system \cite{OptiTrack}.
We place multiple reflective markers on both the robot and the camera.
These can be tracked by the Optitrack system with an accuracy of \SI{1}{mm}.
Given these point observations, we can calculate the robot and the camera poses with respect to the coordinate system of the motion capture system, and then compute the camera pose of interest relative to the coordinate system of the robot.
Based on the deviations we have observed in several test trials, we estimate that this post-processing decreases the accuracy to about \SI{3}{mm}.
In this way, we determine the ground truth of the transformations from the robot frame to the camera frame for three different poses of the camera.
Once horizontally, i.e. parallel to the floor, once vertically, i.e. perpendicular to the floor, and once in an inclined position at about $ 45 $ degrees.
Without moving the camera in between, one of the system's designers performed the calibration 20 times per tilt using the method we presented in \secref{paper02:sec:method}.
For each tilt, we evaluate the translation and rotation components separately.

\tblref{paper02:tab:translation} shows the result of the translation part of our \acs{AR}-based calibration compared to the calibration using OptiTrack.
\begin{table}[b!]
	\centering
	\caption{The translation in meters determined by the calibration with OptiTrack as well as our \acs{AR} interface.}
	\label{paper02:tab:translation}
	\begin{threeparttable}
		\begin{tabular}{lcccccc}
			\toprule
			     & \multicolumn{2}{c}{Vertical} & \multicolumn{2}{c}{Horizontal} & \multicolumn{2}{c}{Inclined}\\
			Axis & OptiTrack & mARC\tnote{a} & OptiTrack & mARC & OptiTrack & mARC \\
			\midrule
			x\tnote{b} & -0.081 & -0.080 & -0.079 & -0.079 & -0.077 & -0.079 \\
			y          & -0.295 & -0.295 & -0.324 & -0.327 & -0.331 & -0.332 \\
			z & \phantom{-}0.973 &\phantom{-}0.973 &\phantom{-}1.072 &\phantom{-}1.071 &\phantom{-}1.033 &\phantom{-}1.035 \\
			\midrule
			\O Dist.\tnote{c} & \multicolumn{2}{c}{0.003}	& \multicolumn{2}{c}{0.004}	& \multicolumn{2}{c}{0.003} \\
			\bottomrule
		\end{tabular}
		\begin{tablenotes}
			\item[a] Average value of our \acs{AR}-based calibration
			\item[b] Coordinate axes refer to the \acs{ROS} coordinate system
			\item[c] Average spatial distance of all runs calculated with the euclidean norm
		\end{tablenotes}
	\end{threeparttable}
\end{table}
In the table, the translation in each direction is given with respect to the \acs{ROS} coordinate system.
The difference between the result of the OptiTrack system and the mean result from our calibration varies, but is not noticeably pronounced with respect to any direction.
The largest difference is observed with \SI{3}{mm} in the direction of the y-axis in the case of horizontal orientation.
All other values do not differ at all or only $ 1 $ to \SI{2}{mm}.
Our analyses have shown that the same is true for the average of all individual differences to the ground truth.

Since the deviation in individual directions is less relevant than the spatial distance, we also want to take this into account.
We measured the euclidean distance of the translation of each individual calibration run from the ground truth translation.
The results are reported in the last row of \tblref{paper02:tab:translation}.
One can see that the spatial error does not exceed \SI{4}{mm} on average.
This is comparable to the accuracy of the extrinsic calibrations evaluated in \cite{zhang2004extrinsic} and \cite{bi2018automatic}.
The distribution of the individual euclidean distances to the ground truth are shown in the box plot in \figref{paper02:fig:TranslationError}.
\begin{figure}
	\centering
    \includegraphics[width=.85\linewidth]{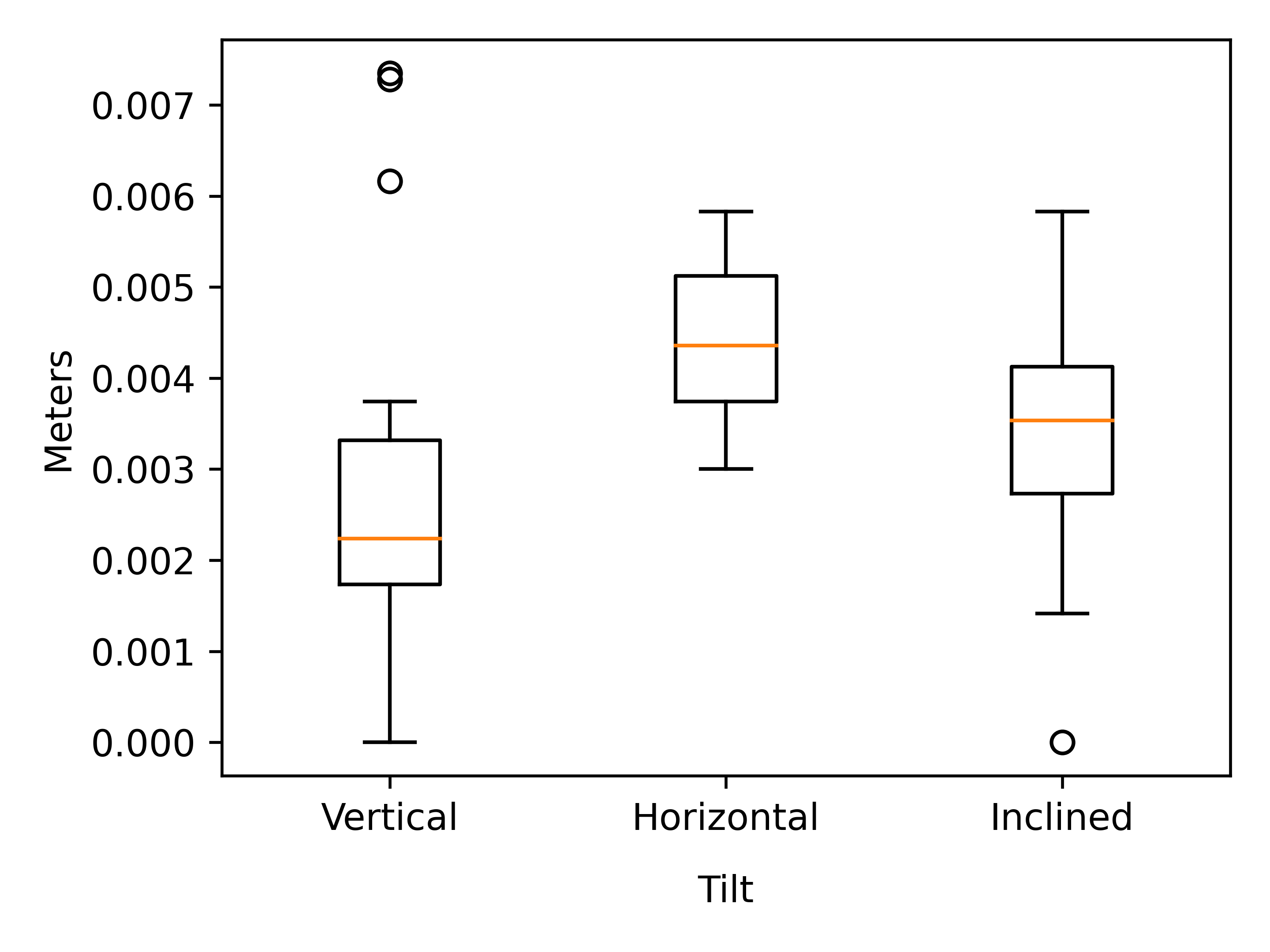}
    \caption{The box plot represents the distribution of the translation errors with respect to the euclidean norm.}
    \label{paper02:fig:TranslationError}
\end{figure}
Although the error in the vertical setting is generally the smallest, there were also some outliers.
Basically, in all three scenarios the vast majority of errors were below \SI{5}{mm}.
The medians lie between \SI{2}{mm} and \SI{4.5}{mm}.
Note that this is only slightly above the accuracy range of the reference ground truth estimated via OptiTrack.

We now examine the rotational error of the transformation.
In general, each rotation can be expressed by an axis of rotation and an angle of rotation.
This rotation angle can be considered a measurement of the similarity of two orientations.
This means that for each rotation component determined by our calibration, we calculate the difference rotation, which transforms the obtained rotation into the ground truth rotation.
The smaller the angle of rotation, the more similar the two rotations.
The angles of all difference rotations are plotted in \figref{paper02:fig:RotationError}.
\begin{figure}
    \centering
    \includegraphics[width=.85\linewidth]{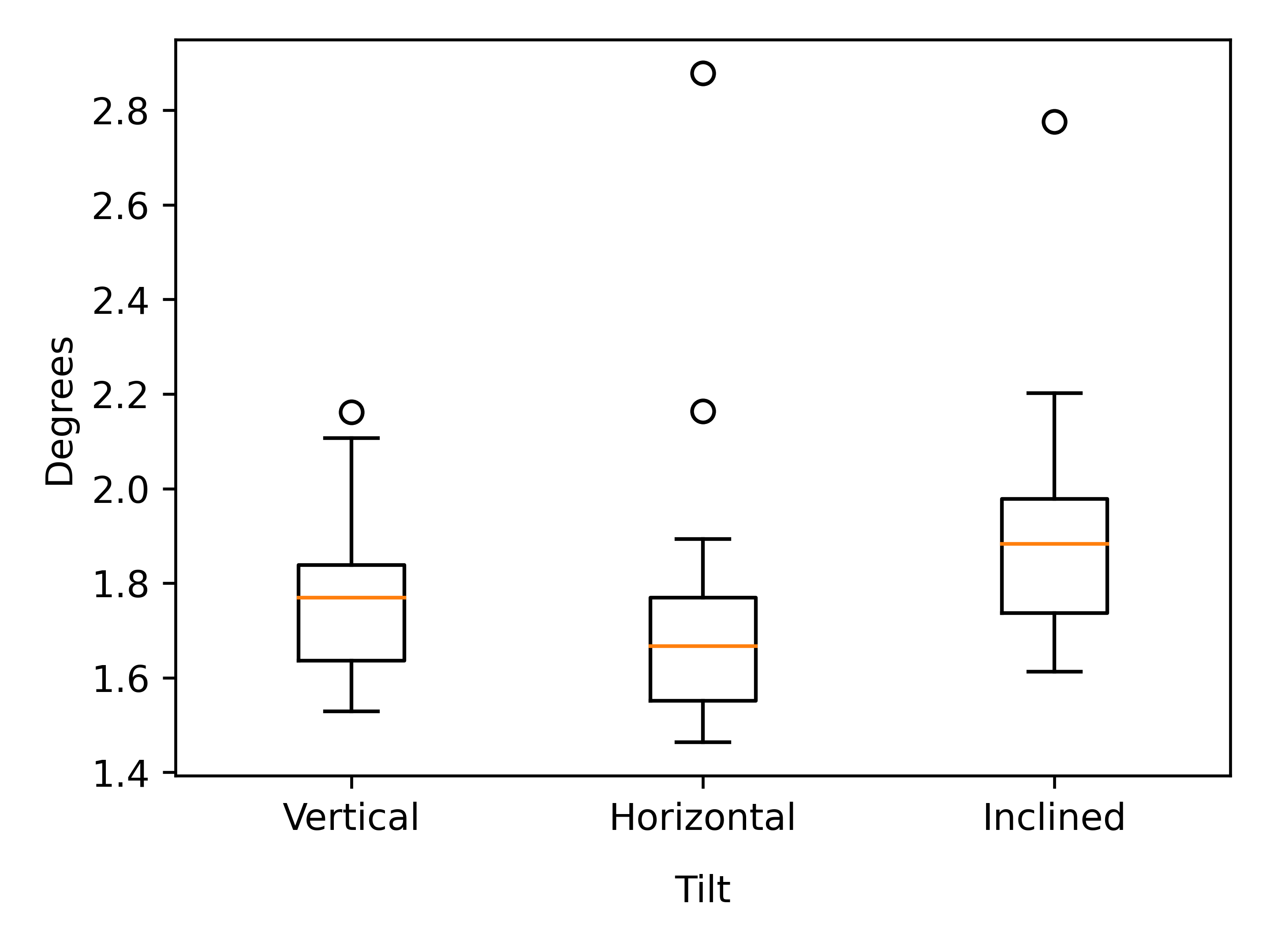}
    \caption{The rotation errors displayed in a box blot.}
    \label{paper02:fig:RotationError}
\end{figure}
Although there are, again, a few outliers, most values are below 2 degrees with medians ranging from 1.6 to 1.9 degrees.
The same applies to the average rotation error.
Thus, the rotation error is of the same order of magnitude as that of classical approaches like \cite{zhang2004extrinsic}.
All in all, the accuracy meets the requirements of most applications, including our gaze segmentation, while being flexible and fast.

Let us now have a closer look at the evaluation of the segmentation part.
Although the performance of current 3D object detectors lags behind the state of the art in 2D object detection, there are 3D object detectors that promise good results on indoor datasets such as SUN RGB-D \cite{song2015sun}.
However, our experiments have shown that the claimed results are difficult to achieve in practical applications.
One possible reason could be that, due to comparability, the evaluations are usually conducted on the same few categories \cite{qi2018frustum}, \cite{wang2019frustum}, \cite{qi2019deep}.
As a result, performance on other classes is often significantly worse or remains unknown.

We trained several neural networks, such as VoteNet \cite{qi2019deep} and Frustum ConvNet \cite{wang2019frustum} on the classes book, bottle, bowl, cup, keyboard, laptop,  mouse, paper, plant, and telephone from the Sun RGB-D dataset.
These objects were more appropriate for our setup although smaller than the furniture used in the original papers.
It turned out that all state-of-the-art networks performed very poorly on our set of objects and could not serve as reasonable reference ground truth.
To put this in numbers: Whereas the mean average precision with a 3D Intersection over Union (\acs{IoU}) threshold of $ 0.25 $ was only \SI{27.8}{\%} for Frustum ConvNet, this value was even less than \SI{1}{\%} for VoteNet.
Thus, almost none of the available test objects were successfully detected by the neural networks and a meaningful comparison was therefore not possible.
For this reason, we devised an alternative evaluation strategy and eventually conducted two different approaches.
In the first one, we labeled the 3D bounding boxes of the objects in the acquired point cloud of the scene by hand and calculated the 3D \acs{IoU} (with regard to the volume) for ten test objects.
In the second one, we used a pretrained 2D object detector to avoid vulnerability regarding a bias in labeling.
While modern 3D detectors are still far from being able to serve as ground truth, 2D detectors certainly are capable of doing so.
Hence, we projected the points segmented by our method onto the 2D image plane and compared the resulting 2D bounding box with Faster~R-CNN \cite{ren2015faster} (ResNet-101 backbone) trained on Microsoft~COCO~\cite{lin2014microsoft}.
This dataset was also the criterion by which the ten test objects were selected.
The results of both evaluations are shown in \tblref{paper02:tab:IoU}.
\begin{table}
	\centering
	\caption{The \acs{IoU} between the bounding boxes obtained by our method and the respective ground truth.}
	\label{paper02:tab:IoU}
	\resizebox{\columnwidth}{!}{%
	\begin{threeparttable}
		\begin{tabular}{r|cccccccccc|c}
			\toprule
			Class name & apple & backpack & book & bowl & clock & cup & keyboard & mouse & remote & tennis ball & mIoU \\
			\midrule
			2D \acs{IoU} & 0.78 & 0.78 & 0.79 & 0.86 & 0.68 & 0.84 & 0.88 & 0.79 & 0.80 & 0.87 & 0.81 \\
			3D \acs{IoU} & 0.70 & 0.66 & 0.72 & 0.84 & 0.62 & 0.73 & 0.66 & 0.59 & 0.64 & 0.71 & 0.69 \\
			\bottomrule
		\end{tabular}
	\end{threeparttable}}
\end{table}
In the 2D case, all values are above $ 0.5 $ and thus all objects can be considered correctly detected \cite{everingham2010pascal}, \cite{zitnick2014edge}.
Furthermore, almost all values are even above $ 0.7 $ with a mean \acs{IoU} of $ 0.81 $.
In contrast, the 3D \acs{IoU} values are naturally smaller.
Nevertheless, all objects are again considered to be detected, using the usual 3D threshold of $ 0.25 $ as reference \cite{qi2018frustum}, \cite{song2015sun}.
The mean 3D~\acs{IoU} is $ 0.69 $.
\figref{paper02:fig:Recall} shows the recall as a function of the \acs{IoU} threshold at which a bounding box is classified as true positive.
\begin{figure}
    \centering
    \includegraphics[width=.9\linewidth]{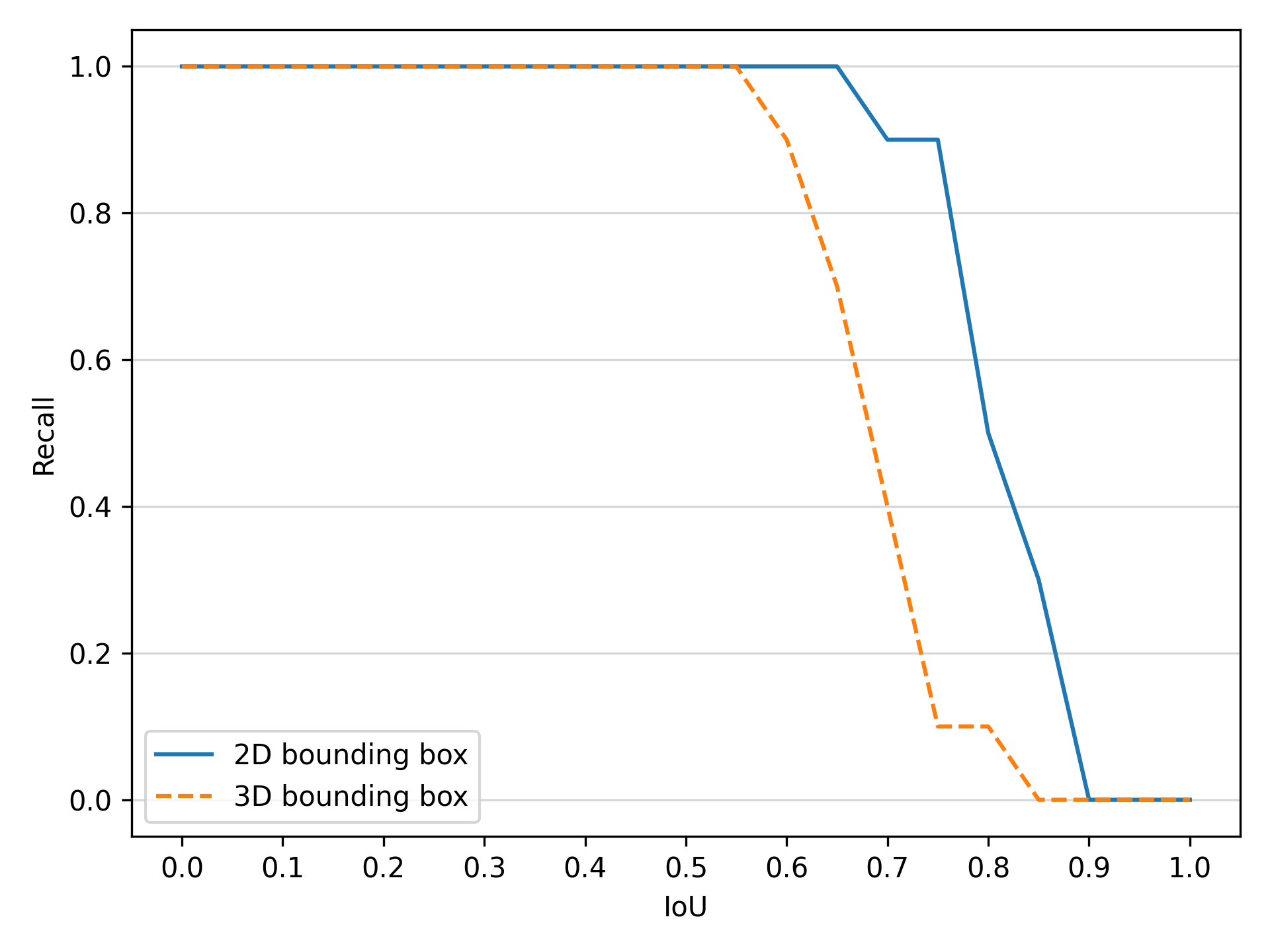}
    \caption{The recall as a function of the \acs{IoU} threshold at which the objects are considered to be detected.}
    \label{paper02:fig:Recall}
\end{figure}
Note that even with a 3D~\acs{IoU} threshold of $ 0.5 $, which is twice as large as that used by the authors of VoteNet and others \cite{wang2019frustum}, \cite{song2015sun}, the recall is still~\SI{100}{\%}.

Overall, our method hints at going far beyond the practical applicability of state-of-the-art neural network-based 3D object detectors, illustrating the importance of diverse solution strategies along with neural networks.

\subsubsection{Limitations}
Although our method of calibration is remarkably fast and user-friendly, the initial setup takes some time. While less in-depth expert knowledge is required compared to other methods, care must be taken to ensure that the markers are positioned accurately and that the distances to the corresponding frames can be determined.
However, since this is a one-time step, this time expenditure is not of any significance compared to the time saved in each subsequent calibration run.

Furthermore, as with any other existing method, our segmentation and the calculated bounding box strongly depend on the quality of the original point cloud provided by the depth sensor.
In this regard, the perspective of the robot's camera on the object also plays a role and whether the depth sensor can correctly determine the distance at the edges of the objects.
However, the fact that the image quality has an influence on the result is in the nature of things and could be resolved by using multiple cameras or additional angles.

For objects that are too close to each other, it is not possible to keep them apart by extracting euclidean clusters.
In this case, one could, for instance, resort to a min-cut based segmentation algorithm, also generally suitable, since a point must be given in the center of the object, which can be provided by the gaze point.
In our tests, min-cut segmentation indicated promising results, but also required the approximate size of the respective object as an additional input argument.

\subsection{Conclusion}
\label{paper02:sec:conclusion}

In this work, we presented a novel method that allows for the deployment of robots under changing and non-predefined conditions. In this course, we combined robotics, augmented reality, and eye tracking to improve human-robot collaboration.
Merely by receiving gaze information from its human partner, the robot was capable of detecting and segmenting unknown objects.

While most existing methods for extrinsic robot calibration are time consuming and often quite complicated to conduct, we have developed a method that is user-friendly, customizable at runtime, and takes only a few seconds to complete.
At the same time, our evaluation has shown that we still achieve competitive accuracy compared to classical methods.

In addition, we bridge the two worlds of human and robot through the use of head mounted augmented reality glasses, giving the robot access to another persistent information channel --- the human gaze. 
Just by having a human look at an object, the robot was able to segment objects it has not seen before and calculate associated three-dimensional bounding boxes.
This goes beyond the capabilities of some state-of-the-art 3D object detectors and we found that our method works in situations where current existing neural networks have failed.
Through direct feedback in the augmented human reality, the human is continuously informed about the results and the initialization of further interactions between the robot and the object is possible.
This could be especially relevant for physically disabled people who are limited to movements in the head or neck area, in combination with a robotic arm that helps them grasp or manipulate objects.

In summary, our proposed method is versatile and facilitates general human-robot collaboration, as well as unknown object detection in the context of such scenarios in particular.
However, there remains a significant amount of future work as we seek to investigate our segmentation in more challenging scenarios and realize a subsequent interaction between the robot and the objects.

\fi

\clearpage
\setcounter{footnote}{0}
\section{Multiperspective Teaching of Unknown Objects via Shared-gaze-based Multimodal Human-Robot Interaction}
\label{app:sec:paper04}

\ifpaper
\subsection{Abstract}
For successful deployment of robots in multifaceted situations, an understanding of the robot for its environment is indispensable.
With advancing performance of state-of-the-art object detectors, the capability of robots to detect objects within their interaction domain is also enhancing.
However, it binds the robot to a few trained classes and prevents it from adapting to unfamiliar surroundings beyond predefined scenarios. In such scenarios, humans could assist robots amidst the overwhelming number of interaction entities and impart the requisite expertise by acting as teachers.
We propose a novel pipeline that effectively harnesses human gaze and augmented reality in a human-robot collaboration context to teach a robot novel objects in its surrounding environment.
By intertwining gaze (to guide the robot's attention to an object of interest) with augmented reality (to convey the respective class information) we enable the robot to quickly acquire a significant amount of automatically labeled training data on its own.
Training in a transfer learning fashion, we demonstrate the robot's capability to detect recently learned objects and evaluate the influence of different machine learning models and learning procedures as well as the amount of training data involved. 
Our multimodal approach proves to be an efficient and natural way to teach the robot novel objects based on a few instances and 
allows it to detect classes for which no training dataset is available.
In addition, we make our dataset publicly available to the research community, which consists of RGB and depth data, intrinsic and extrinsic camera parameters, along with regions of interest.

\subsection{Introduction}
As technology progressed, more and more robots were developed for the industrial sector, and their fields of application became diversified.
Numerous industries, including automotive, electronics, rubber and plastics, cosmetics, pharmaceutical, and food and beverage, benefit from their superior precision, efficiency, working capacity and tolerance to arduous and hazardous environments~\cite{xiao2021robotics}.
In the immediate environment of humans, robots are also increasingly employed in the form of service assistants, for instance in supermarkets such as Walmart~\cite{bogue2019strong,li2022effect} or as tour guides in museums~\cite{thrun1999minerva, velentza2019human}.
This success has also been fueled by recent advances in machine learning, particularly in computer vision, which allows robots to understand their environment and detect objects and people within it.
However, a core assumption is almost always that a large amount of training data with high quality labels exist.
Many large car manufacturers or companies such as Google, Tesla, and Uber have therefore established their own image and video databases, in most cases by outsourcing to crowdsourcing platforms such as Amazon Mechanical Turk~\cite{schmidt2019crowdsourced}.
Regarding service robots operating in warehouses or office environments, there are often no publicly accessible datasets that are tailored to the respective environment, comprise all relevant object classes, and are fully labeled.
Consequently, state-of-the-art object detectors perform excellently given the existence of sufficient training data, but are limited to deployment in previously specified scenarios predefined by the training data \cite{weber2020distilling}.
As soon as an object is not included, it exceeds the capabilities of the object detector and pushes the robot to the limits of its possibilities.
In fact, the proportion of objects covered by data sets is vanishingly small compared to the quantity of objects existing in practice.
For example, ImageNet~\cite{deng2009imagenet}, one of the largest publicly available image datasets, contains just 1000 classes, while the number of classes existing in the real world obviously far exceeds this number.
This fact hinders the deployment of robots in unknown environments and the dynamic adaptation to unfamiliar conditions.

In this work, we aim to make mobile robots not only more capable in terms of the tasks they have to accomplish, but also alleviate data dependency, in the sense that we extend the robot's basic knowledge by building on an existing general understanding of objects and adding new classes.
More specifically, we
teach a robot novel, unknown objects to enable it to redetect said objects within the environment in which the learning process took place.
To this end, we employ fluent and intelligent human-robot interaction, at the intersection of research fields of computer vision, eye tracking, and augmented reality (\acs{AR}).
By means of the latter, we realize a multimodal communication channel, using human gaze to direct attention to an object, and speech or gestures to convey the relevant class information to the robot.
Subsequently, the robot visually segments the object of interest and takes a series of images of the object from slightly different angles.
The data obtained in this user-friendly and convenient process is rich in information and encompasses extrinsic and intrinsic camera parameters, as well as regions of interest, in addition to RGB and depth images.
In conjunction with the class information provided by the human, the robot learns the respective object accordingly.

In summary, our main contributions are as follows:
\begin{enumerate}
    \item We propose a novel pipeline to teach a robot new, yet unknown objects.
    \item Towards this goal, we combine gaze and augmented reality in a human-robot interaction scenario to enable a feasible and swift acquisition of large amounts of labeled training data.
    \item We evaluate the learning process in detail with multiple models, different learning methods, and with varying amounts of data.
    \item We present \acf{OMD}, a versatile dataset, and make it publicly available to the research community under \url{https://cloud.cs.uni-tuebingen.de/index.php/s/2oRPs2o3FZkdBHW}.
    \item We make our system with our entire code base publicly available to the research community under \url{https://github.com/dnlwbr/Multiperspective-Teaching}.
\end{enumerate}

\subsection{Related Work}
Engaging multimodal human-robot interaction to teach a robot unknown objects requires significant multidisciplinary efforts, which we will discuss in this section. From
1) augmented reality in robotics, and
2) previous applications of eye tracking in the context of computer vision, to
3) unknown object detection, to
4) how robots learn and
5) collaborate with humans.

\subsubsection{Augmented Reality in Robotics}
With the increasing popularity of augmented reality devices, industrial applications and research also expanded~\cite{makhataeva2020augmented}.
Especially the combination of \acs{AR},
with robotic-assisted surgeries showed potential~\cite{qian2019review}, as the human remains in control via \acs{AR}, but can take advantage of the precision and consistency of the robots~\cite{vadala2020robotic}.
In terms of robot control and path planning, \cite{krupke2018comparison} controlled an industrial robot arm in pick-and-place tasks using an \acs{AR} device and \cite{quintero2018robot} designed an \acs{AR} interface to plan, preview and execute the trajectory of a robot arm.
In multi-agent systems, \acs{AR} has also been used for visual feedback~\cite{von2016robot} and remote control~\cite{reina2017ark} of robot swarms.
Enhancing the perception of the real world through \acs{AR} has thus proven to be an appealing way to communicate with robots.

\subsubsection{Eye Tracking and Computer Vision}
Eye tracking has also become an important tool in both research and industry~\cite{krafka2016eye}.
For this reason, \cite{krafka2016eye} developed an eye tracking software that works on mobile devices such as mobile phones or tablets and does not require any sensors other than a camera.
The authors of \cite{pfeiffer2014eyesee3d}, analyzed the mobile 3D eye tracking data using computer vision (and augmented reality). This involved tracking 3D markers and aligning them with virtual proxies.
In~\cite{toyama2012gaze}, the class of objects was identified by matching the features of known objects with features in the neighborhood of human fixations.
However, this required the object to appear in the database previously created specifically for this purpose, and it was also not possible to make statements about the position of the objects.
Nevertheless, eye tracking data can help to improve the performance of segmentation algorithms~\cite{mishra2009active}.
Thus, in~\cite{papadopoulos2014training}, fixations were used to train a model to annotate object locations, and in \cite{weber2022exploiting} gaze was incorporated into the segmentation process of a point cloud.
Here in turn, in both cases it was not possible to make a statement about the object classes.

\subsubsection{Unknown Object Detection}
The problem of unknown object detection was also addressed using eye tracking in~\cite{weber2020distilling}.
In this work, the number of candidate bounding boxes of a region proposal method was significantly reduced, but a classification was not possible.
Using heat maps instead of scene frames, \cite{weber2022gaze} categorized video segments based on whether a person was looking at an object and determined the parameters of the associated bounding box.
The detected objects were all unknown, but again the classes were not determined.
The authors of \cite{kabir2022unknown} addressed the problem of unknown object detection using a one-class support vector machine. 
Since the learning process was incremental, multiple robots were involved, connected to each other via a cloud-based station where all the processing took place.
In this approach, unknown objects were only identified as unknown without adding the class information to the learning process.
The purpose was to filter the unknown objects from the classified objects and forward only known objects in order to avoid sending incorrect information to the robots.
In another recent work, \cite{li2022uncertainty} proposed to exploit additional predictions of semantic segmentation models and quantify the uncertainty of the proposed segmentations.
Again, the classification task was binary and only the categories known and unknown were determined without eventually learning the objects.

\subsubsection{Teaching of Robots and Machines}
Robots are employed in a wide range of applications, especially in industry.
These include teaching assembly tasks \cite{wan2017teaching}, where the robot learns from human demonstration:
First the robot observes the human, then it imitates the human's movements.
A different way of learning through user interaction is proposed in \cite{gemignani2015teaching}.
In this approach, natural language is used to explain to the robot, which tasks it should perform.
One such typical task is grasping objects.
Solving this challenge is part of current research such as \cite{karaoguz2019object}, where an object detection approach was used to learn good grasping poses.
Data driven approaches, as stated in \cite{bohg2013data}, are often addressed by providing training data in form of labeled examples, by trial-and-error, or through human demonstration.
This means that communication between human and robot is also important here for a flexible learning progress without prior offline data generation.

\subsubsection{Joint Attention in Collaborative Settings}
Teaching in collaborative scenarios between human and robot has been investigated by \cite{krause2014learning} and \cite{el2021teaching}, among others.
In~\cite{krause2014learning}, natural language context for one-shot learning of visual objects has been used to enable a robot to recognize a described object.
In this proof of concept, however, the objects had to be unambiguously distinguishable by color or spatial relationships, and the component parts also had to be uniquely describable by linguistic expressions.
In~\cite{el2021teaching}, a teaching system for object categorization was proposed.
This system allowed the user to visualize the intermediate states of categorization, that is, to which category the robot would assign an object.
Through interaction, the categorization could be improved and corrected, but all objects had to be marked with AR markers in order to be recognized at all.
In combination with picking tasks, \cite{valipour2017incremental} and \cite{dehghan2019online} also taught a robot new objects.
In both cases, however, exactly the same objects of each class were used for both training and testing, which positively biases the results.
We use several different objects per class, which is more in line with real-world conditions.

In this work, we
combine the different research areas of eye tracking and \acs{AR} for a simple and natural interaction between robot and human to jointly solve a computer vision problem.

\subsection{Method} \label{paper04:sec:method}
The goal of this work is to teach a robot unknown objects in its environment, in such a way that the robot is later capable of detecting these objects in this same environment.
To this end, we will explain below 1) how we communicate with the robot through the modalities vision, gaze, speech, and gestures 2) how the robot identifies the object of interest, 3) how the human teaches the robot, and 4) how the robot eventually manages to learn.

\subsubsection{Augmented Reality Interface}
In order to enable the human to teach the robot anything, a communication channel is mandatory.
As suggested by \cite{weber2022exploiting}, we meet this need in the form of an augmented reality interface.
The entire communication between human and robot takes place via this interface, that is, the robot can be controlled, the gaze information of the human can be transmitted and the respective class information of the objects can be conveyed.
On the human side, we deploy the HoloLens~2, which is a pair of head-mounted augmented reality glasses manufactured by Microsoft with a built-in eye tracker.
We developed the interface, i.e. the HoloLens application, using the game engine and real-time development platform Unity, version~2019.4.36.
In addition, we used assets from the Microsoft Reality Toolkit, MRTK~2.7.2, which Microsoft supplies specifically for this purpose.
The data interconnection between the \acf{UWP} app on the HoloLens and the robot operating system, \acs{ROS}~\cite{quigley2009ros}, takes place via ROS\#~\cite{bischoff2019rossharp}.
The open-source software library ROS\# exchanges JSON based commands with \acs{ROS} through the rosbridge\_suite from within Unity applications.
Both the HoloLens and the robot are continuously connected with each other via WiFi, and data, such as the human's gaze information, can be sent and received in real time.
Finally, gestures and speech serve to operate the \acs{AR} interface and to interact with the robot.
\figref{paper04:fig:interplay} illustrates how the \acs{AR} interface acts as a bridge between human and robot and shows the interplay of the individual components of our teaching pipeline, which we will describe in more detail below.
\begin{figure}
    \centering
    \includegraphics[width=\linewidth]{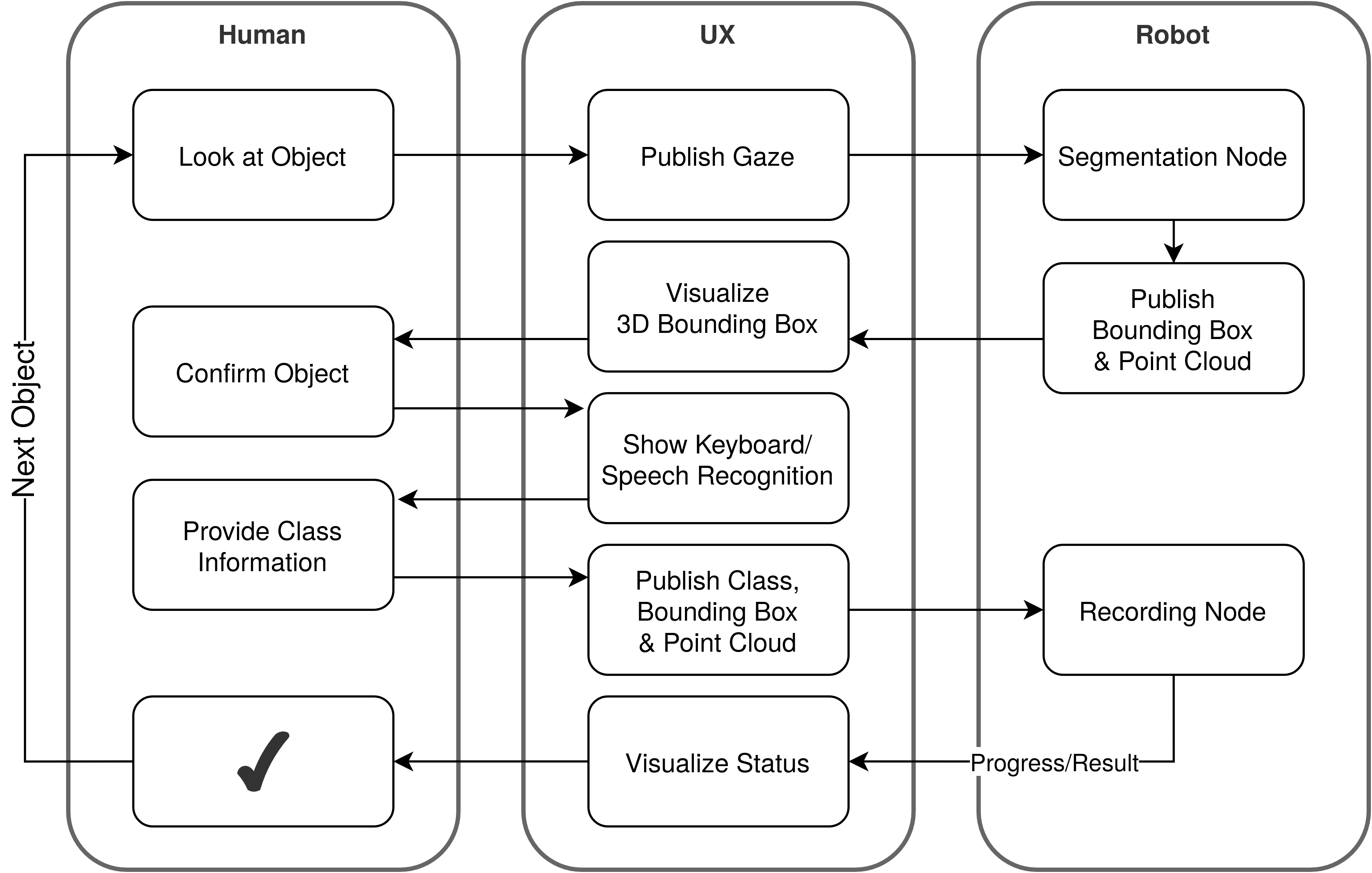}
    \caption{Overview of the entire teaching pipeline. The \acs{AR} user interface (UX) acts as a bridge between human and robot.}
    \label{paper04:fig:interplay}
    \Description{A flow chart showing the intermediate steps of the learning pipeline, each belonging to one of the three main categories, human, user interface or robot. It starts with the human looking at an object (Human category). Then the individual components are processed in the following order: "Publish Gaze" (user interface), "Segmentation Node" (robot), "Publish Bounding Box and Point Cloud" (robot), "Visualize 3D Bounding Box" (user interface), "Confirm Object" (human), "Show Keyboard / Speech Recognition" (user interface), "Provide Class Information" (human), "Publish Class, Bounding Box and Point Cloud" (user interface), "Recording Node" (robot), "Visualize Status" (user interface). Eventually, the human can trigger the pipeline again from the beginning with a new object.}
\end{figure}

\subsubsection{Identifying the Unknown Object of Interest}
In order for the robot to learn a new object, it has to identify it as such in the first place.
This is quite a fundamental problem, as it is, in a sense, a chicken-and-egg problem. 
For the robot, it is difficult to detect the object of interest as it does not know it at this point and it is yet to be taught.
Therefore, since the robot must identify the object before it has learned it, the deployment of neural networks is not possible at this point, and determining where the object begins and where it ends is not trivial.
Instead, we want to incorporate the human's gaze information to help the robot locate the target object. 
This means that the human looks at the object, whereupon the robot can distinguish it from the rest of the environment.
For this purpose, we take the approach of \cite{weber2022exploiting} as a basis, who segmented observed objects using human gaze and the point cloud obtained from the depth sensor of the robot's scene camera.
Thereby, a calibration determines the respective position of the robot and HoloLens, and the HoloLens' motion sensors ensure that the mutual position is tracked during human movements.
In addition, the gaze point, that is, the point at which the human is looking, is continuously tracked by the HoloLens and published as a point in 3D space through our user interface using ROS\#.
Thus, the corresponding \acs{ROS} topic can be subscribed by the \acs{ROS} system of the robot, which means that the gaze point is known to the robot at all times and can be used for segmentation.
In the first instance of the segmentation described in \cite{weber2022exploiting} a pass through filter and a voxel grid filter are applied to reduce the size of the point cloud.
Subsequently, the ground is extracted using \acs{RANSAC} and eventually the object is isolated by means of the gaze point and Euclidean clustering.
We adapt this method with slight adjustments in the last step.
Instead of assigning only the cluster closest to the gaze point to the object, we consider all clusters within a certain distance and with a certain size.
We set the threshold for the maximum distance between cluster and gaze point to 2\,cm and the minimum cluster size to five points.
This way it is possible to even segment very flat objects that do not protrude far from the ground.
Depending on which object the person is currently looking at, the respective object is then segmented in real time.
All the mentioned point cloud processing is accomplished using the open-source Point Cloud Library (PCL) \mbox{\cite{rusu20113d}}.

Owing to the calibration carried out at the beginning, the position of the object of interest is known both from the location of the human wearing the HoloLens as well as from the location of the robot.
The former allows the robot to display its feedback regarding the segmented object as a 3D bounding box on the HoloLens, namely in the human's field of view, using a subscriber, attached to a virtual bounding box, that updates the position of the box depending on the segmentation results.
We can take advantage of the latter during the teaching process described below.

\subsubsection{Teaching through Joint Attention}
The attention of the robot and the human is now jointly directed at one and the same object.
The next step is to confirm to the robot that the framed object is the object of interest and to provide a class information.
By performing a pinching gesture with the index finger and thumb (within the field of view of the HoloLens) during the fixation of the object with the human eyes, we can select the object via the \acs{AR} interface.
Alternatively, it is also possible to just say ``select''.
Thereupon, a virtual keyboard appears in the human's field of vision, on which he can now enter the class name of the object.
Again, it is alternatively possible to simply resort to speech.
\figref{paper04:fig:keyboard} shows the implemented keyboard from the human perspective.
\begin{figure}[b!]
    \centering
    \includegraphics[width=\linewidth, trim={5cm 0 5cm 0}, clip]{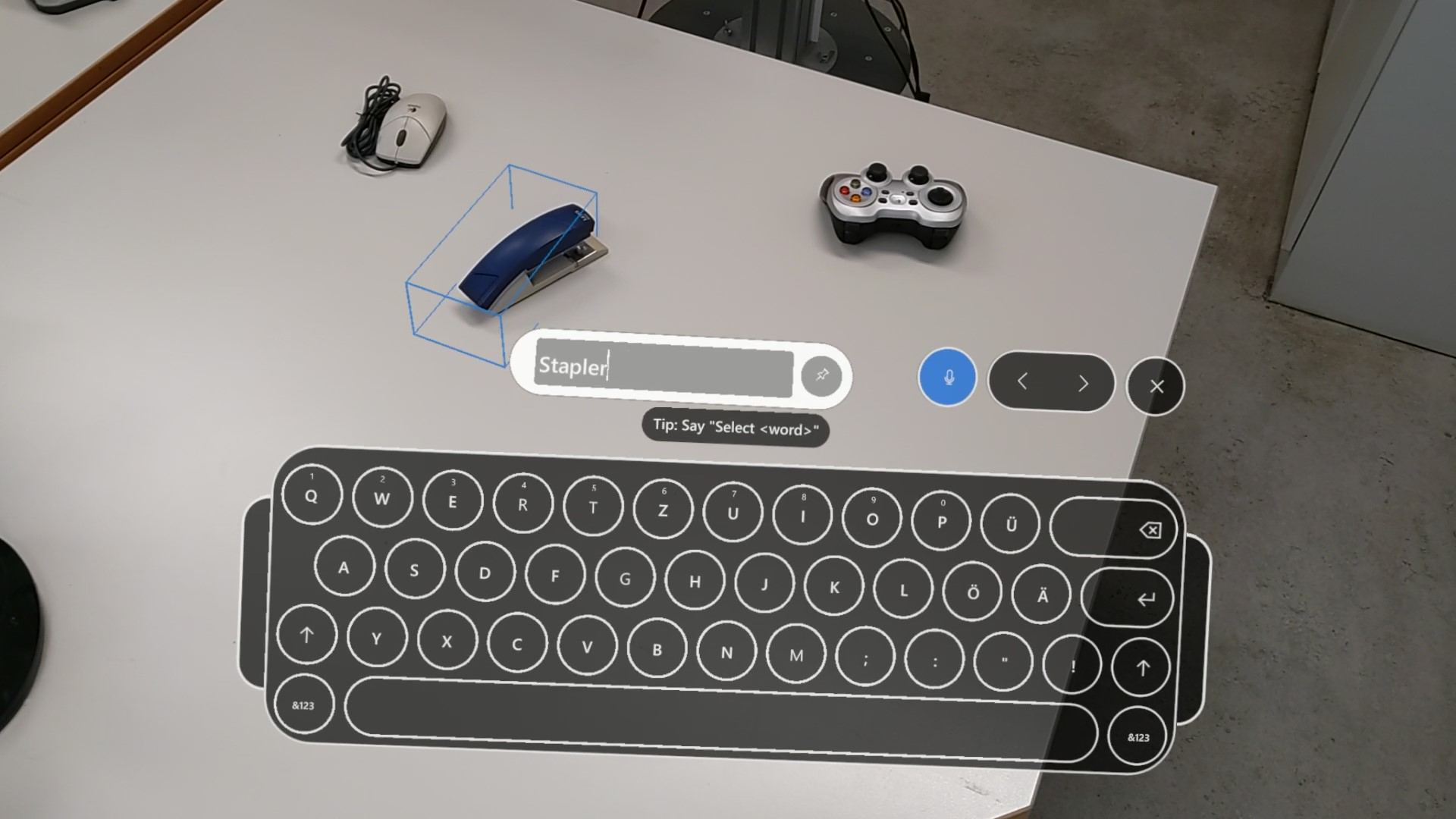}
    \caption{In the human's field of view, a bounding box of the object segmented by the robot is displayed. After the selection it is possible to specify the class name of the respective object using a virtual keyboard or speech.}
    \label{paper04:fig:keyboard}
    \Description{Augmented reality interface displayed in the human's filed of view. In the background, various objects lie on a table. A bounding box is shown around a stapler, which is visible to the human via augmented reality. In the foreground, a virtual keyboard can be seen, through which the class name "stapler" has been specified.}
\end{figure}

Our next goal is to have the robot autonomously capture images of the object of interest, which it can later use as training data.
In order to get as many images from multiple angles as possible, we attach a second camera to the wrist of the robot's arm.
The robot is now supposed to move this camera in a circle around the object.
This means that once the human has transmitted the class name to the robot by means of a \acs{ROS} action, it calculates a circular trajectory of reachable points.
Due to physical limitations, such as the length of the arm, this is usually a partial segment of the circle.
During the movement, the camera is aligned in such a way that it points at a 45 degree angle to the center of the previously determined 3D bounding box.
As the distance between the center and the camera, we use twice the length of the diagonal of the bounding box with a minimum safety distance of twice the distance between the camera and the robot arm end effector.
The recording process is illustrated in \figref{paper04:fig:robotarm}.
\begin{figure}
    \centering
    \includegraphics[width=0.49\columnwidth, trim={23cm 0.2cm 14cm 0}, clip]{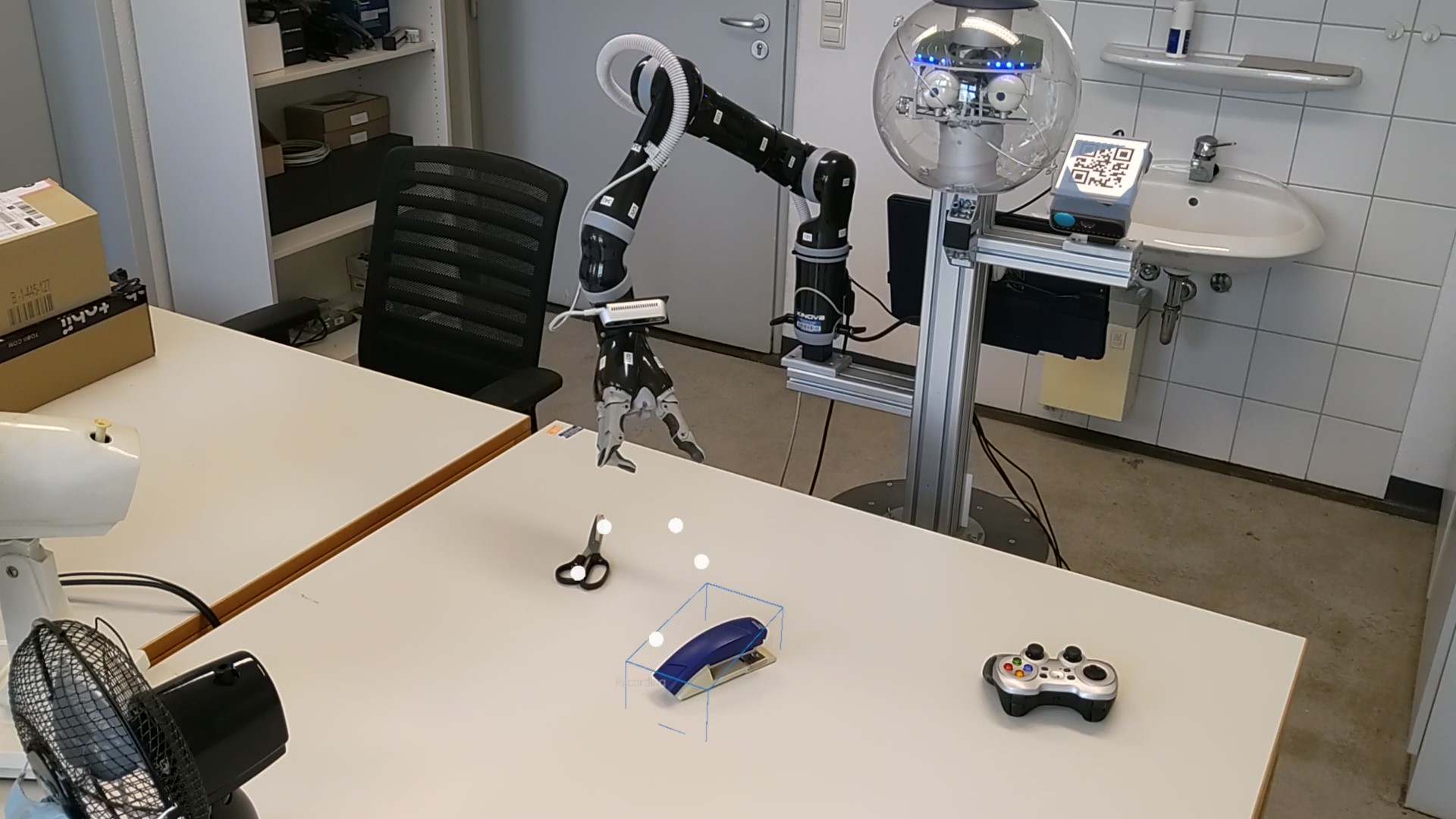}
    \includegraphics[width=0.49\columnwidth, trim={18cm 8cm 26cm 6cm}, clip]{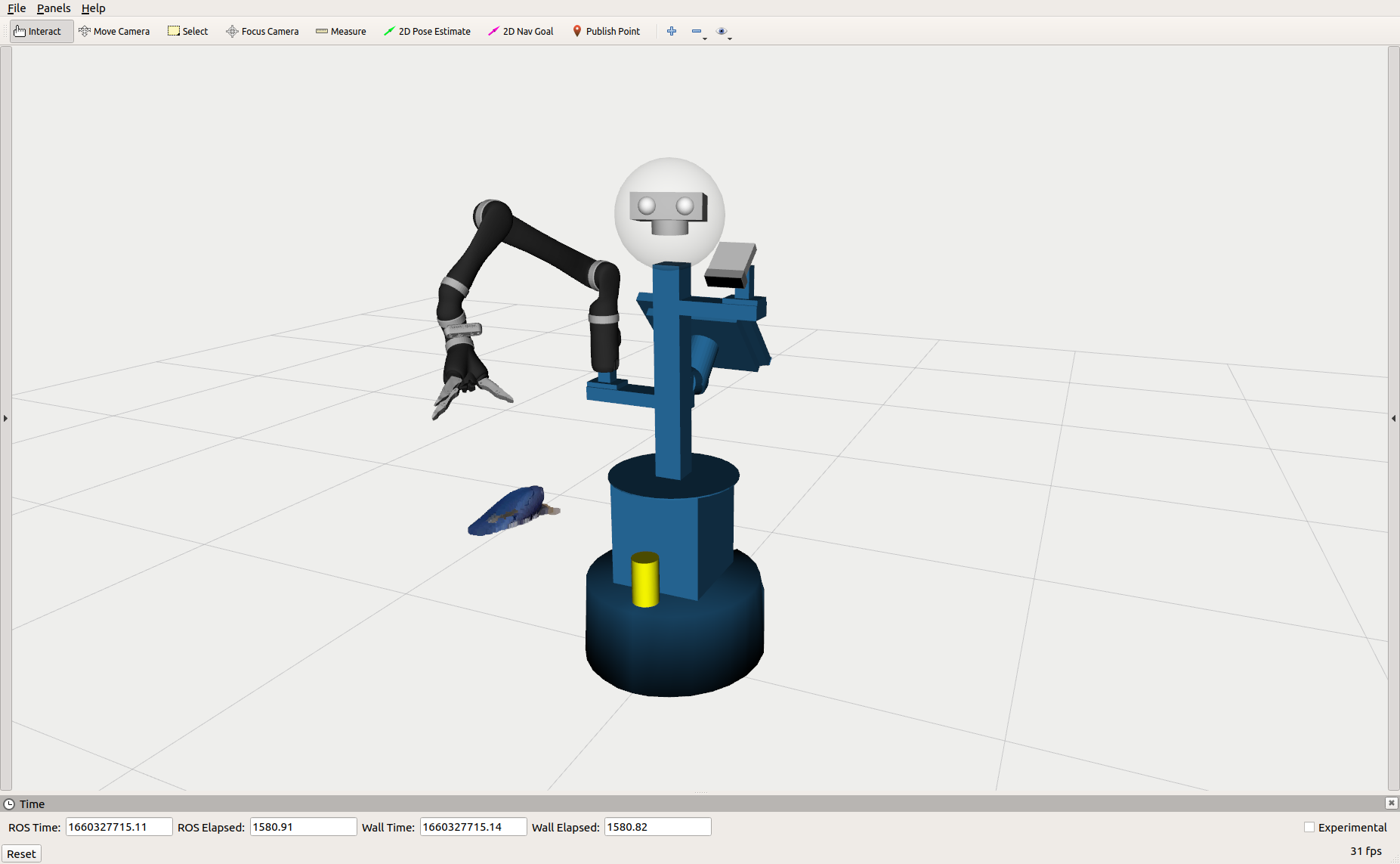}
    \caption{The left side shows the recording process from the human's augmented view and the right side is a visualization in RVIZ with the point cloud of the segmented object. The point cloud is used both to display the bounding box for the human and to label the images captured by the camera on the wrist of the robotic arm.}
    \label{paper04:fig:robotarm}
    \Description{Two images of the recording process of an object. On the left side is a robot with a robotic arm, on whose wrist a camera is attached. The arm is outstretched and the camera is pointed obliquely from above at a stapler on the table. The whole scene is shown from the human's field of view and via the augmented reality interface a bounding box can be seen around the stapler, as well as an progress indicator for the ongoing recording process. On the right side the situation is visualized in RVIZ. Instead of the table with all objects only the pointcloud of the stapler is visible.}
\end{figure}

By virtue of the calibration described in the previous section, the robot is not only aware of the position of the object of interest, but also capable of computing the transformation to all of its coordinate systems, namely the robot frames.
This includes the camera on the robot's arm.
The essential aspect in this step is to continuously transform the point cloud of the segmented object into the coordinate system of the camera.
By projecting the point cloud onto the 2D image plane of the camera, we can derive the region of interest from the boundary points.
Thus, for each captured image, we can additionally store a 2D bounding box calculated in this way.
Overall, the stored synchronized data are RGB and depth images as well as the regions of interest with the 2D bounding boxes.
In addition, we also store the positions of the camera to the captured object.
Eventually, the robot is able to automatically produce hundreds of labeled training images in a remarkably brief period of time.
More precisely, teaching one object takes about one minute and yields about 300 images.
The progress and completion of the recording process is in turn transmitted to the HoloLens via the \acs{ROS} action, visualizing to the human when the robot is ready for a new object.

\subsubsection{Transfer Learning}
Following the teaching part, the learning process of the robot now ensues.
To enable the robot to independently detect the previously seen objects in the future, it must use the information at its disposal in the form of the training data it has created itself.
In other words, the robot, or rather its neural network based object detectors, will be trained on the RGB images obtained.
This will be accomplished by means of transfer learning.
Hence, we assume some prior awareness of objectness, since our method should be seen as an extension rather than a replacement for the training with common large datasets, such as ImageNet~\cite{deng2009imagenet}, PASCAL~VOC~\cite{everingham2010pascal} or MS~COCO~\cite{lin2014microsoft}.
Such an approach is realistic, since most of the existing objects are not part of these datasets.
We aim to extend this state of knowledge with our method.
Consequently, we resort to state-of-the-art object detectors, such as Faster \mbox{R-CNN}~\cite{ren2015faster} and FCOS~\cite{tian2019fcos}.
Starting from one of these pretrained models respectively, we delete the last classification layers, and then reinitialize them with the appropriate number of output neurons for our use case.
Finally, we retrain said layers with our data, freezing all other neurons from the preceding feature layers.
By freezing the feature layers responsible for the general comprehension of objects and fine-tuning only the last few layers, we prevent overfitting~\cite{yosinski2014transferable, zhuang2020comprehensive}.
In fact, since we only retrain the classification heads of the models, even training on the robot itself is possible without relying on a high-performance GPU.
Naturally, the training may take longer.
In \secref{paper04:sec:evaluation}, we will evaluate the performance of the aforementioned models, among others.

\subsection{Dataset: Objects in Multiperspective Detail} \label{paper04:sec:dataset}
The method introduced in the previous section allowed us to create a new type of dataset.
While we publish the validation and test set mainly for the sake of reproducibility of our results, we think that our training set might be especially interesting for further purposes.
In the following, we would like to explain the training data in more detail and provide some statistics.
We will elaborate on the validation and test set within the scope of our evaluation in \secref{paper04:sec:evaluation}.

Our training set is particularly characterized by the fact that it supplies many details about individual objects.
While most object detection datasets often consist of many images with different objects, our data depicts the objects from many different angles.
This is especially attractive for objects that appear different from the front and back, such as a gamepad.

The set consists of 3113 perspectives in total and contains the classes fork, frisbee, gamepad, hole puncher, knife, scissors, shuttlecock, stapler, table tennis ball and toothbrush.
For each class, there are two different entities in the set, differing in color, shape, or both.
\figref{paper04:fig:dataset} shows some sample images and \figref{paper04:fig:num_classes} illustrates the distribution of the viewpoints.
\begin{figure}[b!]
	\centering
	\includegraphics[width=0.49\columnwidth, trim={0cm 0 0cm 0}, clip]{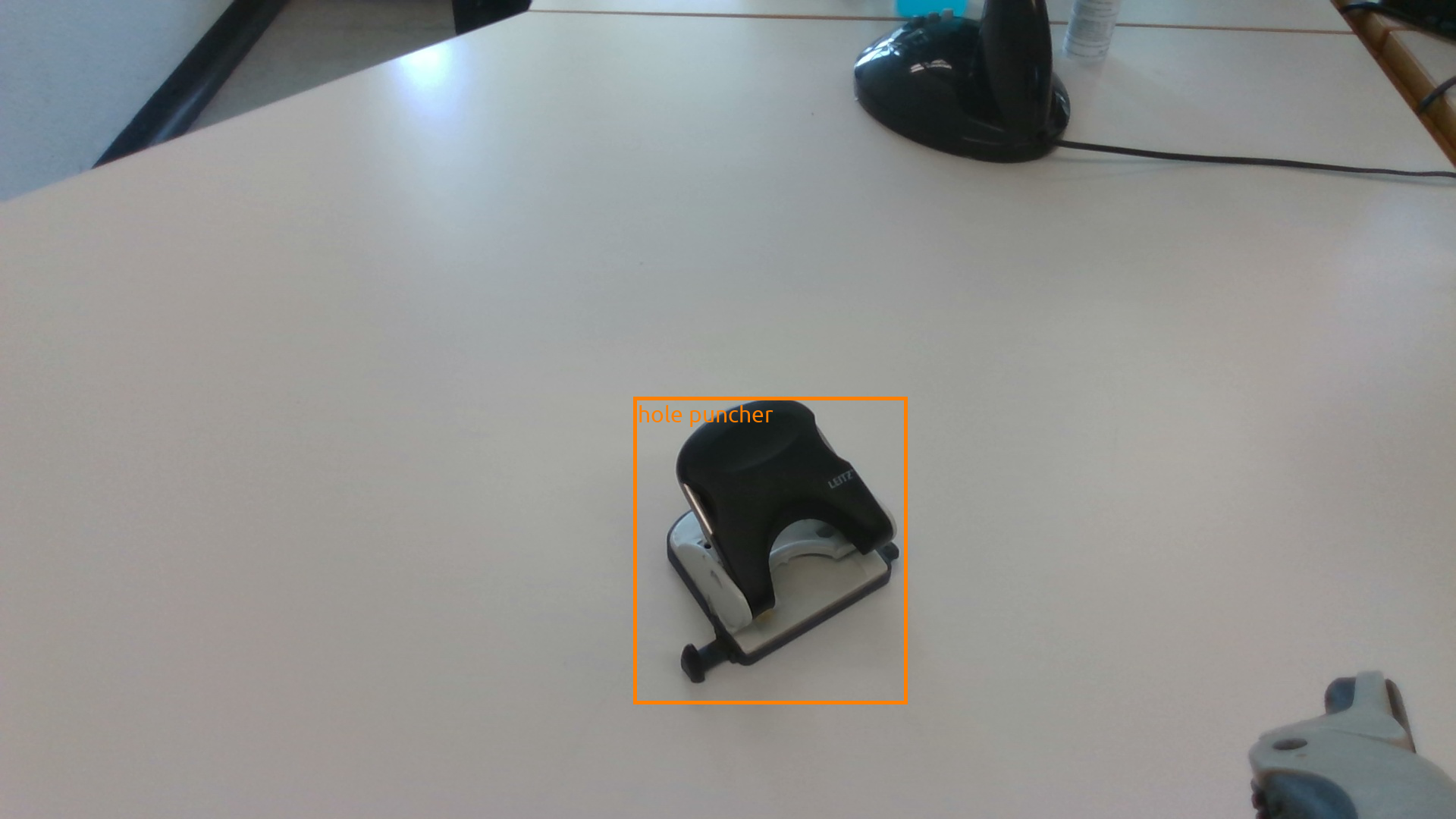}
	\includegraphics[width=0.49\columnwidth, trim={0cm 0 0cm 0}, clip]{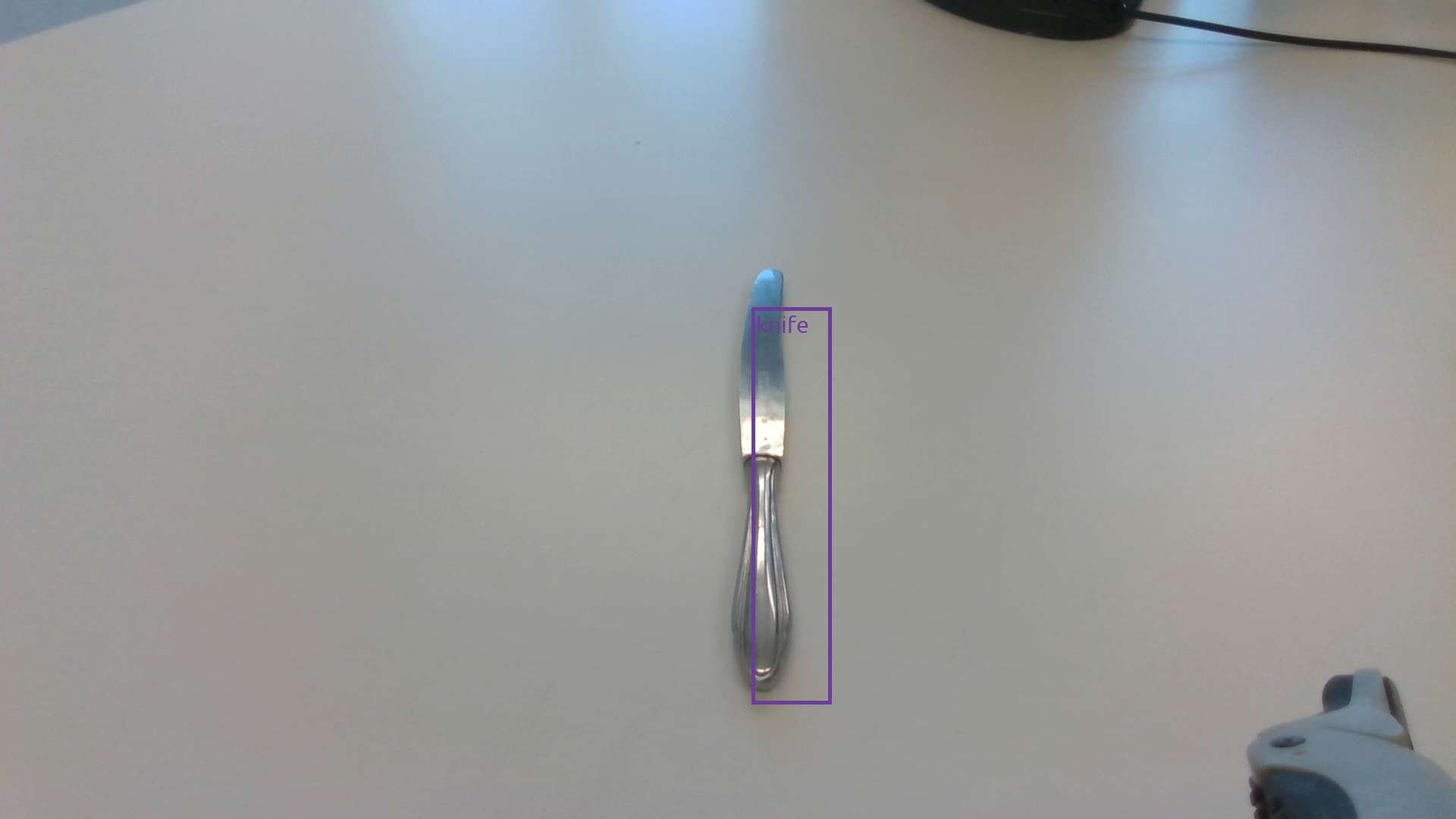}
	\par \vspace{2pt}
	\includegraphics[width=0.49\columnwidth, trim={0cm 0 0cm 0}, clip]{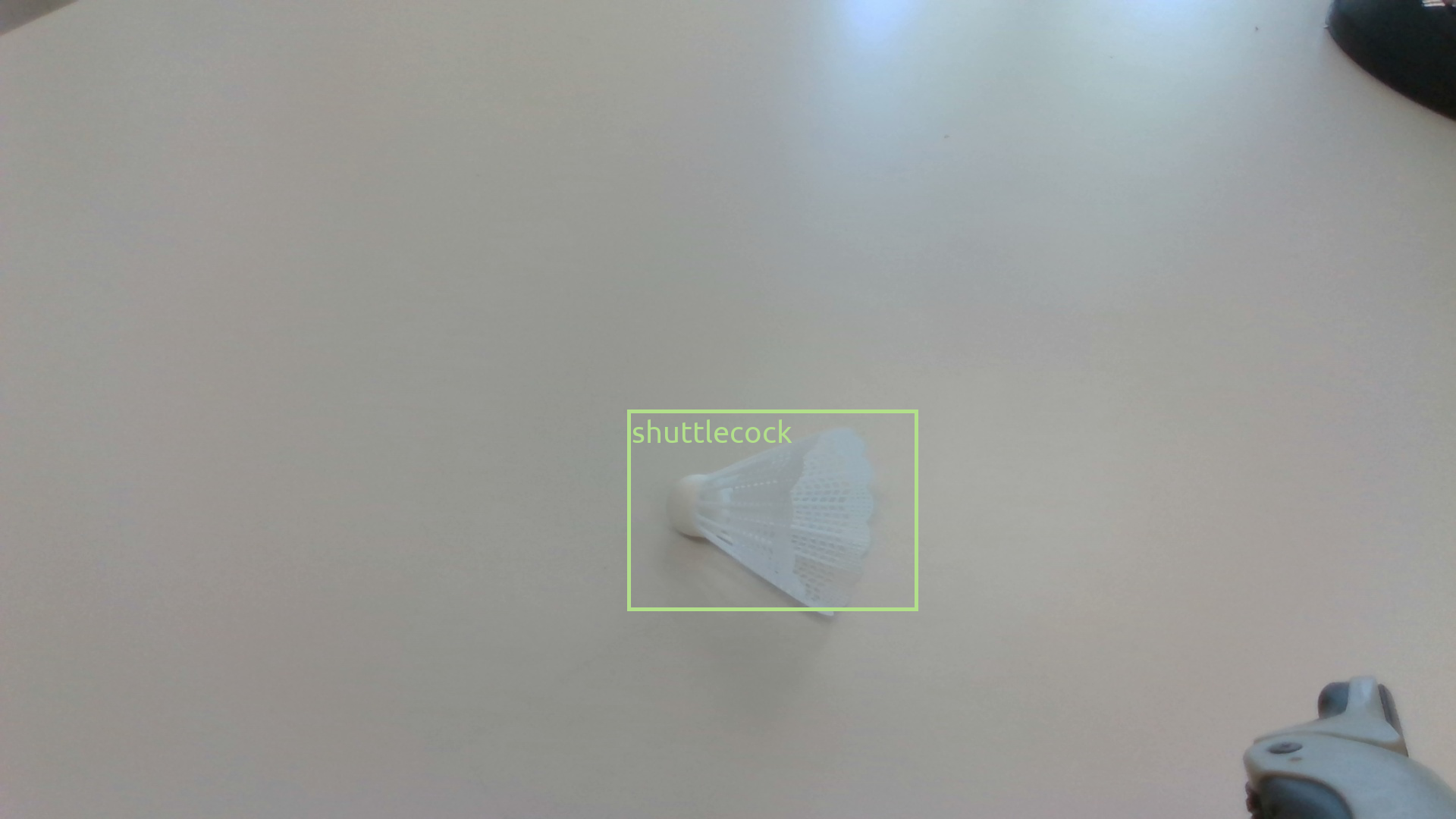}
	\includegraphics[width=0.49\columnwidth, trim={0cm 0 0cm 0}, clip]{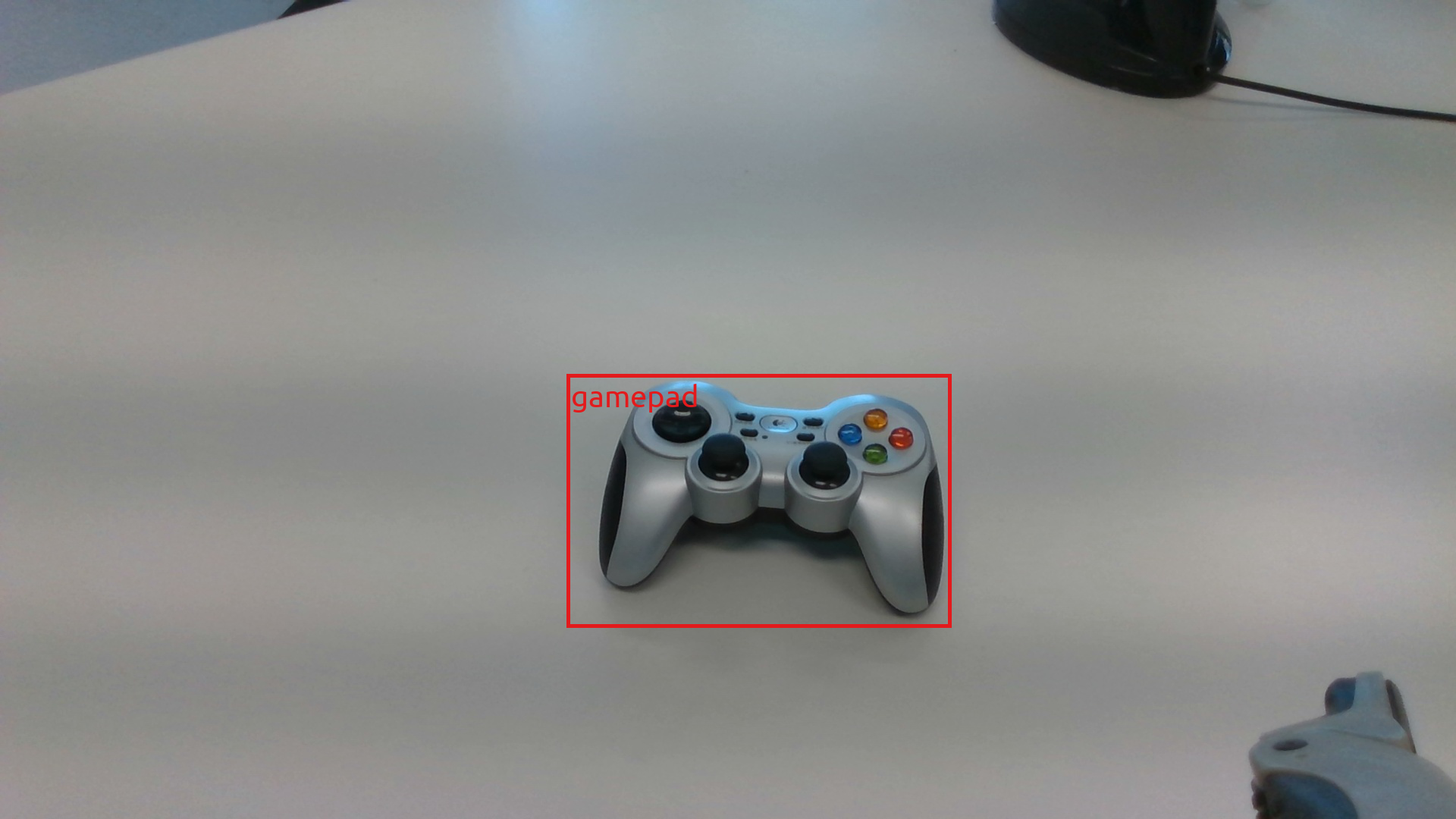}
	\caption{Sample images of a hole puncher, a knife, a shuttlecock and a gamepad from our dataset. The quality of the bounding boxes may vary depending on the point of view and may sometimes be slightly too large, too small or offset. In all images, however, the majority of the box always covers the respective object. The objects contrast differently with the background in terms of flatness and color.}
	\label{paper04:fig:dataset}
	\Description{Four images of four different objects on a table. Around each object is drawn the bounding boxes determined by the robot. The bounding box around the hole puncher and the gamepad are very fitting, while the box around the slim fork has a minimally offset and the one around the shuttlecock is slightly too large.}
\end{figure}
\begin{figure}
    \centering
    \includegraphics[width=.9\linewidth]{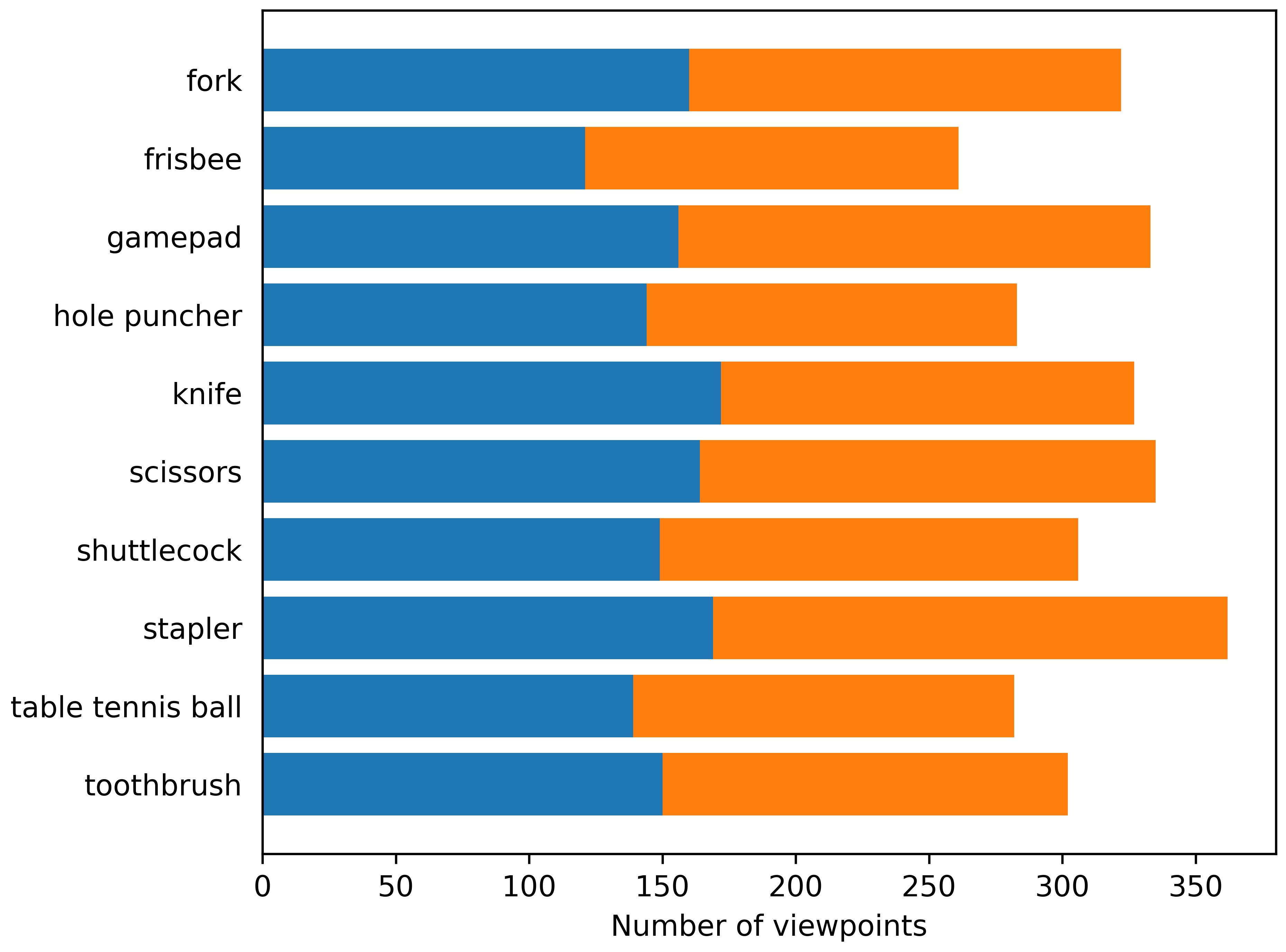}
    \caption{Distribution of the viewpoints across the categories. The colors indicate the two different items within the classes.}
    \label{paper04:fig:num_classes}
    \Description{A bar chart showing the number of viewpoints for each class in the dataset. The ten classes are listed on the y-axis and the number of viewpoints on the x-axis. All classes range from 280 to 380 viewpoints. The class frisbee has the fewest viewpoints and the class stapler has the most. All ten bars consist of two bars in different colors, which represent the two entities within a class. For all classes, the color change is approximately in the middle, which means that there are approximately the same number of viewpoints of both entities in the dataset.}
\end{figure}
The objects are each placed individually on a table and for every camera perspective multiple pieces of information are included.
For each RGB image, alongside the region of interest, there is a corresponding depth image that is aligned to the color image.
All RGB images and depth images have a resolution of $ 1920 \times 1080 $.
In addition, all camera poses are available.
They are specified independently of the robot as a transformation, composed of translation vector and rotation quaternion, from the coordinate system of the camera to the one of the object.
The last component is the meta-information about the camera with the intrinsic parameters of the camera calibration.

Altogether, we believe that the high information density in our data is also interesting for other research areas where camera positions are crucial, such as Neural Radiance Fields~\cite{mildenhall2020nerf, yu2021pixelnerf, lin2021barf, deng2022depth}.
There, either synthetic data must be used or, given real data, the camera positions (and depths) can only be roughly approximated via structure from motion.
For this reason, we make our dataset, \acf{OMD}, publicly available to the research community.

\subsection{Evaluation} \label{paper04:sec:evaluation}
For all our experiments, alongside the aforementioned HoloLens 2 worn by the human, we employed a Scitos G5 from MetraLabs~\cite{MetraLabs} as robot.
The body camera through which the robot observes the scene and performs the segmentation is an Azure Kinect DK from Microsoft.
The robot arm that was additionally installed on the Scitos is a Kinova Jaco2~\cite{KinovaGen2} with 6 DoF.
The camera attached to the wrist of the arm in order to take pictures of the objects is an Intel RealSense D435.

For training we use our own training set, which we have explained in detail in \secref{paper04:sec:dataset}.
Since the objective of this work is to teach the robot its environment, we also had to record a validation and test set located in the environment where the learning process took place.
As mentioned earlier, alongside the training set, we will also publish the validation and test set to ensure the reproducibility of our results.
The validation and test set consist of 1051 and 1410 regular images, respectively, which were manually labeled by hand using DarkLabel~\cite{DarkLabel}.
The classes represented therein are the same ten as in the training set.
For each class, four distinct objects of the respective category were available.
The validation set contains the same two objects of each class as the training set, whereas the test set contains the other two.
That is, the objects differ in shape, color, or both from those used in training.
The objects were photographed randomly grouped (within the set) in the robot's office environment.
We made sure to create challenging scenarios as well, such as items being stacked or the toothbrush still being in its packaging, as shown in \figref{paper04:fig:dataset_test}.
\begin{figure}[hb]
    \centering
    \includegraphics[width=0.49\columnwidth, trim={0cm 0 0cm 0}, clip]{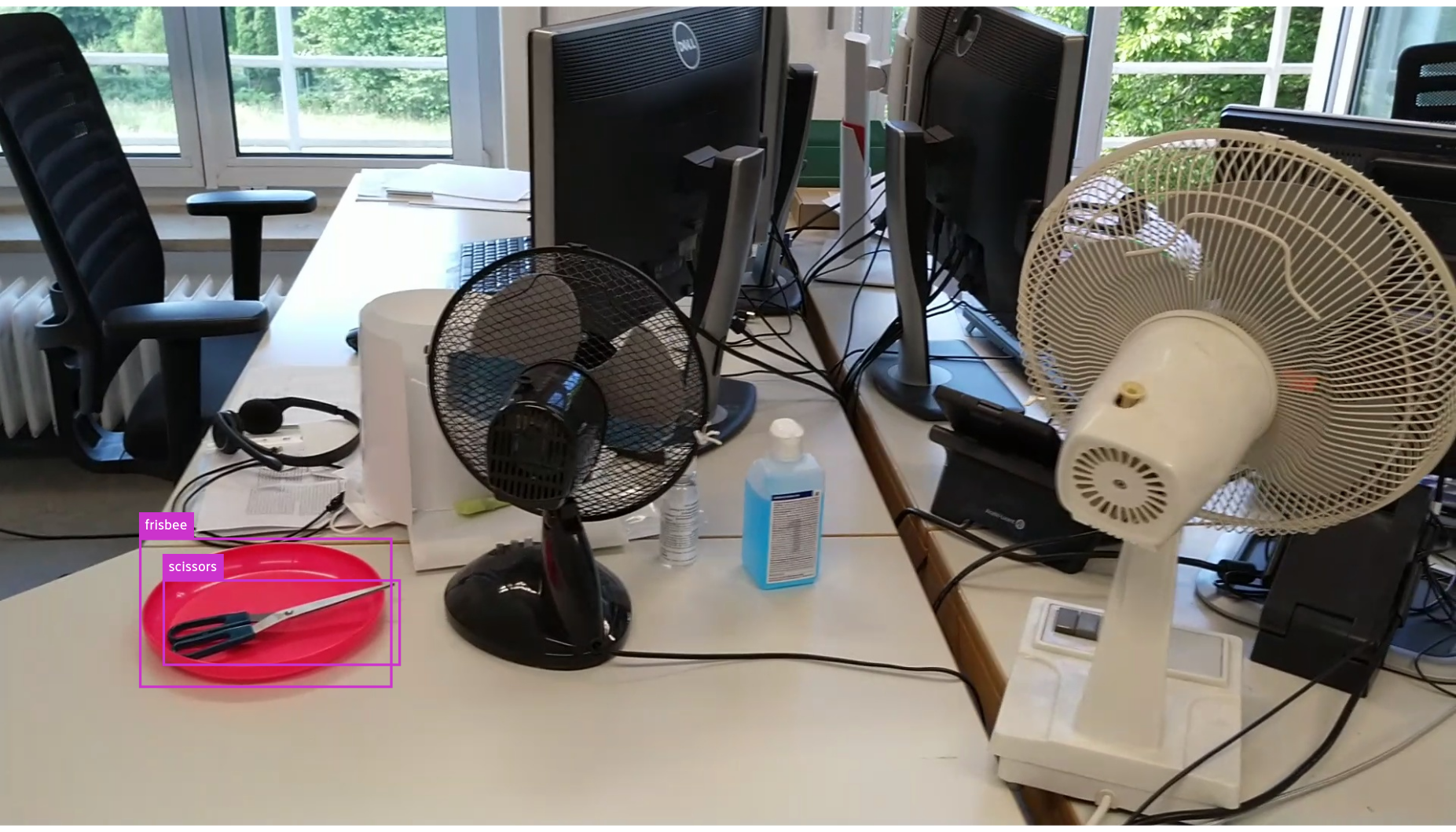}
    \hfill
    \includegraphics[width=0.49\columnwidth, trim={0cm 0 0 0}, clip]{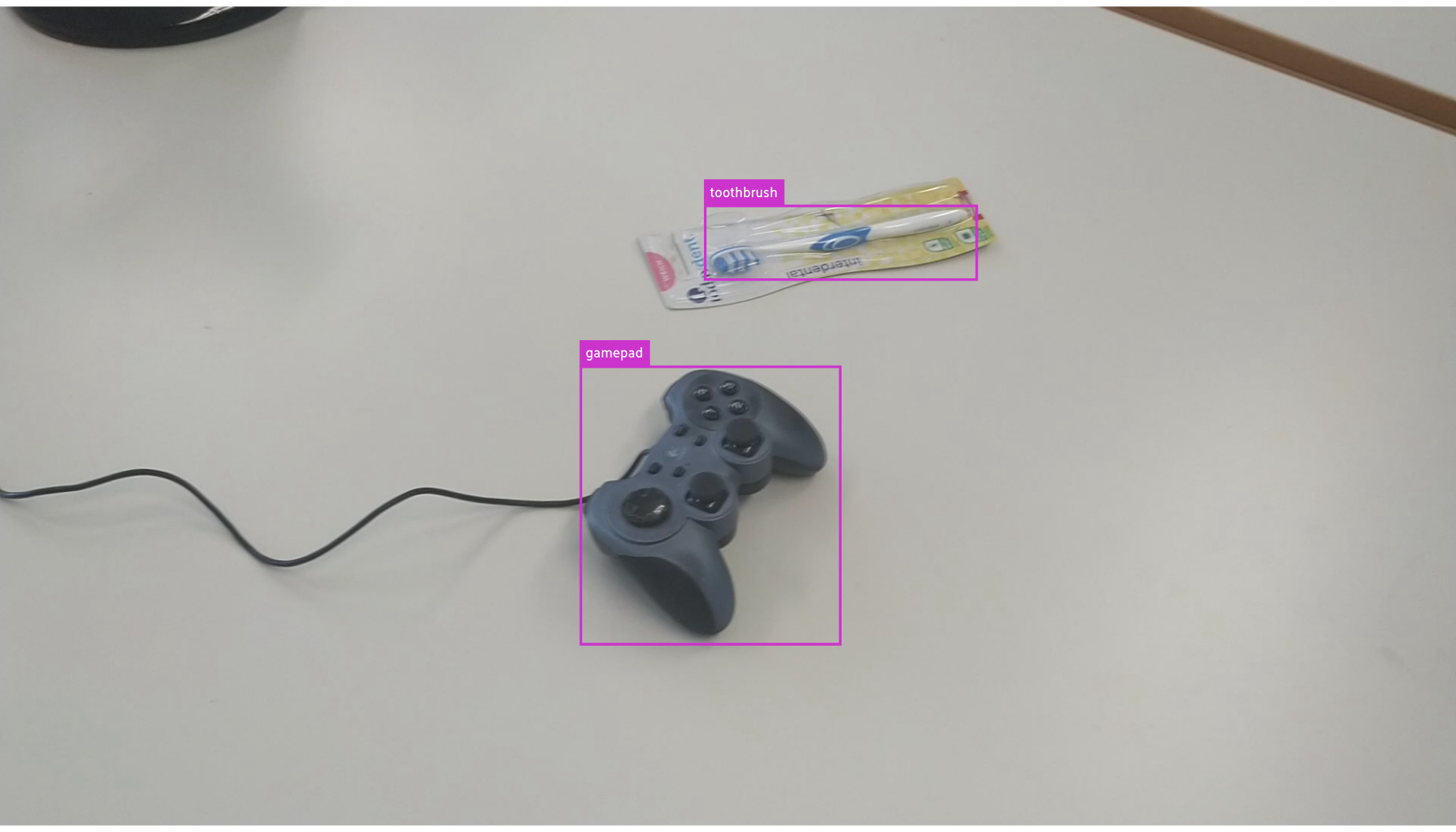}
    \caption{Sample images from the test set. The set of objects is disjoint with the ones from the training set (see gamepad). The set is also diverse in terms of the clutter of the background and the distances to the objects.}
    \label{paper04:fig:dataset_test}
    \Description{Two images of objects in an office environment. On the left is a pair of scissors in an upside-down frisbee surrounded by office items, monitors, fans and other things. On the right side a gamepad and a toothbrush are lying on a white table. The toothbrush is still in its packaging.}
\end{figure}

In the following, we evaluate our learning pipeline using several state-of-the-art object detectors.
Consequently, the object detectors Faster \mbox{R-CNN}~\cite{ren2015faster}, RetinaNet~\cite{lin2017focal}, FCOS~\cite{tian2019fcos}, and SSD300~\cite{liu2016ssd} serve as a foundation.
We complement these with various backbones, such as ResNet-50-FPN~\cite{he2016deep}, VGG16~\cite{simonyan2014very} and MobileNetV3 Large~\cite{sandler2018mobilenetv2, howard2019searching}.
All backbones were trained on ImageNet and can be left as is, since we deliberately picked object classes for our evaluation that had no intersection at all with this dataset.
The reason behind our choice of the ten test objects was as follows.
On the one hand, they must not appear in ImageNet due to the backbones, but on the other hand, at least a part of them ought to be in MS COCO so that we have a comparison later on.
Furthermore, within each class there had to be several different looking objects of that class.
All of this together limited the selection accordingly, especially since we tried to avoid perishable classes like food.
Hence, in terms of the actual object detectors and to ensure that our objects are indeed unknown to the models, we had to train them on a subset of MS COCO.
More precisely, we extracted the classes fork, frisbee, knife, scissors, sports ball, and toothbrush from the dataset using the tool Fiftyone~\cite{moore2020fiftyone} and then trained the above mentioned detectors on the remaining part.
In doing so, we followed the respective training recipe of the original implementation and, for consistency, adhered thereto in all of our subsequent experiments in our own training pipeline.
The only exception in our transfer learning approach was the type of data augmentation applied.
In this case, we used random photometric distortion, random zoom out, random cropping, and random horizontal flipping for all models (not just SSD300) to prevent overfitting.
Subsequently, we trained using the method described in \secref{paper04:sec:method}.

In all our experiments, we evaluate according to the MS~COCO metric~\cite{lin2014microsoft}, namely the average precision for varying intersection-over-union thresholds (\acs{IoU}).
In this context, we use the abbreviations $ \AP{} = \text{AP}^{\text{IoU}=0.5:0.05:0.95} $, $ \AP{50} = \text{AP}^{\text{IoU}=0.5} $, and $ \AP{75} = \text{AP}^{\text{IoU}=0.75} $ within a class and \acs{mAP} as the average over all categories.
Analogously, this applies to the average recall, where we consider the maximum recall given 1, 10, and 100 detections per image, respectively, and use the abbreviations $ \AR{1} = \text{AR}^\text{max=1} $, $ \AR{10} = \text{AR}^\text{max=10} $ and $ \AR{100} = \text{AR}^\text{max=100} $.
Again, \acs{mAR} denotes averaged over all categories.
Unless otherwise stated, average precision and average recall refer to \AP{} and \AR{}, respectively.

A comparison of all tested models is provided in \tblref{paper04:tbl:evalModels}.
\begin{table}[b]
    \centering
    \caption{Comparison of all machine learning models trained in a transfer learning fashion. The best values are highlighted in bold.}
    \label{paper04:tbl:evalModels}
    \resizebox{\columnwidth}{!}{%
    \begin{threeparttable}
    \begin{tabular}{cccccccc}
    \toprule
        Model   & Backbone  & \mAP{} & \mAP{50} & \mAP{75} & \mAR{1} & \mAR{10} & \mAR{100}  \\
        \midrule
        Faster R-CNN~\cite{ren2015faster}   & ResNet-50~\cite{he2016deep}
                                            & \textbf{33.6} & \textbf{66.9} & 31.4 & 43.7 & 50.1 & 50.4 \\
        Faster R-CNN~\cite{ren2015faster}   & MobileNetV3~\cite{sandler2018mobilenetv2,howard2019searching} 
                                            & 15.5 & 38.1 & \pz6.1 & 23.7 & 27.4 & 27.7 \\
        Faster R-CNN~\cite{ren2015faster}   & MobileNetV3~\cite{sandler2018mobilenetv2,howard2019searching}\tnote{$\dagger$}
                                            & 13.0 & 38.4 & \pz3.1 & 22.2 & 25.5 & 25.5 \\
        FCOS~\cite{tian2019fcos}            & ResNet-50~\cite{he2016deep}
                                            & 30.6 & 47.6 & \textbf{35.9} & 44.7 & 53.8 & 55.0 \\
        RetinaNet~\cite{lin2017focal}       & ResNet-50~\cite{he2016deep}
                                            & 31.2 & 52.4 & 34.3 & \textbf{46.2} & \textbf{57.6} & \textbf{59.1} \\
        SSD300~\cite{liu2016ssd}            & VGG16~\cite{simonyan2014very}
                                            & \pz8.0 & 19.1 & \pz5.0 & 21.0 & 31.6 & 34.0 \\
    \bottomrule
    \end{tabular}
    \begin{tablenotes}
			\item[$\dagger$] \footnotesize Tuned for mobile use cases
	\end{tablenotes}
    \end{threeparttable}}
\end{table}
Faster \mbox{R-CNN} with the ResNet-50 backbone generally performed best in terms of average precision.
With a MobileNetV3 backbone, the performance was significantly worse in terms of both precision and recall. 
FCOS and RetinaNet are slightly behind Faster \mbox{R-CNN} in terms of the mean average accuracy.
The latter has the best recall values, while FCOS has the best mean average precision at an intersection-over-union of $ 0.75 $.
SSD300 clearly lags behind all other models in terms of precision.

As we proceed, we will continue with Faster R-CNN for further analysis, since MS~COCO \cite{lin2014microsoft} considers \acs{mAP} as the single most important metric.
In general, the model seems to detect the objects quite well, but some classes cause more difficulties than others.
This also becomes apparent by looking at the curve in \figref{paper04:fig:PRCurve}.
\begin{figure}
    \centering
    \includegraphics[width=\linewidth, trim={0cm 0 0 0}, clip]{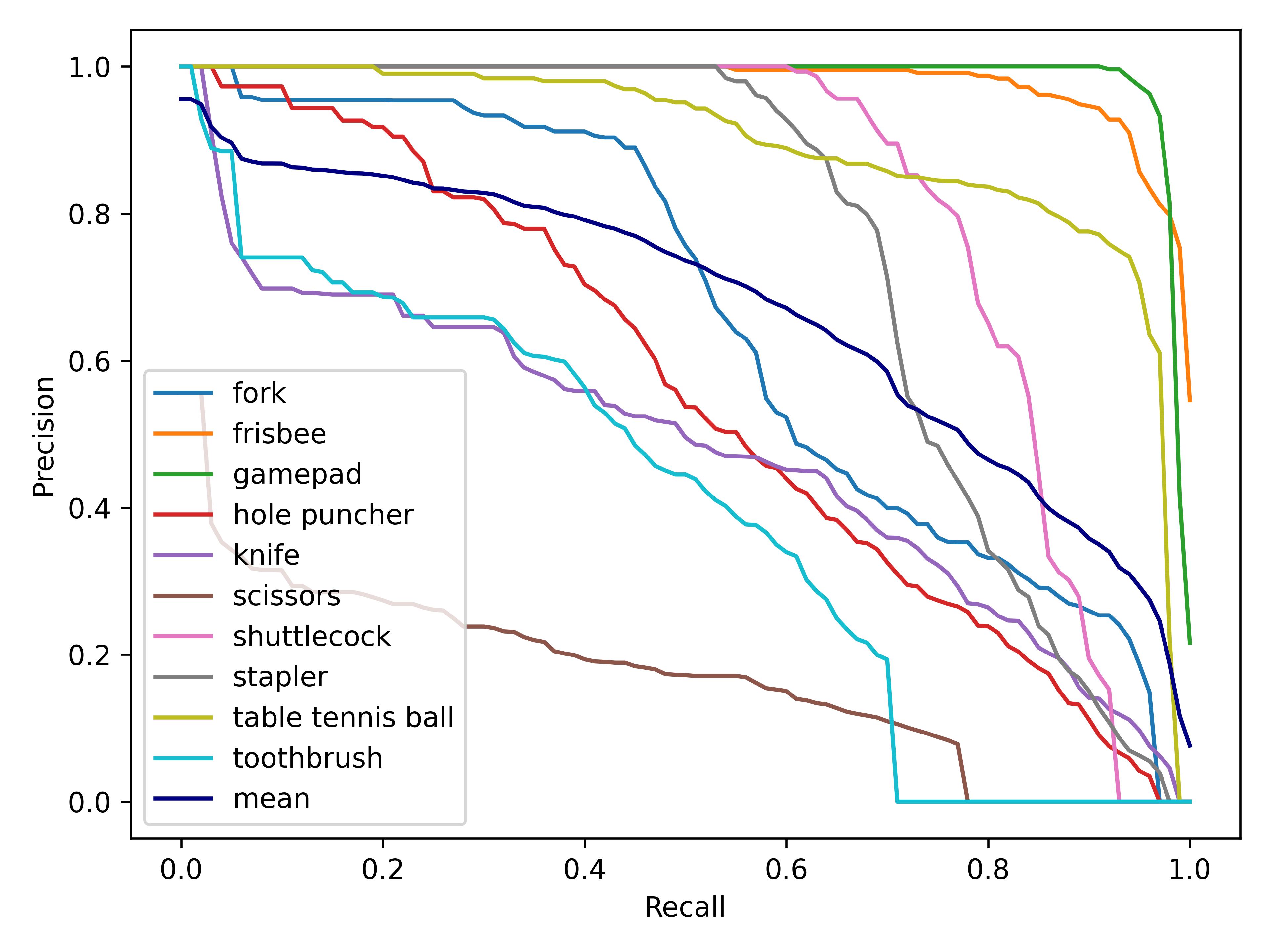}
    \caption{Precision-recall curve of Faster \mbox{R-CNN} at an \acs{IoU} of~0.5. Above this value, objects can be considered as detected~\cite{everingham2010pascal, zitnick2014edge}.}
    \label{paper04:fig:PRCurve}
    \Description{A line graph of the precision recall curves of the ten classes. The recall is shown on the x-axis and the precision on the y-axis. The curves start at a recall of 0.0 with precision values in the range of 1.0 and decrease with increasing recall. The curve of the scissors is the only one that starts in the range of 0.4 and the precision decreases evenly until a recall of 0.8 to zero. The toothbrush reaches zero a bit earlier, but is overall far above the curve of the scissors, since it starts at a precision of 1.0.  The gamepad and the frisbee are still almost at a precision of 1.0 up to a recall of 0.9. All other curves lie between the toothbrush and the gamepad.}
\end{figure}
Even at low recall values, the precision of the scissors is below $ 0.5 $.
The class toothbrush also decreases early.
On the other hand, the precision for the classes gamepad and frisbee is consistently excellent, even for a high recall.

In \tblref{paper04:tbl:evalClasses}, we compare different training variants.
\begin{table}
    \centering
    \caption{Comparison of the average precision (\AP{}) for different training types of Faster \mbox{R-CNN} on our test set. Namely, apart from the backbone, trained from scratch (S) or trained in the sense of transfer learning with frozen non-classification layers (TL-F) or completely unfrozen (TL-U), respectively. All three on the data collected by the robot. The best values are highlighted in bold. The last column (COCO) serves as an orientation and reports the results of Faster \mbox{R-CNN} trained on the entire MS COCO training set.}
    \label{paper04:tbl:evalClasses}
    \begin{tabular}{rccc|c}
    \toprule
        Class               & S     & TL-U  & TL-F  & COCO \\
        \midrule
        fork                & \textbf{21.7}  & \textbf{21.7}  & 18.6  & 63.8 \\
        frisbee             & 19.5  & 47.6  & \textbf{58.8}  & 65.9 \\
        gamepad             & 38.4  & 24.8  & \textbf{62.6}  & -    \\
        hole puncher        & \textbf{42.5}  & 26.3  & 23.0  & -    \\
        knife               & 20.4  & 19.1  & \textbf{27.6}  & 50.0 \\
        scissors            &\pz6.5 &\pz\textbf{7.3} &\pz5.1 & 70.9 \\
        shuttlecock         & 24.9  & 27.8  & \textbf{51.5}  & -    \\
        stapler             & 29.3  & 24.8  & \textbf{38.9}  & -    \\
        table tennis ball   &\pz3.4 & 27.6  & \textbf{44.0}  & 17.3 \\
        toothbrush          & \textbf{21.6}  & 8.8   & 6.2   & 37.4 \\
        \midrule
        \mAP{50}            & 62.8  & \textbf{68.8}  & 66.9  & 84.4 \\
        \mAP{75}            &\pz9.8 & 7.5   & \textbf{31.4}  & 55.4 \\
        \mAP{\phantom{00}}  & 22.8  & 23.6  & \textbf{33.6}  & 50.9 \\
    \bottomrule
    \end{tabular}
\end{table}
If we ignore the baseline variant (COCO) for a moment, we find that the transfer learning method, in which only the last layers had been trained (TL-F), has the best \AP{} for the majority of classes.
In particular, \mAP{75} is significantly higher.
It is worth mentioning that the baseline values (COCO) are naturally superior.
The difference between the full MS COCO training set and the part we used for pretraining remains still $ 25\,713 $ objects in $ 14\,296 $ images, which is five times as many images as we used.
In addition, our images are distributed among all ten classes, while MS COCO does not contain four of them and the baseline thus does not recognize them at all.
This demonstrates the strength of our pipeline, which is designed to enable the learning of additional, as yet unknown classes, that is, to extend existing knowledge.
The comparison with the baseline, which is the ideal case, namely 1) a suitable data set exists 2) it is accessible and 3) the object is part of it, serves primarily to better classify our results into the overall picture.
It is not intended to outperform a model trained on such a large data set, but rather to determine an upper bound and test how close we can get with our method.
Taking this into account, it is remarkable how well our pipeline has learned especially the classes gamepad or shuttlecock, whose \AP{} is even higher than the \acs{mAP} of the baseline.
Furthermore, since the table tennis ball occurs in MS COCO only as a subset of the class sports ball, we can see how our system becomes more attentive to table tennis balls.
In contrast, the class frisbee does not quite reach the baseline as the corresponding MS COCO class contains exclusively frisbees.
In the case of the category hole puncher, the results are satisfying even without pre-existing basic knowledge (S).

Considering the amount of time needed for the respective training, major differences become apparent.
Training of the entire MS COCO training set lasted the longest, at more than two days on two NVIDIA RTX A4000 deployed in parallel.
The other three variants were trained on our dataset on only one of the GPUs and took 4 hours for the entire model (S, TL-U) and 2.5 hours for the freezed variant (TL-F), respectively.
As mentioned above, although for consistency reasons we trained 26 epochs as in the original recipe, the weights with the best validation accuracy that we eventually used for testing were often reached earlier.
For Faster \mbox{R-CNN} trained via TL-F, this was even the case after three epochs (starting at $ \mAP{}=0.0 $ before training), which corresponds to a training time of about 40 minutes on our Scitos equipped with an NVIDIA GeForce GTX 1050 Ti.
This makes the entire pipeline also suitable for stand-alone learning directly on the robot.

Finally, we analyze the influence of the amount of images used for training.
\tblref{paper04:tbl:evalSubset} lists the results of Faster R-CNN trained using TL-F for varying dataset sizes.
\begin{table}
    \centering
    \caption{Results of Faster \mbox{R-CNN} trained via TL-F on different sized subsets of our dataset. The best values are highlighted in bold.}
    \label{paper04:tbl:evalSubset}
    \begin{tabular}{ccccc}
    \toprule
                            & 25\%  & 50\%  & 75\%  & 100\%\\
        \midrule
        \mAP{\phantom{000}} & 31.5  & 32.9  & 31.7  & \textbf{33.6} \\
        \mAP{50\phantom{0}} & 64.7  & 66.7  & 66.7  & \textbf{66.9} \\
        \mAP{75\phantom{0}} & 28.6  & 28.4  & 26.0  & \textbf{31.4} \\
        \mAR{1\phantom{00}} & 41.1  & 42.9  & 41.4  & \textbf{43.7} \\
        \mAR{10\phantom{0}} & 46.9  & 49.9  & 46.9  & \textbf{50.1} \\
        \mAR{100}           & 47.2  & 50.2  & 47.1  & \textbf{50.4} \\
    \bottomrule
    \end{tabular}
\end{table}
The images were removed from the sequence of perspectives with equidistant spacing.
Apart from the case where 75\% of the data is used, the tendency emerges that as the number of images increases, so does the average precision and average recall.
The best result is obtained using all the data.

\subsection{Limitations \& Discussion}
Similar to all supervised machine learning methods, we depend on the quality of the training data.
In our case, this can vary depending on the preceding segmentation of the point cloud.
This in turn is naturally dependent on the quality of the data obtained by the depth sensor of the robot's scene camera.
Especially with very dark or glossy surfaces, we noticed that the depth sensor had problems determining the depth accurately.
As a result, the accuracy of the bounding boxes suffers, which eventually has an impact on performance.
However, this problem can be compensated with an even larger number of objects and our tests have shown, moreover, that the robot is still capable of detecting the learned objects in its environment despite such difficulties.

One further point is that while our approach generalizes well even to other objects of the learned classes, our tests were inferior on popular datasets such as MS COCO.
This is due to the diversity of the images and the versatile situations depicted in them.
For instance, fruits such as bananas and apples can be found in their natural form as well as cut into small pieces in a fruit salad.
This task can only be solved with an enormous amount of training data.
Our method, on the other hand, although it cannot achieve the performance of training on large datasets, is primarily designed to teach the robot objects for which data does not yet exist.
In such scenarios, a semantic scene understanding is necessary and humans can assist in gaining this understanding by means of our method.
While an extension of existing datasets would also be conceivable, our method, in contrast, does not require tremendous labeling resources and can be used spontaneously in the respective situation.
Our evaluations (\mbox{\tblref{paper04:tbl:evalClasses}}) show that our method is capable of learning unknown objects that are not detected by the baseline.
It can therefore be used more flexibly without the need to know the situation in advance and rely on the existence of appropriate datasets.
Moreover, teaching through two-way interaction is extremely natural, especially since the \acs{AR} system enables real-time communication between human and robot by directly connecting both worlds, the analog world of the human and the digital world of the robot, so that the human can supply information (gaze, class) to the robot, thus initiating the recording process, and the robot can in turn communicate feedback visually.
Pointing to objects using gaze completes the interplay, as it is intuitive, less ambiguous than pointing with a finger and, unlike speech, can be applied before the object is known to the robot.
All in all, we therefore consider our approach to be less of a replacement and more of a supplement.
\subsection{Conclusion}
In this work, we presented a novel pipeline towards the deployment of robots in non-predefined scenarios.
To this end, we leveraged human gaze and augmented reality in the interaction between robot and human to successfully teach the robot new, yet unknown objects in its environment.

In order for robots equipped with machine learning based object detectors to detect their environment and the objects contained therein, a lot of training data is usually required.
In practice, however, under unpredictable conditions and due to the wide range of existing objects, the availability of a suitable data set can not always be guaranteed.
Our approach can complement popular datasets in exactly such situations and produce large amounts of automatically labeled, non-synthetic training data in a user-friendly manner and in a short period of time.
On the basis of such data, we have trained state-of-the-art object detectors in several different ways and shown that it is possible to learn and detect new objects in this manner.
In fact, the training can even take place standalone on the robot due to transfer learning, without the need for tremendous computational resources.
Further, with a few instances, it was possible for the robot to generalize to unseen objects in the given class and to detect classes that could not be detected by the baseline due to an unsuitable underlying training dataset.
This makes our teaching pipeline a valuable extension to training exclusively on standard datasets.
Overall, our approach is supremely natural and intuitive by virtue of its multimodality, including \acs{AR} and the shared gaze of human and robot.
The dataset we have recorded in the course of our evaluation is also made publicly available and is characterized by a high level of information density owing to the many different perspectives on the respective object and the data gathered in this process.
As a result, it has the potential to be relevant for a variety of purposes, aside from ours.

However, a significant amount of work remains for the future as we plan to investigate the usability of our system in a user study as well as to extend our approach to enable the robot to successfully detect objects outside of its trained environment and to further leverage the acquired knowledge through active learning in another human-robot interaction scenario.

\subsection*{Acknowledgments}
This research was funded by the Deutsche Forschungsgemeinschaft (DFG, German Research Foundation) under Germany's Excellence Strategy -- EXC number 2064/1 -- Project number 390727645.
We also thank Julia Dietl for her valuable efforts in labeling.
\fi

\clearpage
\setcounter{footnote}{0}
\section{Leveraging Saliency-Aware Gaze Heatmaps for Multiperspective Teaching of Unknown Objects}
\label{app:sec:paper05}

\ifpaper
\subsection{Abstract}
As robots become increasingly prevalent amidst diverse environments, their ability to adapt to novel scenarios and objects is essential.
Advances in modern object detection have also paved the way for robots to identify interaction entities within their immediate vicinity.
One drawback is that the robot's operational domain must be known at the time of training, which hinders the robot's ability to adapt to unexpected environments outside the preselected classes.
However, when encountering such challenges a human can provide support to a robot by teaching it about the new, yet unknown objects on an ad~hoc basis.
In this work, we merge augmented reality and human gaze in the context of multimodal human-robot interaction to compose saliency-aware gaze heatmaps leveraged by a robot to learn emerging objects of interest.
Our results show that our proposed method exceeds the capabilities of the current state of the art and outperforms it in terms of commonly used object detection metrics.

\subsection{Introduction}
With all the advancements in technology, the industrial sector witnessed a proliferation of robots and an expansion of their areas of application, spanning a wide range of industries and use cases.
From automotive and electronics over food and beverage to rubber and plastics, cosmetics, and pharmaceutical industries, an array of domains profit from the exceptional precision, work capacity, efficiency and high tolerance in demanding and hazardous environments of these robots~\cite{xiao2021robotics}.
Robots are also becoming more prevalent in close proximity to humans, serving as tour guides in museums~\cite{thrun1999minerva, velentza2019human} or as service assistants in places like supermarkets, such as Walmart~\cite{bogue2019strong,li2022effect}.
The recent advancements in machine learning, especially in computer vision, have been a major catalyst for this success, as they enabled robots to identify objects and individuals within their environment with great accuracy and speed.
Nevertheless, a fundamental premise underlying most cases is the existence of abundant training data with high quality labels.
In practice, the assumption that a fully labelled data set is available, is suited to the field of application, and encompasses all relevant objects, is not inevitably valid.
While state-of-the-art object detectors can achieve outstanding performance, given sufficient training data, their deployment is restricted to predefined scenarios imposed by the available training data~\cite{weber2020distilling}.
An unknown object that was not initially already part of the training set, but appears in front of a robot, cannot be detected by definition.
As a result, object detectors and, consequently, robots reach their capability limits.
For the latter, however, scene understanding is indispensable for all kinds of interaction.

In this work, we aim at minimizing such constraints towards the deployment of robots in unfamiliar environments by promoting the adaptation to new conditions involving unknown objects, while simultaneously reducing data dependency.
Within a human-robot interaction setting, we teach a robot novel objects using both multimodal and natural communication channels, such as augmented reality (see \figref{paper05:fig:keyboard}), speech, and human gaze.
The human looks at an object and provides the class information, whereupon the robot, equipped with an object detector, learns it.
\begin{figure}
    \centering
    \includegraphics[width=\linewidth, trim={1cm 2cm 0cm 0cm}, clip]{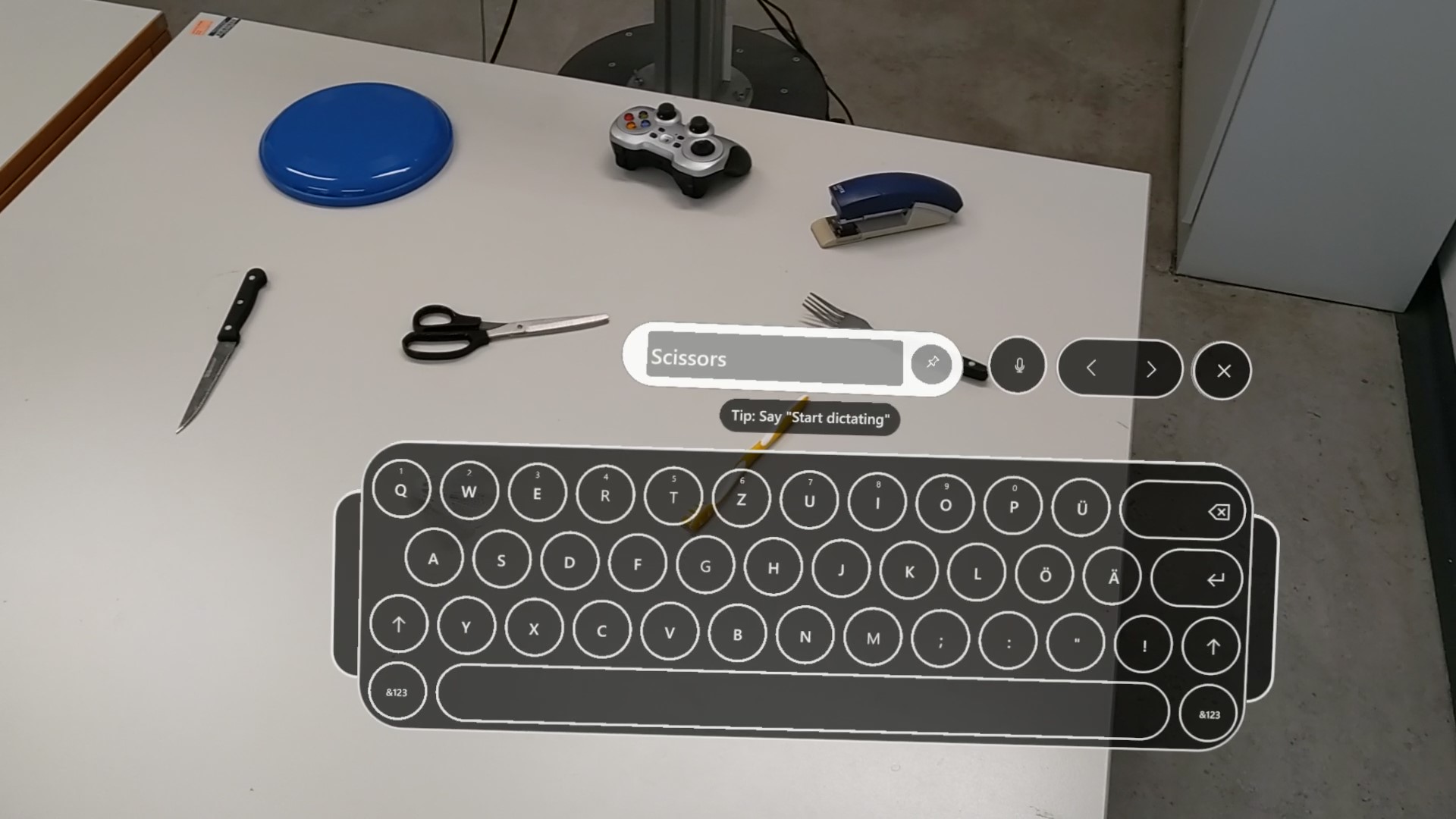}
    \caption{The augmented reality interface through which the human can teach the robot the class of an object using a virtual keyboard or speech.}
    \label{paper05:fig:keyboard}
\end{figure}
To this end, the robot autonomously acquires a series of images using a robot arm and determines the area of interest based on the gaze data.
Following up on the work of~\cite{weber2023multiperspective}, where the robot had to rely on point cloud segmentation, we streamline the teaching process by using saliency~aware gaze heatmaps.
The core idea is to leverage the \acf{GBVS} algorithm~\cite{harel2006graph} in combination with gaze~\cite{geisler2020exploiting} to refine the human gaze points and guide the robot's attention to the salient parts of the images that are of interest.
As a result, the teaching process becomes more lightweight, yet more efficient.

Overall, the contribution of our work is twofold:
\begin{enumerate}
    \item In a human-robot interaction setting, we utilize augmented reality to encode 3D gaze points of a human as saliency-aware 2D~gaze heatmaps, applying a single and a dual gaze-assisted approach.
    \item Having taught a robot unknown objects, our evaluations demonstrate that we outperform the current state of the art with regard to the common object detection metrics.
\end{enumerate}

\subsection{Related Work}
Teaching a robot to detect unknown objects through multimodal human-robot interaction is a complex task that necessitates substantial collaboration between different research fields, such as computer vision, eye tracking, and robotics.

In \cite{kabir2022unknown}, the authors approached the task of detecting unknown objects by employing a one-class support vector machine.
As the learning process was carried out incrementally, a series of robots were operating simultaneously, connected to a cloud-based system where all the computations were performed.
In this approach, the classification of unknown objects was limited to the property of being unknown and the learning process did not incorporate any specific class information.
By separating known and unknown items and exclusively relying on the familiar objects, the amount of incorrect data transmitted to the robots ought to be reduced.
The awareness that an object is unknown was also the basis in~\cite{du2022unknown}.
This work addressed the problem that datasets often contain unlabelled objects that are misinterpreted by object detectors as known classes.
The authors identified such out-of-distribution objects in videos to develop an unknown-aware object detector.
This model, however, used that information solely to reduce the number of false positives.
The unknown objects were neither learned nor classified.
Another recent work, as presented in~\cite{li2022uncertainty}, proposed to assess the uncertainty of the predictions of semantic segmentation models to binarily classify whether an object was known or unknown.
Again, no attempt was made to learn the individual categories.

In the field of eye tracking, the problem of unknown object detection was investigated in~\cite{weber2020distilling}.
This proof of concept aimed to decrease the number of candidate bounding boxes generated by a region proposal method, without being able to classify them.
The same authors ascertained in~\cite{weber2022gaze} whether an object was being observed by a person and determined the corresponding bounding box parameters using simple heatmaps instead of a scene image.
All the objects detected in the video segments were unknown, but their categories were not identified.
In~\cite{mazzamuto2022weakly}, several egocentric videos of museum visitors we taken to train a model to identify, among $15$ different objects, which specific object a subject was currently looking at.
However, the experimental setting was fixed and the set of $15$ objects were specified in advance.
Furthermore, the model was only capable of identifying the attended object, and none of the other known objects in the scene.
Hence, the model did not have the ability to detect objects in a general sense.
A similar objective was analysed in~\cite{higgins2022head}.
In a virtual reality study the set of objects that a user's attention was directed towards, was narrowed down based on the combination of head pose and linguistic description.
Nevertheless, neither was the class of the attended object learned, nor could the object itself be detected at a later point in time.

Several researchers, including \cite{krause2014learning} and \cite{el2021teaching}, have also investigated the potential of collaborations involving both humans and robots for teaching purposes.
The work presented in~\cite{krause2014learning} utilizes a natural language context to enable a robot to recognize objects through one-shot learning based on visual descriptions.
Nevertheless, in this proof of concept, the objects were required to be easily distinguishable through color or spatial relationships, and their component parts needed to be precisely describable using linguistic expressions to avoid ambiguity.
The authors of~\cite{el2021teaching} introduced a teaching system for object categorization that provided users with the ability to visualize the intermediate categorization stages, meaning the class into which the robot would categorize an object.
While the system allowed for interactive improvement and correction of the categorization, it required all objects to be marked with fiducial markers to be recognized at all.
Previous works, such as \cite{valipour2017incremental} and \cite{dehghan2019online}, have taught robots new objects in conjunction with picking tasks.
However, both studies used identical objects for both training and testing, which creates a positive bias in the results. 
In order to reflect the real-world conditions more accurately, we use distinct objects per class in our experiments.

\subsection{Method} \label{paper05:sec:method}
The objective of this work is to teach a robot unknown objects within its environment when training data does not yet exist or may not be available.
In other words, the human looks at an object that is unknown to the robot and provides the respective class information.
The robot then autonomously records the object from several different viewpoints and labels the data.
Finally, the robot uses the obtained data to learn the respective class.
Building upon the work of~\cite{weber2023multiperspective}, we effectively extend their method, resulting in a significant improvement of the achieved results.

The interaction between the human and the robot requires a way of communication, which we meet by means of the augmented reality interface introduced in~\cite{weber2023multiperspective}.
This user interface operates on a HoloLens~2 worn by the human and enables real-time transmission of the human's gaze data to the robot.
In the course of the teaching process, however, we eliminate the assumption that the robot must identify the object prior to recording the data.
Instead of a segmentation that provides a 3D bounding box with the size and position of the object, our method is based on successive gaze points of a given time interval.
The human selects the object of interest by saying ``select'' and then observes it for a few seconds (10 seconds in our experiments), seeking to capture the entire surface of the object as completely as possible.
A visualization is shown in \figref{paper05:fig:gaze_and_arm}.
\begin{figure}[b!]
    \centering
    \includegraphics[width=0.49\columnwidth, trim={3.2cm 1.5cm 13cm 1cm}, clip]{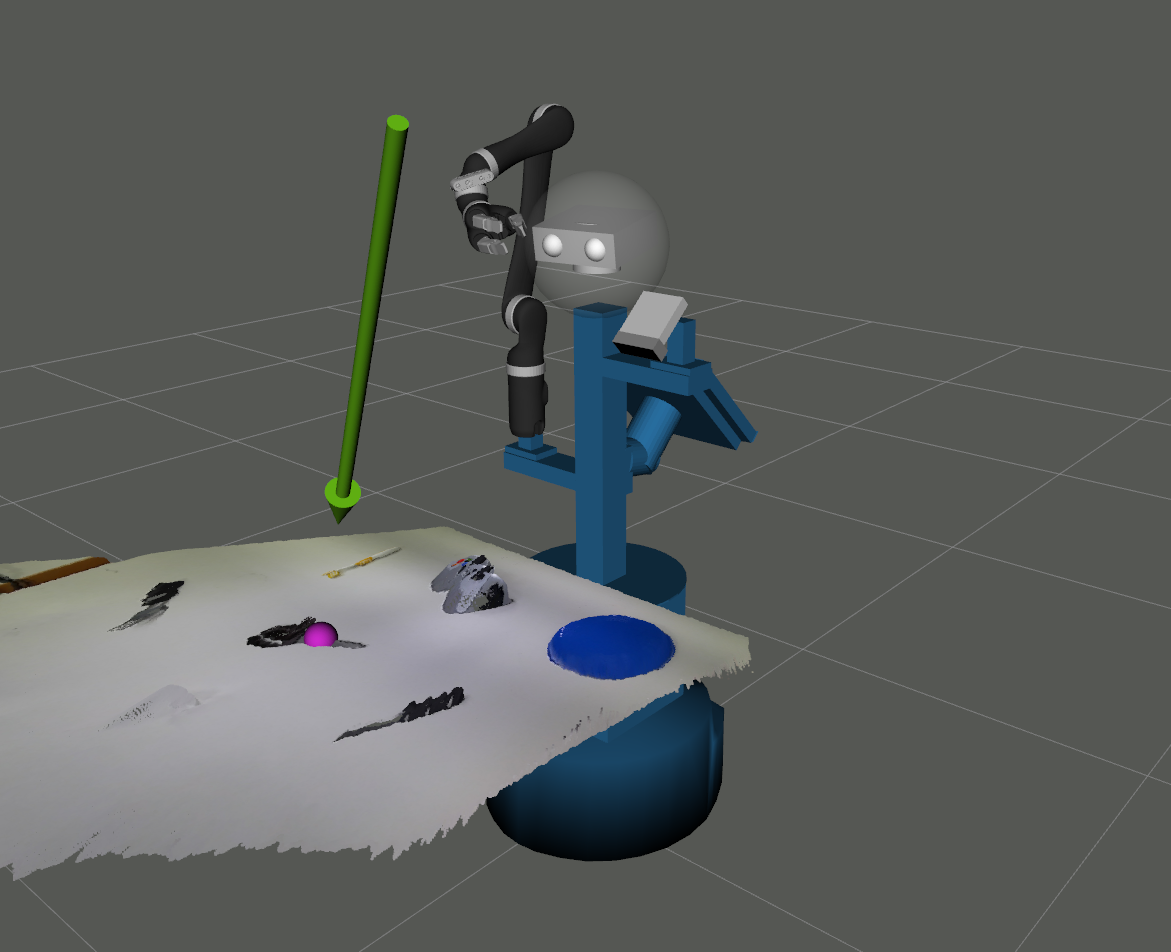}
    \includegraphics[width=0.49\columnwidth, trim={19cm 0 18cm 0}, clip]{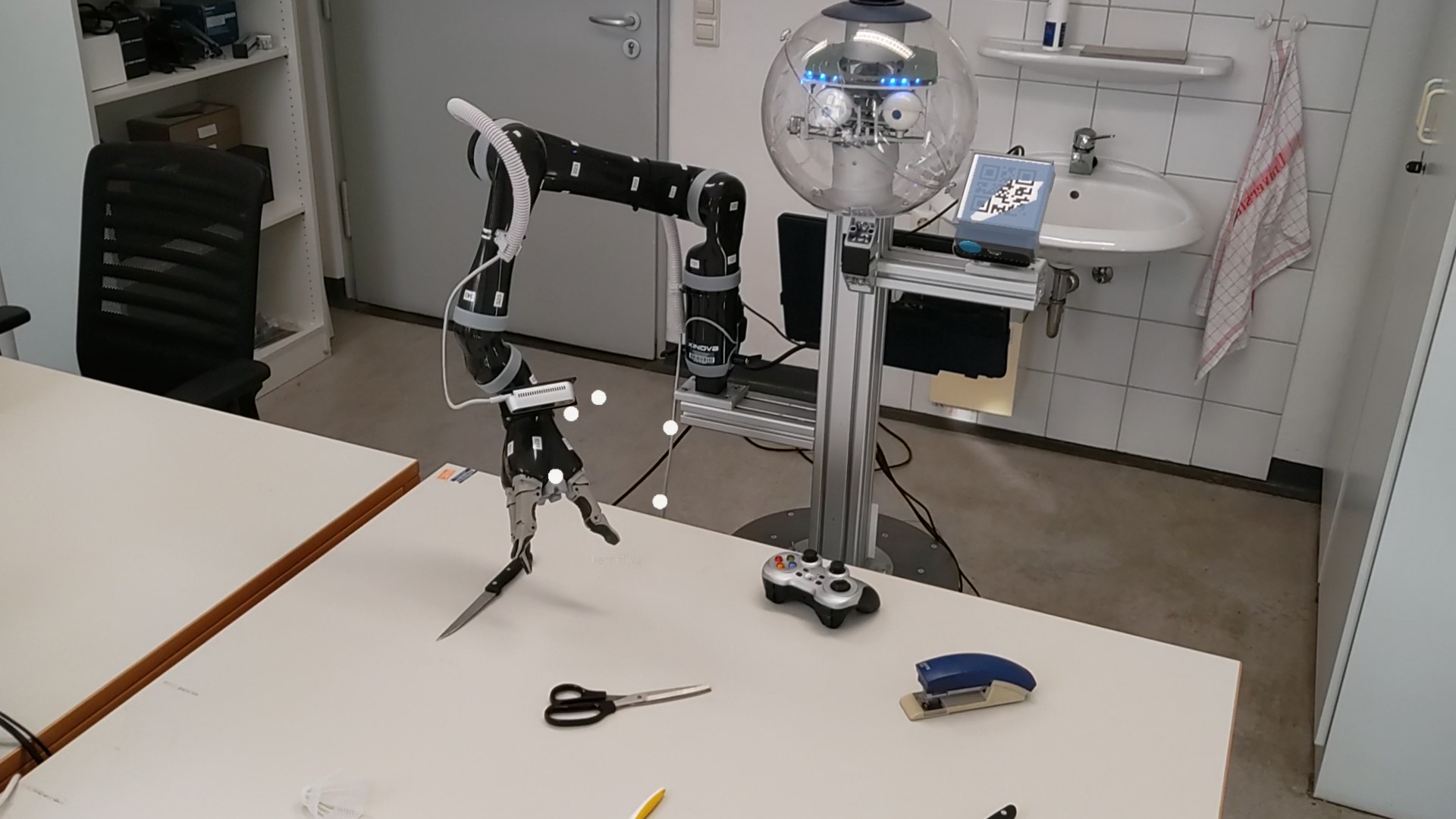}
    \caption{On left side, the human gaze ray (green arrow) and the ensuing gaze point (purple sphere) are visualized in RVIZ. The right side shows the recording process as seen from the augmented view of the human.}
    \label{paper05:fig:gaze_and_arm}
\end{figure}
Based on the resulting gaze points the position and size of the object is approximated in the world space.
Thereby, the median of all gaze points serves as the center of a 3D bounding box and three times the interquartile range as its size, component-wise.
One limitation of~\cite{weber2023multiperspective} is the dependence on the segmentation presented in~\cite{weber2022exploiting} and, consequently, on the depth sensor data of the robot's body camera, which leads to problems with very flat, dark or glossy objects.
By directly determining the position of the object using the gaze points, these types of objects do not pose a challenge for our approach.
Moreover, since we will later also refer to the gaze data to automatically label the recorded perspectives with a bounding box each, we achieve better overall results with less error proneness due to fewer system components.

After the human has looked at the object for the specified time interval, a virtual keyboard automatically appears in the human's field of view.
Using this keyboard or alternatively by speech, the class of the object can now be specified (see \figref{paper05:fig:keyboard}).
Once the human has submitted the input, it is transmitted to the robot, which then calculates a circular path around the object from which it can be recorded by means of a robot arm.
Due to physical constraints, such as the robot arm length, the final reachable trajectory of the arm is usually a sub-segment of the circle.
The robot then automatically brings its arm to the start of the trajectory and moves along it.
Using a camera, which is attached to the wrist of the robot arm, the robot records the object from various angles.
This process is illustrated in \figref{paper05:fig:gaze_and_arm}.
For each perspective, the gaze points gathered at the beginning are continuously mapped from the world space to the respective coordinate system of the moving camera.
Subsequently, these transformed 3D gaze points are projected onto the 2D image plane of the camera.
In principle, a 2D area of interest could now already be determined for each image from the boundary points.
However, in practice, many small inaccuracies are involved.
These include among others the eye tracking and gaze determination, the calibration of the robot with body, arm and camera, as well as the transformation between the individual robot frames. 
Although these are negligible individually, they add up in combination to an offset that corrupts the result. This offset can also be seen in \figref{paper05:subfig:roi:gaze}.

In order to determine the region of interest more precisely, we incorporate saliency in addition to the gaze data.
A saliency map is an image that highlights the most visually prominent regions in an input image that are likely to attract the attention of a human observer.
Inspired by~\cite{geisler2020exploiting}, we leverage the \acf{GBVS} algorithm~\cite{harel2006graph} for saliency-aware gaze heatmaps.
\acs{GBVS} refers to a computational method to determine saliency maps based on principles of graph theory.
It uses the visual and spatial relationships between image pixels to estimate their degree of importance.
The \acs{GBVS} algorithm can be structured into three steps: 
\begin{enumerate}
    \item Extraction of a feature map $ \FeatureMap \in \mathbb{R}^{m\times n} $ based on a given Image~$ \Image \in \mathbb{R}^{m\times n} $.
    \item Generation of an activation map $ \ActivationMap \in \mathbb{R}^{m\times n} $ based on~$ \FeatureMap $.
    \item Normalization of the activation map~$ \ActivationMap $ (and combination with the activation maps of all other feature maps).
\end{enumerate}
In the first step, low-level visual features such as color, luminance, and orientation are extracted from the image~$ \Image $.
The creation of such feature maps $ \FeatureMap $ is a well-known task and can be found in the literature~\cite{itti1998model, liu2018visual}, therefore we refer the readers to these sources for further details.

The second step is to calculate the saliency of each feature map based on the concept that a pixel is more salient if it is different from its surroundings. 
This step is modeled on~\cite{geisler2020exploiting}, except that we omit the temporal domain as we merge the series of successive gaze points into a single heat map per perspective.
The idea is to construct a graph representation of the image, where each node represents an image pixel, and the edges between nodes reflect the relationships between the pixels in terms of their visual dissimilarity or spatial proximity.
The activation map $ \ActivationMap $ can then be interpreted as a state vector of a Markov chain on this graph.
So let the visual dissimilarity of two nodes $ s_i = (p,q) $ and $ s_j = (u,v) $, where $ p,u \in \{1,2,\dots,m\} \eqqcolon [m] $, $ q,v \in [n]$ and $ i,j \in [mn]$, be defined as
\begin{equation}
    \VisualDissimilarity(s_i, s_j) = \abs{\log \dfrac{\FeatureMap(s_i)}{\FeatureMap(s_j)}}.
\end{equation}
Further, let $ \pi_1\colon [m]\times[n] \to [m], (x,y) \mapsto x $ and $ \pi_2: [m]\times[n] \to [n], (x,y) \mapsto y $ be the projection onto the first and second coordinate, respectively.
Then $ \SpatialDissimilarity \colon \left([m]\times[n]\right)^2 \to \mathbb{R} $ given by
\begin{equation}
    \SpatialDissimilarity(s_i, s_j) = \exp\left( -\dfrac{\hat{\SpatialDissimilarity}(s_i,s_j)}{2\sigma^2} \right),
\end{equation}
where
\begin{equation}
    \hat{\SpatialDissimilarity}(s_i,s_j) = \left(\pi_1(s_i) - \pi_1(s_j)\right)^2 + \left(\pi_2(s_i) - \pi_2(s_j)\right)^2,
\end{equation}
is the exponential weighted square distance of the two nodes $ s_i $ and $ s_j $.
The variable~$\sigma$ is a free parameter that is usually set to one tenth to one fifth of the mean value between the width and height of the feature map~\cite{harel2006graph}.

We now consider a fully-connected, directed graph that links all nodes of the lattice $ \FeatureMap $.
The weight assigned to the directed edge from node $ s_i $ to node $ s_j $ is the product of the visual dissimilarity  $ \VisualDissimilarity $ in the domain of $ \FeatureMap $ and the spatial proximity $ \SpatialDissimilarity $:
\begin{equation}
    \Weight(s_i, s_j) = \VisualDissimilarity(s_i, s_j) \cdot \SpatialDissimilarity(s_i, s_j).
\end{equation}
By normalizing the weights of each node's outbound edges to $ 1 $, we can define a Markov chain.
This allows us to establish an equivalence between nodes and states, and between edge weights and transition probabilities.
The Markov transition matrix between the $ mn $ states is
\begin{equation}
    \TransMat = (t_{i,j})_{1\leq i,j \leq mn} \in \mathbb{R}^{mn \times mn},
\end{equation}
where $ t_{ij} = \Weight(s_i, s_j) $.
The final activation $ \ActivationMap $ results from the equilibrium distribution, that is, $ \hat{\ActivationMap}\TransMat = \hat{\ActivationMap}$, where $ \hat{\ActivationMap}\in\mathbb{R}^{1\times mn}$ is the flattened version of $ \ActivationMap $.
This boils down to an eigenvector problem.
In practice, a common method for determining the equilibrium distribution involves repeatedly multiplying the Markov transition matrix with a vector $ \nu \in \mathbb{R}^{1\times mn} $ that is initially uniformly distributed.
Consequently, given $ \lim_{k\rightarrow\infty} \nu\TransMat^k = \hat{\ActivationMap} $, $ \hat{\ActivationMap} $ can be estimated as $ \hat{\ActivationMap} = \nu\TransMat^k $, using a sufficiently large $ k \in \mathbb{N} $.

Motivated by~\cite{geisler2020exploiting}, we incorporate gaze into this step.
Rather than distributing the initial (flattened) activation map $ \nu $ uniformly, we initialize it based on the gaze points:
\begin{equation}
    \label{paper05:eq:nu}
    \nu = \sum_t \frac{\nu_t}{\Vert \nu_t \Vert_1}, \quad \nu_t = \left(\SpatialDissimilarity(s_0, g_t)\; \dots \; \SpatialDissimilarity(s_{nm}, g_t)\right),
\end{equation}
where $ g_t $ is the recorded gaze position at time $ t $.
In this way, $ \nu $ is influenced by the weighted distance of the pixels to the individual gaze points.
Since the equilibrium state remains the same regardless of the initialization, we do not perform the multiplication with the transition matrix multiple times, but only once, that is $ k = 1$.
The greater the value of $ k $, the more significant the impact of saliency, while the influence of the gaze decreases proportionally.
Depending on the quality of the gaze data, this value can be adjusted accordingly.

\begin{figure}[b]
	\centering
	\begin{subfigure}[b]{0.33\columnwidth}
		\includegraphics[width=.98\linewidth]{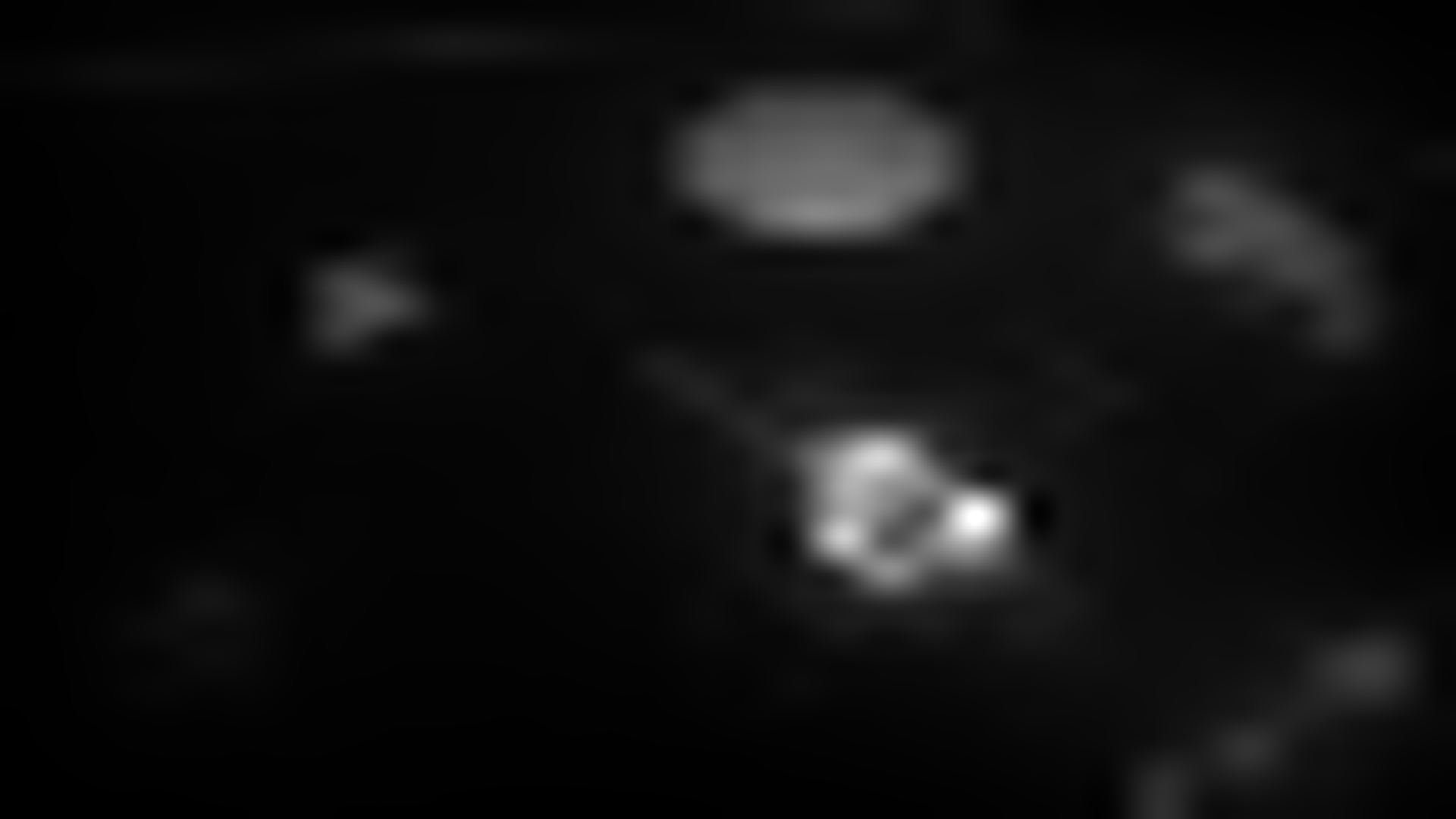}
		\caption{}
		\label{paper05:subfig:SaliencyComparison:GBVS}
	\end{subfigure}%
	\begin{subfigure}[b]{0.33\columnwidth}
		\includegraphics[width=.98\linewidth]{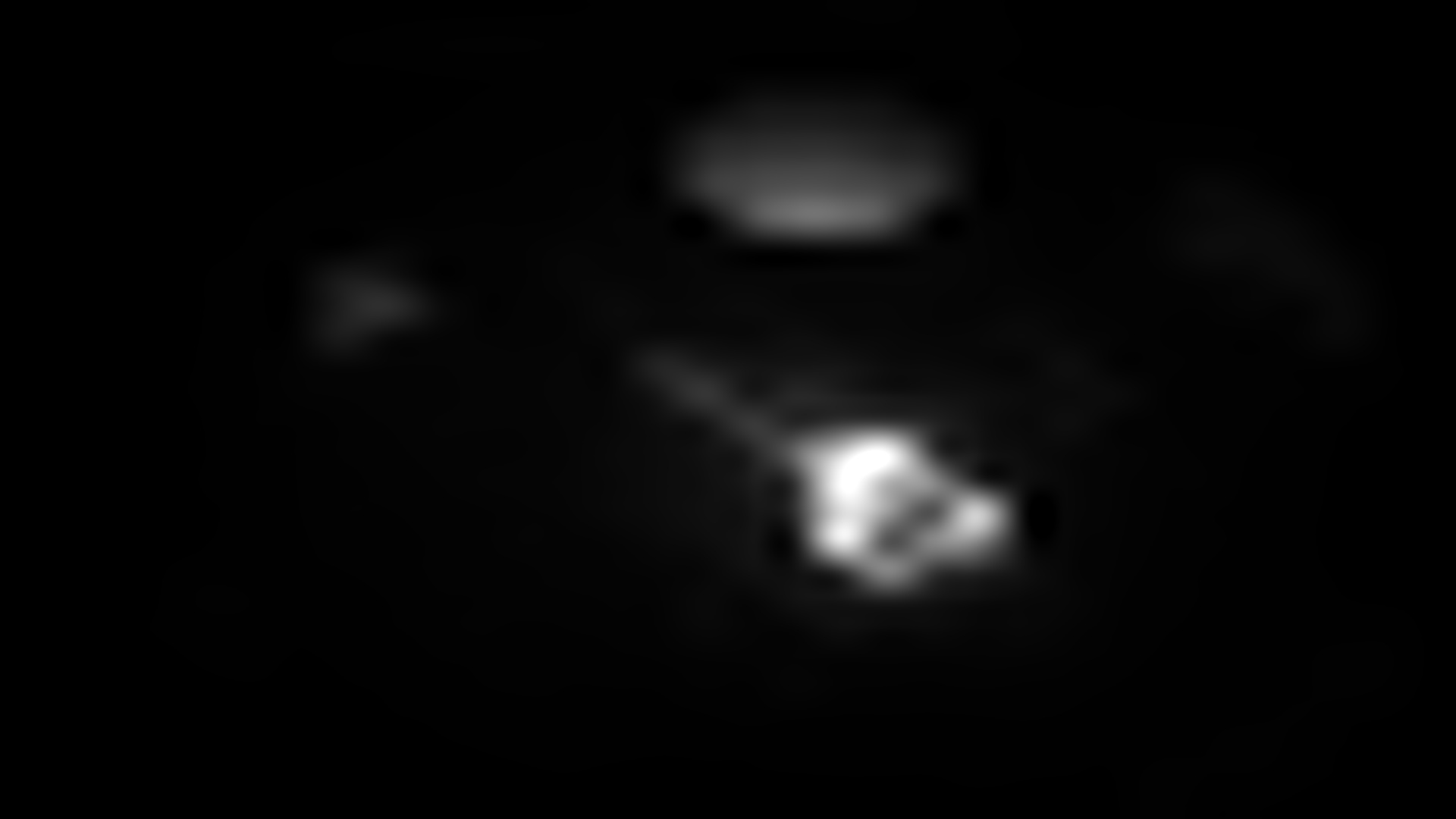}
		\caption{}
		\label{paper05:subfig:SaliencyComparison:GA-GBVS}
	\end{subfigure}%
	\begin{subfigure}[b]{0.33\columnwidth}
		\includegraphics[width=.98\linewidth]{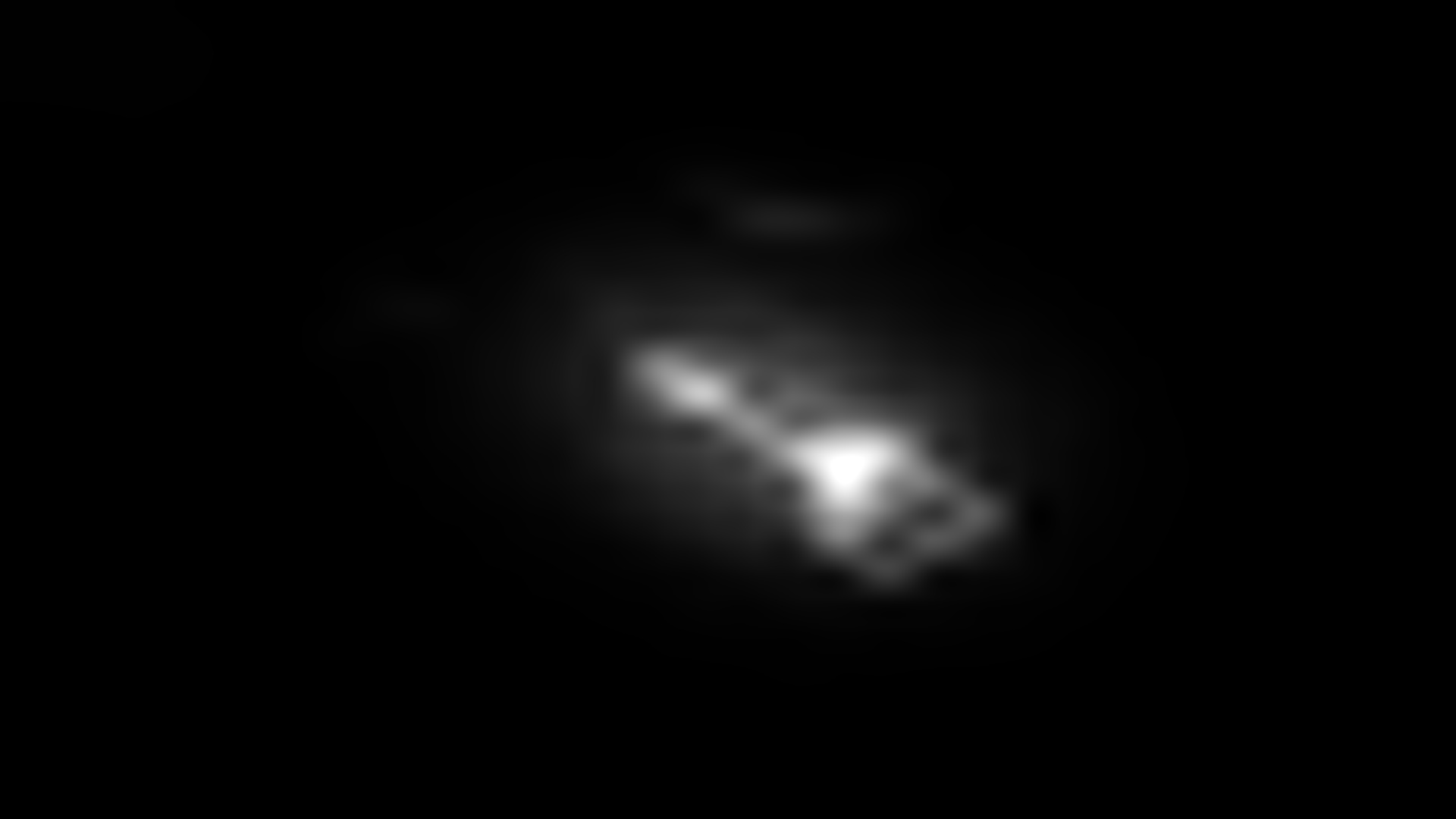}
		\caption{}
		\label{paper05:subfig:SaliencyComparison:DGA-GBVS}
	\end{subfigure}%
	\caption{A comparison of the saliency maps obtained by (\subref{paper05:subfig:SaliencyComparison:GBVS}) \acs{GBVS}, (\subref{paper05:subfig:SaliencyComparison:GA-GBVS}) \acl{GA-GBVS}, and (\subref{paper05:subfig:SaliencyComparison:DGA-GBVS}) \acl{DGA-GBVS}.}
	\label{paper05:fig:SaliencyComparison}
\end{figure}

The next and last \acs{GBVS} step is to ``normalize'' and combine the activation maps.
The aim is to further accentuate salient areas, in order to produce an informative saliency map that is not overly uniform.
This can be achieved with an analogous Markovian approach as in step 2, selecting the weights of the transition matrix $ \TransMatN $ as follows:
\begin{equation}
    \WeightN(s_i, s_j) = \ActivationMap (s_j) \cdot \SpatialDissimilarity(s_i, s_j).
\end{equation}
While the authors of~\cite{geisler2020exploiting} only included gaze data in the second step, we would like to point out that this is additionally possible in this third step, if the influence of the gaze needs to be increased further.
In that case, the initial normalized activation map $ \nu' $ has to be initialized according to \eqref{paper05:eq:nu}.
Under certain circumstances, such as when an input image contains multiple objects, our experiments have shown that this \acl{DGA-GBVS} approach (\acs{DGA-GBVS}), where gaze data is taken into account in the normalization step as well, can be beneficial and may lead to improved performance.
Nevertheless, when it comes to the overall learning process, the consideration of gaze data only in the second step (\acs{GA-GBVS}) outperforms the method proposed in~\cite{weber2023multiperspective} even more apparent.

The finalization is done by combining the activation maps of all extracted feature maps into one single saliency map.
This can be achieved by summing up and then normalizing the outcome to the image value range.
The ultimate resulting saliency-aware gaze heatmap contains less noise than the standard \acs{GBVS} saliency map and more intensely concentrates attention on the area of interest.
A comparison of the final saliency maps is shown in \figref{paper05:fig:SaliencyComparison}.

\begin{figure}[b!]
	\centering
	\begin{subfigure}[b]{0.49\columnwidth}
		\includegraphics[width=.98\linewidth]{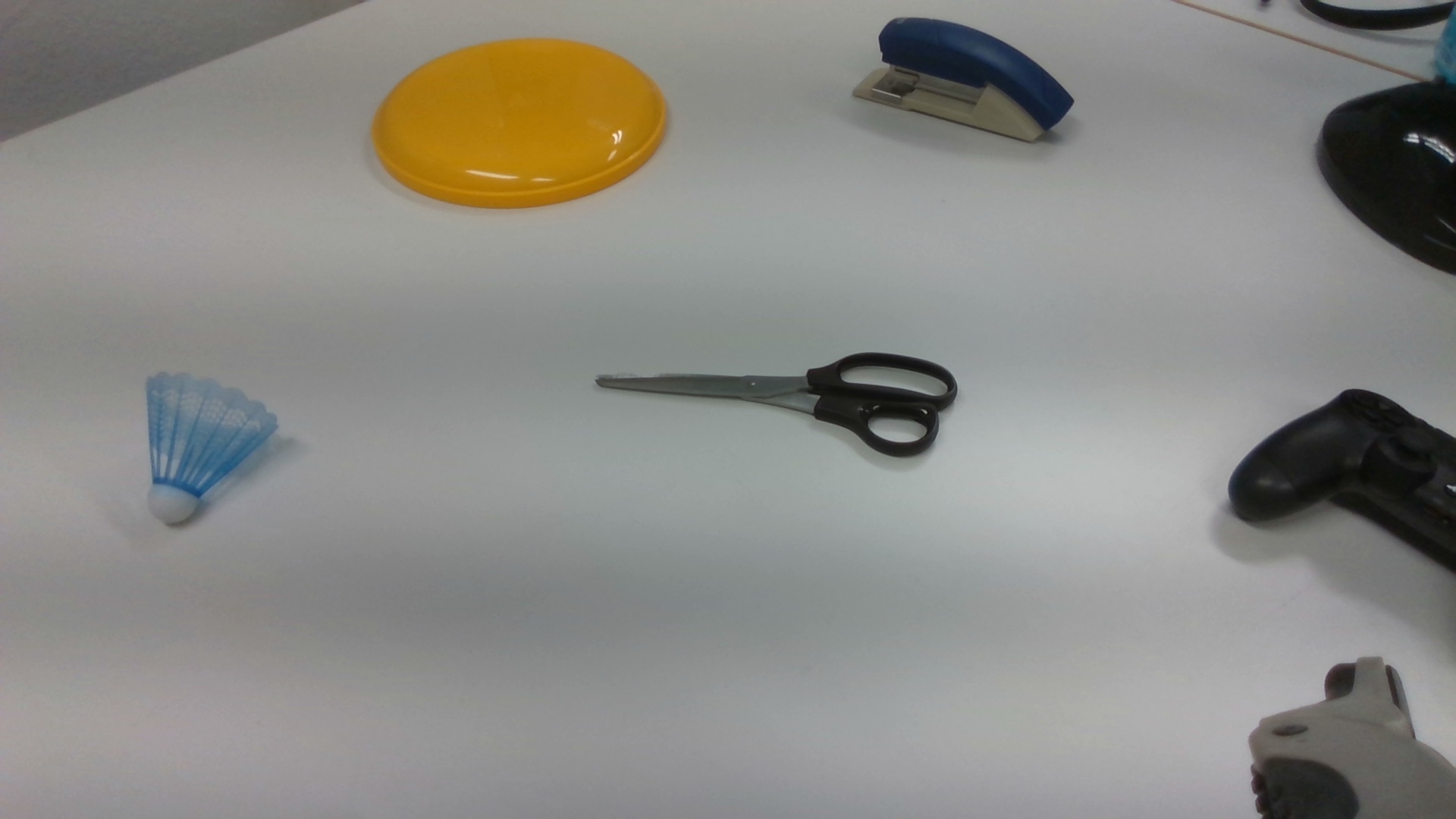}
		\caption{}
		\label{paper05:subfig:roi:stimulus}
	\end{subfigure}%
	\begin{subfigure}[b]{0.49\columnwidth}
		\includegraphics[width=.98\linewidth]{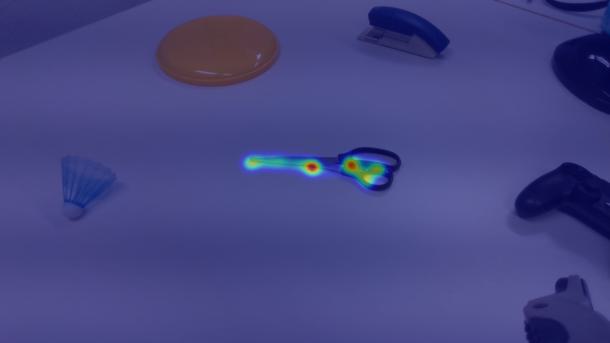}
		\caption{}
		\label{paper05:subfig:roi:gaze}
	\end{subfigure}%
	\par\smallskip
	\begin{subfigure}[b]{0.49\columnwidth}
		\includegraphics[width=.98\linewidth]{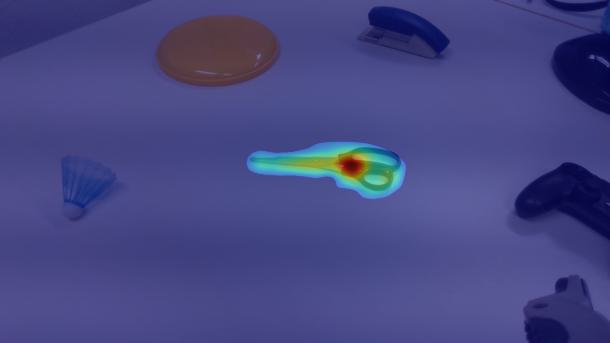}
		\caption{}
		\label{paper05:subfig:roi:GBVS}
	\end{subfigure}%
	\begin{subfigure}[b]{0.49\columnwidth}
		\includegraphics[width=.98\linewidth]{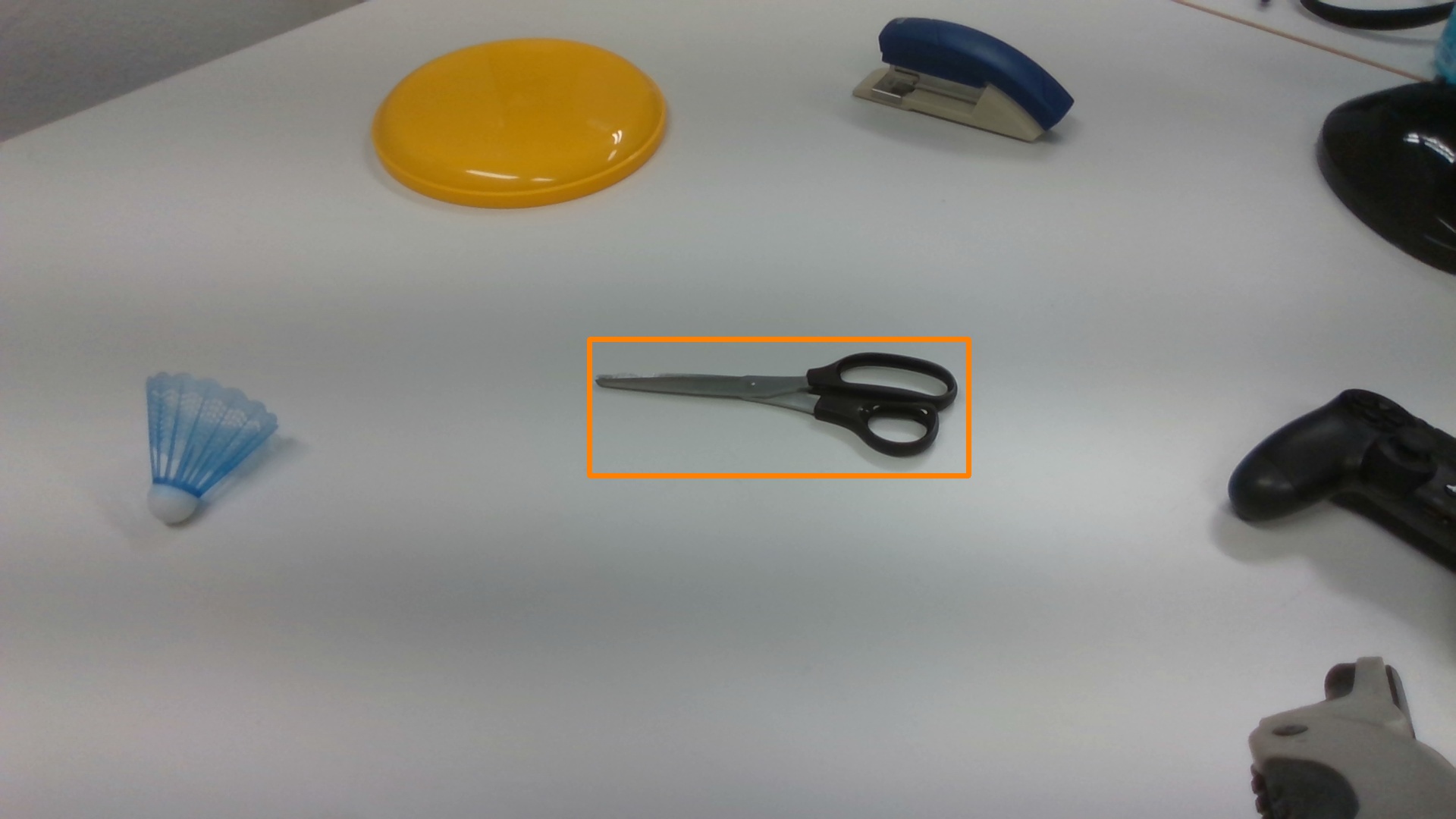}
		\caption{}
		\label{paper05:subfig:roi:box}
	\end{subfigure}%
	\caption{The intermediate stages of the bounding box determination. (\subref{paper05:subfig:roi:stimulus}) Various images of the object are taken by means of the robot arm. (\subref{paper05:subfig:roi:gaze}) The 3D gaze points are mapped to each image obtained. (\subref{paper05:subfig:roi:GBVS}) The heatmap is refined using \acs{GBVS} in combination with gaze. (\subref{paper05:subfig:roi:box}) Eventually, after Otsu's binarisation~\cite{otsu1979threshold}, the boundary points lead to the desired bounding box.}
	\label{paper05:fig:roi}
\end{figure}

Owing to less noise, a bounding box can now be determined on the basis of the boundary points, whereby a threshold value for the points under consideration is set beforehand using Otsu's binarization~\cite{otsu1979threshold}.
An example with the intermediate stages of the entire bounding box estimation is shown in \figref{paper05:fig:roi}.

In the final step, each image taken with the robot arm can now be automatically labeled with a bounding box and the robot can be trained using the transfer-learning approach proposed in~\cite{weber2023multiperspective}.
This means that we follow the same training routine and also pretrain the robot on a subset of MS COCO that does not contain any objects that will be taught later in the course of the evaluation.

\subsection{Evaluation} \label{paper05:sec:evaluation}
Our evaluation setup comprises of a Scitos~G5 robot manufactured by MetraLabs~\cite{MetraLabs}, which serves as the base platform.
Attached to this platform is a Kinova Jaco2~\cite{KinovaGen2} robotic arm featuring six degrees of freedom.
The camera, which is attached to the wrist of the arm and is guided around the objects by the robot to capture images from multiple perspectives, is an Intel RealSense~D435~\cite{IntelRealSense}.

For the comparability of our results, we evaluate on the publicly available \acf{OMD} dataset~\cite{weber2023multiperspective} and use the approach presented there as a baseline.
The dataset contains ten classes, which we teach to the robot following the methodology described in detail in \secref{paper05:sec:method}. 
For this purpose, conforming to~\cite{weber2023multiperspective}, we place the two training items of each class individually on the table in front of the robot, which differ in shape, color or both from the two items in the test set.
To replicate their results, we employ Faster \mbox{R-CNN}~\cite{ren2015faster} with the ResNet-50-FPN~\cite{he2016deep} backbone as underlying machine learning model and we also pretrained the backbone on Imagenet~\cite{deng2009imagenet} and Faster \mbox{R-CNN} on a subset of MS~COCO~\cite{lin2014microsoft}, excluding the classes from the \acs{OMD} dataset, which we intend to teach.

Upon completion of the training, we evaluate based on the MS~COCO metrics~\cite{lin2014microsoft}.
That means we examine the average precision and the average recall for a variety of intersection-over-union (\acs{IoU}) thresholds.
To simplify the discussion, we adopt the following abbreviations: \AP{} denotes the average precision across all \acs{IoU} thresholds from $0.5$ to $0.95$, with a step size of $0.05$. 
The abbreviations \AP{50} and \AP{75} correspond to the average precision at \acs{IoU} thresholds of $0.5$ and $0.75$, respectively;
All are calculated separately for each individual class.
The analog notation applies to the average recall, where we use the abbreviations \AR{1}, \AR{10}, and \AR{100} to refer to the average recall when allowing up to $1$, $10$, and $100$ detections per image, respectively.
Additionally, we use \acs{mAP} and \acs{mAR} to represent the mean average precision and the mean average recall over all classes.

\figref{paper05:fig:pr_curve} compares the precision-recall curves of the baseline method and our approach at an \acs{IoU} of $0.5$.
\begin{figure}
	\centering
	\begin{subfigure}[b]{.8\textwidth}
		\captionsetup{skip=0pt}
		\includegraphics[width=\linewidth]{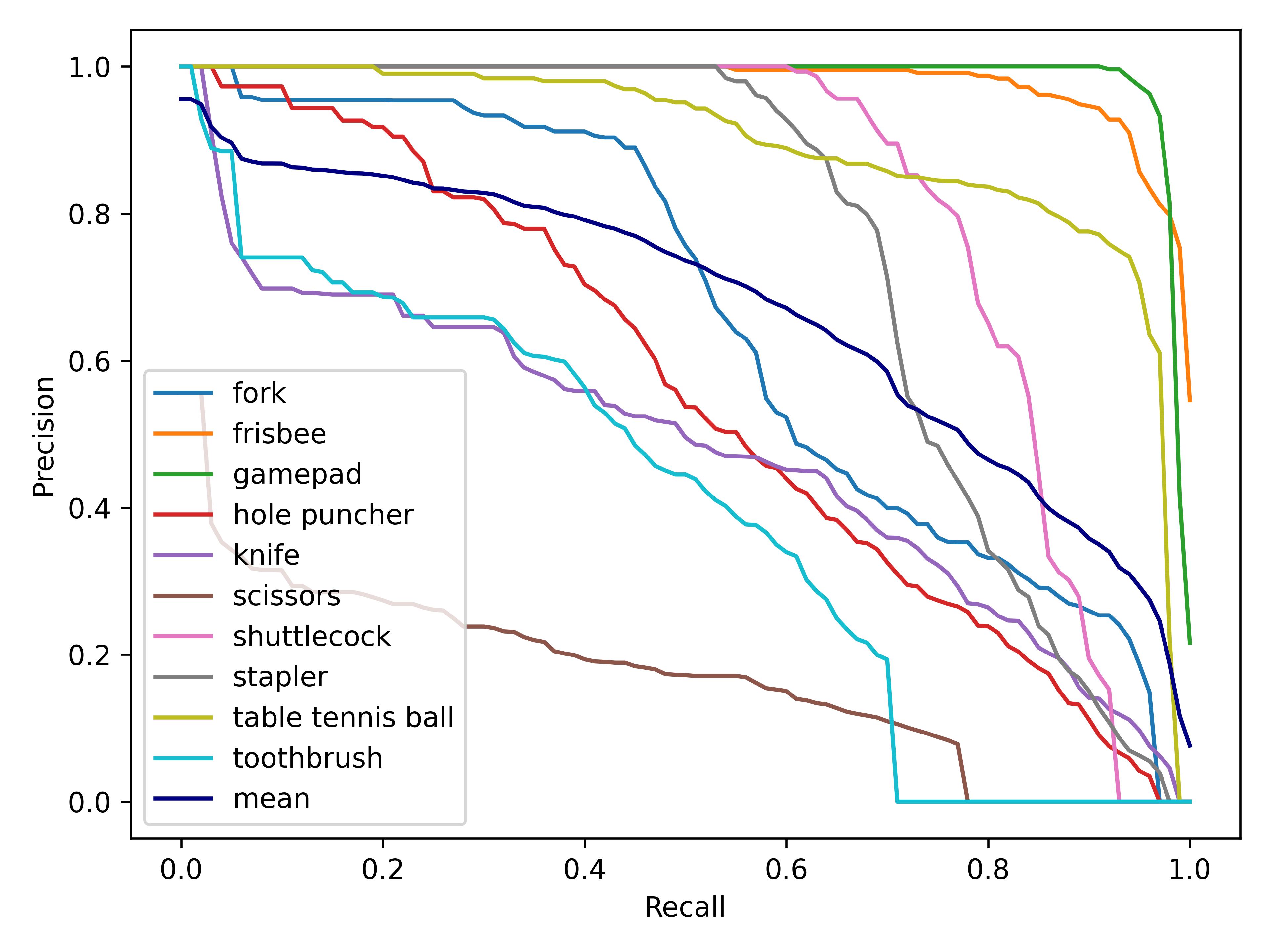}
		\caption{Baseline}
		\label{paper05:subfig:pr_curve:segmentation}
	\end{subfigure}%
	\par\medskip
	\begin{subfigure}[b]{.8\textwidth}
		\captionsetup{skip=0pt}
		\includegraphics[width=\linewidth]{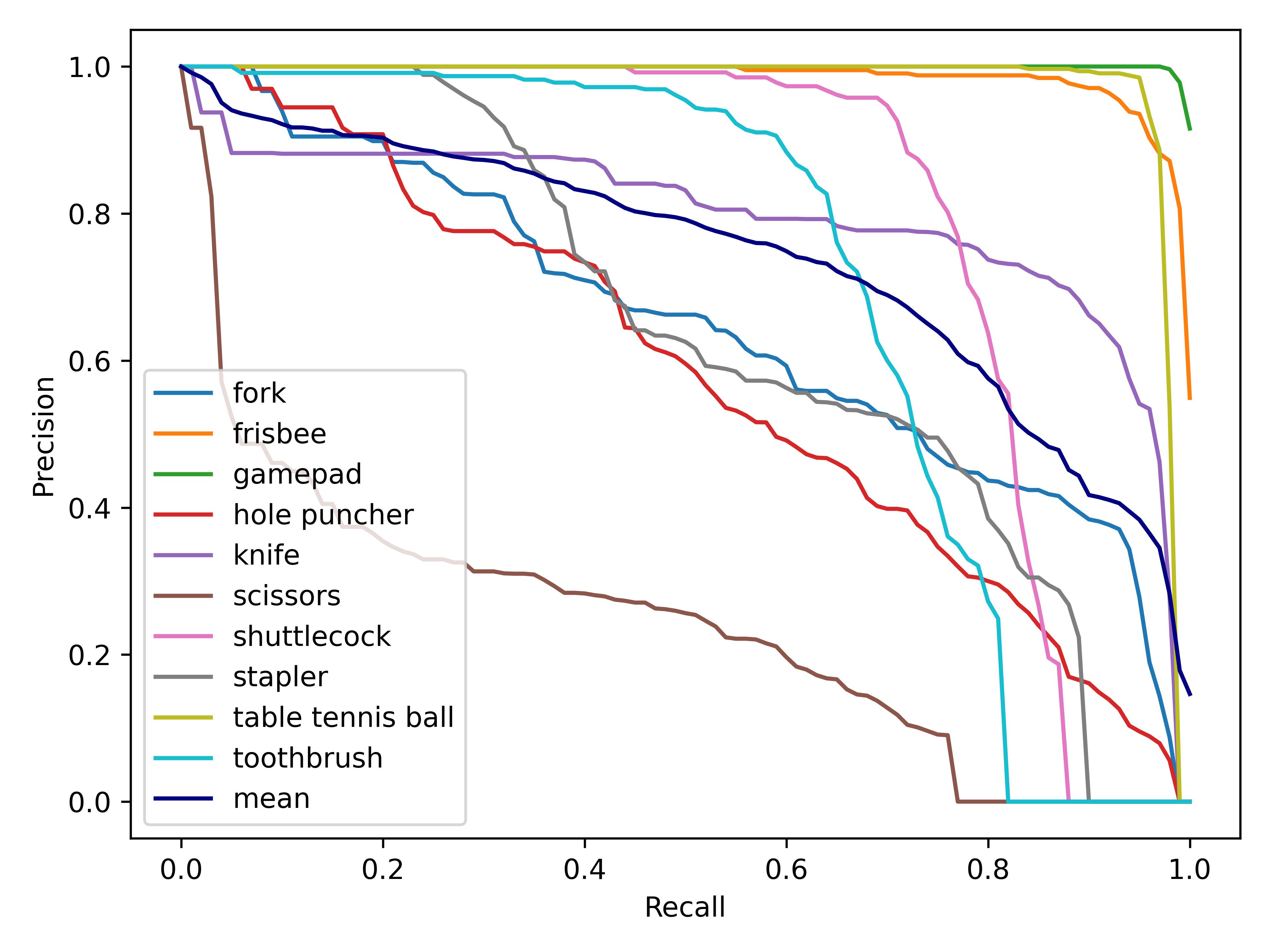}
		\caption{\acs{GA-GBVS}}
		\label{paper05:subfig:pr_curve:GA-GBVS}
	\end{subfigure}%
	\caption{Precision-recall curves at an \acs{IoU} of 0.5 of (\subref{paper05:subfig:pr_curve:segmentation}) the baseline~\cite{weber2023multiperspective} and (\subref{paper05:subfig:pr_curve:GA-GBVS}) our teaching approach using \acs{GA-GBVS}.}
	\label{paper05:fig:pr_curve}
\end{figure}
This is the threshold above which objects can be considered as detected~\cite{everingham2010pascal, zitnick2014edge}.
In general, the mean curve shows a tendency towards better precision at higher recall applying our method.
Although the stapler class has deteriorated slightly, the overall improvement is particularly apparent for the knife, the toothbrush, and the table tennis ball.
While, for example, in the latter the curve of the baseline decreases continuously beginning with a recall value of $0.3$, our method still reaches a precision of almost $1$ at a recall value of over $0.9$.
The knife and toothbrush reveal an even more pronounced improvement over the baseline, as both curves in \figref{paper05:fig:pr_curve}(\subref{paper05:subfig:pr_curve:GA-GBVS}) are much more concave and lie further above the corresponding baseline curves in \figref{paper05:fig:pr_curve}(\subref{paper05:subfig:pr_curve:segmentation}).

The same picture emerges when the average accuracy is considered as a function of the \acs{IoU}.
A comparison can be seen in \figref{paper05:fig:ap_iou_curve}.
\begin{figure}
	\centering
	\begin{subfigure}[b]{.8\textwidth}
		\captionsetup{skip=0pt}
		\includegraphics[width=\linewidth]{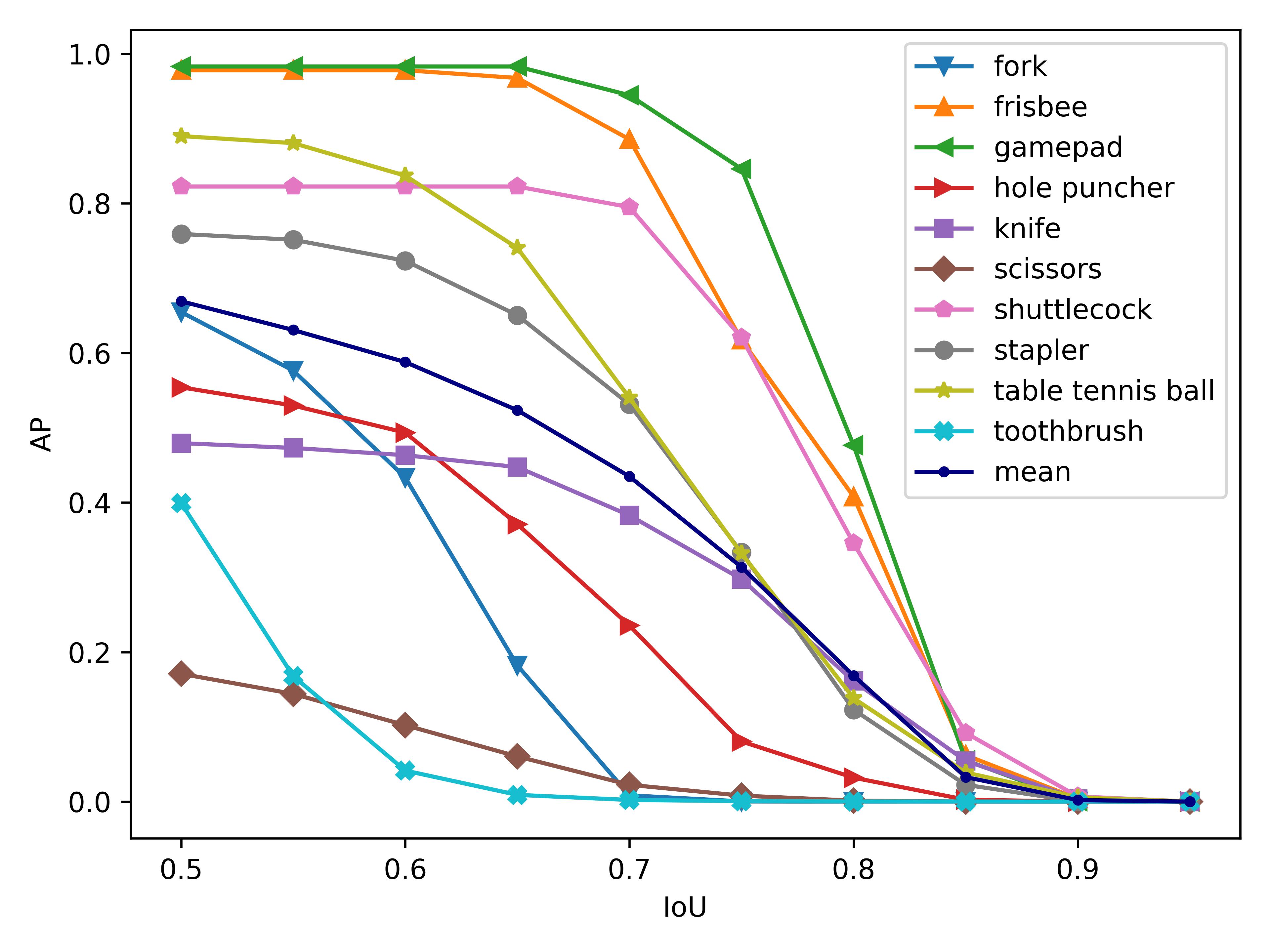}
		\caption{Baseline}
		\label{paper05:subfig:ap_iou_curve:segmentation}
	\end{subfigure}%
	\par\medskip
	\begin{subfigure}[b]{.8\textwidth}
		\captionsetup{skip=0pt}
		\includegraphics[width=\linewidth]{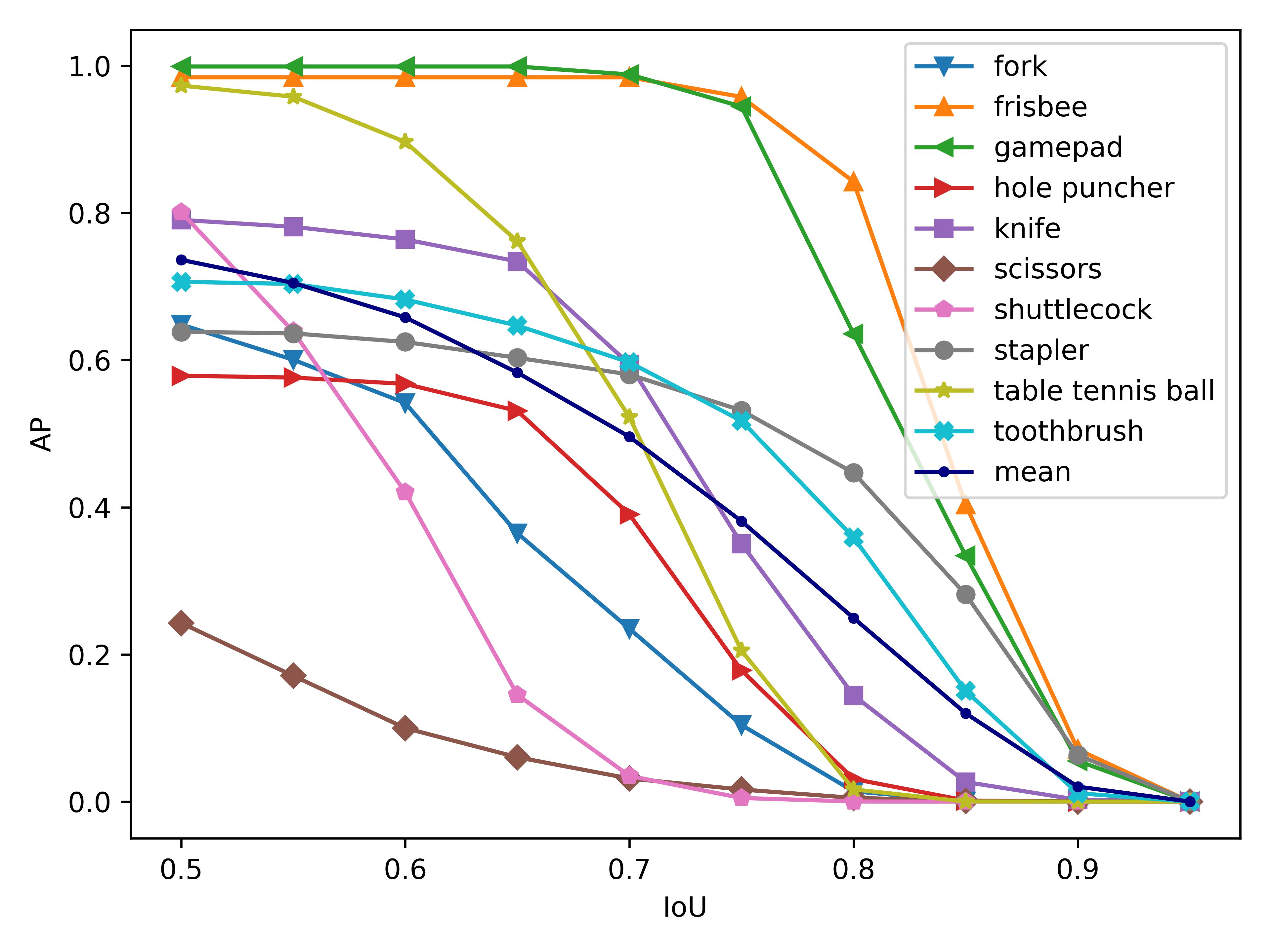}
		\caption{\acs{GA-GBVS}}
		\label{paper05:subfig:ap_iou_curve:GA-GBVS}
	\end{subfigure}%
	\caption{AP-\acs{IoU} curves of (\subref{paper05:subfig:ap_iou_curve:segmentation}) the baseline~\cite{weber2023multiperspective} and (\subref{paper05:subfig:ap_iou_curve:GA-GBVS}) our teaching approach using \acs{GA-GBVS}.}
	\label{paper05:fig:ap_iou_curve}
\end{figure}
The results of the toothbrush show the most improvement.
Whereas with the baseline the average precision drops steeply right at the beginning, our method demonstrates a slower, more gradual decline, remaining almost constant at first.
In contrast, the shuttlecock has deteriorated, however, the overall result across all classes is still better.
This is supported by the curve representing the mean of all classes, which proves that our method outperforms the baseline in terms of mean average accuracy at every \acs{IoU} value.
Specifically, our method achieves $ \mAP{60} \approx 0.71$ and $ \mAP{80} \approx 0.25 $, respectively, while the baseline achieves only around $0.63$ and $0.17$ at the same \acs{IoU} values.

\figref{paper05:fig:recall_iou_curve} displays the recall as a function of the \acs{IoU}.
\begin{figure}
\centering
    \begin{subfigure}[b]{.8\textwidth}
        \captionsetup{skip=0pt}
	    \includegraphics[width=\linewidth]{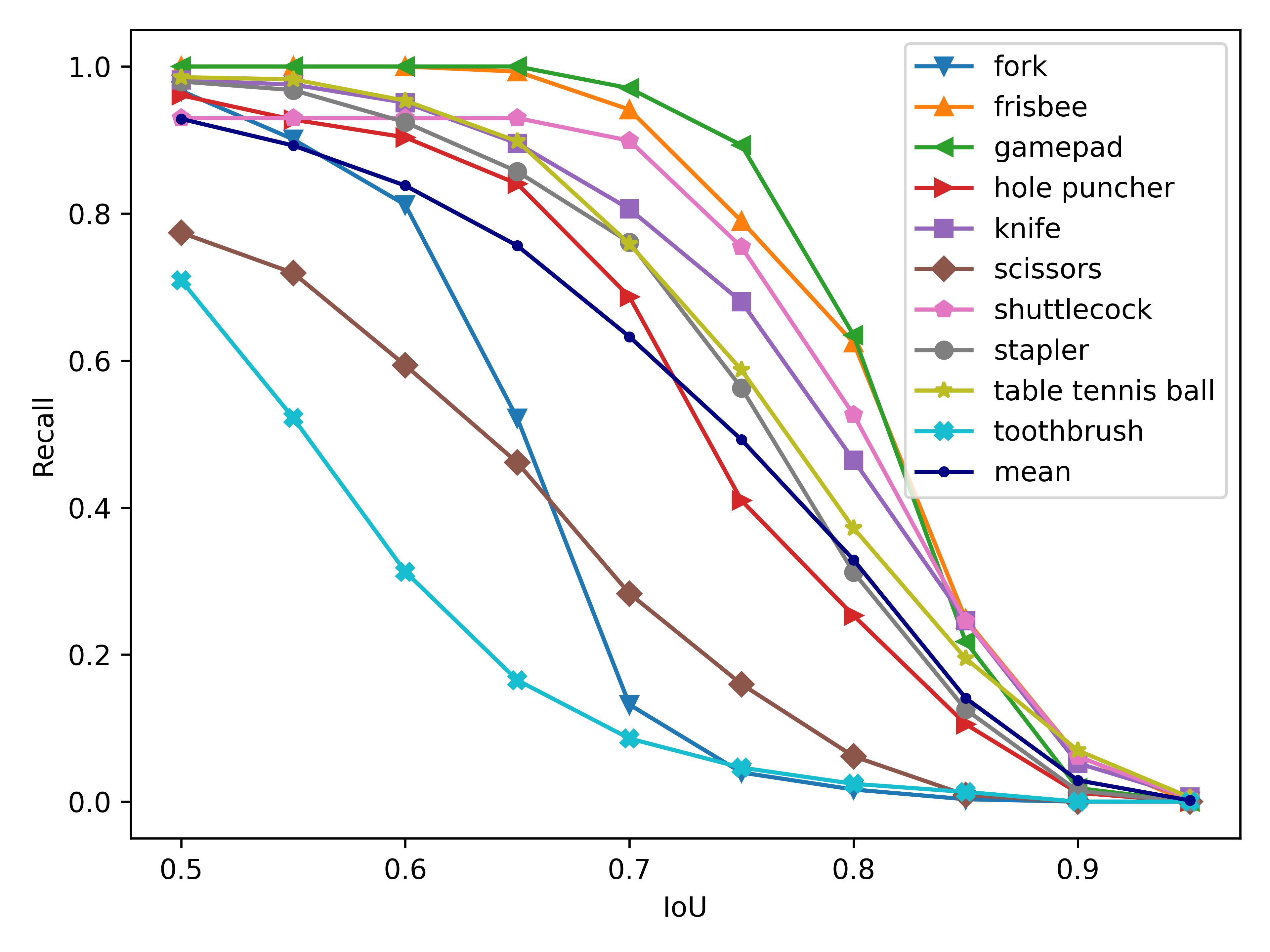}
		\caption{Baseline}
		\label{paper05:subfig:recall_iou_curve:segmentation}
	\end{subfigure}%
    \par\medskip
    \begin{subfigure}[b]{.8\textwidth}
        \captionsetup{skip=0pt}
	    \includegraphics[width=\linewidth]{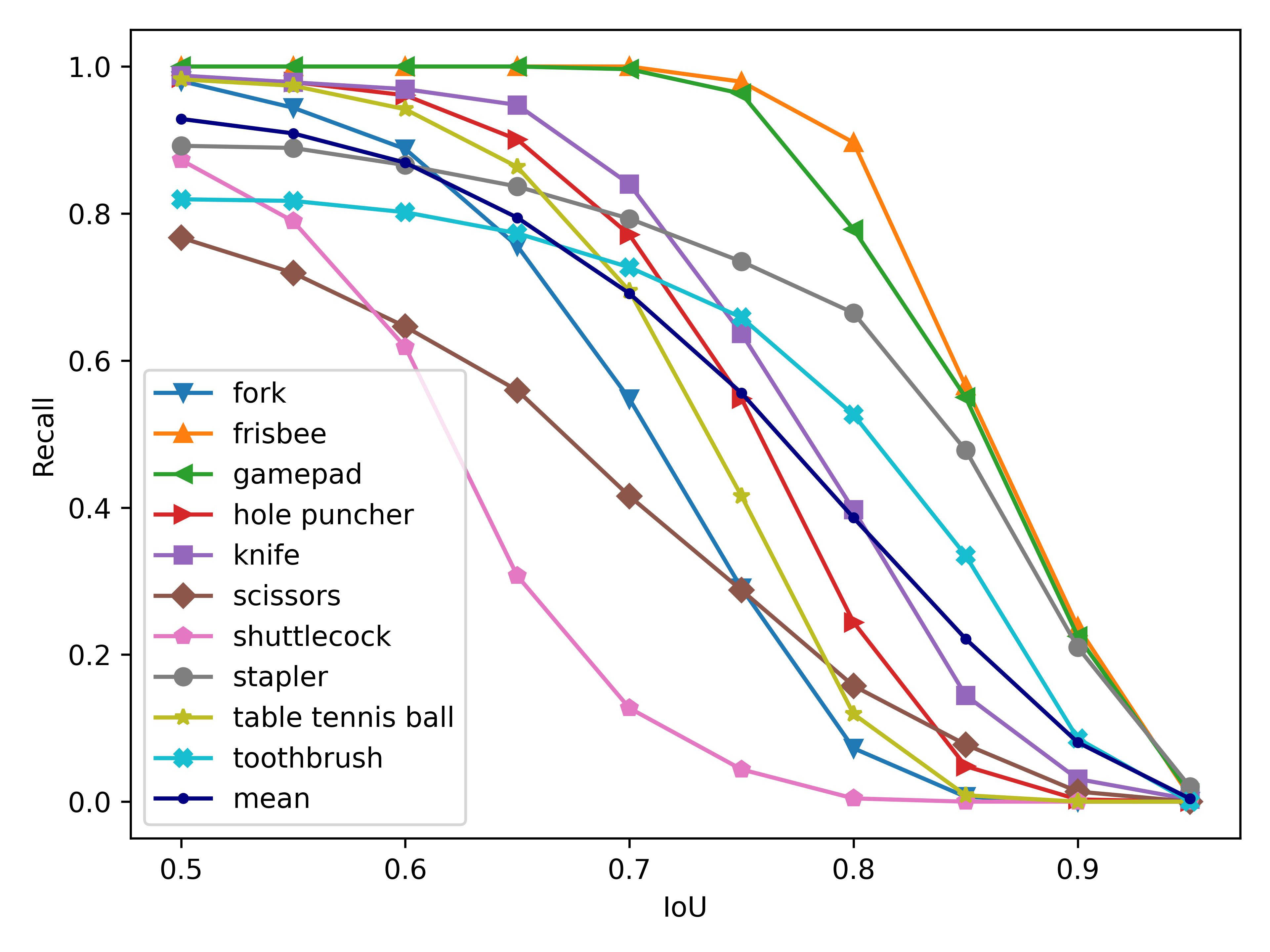}
		\caption{\acs{GA-GBVS}}
		\label{paper05:subfig:recall_iou_curve:GA-GBVS}
	\end{subfigure}%
    \caption{Recall-\acs{IoU} curves of (\subref{paper05:subfig:recall_iou_curve:segmentation}) the baseline~\cite{weber2023multiperspective} and (\subref{paper05:subfig:recall_iou_curve:GA-GBVS}) our teaching approach using \acs{GA-GBVS}.}
    \label{paper05:fig:recall_iou_curve}
\end{figure}
This comparison also shows that our method outperforms the baseline in terms of recall averaged across all classes.
Although the \mAR{50} is approximately the same for both methods, the baseline values decrease faster as the \acs{IoU} threshold increases.
As an example, our method yields an \mAR{80} of approximately $0.39$, while the baselines value is already $0.32$ at the same \acs{IoU} threshold.

In \tblref{paper05:tbl:evalClassesAP} an extract of the detailed average precision values is listed.
\begin{table*}
    \centering
    \caption{The average precision on the \acs{OMD} test set for the different training methods. The best values are printed in bold.}
    \label{paper05:tbl:evalClassesAP}
    \resizebox{\columnwidth}{!}{%
    \begin{threeparttable}
    \begin{tabular}{rcccccccccccc}
    \toprule
        \multirow{2}{*}[-0.5\belowrulesep]{Class} & \multicolumn{4}{c}{\AP{50}} & \multicolumn{4}{c}{\AP{75}} & \multicolumn{4}{c}{\AP{\phantom{00}}}\\
        \cmidrule(lr){2-5}\cmidrule(lr){6-9}\cmidrule(lr){10-13}
        & \cite{weber2023multiperspective} & GA\tnote{$\dagger$} & DGA\tnote{$\ddagger$} & COCO & \cite{weber2023multiperspective} & GA & DGA & COCO & \cite{weber2023multiperspective} & GA & DGA & COCO \\
        \midrule
        fork              & 65.5 & 64.8 & 68.7 & \textbf{98.3} &\pz0.0& 10.4 &\pz2.3& \textbf{77.6} & 18.6 & 25.1 & 21.8 & \textbf{63.8}\\
        frisbee           & 97.8 & \textbf{98.4} & 97.6 & 87.1 & 61.7 & \textbf{95.8} & 23.0 & 79.8 & 58.8 & \textbf{72.0} & 46.3 & 65.9\\
        gamepad           & 98.3 & \textbf{99.9} & 99.5 &\pz0.0& 84.6 & \textbf{94.4} & 32.3 &\pz0.0& 62.6 & \textbf{69.5} & 49.0 & \pz0.0\\
        hole puncher      & 55.4 & 57.9 & \textbf{59.5} &\pz0.0&\pz8.0& \textbf{17.8} &\pz0.1&\pz0.0& 23.0 & \textbf{28.5} & 18.0 & \pz0.0\\
        knife             & 47.9 & 79.0 & 78.9 & \textbf{87.9} & 29.8 & 35.1 & 48.3 & \textbf{50.7} & 27.6 & 41.9 & 44.7 & \textbf{50.0}\\
        scissors          & 17.1 & 24.3 & 50.1 & \textbf{94.9} &\pz0.8&\pz1.6& 24.0 & \textbf{87.1} &\pz5.1&\pz6.3& 27.4 & \textbf{70.9}\\
        shuttlecock       & \textbf{82.3} & 80.1 & 32.8 &\pz0.0& \textbf{62.1} &\pz0.5&\pz0.0&\pz0.0& \textbf{51.5} & 20.5 &\pz5.2& \pz0.0\\
        stapler           & \textbf{75.9} & 63.8 & 61.6 &\pz0.0& 33.3 & \textbf{53.2} & 52.3 &\pz0.0& 38.9 & \textbf{44.1} & 41.0 & \pz0.0\\
        table tennis ball & 89.0 & \textbf{97.3} & 91.0 & 67.4 & \textbf{33.2} & 20.5 & 27.3 &\pz0.8& \textbf{44.0} & 43.3 & 38.1 & 17.3\\
        toothbrush        & 39.9 & 70.6 & 66.0 & \textbf{70.9} &\pz0.1& \textbf{51.7} & 13.6 & 36.6 &\pz6.2& \textbf{43.7} & 27.7 & 37.4\\
        \midrule
        mean              & 66.9 & \textbf{73.6} & 70.5 & 50.7 & 31.4 & \textbf{38.1} & 22.3 & 33.3 & 33.6 & \textbf{39.5} & 31.9 & 30.5\\
    \bottomrule
    \end{tabular}
    \begin{tablenotes}
			\item[$\dagger$] \footnotesize \acf{GA-GBVS}
            \item[$\ddagger$] \footnotesize \acf{DGA-GBVS}
	\end{tablenotes}
    \end{threeparttable}}
\end{table*}
The results of both of our teaching variants, \acs{GA-GBVS} and \acs{DGA-GBVS}, are compared with the baseline approach~\cite{weber2020distilling}.
Furthermore, in accordance with their evaluation methodology, we also compare our method with a model trained on the entire MS~COCO dataset, that is, including the classes present in the \acs{OMD} dataset.
However, the authors have pointed out that this comparison has to be treated with caution and is only conditionally meaningful, as it involves fundamentally disparate starting points.
On the one hand, the model that has been trained on the entire MS~COCO dataset naturally has an advantage, as it has seen far more and also more diverse images.
On the other hand, only six out of the ten \acs{OMD} classes are part of MS~COCO, which means that more images are distributed among fewer classes and, in addition, the four other classes cannot be detected at all.
This lack of coverage entails that there is no flexibility to detect unknown objects.
Nevertheless, this is exactly the task.
The intention is to make it possible to detect all objects, irrespective of whether a corresponding data set is available, and to teach the unknown objects ad~hoc if necessary, expanding the present knowledge of the robot.
In fact, this is precisely the strength of our method and what differentiates the starting points.
Still, we want to consult COCO as an additional comparison to the baseline~\cite{weber2023multiperspective} to get a kind of ``upper bound'' and to assess which values are realistic in case a suitable dataset exists, is available and contains the relevant objects.
The values for \acs{mAP} and \acs{mAR}, however, refer to all ten classes to reflect the result on the overall task.

Our method outperforms the state-of-the-art approach proposed by~\cite{weber2023multiperspective} across nearly all classes.
For example, while the average accuracy at an \acs{IoU} of $0.5$, the threshold at which an object is considered to be detected~\cite{everingham2010pascal, zitnick2014edge}, slightly deteriorated for the shuttlecock with \acs{GA-GBVS} compared to~\cite{weber2023multiperspective}, the toothbrush showed significant improvements.
In terms of the scissors, \acs{DGA-GBVS} is superior to the more basic \acs{GA-GBVS} approach.
It is remarkable that our training method utilizing \acs{GA-GBVS} has surpassed the \AP{} values of MS~COCO in three classes, which is especially noteworthy given that MS~COCO considers \AP{} to be the primary metric for evaluation~\cite{lin2014microsoft}.
Overall, our approach utilizing \acs{GA-GBVS} outperforms all other alternatives across all \acs{IoU} values in terms of the mean average precision~(\acs{mAP}).
Moreover, a central aspect to emphasize is that teaching the robot via \acs{GA-GBVS} from scratch yields also promising performance, as indicated by our achieved \mAP{50}=71.9, \mAP{75}=32.2, and \mAP{}=34.6 scores.
That means even without any pretraining our method is superior to the baseline.

\begin{table*}
	\centering
	\caption{The average recall on the \acs{OMD} test set for the different training methods. The best values are printed in bold.}
	\label{paper05:tbl:evalClassesAR}
	\resizebox{\columnwidth}{!}{%
		\begin{threeparttable}
			\begin{tabular}{rcccccccccccc}
				\toprule
				\multirow{2}{*}[-0.5\belowrulesep]{Class} & \multicolumn{4}{c}{\AR{1}} & \multicolumn{4}{c}{\AR{10}} & \multicolumn{4}{c}{\AR{100}}\\
				\cmidrule(lr){2-5}\cmidrule(lr){6-9}\cmidrule(lr){10-13}
				& \cite{weber2023multiperspective} & GA\tnote{$\dagger$} & DGA\tnote{$\ddagger$} & COCO & \cite{weber2023multiperspective} & GA & DGA & COCO & \cite{weber2023multiperspective} & GA & DGA & COCO \\
				\midrule
				fork              & 28.4 & 40.4 & 34.3 & \textbf{72.3} & 33.9 & 44.9 & 36.9 & \textbf{72.3} & 33.9 & 44.9 & 36.9 & \textbf{72.3}\\
				frisbee           & 66.3 & \textbf{76.7} & 53.7 & 71.6 & 66.6 & \textbf{76.9} & 53.9 & 71.6 & 66.6 & \textbf{76.9} & 53.9 & 71.6\\
				gamepad           & 67.3 & \textbf{75.2} & 57.9 & \pz0.0   & 67.3 & \textbf{75.2} & 58.1 & \pz0.0   & 67.3 & \textbf{75.2} & 58.1 &\pz0.0\\
				hole puncher      & 34.2 & \textbf{42.5} & 26.7 & \pz0.0   & 49.2 & \textbf{54.0} & 37.2 & \pz0.0   & 51.0 & \textbf{54.4} & 37.6 &\pz0.0\\
				knife             & 41.4 & 57.8 & 59.8 & \textbf{63.2} & 60.4 & 59.4 & 62.0 & \textbf{63.4} & 60.6 & 59.4 & 62.0 & \textbf{63.4}\\
				scissors          & 23.4 & 20.5 & 44.2 & \textbf{74.8} & 30.6 & 36.4 & 51.5 & \textbf{75.4} & 30.6 & 36.4 & 51.5 & \textbf{75.4}\\
				shuttlecock       & \textbf{61.0} & 27.1 &\pz9.6& \pz0.0   & \textbf{62.1} & 27.6 & 14.8 & \pz0.0   & \textbf{62.1} & 27.6 & 15.0 &\pz0.0\\
				stapler           & 45.6 & \textbf{57.6} & 53.8 & \pz0.0   & 54.0 & \textbf{63.8} & 60.3 & \pz0.0   & 55.0 & \textbf{63.8} & 60.3 &\pz0.0\\
				table tennis ball & \textbf{56.1} & 49.8 & 49.0 & 23.8 & \textbf{58.1} & 50.0 & 51.9 & 23.8 & \textbf{58.1} & 50.0 & 51.9 & 23.8\\
				toothbrush        & 13.6 & \textbf{54.0} & 35.4 & 45.9 & 18.8 & \textbf{55.4} & 37.7 & 45.9 & 18.8 & \textbf{55.4} & 37.7 & 45.9\\
				\midrule
				mean              & 43.7 & \textbf{50.1} & 42.4 & 35.2 & 50.1 & \textbf{54.4} & 46.4 & 35.2 & 50.4 & \textbf{54.4} & 46.5 & 35.2\\
				\bottomrule
			\end{tabular}
			\begin{tablenotes}
				\item[$\dagger$] \footnotesize \acf{GA-GBVS}
				\item[$\ddagger$] \footnotesize \acf{DGA-GBVS}
			\end{tablenotes}
	\end{threeparttable}}
\end{table*}
With respect to the average recall, our method can also prevail.
The results in \tblref{paper05:tbl:evalClassesAR} demonstrate that we not only obtain a higher \AR{} than the baseline for almost every class regardless of the number of detections per image, but also partially exceed the ``upper bound'' of MS~COCO.
Eventually, our approach based on \acs{GA-GBVS} also outperforms all other alternatives in terms of the mean average recall~(\acs{mAR}).

\subsection{Limitations}
While our approach does partially mitigate inaccuracies arising from factors such as gaze tracking, robot-to-arm calibration, and human-to-robot transformations, the extent of the ensuing deviations varies depending on the arm's position.
This means that the offset between the projected gaze points on the recorded images is also varying, which in turn causes the quality of the final bounding boxes to differ slightly.
Nonetheless, as our results support, the quality is sufficient for the model to learn the objects successfully.

\subsection{Conclusion}
In this paper, we introduced a novel technique enabling robots to adapt to unfamiliar environments, along with the objects contained therein, beyond the predefined ones.
Rather than having the potential interaction entities dictated by fixed datasets, we expand existing knowledge in an ad~hoc manner.
Such approaches are inevitable in the long run, as the world is multifaceted and dynamic and the number of existing objects -- at least for all practical purposes -- is infinite, precluding the compilation of comprehensive datasets that cover them all.
The existence of a suitable dataset that contains the required classes and is also available can therefore not be inevitably assumed.
For this reason, we tasked the human with the role of a teacher, educating the robot about unknown objects of interest.
To this end, we transformed human gaze data acquired by means of an augmented reality interface into saliency-aware gaze heatmaps.
This process involved two different gaze-assisted approaches, which eventually allowed the robot to precisely perceive the region of interest.
Based on the class name provided by the human, the robot was capable of learning new objects in a flexible way.
The results of our evaluation have shown that our proposed method is superior to the current state of the art in terms of commonly used object recognition metrics.
This remains the case even if we omit the pretraining used by the baseline.
Therefore, our findings suggest that our approach has the potential to significantly enhance the adaptability of robots to novel scenarios and objects.
Altogether, we hope that our approach will offer new avenues for future research in human-robot interaction, leading to more dynamic and versatile robotic systems in non-predefined scenarios.
Despite the progress made, a considerable amount of work remains to be addressed as we intend to conduct a user study to asses the usability of our system as well as to further broaden the interaction between human and robot in an active learning context.

\subsection*{Acknowledgment}
This research was funded by the Deutsche Forschungsgemeinschaft (DFG, German Research Foundation) under Germany's Excellence Strategy -- EXC number 2064/1 -- Project number 390727645.
\fi

\backmatter

\cleardoublepage
\phantomsection
\bibliographystyle{unsrtnat}
\bibliography{misc/references}

\end{document}